\documentclass[rog]{AGUTeX}

\usepackage[pdftex]{graphicx}

\newcommand{\sizeA}{\rule{0mm}{6mm}}
\newcommand{\LSBA}{\mbox{$\left[\sizeA\right.$}}  
\newcommand{\RSBA}{\mbox{$\left.\sizeA\right]$}}  

\authorrunninghead{Polzin and Lvov}

\titlerunninghead{Regional Characterizations}

\authoraddr{K. L. Polzin
MS\#21, Woods Hole Oceanographic Institution, Woods Hole, MA 02543, USA.  
(kpolzin@whoi.edu)}

\authoraddr{Y. V. Lvov
Department of Mathematical Sciences, Rensselaer Polytechnic Institute,Troy NY 12180, USA.  
(lvovy@rpi.edu)}

\begin{document}

%
%

\title{Toward Regional Characterizations of the Oceanic Internal Wavefield}
%

%
%


\author{K. L. Polzin}
\affil{Department of Physical Oceanography, Woods Hole Oceanographic Institution, Woods Hole, Massachusetts, USA}

\author{Y. V. Lvov}
\affil{Department of Mathematical Sciences, Rensselaer Polytechnic Institute,Troy, New York, USA}  



%
%
%

%
%


\begin{abstract}
Many major oceanographic internal wave observational
programs of the last four decades are reanalyzed in order to characterize variability of 
the deep ocean internal wavefield.  The observations are 
discussed in the context of the universal spectral model proposed 
by Garrett and Munk.  The Garrett and Munk model is a good description 
of wintertime conditions at Site-D on the continental rise north of 
the Gulf Stream.  Elsewhere and at other times, significant deviations 
in terms of amplitude, separability of the 2-D vertical wavenumber - frequency
spectrum, and departure from the model's functional form are reported.  


Specifically, the Garrett and Munk model overestimates annual 
average frequency domain spectral levels both at Site-D and in general.  
The bias at Site-D is associated with the Garrett and Munk model 
being a fit to winter time data from Site-D and the 
presence of an annual cycle in high frequency energy in the 
western subtropical North Atlantic having a maximum in winter.  
The wave spectrum is generally non-separable, with near-inertial waves typically having greater bandwidth (occupying smaller vertical scales) than continuum frequency waves.  Separability is a better approximation for more energetic states, such as wintertime conditions at Site-D.  
Subtle geographic differences from the high frequency and high vertical wavenumber power laws of the Garrett and Munk spectrum are apparent.  Such deviations tend to co-vary: whiter 
frequency spectra are partnered with redder vertical wavenumber spectra.  

We review a general theoretical framework of statistical radiative balance equations and interpret the observed variability in terms of the interplay between generation, propagation and nonlinearity.  First:  Nonlinearity is a fundamental organizing principle in this work.  The observed power laws lie close to the Induced Diffusion stationary states of the resonant kinetic equation describing the lowest order nonlinear transfers.  Second:  Eddy variability and by implication wave-mean interactions are also an organizing principle.  Observations from regions of low eddy variability tend to be outliers in terms of their parametric spectral representation; other data tend to cluster in two regions of parameter space. More tentatively, the seasonal cycle of high frequency energy is in phase with the near-inertial seasonal cycle in regions of significant eddy variability.  In regions of low eddy variability, the seasonal cycle in high frequency energy lags that of near-inertial energy.  

The Induced Diffusion stationary states are approximate analytic solutions to the resonant kinetic equation and the Garrett and Munk spectrum represents one such analytic solution.  We present numerical solutions of the resonant kinetic equation, though, that are inconsistent with the Garrett and Munk model representing a stationary state, either alone or in combination with other physical mechanisms.  We believe this to be the case for other regional characterizations as well.  We argue that nonstationarity of the numerical solutions is related to local transfers in the horizontal wavenumber domain whereas the analytic Induced Diffusion solutions consider only nonlocal transfers in the vertical wavenumber domain.  Consequences for understanding the pathways by which energy is transfered from sources to sinks are considered.  Further progress likely requires self-consistent solutions to a broadened kinetic equation.  

\end{abstract}

%
%

%

\begin{article}

\setcounter{errorcontextlines}{999}

\section{Introduction}

Internal waves are ubiquitous features of geophysical fluids, contain a significant fraction of the total variance and constitute an effective mechanism for transferring energy and momentum across large distances and across different scales. In particular, the drag associated with
internal wave breaking needs to be included in order to obtain
accurate simulations of the atmospheric Jet Stream \citep{FA03} and it
has been argued that the ocean's Meridional Overturning Circulation
\citep{WF} is forced by the diffusion of mass \citep{L00} associated
with internal wave breaking \citep{P97} rather than by the convective production
of cold, dense water at high latitudes.  Both circulations represent fundamental pieces of the Earth's climate system.

The oceanic internal wavefield is "complex".  It spans a vast range of scales:  
in the time domain periods range from about one day for inertial waves
to a few minutes for buoyancy frequency waves in the pycnocline.
Horizontal wavelengths range from a few meters to hundreds of
kilometers.  The internal wave field's complexity arises not just from
its extended range of scales, but also from a multiplicity of possible sources 
at the surface (atmospheric forcing associated with wind stress and buoyancy fluxes)
and bottom (tides and mean currents impinging upon nonuniform
topography); the potential for scale transformations (reflection and scattering from
topography) and dissipation at the bottom boundary; and interior
transfers of energy and momentum through wave-mean flow interactions.
A significant sink is wave breaking, which is believed to be product
of nonlinearity resulting in downscale transfers of energy.  

Despite, or perhaps because of, this complexity, a universal character is usually
ascribed to the oceanic internal wave energy spectrum.  In a classic
work, \cite{GM72} demonstrated how observations from
various sensor types could be synthesized into a combined
wavenumber-frequency spectrum, now called the Garrett-and-Munk (GM)
spectrum of internal waves. Consistent only with linear internal wave
kinematics, the GM spectrum was developed as an empirical curve fit to
available data.  Deviations have been noted near boundaries
\citep{WW79, P04b} and at the equator \citep{E85}.  However, our notion of what
constitutes the background wavefield has remained static since the
last significant model revision provided by \cite{CW76}.  

The authors' sentiments are captured by the first paragraph of \cite{Briscoe75a}'s Survey section:  
``The single paper motivating the most comments, experiments, and disquiet in a lot of readers was \cite{GM72}, now called familiarly GM72.  (The updated version is \cite{GM75}, called GM75.)  The paper is a virtuoso orchestration of synthesis, approximation, boldness, normalization, and implication.  Starting from the observation that the wavenumber-frequency ($\omega-\kappa$) spectrum of a multimodal dispersive system is not uniquely determined by pure wavenumber and pure frequency spectra, the paper goes onto construct from linear dynamics and WKBJ methods an $\omega-\kappa$ spectrum that is astonishingly consistent with many kinds of data from all kinds of instruments.  \cite{Wunsch75} iconoclastically discusses the pitfalls of such an approach and correctly emphasizes the fact that the dynamically important trait is the inconsistency of various data sets with some base state model.''    With immense inspiration provided by Garrett and Munk, we take up, after 35 years, the difficult task posed by Wunsch.  We document variability of the oceanic internal wavefield and then review the field to assess the dynamical underpinnings of that variability.  While heavily referenced, this work does not attempt to be a literature review.  

In previous work \citep{LPT} we noted that deviations of frequency/vertical wavenumber power laws from the universal model tend to co-vary: whiter frequency spectra are partnered with redder
vertical wavenumber spectra.  Here we provide a more systematic study of such deviations.  We define spatial and temporal patterns of spectral parameters (amplitude, power laws, bandwidth and separability) in a review of many observational programs of the last four decades.  Having documented variability in the background wavefield, we interpret the spatial patterns in terms of generation, nonlinearity, wave-mean interactions and dissipation.  

Nonlinearity plays a key role.  In a companion manuscript \citep{theory} we discuss the application of wave turbulence formalisms to this problem and rigorously demonstrate the validity of conventional wisdom regarding a truncated analytic approximation to nonlinear spectral transfers described in \cite{MMa, MMb}.   The observed power law combinations in frequency and vertical wavenumber documented here are largely consistent with the stationary states associated with the truncated analytic approximations.  While this appears to be a stunning affirmation of this approximation to the wave turbulence transfer integral, it comes with caveats, disclaimers and possible contradictions.  First, the theory is not formally valid at high wavenumber for realistic wave amplitudes.  Second, the identification of GM76 being a stationary state is only true after integration over the frequency domain and invocation of dissipation.  Finally, numerical evaluations of the nonlinear transfers are inconsistent with the high frequency, high vertical wavenumber domain of GM76 being a stationary state.  This tension also applies to other stationary states identified by the truncated analytic theory having power law combinations associated with energy transfers from larger to smaller vertical scales and from higher to lower frequencies.  Such stationary states require an energy source at high frequencies.  But our review of the remaining generation and spectral transport mechanisms does not yet reveal what this mechanism might be.  

The article is organized as follows. In Section \ref{Background} we summarize
the development of the GM model through its various incarnations. We
then review and re-analyze data obtained by major internal wave
observational programs of the last four decades in Section \ref{obs}.
These data clearly demonstrate that deviations from the GM spectrum
are not random, but rather form a distinct pattern which we summarize in Section \ref{Summary}.  We discuss the theoretical framework which may be used to interpret these
observations in Section \ref{Theory} and conclude in Section 
\ref{Conclusions}.  An appendix contains additional useful information relating to the interpretation of oceanographic data.

\section{Background }\label{Background}

\subsection{Kinematic Structure}

\subsubsection{Dispersion Relation} 

Internal waves arise through the restoring force of gravity in a stably stratified fluid.  In a rotating system they comprise a 'fast' mode of oscillation with frequencies between inertial $f$ and buoyancy $N$ frequencies.  In terms of an Eulerian description with velocity vector ${\bf u} = (u, v, w)$, the linearized equations of motions are
\begin{eqnarray}
u_t - fv = - \pi_x \nonumber \\
v_t + fu = - \pi_y \nonumber \\
w_t  = -\pi_z - b \nonumber \\
\rho_t + w \overline{\rho}_z = 0 \nonumber \\
u_x + v_y + w_z = 0 \nonumber
\end{eqnarray}
in which kinematic pressure $\pi = p/\rho_o$ and buoyancy $b=-g\rho / \rho_o$ are perturbations from a hydrostatically balanced equilibrium state, $\overline{\pi}_z = - \overline{b}$, $g$ is gravity and the fluid is assumed to be incompressible ($\nabla \cdot {\bf u}=0$).  We use ${\bf r} = (x,y,z)$ as a position vector.  An over-bar $(\overline{\phi})$ indicates a mean on time scales much longer than a wave period.  The constant $\rho_o$ is the mean density and arises from assuming density to be constant in computing momentum changes associated with accelerations, but taking density changes into account when they give rise to buoyancy forces (the Boussinesq approximation).  The buoyancy gradient relates to oscillations  of frequency $N=\sqrt{\overline{b}_z} = \sqrt{-g\overline{\rho} / \rho_o} $ corresponding to periods of minutes in the upper ocean to hours in the abyss.  This work is concerned almost exclusively with horizontal scales small enough that the Earth's rate of rotation about the local vertical, $f=2 \Omega \sin ({\rm latitude})$, can be considered as constant (the f-plane approximation).  The rate of rotation of the Earth about its axis is $\Omega$.  The Coriolis frequency $f$ thus corresponds to periods of 1/2 day at the poles to several days in the tropics and vanishes at the equator.  The foundation for interpreting observations lies in linear polarization relations (Section \ref{ConsistencyRelations}) obtained from these equations of motion.  

The equations of motion can be manipulated to obtain a wave equation  in vertical velocity (e.g., \cite{Gill}):
\begin{equation}
(\partial_t^2 + f^2)\partial_z^2 w+[N^2(z)+\partial_t^2](\partial_x^2 + \partial_y^2) w = 0
\label{wave-eq}
\end{equation}
for arbitrary stratification profile $N^2(z)$.  Assuming the background stratification is constant, plane wave solutions $w = w_{o} e^{i[{\bf r} \cdot {\bf p} - \sigma t]}$ with three dimensional wavevector ${\bf p}=(k,l,m)$ having vertical component $m$ and horizontal magnitude $k_h = \mid {\bf k} \mid$ return 
\begin{equation}
\sigma_{\bf p}= \pm \sqrt{ \frac{m^2f ^2 + N^2 k_h^2 }{m^2 + k_h^2} },
\label{DR}
\end{equation}
which can be rewritten as 
$$
\frac{\sigma^2-f^2}{N^2-\sigma^2} = \frac{k_h^2}{m^2} . 
$$

Many of the important properties of internal waves are immediately available:
\begin{itemize}
\item Incompressibility $(\nabla \cdot {\bf u} = 0)$ and a plane wave solution imply particle velocities are normal to the wavevector, ${\bf p} \cdot {\bf u} = 0$.  
\item Freely propagating waves, for which ${\bf p}$ is real, require $f < \mid \sigma \mid < N$.  
\item Wave frequency solely determines the angle of the wavevector ${\bf p}$ relative to the vertical and, since ${\bf p} \cdot {\bf u} = 0$, the ratio of horizontal to vertical velocity.  
\item The group velocity ${\bf C_g} = \nabla_{\bf p} \sigma$  $(=[\sigma k /k_h^2, $ $\sigma l /k_h^2, -\sigma/m]$ if $f^2 \ll \sigma^2 \ll N^2$) has the 'odd' property that crests and troughs (i.e., lines of constant phase) propagate in the same direction as a wave packet (group) in the horizontal but the signs of phase and group velocity differ in the vertical coordinate.  Upward phase propagation implies downward energy propagation.   
\end{itemize} 

\subsubsection{Polarization Relations}\label{ConsistencyRelations}
A suite of diagnostic relations can be obtained by assuming constant stratification rate, plane wave solutions in which all variables are proportional to $e^{i({\bf p} \cdot {\bf r} - \sigma t)}$and expressing the dependent variables in terms of wave amplitude $a$:
\begin{eqnarray}
u & = & \LSBA \frac{k_h^2}{m^2 \mid {\bf p} \mid^2 } \RSBA^{1/2}
\frac{m^2(k-ifl/\sigma)}{k_h^{2}} ~ a ~ {\rm e }^{i({\bf p \cdot r} -\sigma t)}\label{PR1}
 \\
v & = & \LSBA \frac{k_h^2}{m^2 \mid {\bf p} \mid^2 } \RSBA^{1/2}
\frac{m^2(l+ifk/\sigma)}{k_h^{2}} ~ a ~ {\rm e }^{i({\bf p \cdot r} -\sigma t)}
\\
w & = & \LSBA \frac{k_h^2}{m^2 \mid {\bf p} \mid^2 } \RSBA^{1/2} -m ~ a ~ {\rm e
}^{i({\bf p \cdot r} -\sigma t)} \\
b & = & \LSBA \frac{k_h^2}{m^2 \mid {\bf p} \mid^2 } \RSBA^{1/2} -\frac{imN^2}{\sigma}
~ a ~ {\rm e }^{i({\bf p \cdot r} -\sigma t) } \\
\pi & = & \LSBA \frac{k_h^2}{m^2 \mid {\bf p} \mid^2 } \RSBA^{1/2}
-\frac{(N^2-\sigma^2)}{\sigma} ~ a ~ {\rm e }^{i({\bf p \cdot r} -\sigma t)}\label{PR5}
\end{eqnarray}
The prefactor in these polarization relations is such that the wave amplitude
$a$ is normalized to represent the total energy:
\begin{equation}
E_k + E_v + E_p = aa^{\ast} ~.
\end{equation}
The total energy is the sum of horizontal kinetic $E_k$, vertical kinetic $E_v$
and potential $E_p = \frac{1}{2} N^2 \eta^2$ energy in which $\eta$ represents
the vertical displacement.

Two important diagnostics utilized here are: 
First, assuming constant stratification, the ratio of horizontal kinetic to potential energy for a single internal wave is: 
\begin{equation}
\frac{E_k}{E_p} = \frac{uu^{\ast}+ vv^{\ast}}{bb^{\ast} N^{-2} }= \frac{\sigma^2+f^2}{\sigma^2-f^2} \frac{N^2-\sigma^2}{N^2}~.
\label{Ek_on_Ep}
\end{equation}
Second, the horizontal velocity trace for high frequency internal waves is rectilinearly polarized.  Rotation alters this so that the velocity trace at inertial frequency inscribes a circle with time.  
With depth, the near inertial velocity vector traces out an elliptical helix in which 
the ratio of velocity variance in the clockwise ($E_{cw}$) and counter-clockwise 
($E_{ccw}$) rotating components is given by:  
\begin{equation}
\frac{E_{cw}}{E_{ccw}}= \frac{(u-iv)(u-iv)^{\ast}}{(u+iv)(u+iv)^{\ast}} = \frac{(\sigma+f)^2}{(\sigma-f)^2}~,
\end{equation}
with the sign convention that positive frequency $f < \sigma < N$ implies 
upward phase propagation (and hence downward energy propagation) in 
the northern hemisphere.  

\subsubsection{Weak Spatial Inhomogeneities}

If the stratification profile varies much more slowly than the
wave phase, a WKB approximation for vertically propagating waves
provides the approximate solution to (\ref{wave-eq}):
\begin{equation}
 w \propto N(z)^{-1/2} e^{ \displaystyle{ \frac{\pm i k_h}{\sqrt{\sigma^2-f^2}} \int N(z) dz} }
\end{equation}
and so the effects of a variable buoyancy profile can be accounted
for by stretching the depth coordinate by $N$.  This amounts to scaling the 
vertical wavenumber by $N/N_o$, in which $N_o$ is a
reference stratification.  The value $N_o=$ 3 cph is often used.  
The buoyancy scaling of other dependent variables follows from the polarization relations (\ref{PR1})-(\ref{PR5}).  The use of the WKB approximation requires $\sigma^2 \ll N^2$.  If this
relation is not satisfied, solutions can be found by
treating (\ref{wave-eq}) as an eigenvalue problem with appropriate
boundary conditions.  If $\sigma^2 \ll N^2$ but the wave-phase is not
slowly varying, the boundary conditions are that $w=0$ at the top
($z=H$) and bottom ($z=0$), which then implies the horizontal
velocities $[u(z), v(z)]$ are proportional to:
\begin{equation}
N(z)^{1/2} cos(n \pi \int_0^z N(z^{\prime}) dz^{\prime}/ \int_0^H
N(z^{\prime}) dz^{\prime}),
\end{equation}
for integer values of n.  

Similarly, one can investigate interactions of the internal wavefield with a mean velocity field by assuming the internal wavefield has small amplitude and small spatial scales relative to a time independent geostrophically balanced flow, Section \ref{WaveMean}.  This problem is amenable to WKB and ray-tracing techniques.  In this slowly varying limit, the linear polarization relations reported here are modified by replacing the Eulerian frequency $\sigma$ with the intrinsic frequency $\omega = \sigma - {\bf p} \cdot \overline{{\bf u}}$.  

\subsubsection{Slow Oscillations}\label{SlowOsc}
Internal waves are to be distinguished from a 'slow' mode consisting of, at lowest order, the {\bf steady} geostrophic balance.  
\begin{eqnarray}
fv = \frac{1}{\rho_o} p_x \nonumber \\
fu = -\frac{1}{\rho_o} p_y \nonumber \\
0 = -\frac{1}{\rho_o} p_z - g \nonumber \\
w \overline{\rho}_z = 0 \nonumber \\
u_x + v_y = 0 \nonumber
\end{eqnarray}
Higher order contributions are time dependence, nonlinearity and the effects of a variable rate of planetary rotation ($f$).  Substitution of a plane wave solution into the geostrophic balance returns primarily the information that small aspect ratio fluctuations ($k_h/m \ll f/N$) have $E_k/E_p$ ratios of less than one.  

There are two main paradigms for considering these slow motions {\it vis-a-vis} the fast internal wavefield.  The first, already alluded to, is the limit that geostrophic velocities are larger and have significantly larger spatial scales.  The second paradigm comes about as the two share similar spatial scales.  In this instance the slow motions have typically smaller amplitudes and hence appear as a 'contamination' to the wavefield.  Beyond the issue of aspect ratio, the consistency relations for this non-propagating mode have little utility and this  'contamination field' has the character of uncorrelated noise, which is the content of several diagnostic studies \citep{DAM91, KS93}.  The characterization of this slow mode as 'noise' should not surprise.  The dynamics in this limit are those of stratified rotating turbulence in which motions at different isopycnals are uncoupled at scales larger than those over which shear instabilities can develop \citep{RL00}.  

While diagnostic models of the slow mode usually assume an incoherent field \citep{M78}, the interaction of the internal wavefield with quasi-permanent density finestructure can be exploited by assuming the wavefield buoyancy scales on the density finestructure \citep{P03}.  This buoyancy scaling results in a correlation between internal wave shear and quasi-permanent density gradients when such a correlation is inconsistent with fields composed of either pure waves or pure finestructure.  This enables an estimate of the quasi-permanent spectrum, \cite{P03} and Section \ref{SE_ST_NA}.  

One finestructure consistency relation has great importance.  The plane wave solutions of internal waves have no Ertel potential vorticity signature, whereas the non-waves have non-zero Ertel potential vorticity \citep{M86}.  The distinction is important for issues of isopycnal dispersion \citep{PF04} and for dynamics \citep{M95}.  

Reliance upon the plane wave solutions, however, has its limitations.  While internal waves and slow modes can be distinguished in terms of Ertel potential vorticity content, this distinction disappears when one considers a slowly varying wavepacket with spatial structure in {\bf both} horizontal directions, i.e., solutions of the form 
$$
a \rightarrow a(x,y)e^{i[{\bf r} \cdot {\bf p} - \sigma t]}~. 
$$
See \cite{BM05} and \cite{P08a} for further details.  

\subsection{Parametric Spectral Representations}

Much of the original \cite{GM72} paper is a demonstration of how observations from various sensor types could be synthesized into a combined wavenumber-frequency spectrum consistent with linear internal wave kinematics.  The toolbox contains:  
\begin{itemize}
\item a linear internal wave dispersion relation (\ref{DR}) to transfer from one domain to another, e.g.
$$
E(k_h) ~ d k_h  ~ = E(m) ~ d m 
$$
in which $dk_h/dm$ follows from (\ref{DR}), 0
\item diagnostic relations (Section \ref{ConsistencyRelations}) to test consistency with linear wave kinematics,  and 
\item the choice of a vertical spectral rather than modal representation implies the invocation of an `equivalent continuum' in which the waves are assumed to be vertically propagating.  Boundary conditions and turning points are neglected.  See \cite{GM72} for a discussion of modal properties.  
\end{itemize}
Assuming that the buoyancy profile varies much more slowly than the wave phase implies the total energy $E$, obtained as an integral of the energy density over the spectral domain,  
\begin{equation}
E=\int E({\bf k},\sigma) ~ d {{\bf k}} ~ d \sigma  ~ =\int E(m,\sigma) ~ d {m} ~ d \sigma  ~ 
\end{equation}
varies as
\begin{equation}
E=E_o \frac{N}{N_o}  ;
\label{EN}
\end{equation}
i.e., $E/N$ is an adiabatic invariant.
The energy spectrum is denoted as $E$ with following arguments [e.g., $E(\sigma)$].     
The variable $E$ without following arguments represents total energy.)  The factors $E_o$ and $N_o$ represent reference values for the total energy of the internal wavefield and stratification.  

Garrett and Munk proposed that the spectral energy density can be represented as a {\bf separable} function, i.e., the product of a function $A$ of vertical wavenumber only and function $B$ of wave frequency only:
\begin{equation}
E(\sigma,m)= E ~ A(\frac{m}{m_*})B(\sigma),
\label{Separability}
\end{equation}
where $m_*$ is a fixed reference number.  
This reference wavenumber is conveniently expressed as 
\begin{equation} m_* = \pi j_* / b,
\label{Jstar}
\end{equation} 
in which the variable $j$ represents the mode number of an ocean with an exponential buoyancy frequency profile having a scale height of $b$ [$b$ = 1300 m in the GM model].  Separability in vertical wavenumber
and frequency was invoked as the simplest representation not inconsistent with available data.  The spectral amplitude was regarded as being independent of spatial coordinate apart from the dictates of buoyancy scaling.  Use of non-separable and anisotropic parametric spectral representations can be found in \cite{M78}.  
  
Functions $A(x=\frac{m}{m_*})$ and $B(\sigma)$ should be normalized to unity, so that the integrals
$$\int\limits_{m_1/2m_{\ast}}^{m_c/m_{\ast}} A(x) d x \cong \int\limits_{0}^{\infty} A(x) d x=1, $$  
and
 $$\int\limits_f^N B(\sigma) d \sigma \cong  \int\limits_f^{\infty} B(\sigma) d \sigma =1 $$ 
are dimensionless and in practice the limits of integration ($m_1$ represents vertical mode-1, $m_c$ a high wavenumber cutoff of approximately 0.1 cpm) are approximated for analytic convenience.  The high wavenumber cutoff, defined by
\begin{equation}
\displaystyle \int_0^{m_c} 2 m^2 E_k(m) dm = \frac{2\pi}{10} N^2
\label{MC}
\end{equation}
represents an observed transition to steeper spectral slopes and likely signifies a dynamical transition to a more strongly nonlinear, yet not turbulent, regime.  

The GM model evolved over time, resulting in three major versions, denoted GM72, GM75 and GM76:  
\begin{eqnarray}
{\rm \bf GM72}: \hspace{0.75cm}&& \nonumber \\
A(\frac{m}{m_*}) & = &   \left[
                                   \begin{array}{l} m_*^{-1}, \ {\rm if}\  m<m_*\nonumber \\ 
                                               0,    \ \ \ \ \       {\rm if}\  m<m_*\end{array}
\right.
\nonumber \\
{\rm \bf GM 75,76}: \\
A(\frac{m}{m_*}) & = & 
\frac{s\Gamma(\frac{t}{s})}{\Gamma(\frac{1}{s})\Gamma(\frac{t-1}{s})} \frac{m_*^{-1}}{[1+(m/m_*)^s]^{-t/s}}
\nonumber\\
{\rm with }\nonumber\\
&  & (s,t,j_{*})=\left[\begin{array}{l}
(1,2.5,6) ~~~~~~ \ \ \ {\rm GM75} \nonumber \\
(2,2,3) ~~~~~~\ \ \ \ \ {\rm GM76}.  \nonumber \\\end{array}\right. 
\label{vertical_spectrum}
 \end{eqnarray}
and
\begin{equation}
B(\sigma)   \propto  \sigma^{-r+2q}(\sigma^2 - f^2)^{-q} ~. 
\label{frequency_spectrum}
\end{equation}
The gamma function ($\Gamma$) appears in the normalization for the vertical spectrum.  
All versions of GM utilize $r=1$ and $q=1/2$, so that
$$
B(\sigma ; r=1, q=1/2) = \frac{2 f}{\pi} \frac{1}{\sigma \sqrt{\sigma^2-f^2} }  \, .
$$
Both functions $A(x=\frac{m}{m_*})$ and $B(\sigma)$ behave as power-laws at high wavenumber and
frequency; $B$ has an integrable singularity at the inertial frequency $f$ (which constrains $q < 1$), and $A$ has a plateau for small value of its argument.  As explained below these features were
found to be representative of the oceanic internal wave field.

The total energy in the model, 
\begin{equation}
E_{o}^{GM} = 30 \times 10^{-4} \, {\rm m}^2 {\rm ~ s}^{-2} , 
\end{equation}
is based upon fits to observed high frequency spectra rather than estimates of total observed internal wave band energy.

\subsection{The GM model's three incarnations}\label{GM}
The GM model evolved over time as more and better data became available.  This section documents reasons behind the choice of specific parameters.    

\subsubsection{Frequency Domain}
While \cite{GM72} noted variability in the tidal peak, the
tides are not part of the GM model.  The specific shape of the
inertial peak was simply chosen out of analytic convenience.
Variability in the frequency domain power law was apparent even then,
with a noted alternative being $\sigma^{-5/3}$ \citep{Webster69}.  With many possible 'noise' sources such as mooring motion and finestructure contamination serving to whiten the frequency spectrum,
the judicious choice for a model of the background spectrum
is a redder $\sigma^{-2}$ (Section \ref{SiteD}).  This choice remains stable throughout the
various model versions.

\subsubsection{Vertical Wavenumber Domain}
The spatial domain information available in the late-1960's was limited to stationary current meters and horizontally towed thermistor chains.  The direct information available from current meters was through spatial lag coherences\footnote{Estimates of vertical bandwith $j_{\ast}$ can be obtained by assuming a parametric spectral representation and fitting lagged coherence estimates to the inverse Fourier transform of the spectrum, \citep{GM72, Cairns75, M78}.  This technique can be found in many textbooks as the Wiener-Khinchine relation.}.
The towed data return both a direct estimate of the horizontal spectrum and vertical lag coherence information.  These data supported only a crude representation of the vertical wavenumber spectrum as a top-hat model with about 20
equivalent modes excited ($j_* = 20$ in GM72).

The quality of the information improved dramatically in the early 1970's with the introduction of near-continuous vertically profiling instrumentation such as the Neil Brown CTD \citep{Brown} and Tom Sanford's electric field sensing device \citep{Sanford75}.  The choice of which data to rely upon, though, is quite subjective and the perception is that finestructure contamination of the density field is more problematic than contamination of the velocity field, e.g., \cite{P03}.  See Section \ref{SargassoSea} for a discussion of the \cite{Sanford75} data set.  Extant temperature gradient spectra tended to be white at an intermediate range of vertical wavenumbers (roughly 0.01 $\leq m \leq$ 0.2 cpm) and roll off thereafter.  GM75, however, relied upon even higher wavenumber CTD data from \cite{Millard72}, ignored possible inconsistencies with CTD data in \cite{Hayes75}, and heavily weighted the velocity profile data in \cite{Sanford75} to argue for a high wavenumber slope $t=5/2$.  Estimates of isotherm fluctuations from a vertically profiling float \citep{Cairns75} suggested a much lower bandwidth ($j_{*}=6$) than in the GM72 model.  

GM75 was soon replaced by a more refined model.  Further analysis of float data by \cite{CW76} suggested revision of the high wavenumber power law to $m^{-2}$ and a change to the functional form $1/(m_{*}^2 + m^2)$.  The rationale given for choosing such a functional form is simply its analytic convenience: $1/(m_{*}^2 +
m^2)$ has a simple cosine transform, see also \cite{D76}.  \cite{M81}
follows suit.  This revision is labeled GM76.  The distinction between
$1/(m_*+m)^2$ and $1/(m_*^2+m^2)$ is in the rapidity of the roll-off
from the low-wavenumber plateau region to the high-wavenumber
asymptote.  GM76 has a more rapid transition than GM75 for equivalent
power laws.

\section{Observations of Internal Wave Spectra}\label{obs} 
\subsection{Observational framework/preliminary notes}

Observations of the oceanic internal wave field are analyzed in this
section in order to define the extent to which the deep ocean internal
wavefield is indeed universal.  We follow the lead of GM by characterizing observed spectra with the parametric power law representations (\ref{vertical_spectrum}) and (\ref{frequency_spectrum}) and also allow for variable spectral levels with the focus upon identifying spatial/temporal variability of the spectral parameters.  

Variability in the frequency domain is 
quantified by fitting
\begin{equation}
E(\sigma) = E B(\sigma ; r,q=1/2)  \propto e_o~ \sigma^{-r+1}(\sigma^2 - f^2)^{-1/2},
\label{shape}
\end{equation}
to the high frequency portion of the observed spectra.  'High'
frequency refers to periods less than 10 hours, thereby eliminating
the inertial peak and semi-diurnal tides from consideration.  The amplitude $e_o$ is reported as a fraction 
of the energy associated with the GM spectrum:
\begin{equation}
e_o = \frac{ \int_f^N EB(\sigma ; r,q=1/2) ~ d\sigma }{E_{o}^{GM} N / N_o} ~ .
\label{percent}
\end{equation}

An additional reported statistic is the total internal wave
band energy in the observed spectrum.  Internal wave band energy
is estimated by integrating the observed frequency spectra over
frequencies greater than approximately 0.7$f$, thereby including
broadening of the inertial cusp.

Variability in the vertical wavenumber domain is quantified by fitting
variants of (\ref{vertical_spectrum}) to the observed spectra.  The
resulting statistics are less satisfactory than those in the frequency
domain as low wavenumbers are not well resolved in most instances.

\subsection{Some words of caution}
Power laws are an attractive and simple way to describe
complex systems in general and internal waves in particular.  
Some details of the observed spectra may defy this simple approach.  Our view 
is to treat the parametric representation as a general tendency to help summarize and interpret otherwise complicated data.  With that caveat, there is no shortage of reasons for the observed variability:  nonlinearity, instrumentation, the
buoyancy profile, low frequency flows, tides, bathymetry, horizontal
inhomogeneity, vertical asymmetry and contamination by either
quasi-permanent finestructure or self advection (Doppler shifting)
within the internal wavefield:  

\begin{itemize}
\item It matters how you look at the ocean.  Instrument response and
processing methods may affect the interpretation.  An appendix contains remarks
to aid the reader in negotiating this maze.

\item All that wiggles is not necessarily an internal wave.  Internal
waves typically dominate the small scale variability and
geostrophically balanced motions dominate the largest scales.  But
quasi-permanent finestructure has been identified as making an
increasing contribution to the total spectrum at high wavenumber
(e.g., \cite{M78, P03}, Section \ref{SlowOsc}).

\item In many cases, the quoted spectral power laws are derived from
one-dimensional (vertical wavenumber and frequency) spectra.  Thus there is an 
implicit assumption that the spectra are separable, (\ref{Separability}).   Since 
both 1-d spectra are red, frequency spectra are typically dominated by low vertical wavenumber
motions and vertical wavenumber spectra are dominated by low
frequencies.  The one-dimensional power laws are not identical to the 
2-d power laws if the spectra are non-separable.   

\item The sampling strategy matters.  Vertical wavenumber-frequency
domain information are seldom obtained from a single instrument, nor
are the measurements taken instantaneously.  Vertical profile data
obtained as a spatial survey can be used to define a spatial mean and
residuals, and the internal wave contribution identified as the
residual field.  However, this will alias some low-frequency
variability into the residual fields.  Vertical profile data obtained
as a time series can effectively separate the internal wave
variability from lower frequency motions.  But the resulting residuals
may not be representative of the background wavefield: that background
may be temporarily distorted, for example, by wave-mean interactions.

\item Anisotropic propagation in a region of variable stratification
or into/away from a critical layer can imply significant transport of
energy in the vertical wavenumber domain and thereby create apparent
structure unrelated to nonlinearity.

\item The results presented here are primarily from analyses in depth
or pressure coordinates, rather than in an isopycnal coordinate system.  
High frequency Eulerian spectra are prone to Doppler shift contamination at high vertical wavenumber.  Use of an isopycnal coordinate system limits the Doppler shift contamination 
\citep{SandP91}.

\end{itemize}

Despite these limitations, we believe
that the patterns of spatial and temporal variability discussed below reflect 
physical reality and are not an artifact of processing or interpretation. 

\subsection{A Regional Catalog}

%
\begin{figure*}[b]
\begin{tabular}{lr}
\begin{minipage}[t]{39pc}
\noindent\includegraphics[width=15cm]{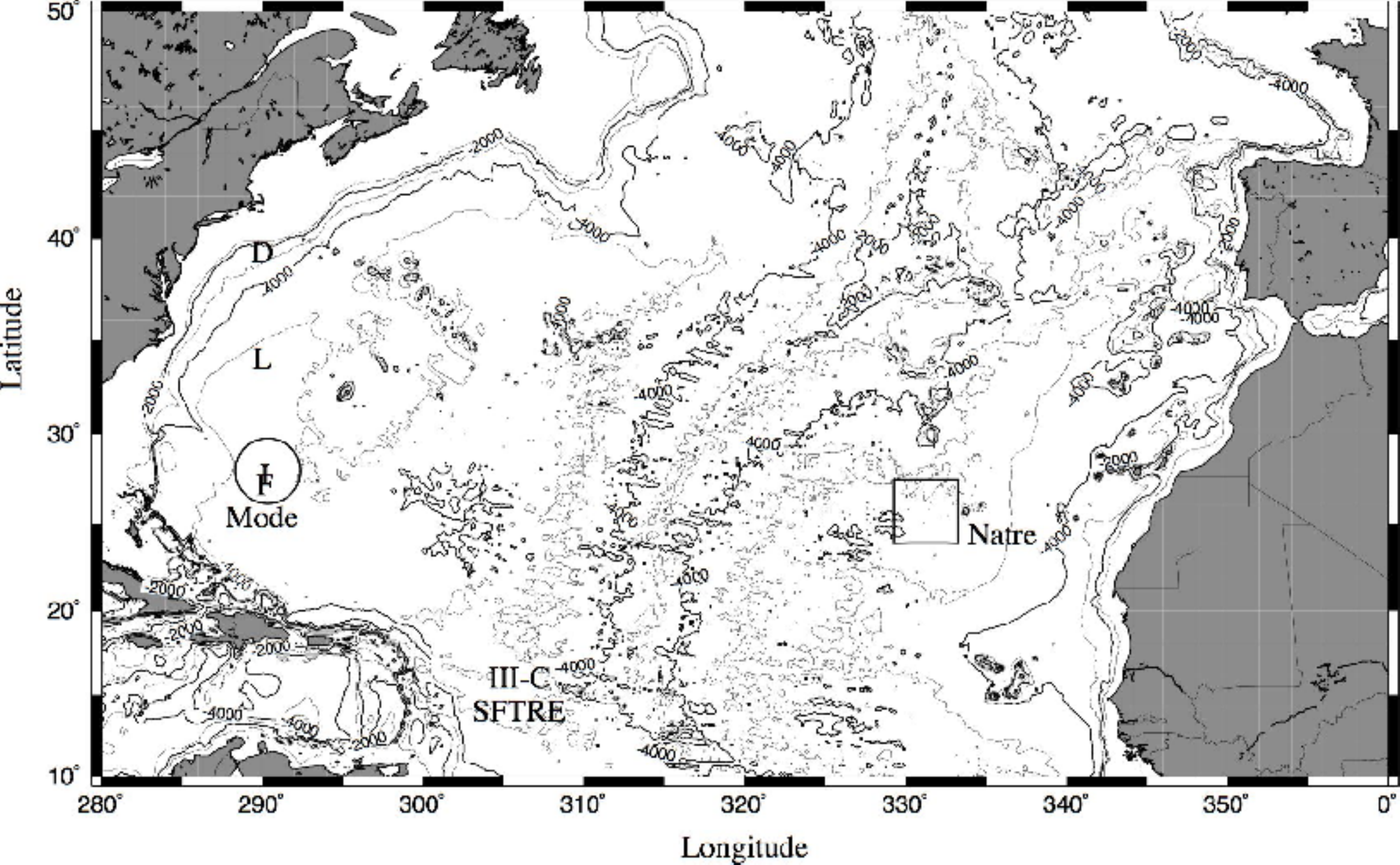}
\end{minipage} & \\
\begin{minipage}[h]{20pc}
\noindent\includegraphics[width=7.5cm]{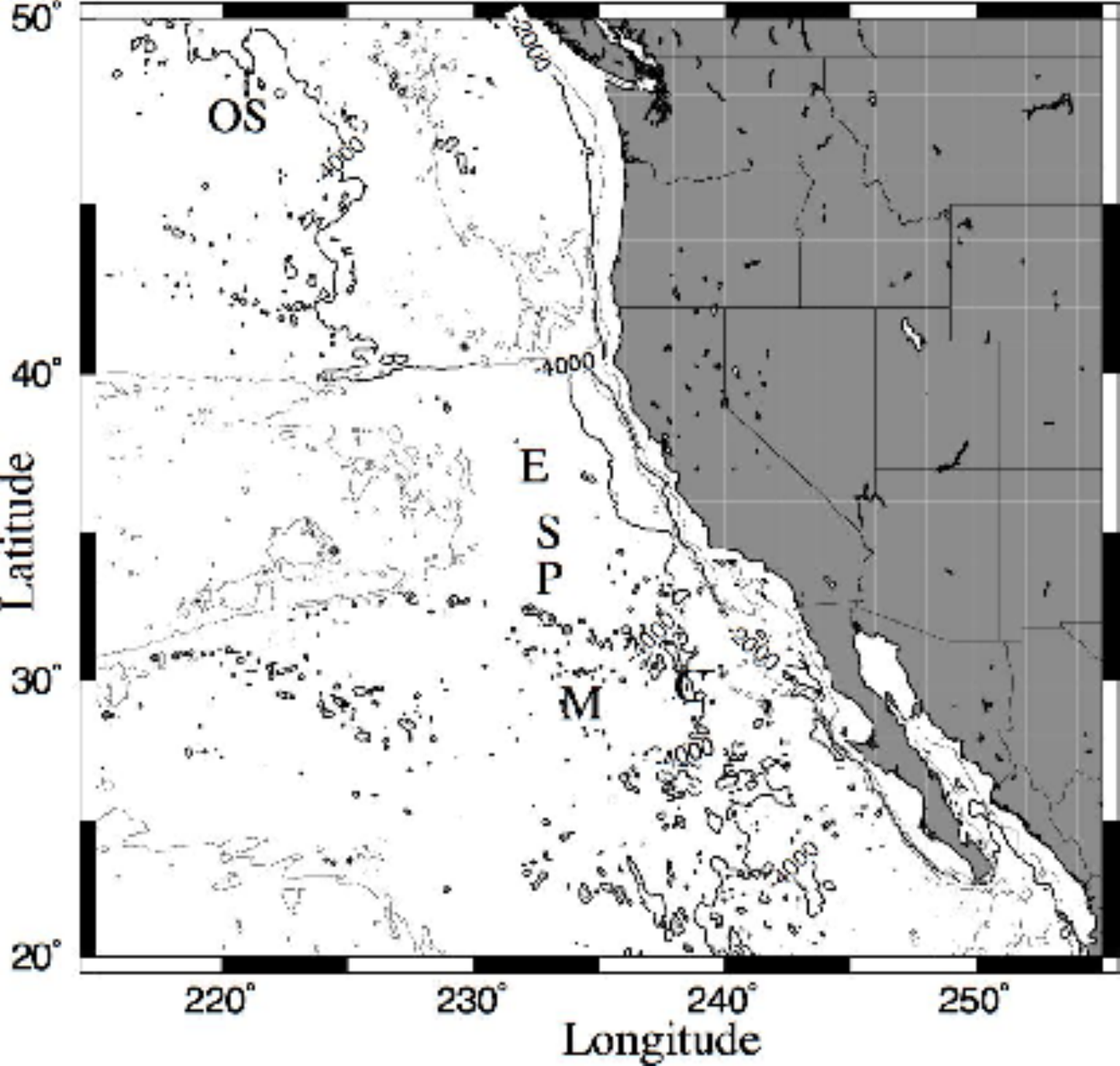}
\end{minipage} & 
\begin{minipage}[b]{20pc}
\caption{Geographic depictions of the
data sets referred to in this study.  Surveys for the Mid-Ocean
Dynamics Experiment (MODE) and the North Atlantic Tracer Release
Experiment (NATRE) are enclosed by the circle and square,
respectively.  This study uses current meter data from near the center
of these survey regions.  The alphabetical keys are: D (Site-D), I
(Internal Wave EXperiment), L (the Long-Term Upper-ocean Study), F
(Frontal Air-Sea Interaction EXperiment), III-C (PolyMode IIIc) and SFTRE 
(the Salt Finger Tracer Release Experiment) in the North Atlantic; S
(SWAPP), P (Patchex), M (Misery 1 and 3 as referred to in
\cite{CW76}), C \cite{Cairns75}, E (The offshore array from the
Eastern Boundary Currents field program, \cite{Ch2000}), MATE 
(the MidOcean Acoustic Transmission Experiment, \cite{L86}) and OS 
(Ocean Storms, \cite{DAsaro95}).  Bathymetric
contours are every 1000 m, alternately bold and thin.  The Gulf Stream
lies shoreward of the 800 m isobath south of Cape Hatteras, cutting
between Site-D and the Lotus region. 
\label{fig-obsmap} } 
\end{minipage}
\end{tabular}\newpage\end{figure*}

Variability of the oceanic internal wave spectrum will likely exist on
multiple temporal and spatial scales, e.g., those scales associated
with individual wave packets and isolated forcing events.  In the present 
study we assume that differences in the {\bf background} spectrum
are a product of regional differences in forcing and boundary
conditions averaged over the characteristic time scale to dissipate
the energy resident in the internal wavefield (about 50 days) and
averaged over the characteristic spatial scale that internal waves can
propagate in that time span.  The forcing mechanisms and propagation
scales are, in general, not regionally well defined.

Thus multiple data sets from Site-D (north of the Gulf Stream), the
Sargasso Sea and the California Current System are grouped together 
(Fig. \ref{fig-obsmap}).  In terms of low frequency variability, Site-D
exhibits the largest eddy energy, e.g., \cite{Wunsch97}.  Eddy energies
in the Sargasso Sea and California Current System exhibit spatial
trends but are comparable to each other.  Short narratives are used to
document regional characterizations of eddy amplitudes, tides and
bottom topography.

\subsubsection{Site D - $m^{-2.0}$ {\rm and} $\sigma^{-2.0}$.\label{SiteD}}

%
\begin{figure}[t]\vspace{-0.75cm}
\noindent\includegraphics[width=20pc]{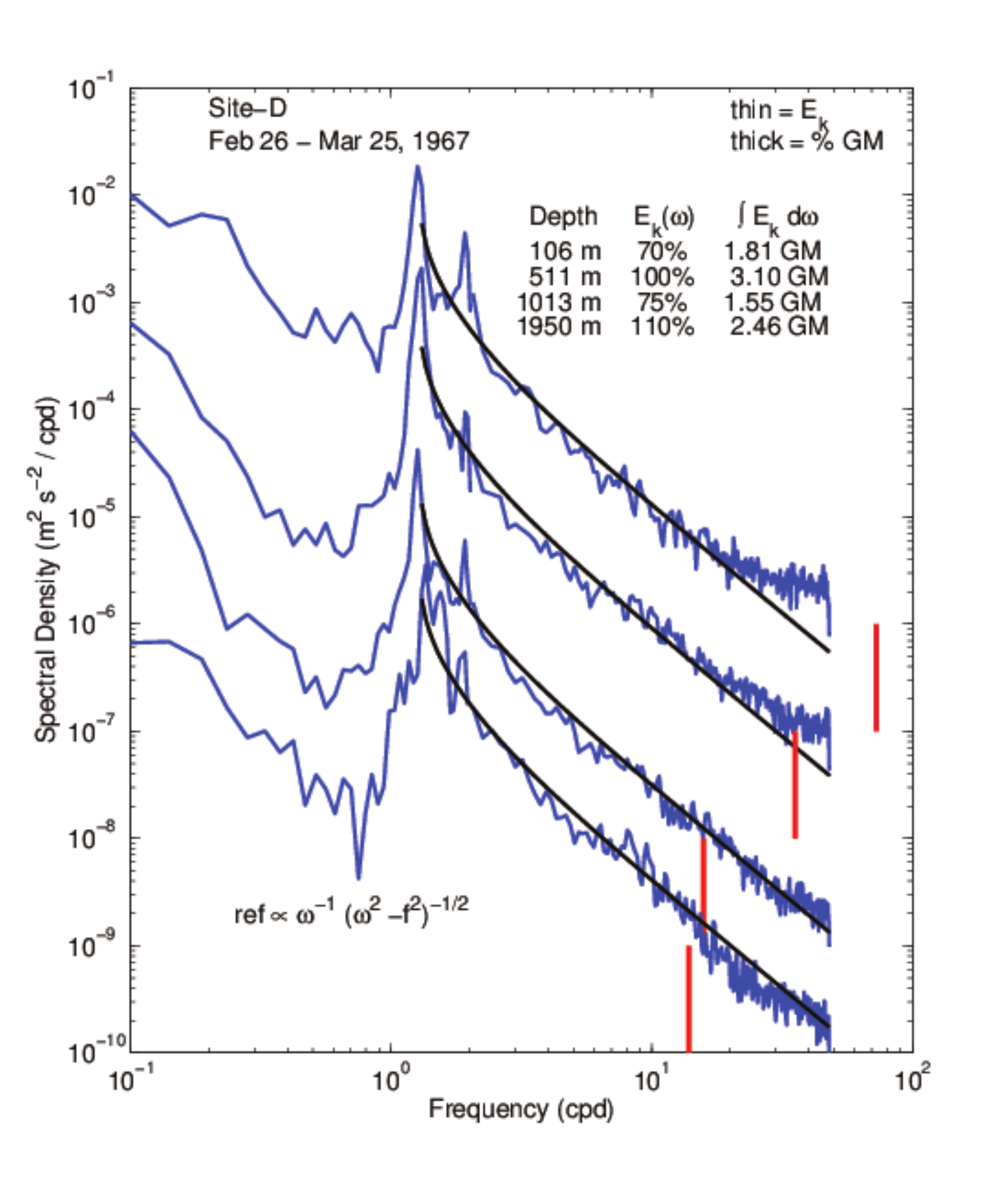}\vspace{-1.0cm}
\caption{ Site-D frequency spectra of
horizontal kinetic energy (blue lines).  These are the
Site-D data that appeared in the original GM72 paper.  Black curves
represent fits of (\ref{shape}) with $r=2$.  The thick vertical lines
represent the buoyancy frequency cut-off. The spectra have been offset 
by one decade for clarity.  \label{D-freq} }
\end{figure}
Site-D was fourth of a series of stations on a section from Woods Hole
to Bermuda.  Historically, the technology for long term current meter
deployments was developed in Woods Hole, and Site-D served as the test
bed for much of that instrumentation.  Much of the data in the
original Garrett and Munk paper (GM72) came from Site-D, if for no
other reason than not much else was available.  Site-D is characterized by a relatively small amplitude tide and large low-frequency (predominantly Topographic Rossby Wave) activity,
e.g., \cite{Hogg81}.  Site-D is situated under the Eastern Seaboard storm track, resulting in an enhanced potential for the resonant coupling between windstress and mixed layer at near-inertial frequencies \citep{DA85}.  Site-D may be relatively unique because of its geographic location.

\begin{figure}[h]\vspace{-0.75cm}
\noindent\includegraphics[width=20pc]{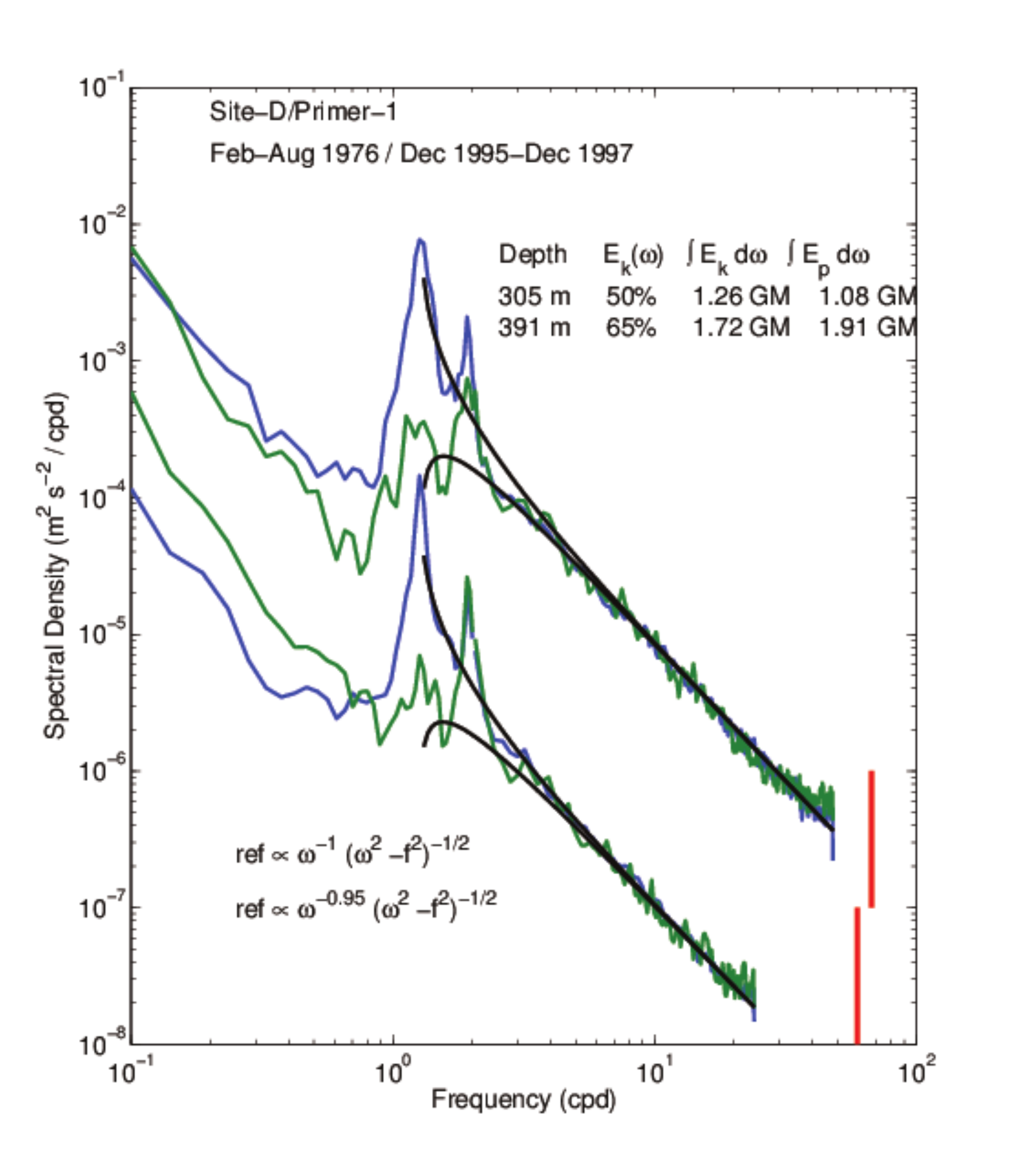}\vspace{-1.0cm}
\caption{ Frequency spectra of
horizontal kinetic energy and potential energy (blue and green lines) from
nearby Site-D.  Black curves represent fits of (\ref{shape}) with
$r=2$.  Thick vertical lines represent the buoyancy frequency cut-off.
The shape of the fit is stable throughout the three decades separating
the data in the original GM72 work.  Note that the amplitude is
somewhat lower.  The Primer-1 spectra have been offset by one decade
for clarity. \label{D-time} }
\end{figure}

It is apparent that, although many data were under consideration in
\cite{GM72}, the GM72 model is a curve fit to the original Site-D data
presented in \cite{Foff69} in both amplitude and $\sigma^{-2}$ power
law (Fig. \ref{D-freq})\footnote{Departures from a simple power law
behavior are apparent in the thermocline data at high frequencies and
spectra from the deeper current meters do not roll-off at frequencies
exceeding the local $N$.  Data from the very stable trimoored IWEX
mooring roll-off nicely (see below) and thus \cite{Briscoe75b}
suggests the super-buoyancy extension is an artifact of mooring
motion.  On the other hand, this does not mean that the departures
from a simple power law behavior apparent in the upper current meters
is an artifact of mooring motion.  See Appendix \ref{CurrentMeters} for further
details.}.  The GM72 model is a good characterization of the
spectral shape at Site-D throughout the decades (Fig. \ref{D-time}).
However, variability is apparent in spectral amplitude.

\begin{figure}[h]\vspace{-1.00cm}
\noindent\includegraphics[width=20pc]{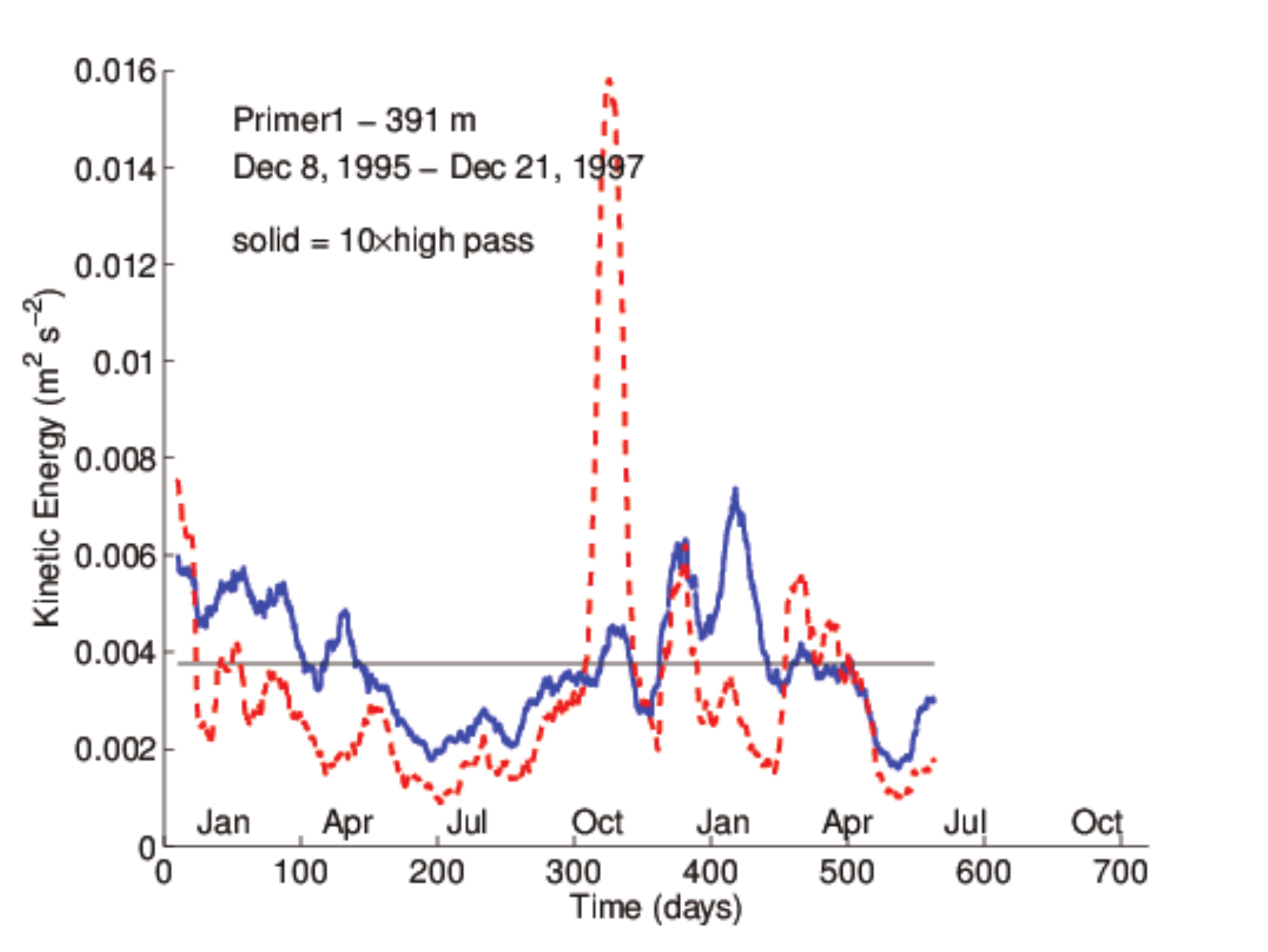}\vspace{-0.75cm}
\caption{ Site-D time series of horizontal kinetic energy.  
The solid line represents 10 times the high frequency energy.  The dashed line represents 
the entire internal wave band energy estimate.  The last 150 days of this record were not 
included in this analysis as they were contaminated by biological fouling of the rotor.  
\label{Primer-time} }
\end{figure}

The original (1967) Site-D data are characterized by larger spectral
levels than the other, longer term, estimates.  \cite{BW84} document
an annual cycle in high frequency internal wave energy using data from
the Lotus region south of the Gulf Stream, with maximum values in
winter.  This annual cycle is also apparent in the Primer data from
Site-D (Fig. \ref{Primer-time}).  A seasonal cycle in near-inertial energy for this location is discussed in detail in \cite{ST09}.  

\begin{figure}[h]\vspace{-0.75cm}
\noindent\includegraphics[width=20pc]{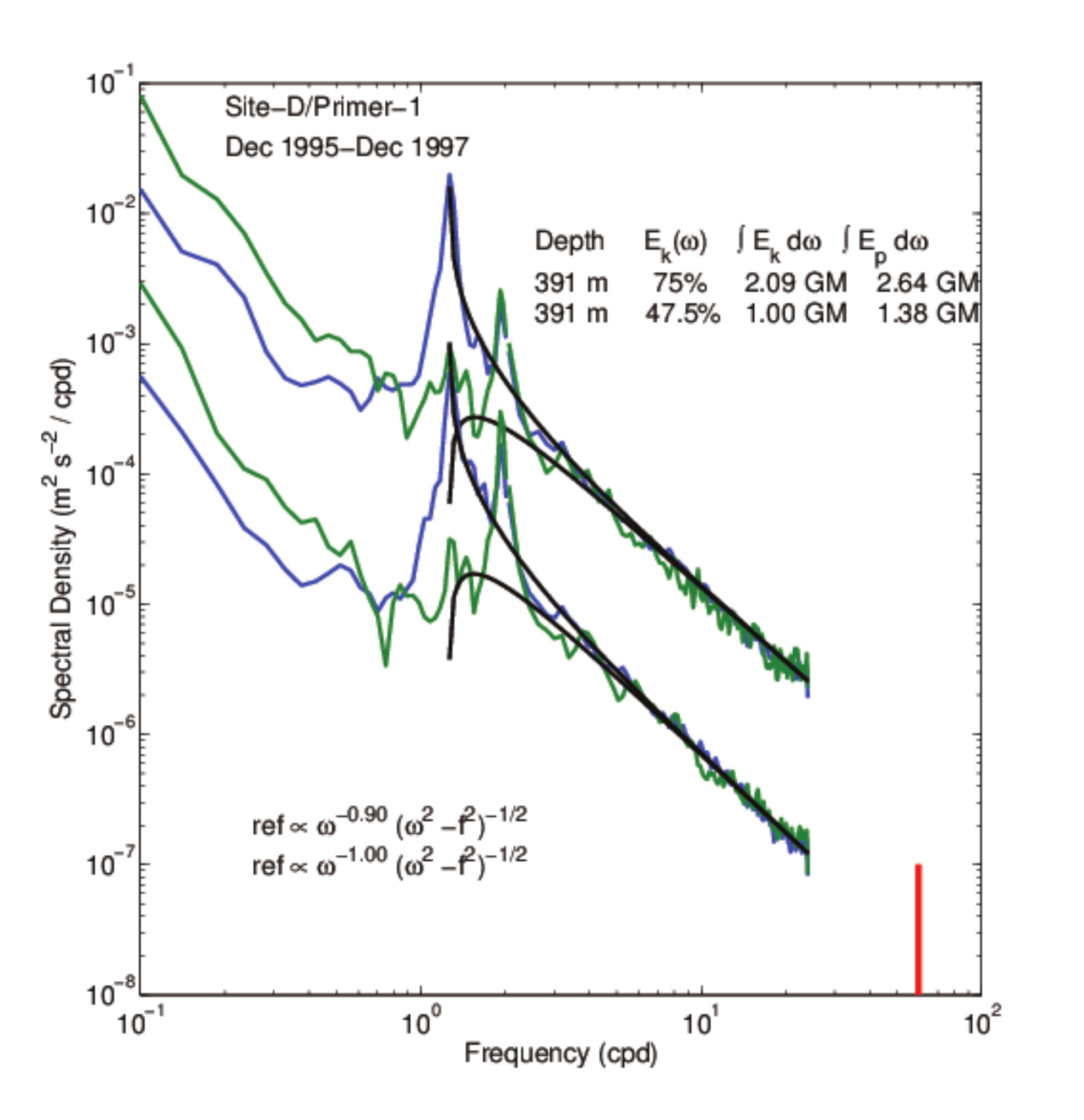}\vspace{-0.75cm}
\caption{ Frequency spectra of
horizontal kinetic and potential energy (blue and green lines) from nearby
Site-D.  The 2-year Primer data set has been divided into high and low
amplitude states.  The low amplitude state has been offset one decade
for clarity.  Black curves represent fits of (\ref{shape}) with
$r=1.90$ (high amplitude state) and $r=1.95$ (low amplitude state).
The thick vertical line represents the buoyancy frequency cut-off.  \label{D-HiLo} }
\end{figure}

With the GM spectrum being a curve fit to the early (1967) data, and with those data being obtained in February-March, the GM model is essentially a description of the winter time spectrum at Site-D.  The summer time spectrum has lower amplitude with 
little change in spectral slope at high frequency (Fig. \ref{D-HiLo}).  
We anticipate results presented below 
by stating here that annual average spectral levels at Site-D tend to be larger than 
other places and thus the Garrett and Munk model is a poor 
description of the background internal wavefield in much of the World Ocean.  


\begin{figure*}[t]\vspace{-0.5cm}
\noindent\includegraphics[width=20pc]{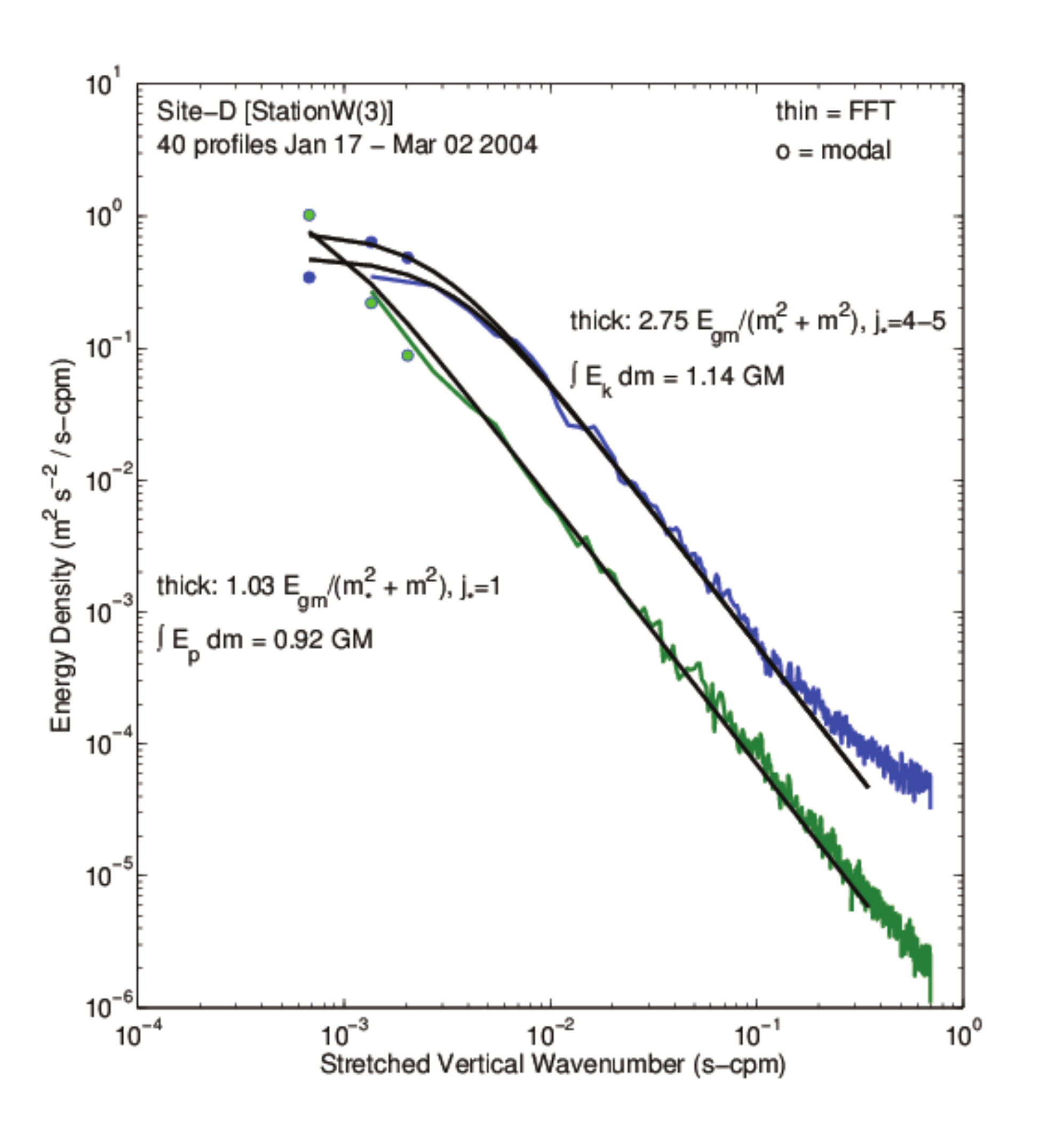}
\noindent\includegraphics[width=20pc]{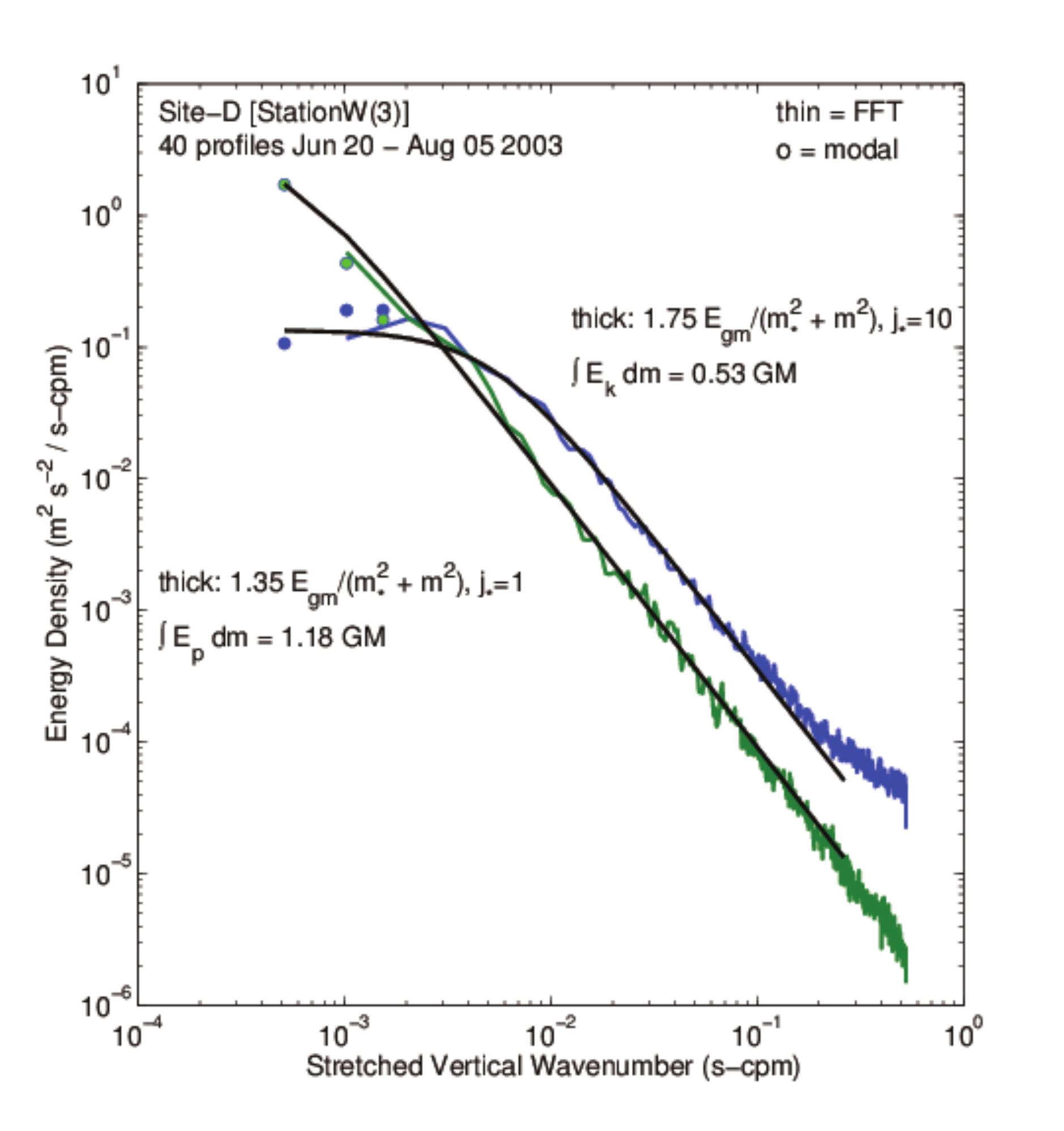}\vspace{-0.50cm}
\caption{ Site-D vertical wavenumber
spectra of horizontal kinetic and potential energy. The black line
represents a fit of the GM76 spectrum.  Spectral estimates at the
lowest wavenumber (enclosed circles) were made using a modal fit.
These data were recently acquired as part of a long term climate
monitoring project, Station-W (unpublished data from J. Toole,
personal communication 2004).  Velocity and density profiles were
obtained with the Moored Profiler.  Information regarding the internal
wavefield is returned by burst sampling 4 times using a 9.5 hour
sampling interval, then waiting 5 days before repeating.  The spectrum
presented here represents departures from the burst means.  See the
appendix for further details.  Departure from the curve fit at
vertical wavelengths of 10 m and smaller is interpreted as noise.  \label{D-vert} }
\end{figure*}

Despite the vertical spectrum being defined using data obtained elsewhere, recent vertical profile data data from Site-D are remarkably consistent with the GM76 model [$1/( m_*^2 + m^2)$] (Fig. \ref{D-vert}).  Wintertime conditions exhibit both enhanced spectral levels (amplitude factors of 2.75 vs 1.75)
and relatively more variance at low modes ($j_* = 4-5$ vs $j_* = 10$) than summertime data.
Winter time conditions also exhibit larger ratios of
kinetic-to-potential energy at high wavenumber, implying an increased
input of near-inertial energy during winter time and relaxation to higher frequencies.

\subsubsection{The Sargasso Sea \label{SargassoSea}}

A large number of experiments have been located in the Sargasso Sea
over the Hatteras Abyssal Plain.  On the southern side of the Gulf
Stream, this region exhibits an energetic eddy field having significant north-south gradients.  
Eddy energy levels are typically less than noted at Site-D.  A tidal ($M_{2}$) peak is apparent in the
temperature and velocity spectra.  \cite{M78} find that fluctuations
at this frequency have larger characteristic vertical scales than the
internal wave continuum, and there is evidence of similar features at
the first several harmonics.  From current meter data at 28$^{\circ}$N, 70$^{\circ}$W \cite{N75} and \cite{H77} estimate net fluxes at $M_2$ to be to the southeast and infer the
source to be the Blake Escarpment, near the western boundary.  
\cite{AZ07}, on the other hand, document net semidiurnal fluxes to the north-northwest (at 31$^{\circ}$N, 69$^{\circ} 30^{\prime}$W ) and southwest (at 34$^{\circ}$N, 70$^{\circ}$W ).  

The bottom near mid-basin is well sedimented and smooth at 28$^{\circ}$N, 70$^{\circ}$W, the locus of the Mid-Ocean Dynamics Experiment and the Internal Wave Experiment.  Rougher topography is noted to the east\footnote{One also finds mud waves.  Mud waves are sedimentary features of 1-10 km horizontal wavelength having amplitudes of 10's to 100's of meters.  These horizontal scales are appropriate for the generation of freely propagating internal lee waves (with Eulerian frequency $\sigma = 0$) if the intrinsic frequency $\sigma = \sigma - {\bf p \cdot \overline{u}}$ lies between the Coriolis and buoyancy frequencies: $f \leq {\bf p \cdot \overline{u}} \leq N$.  Significant coupling between the 'mean' and internal wavefield is anticipated at mean flow rates of 0.1-0.2 m s$^{-1}$.  Sediment transport is an issue at such flow rates and the possibility exists that the lee wave velocity perturbations affect the deposition/erosion process so as to reinforce the mud-waves \citep{B93}.  But this gets us off the topic of the background internal wavefield.}.  The PolyMode Local Dynamics Experiment current meter array (at 31$^{\circ}$N, 69$^{\circ} 30^{\prime}$W ) is likely situated above relatively rough terrain.  Bathymetry is relatively smooth at 34$^{\circ}$N, 70$^{\circ}$W, the locus of the Long Term Upper Ocean Study.  

The buoyancy frequency profile has a relative minimum in $N(z)$ associated with the 18$^{\circ}$ water thermostad at
about 300 m water depth.  The main thermocline exhibits nearly
constant stratification between 500-1000 m.\newline

\noindent{{\bf MODE}} 

Conducted during March-July of 1973, the Mid-Ocean Dynamics Experiment
(MODE) was one of the first concentrated studies of mesoscale ocean
variability.  The experiment featured arrays of moored current meters,
neutrally buoyant floats, standard hydrographic station techniques and
the use of novel vertically profiling instrumentation.  An extensive
array of current meter moorings was deployed in a 300 km radius
centered about ($28^{\circ}$ N, $69^{\circ} ~ 40^{\prime}$ W).
Data return from the current meters was limited \citep{Mode_Atlas}. 

While designed primarily to investigate low frequency motions, the
experiment returned a great deal of information about internal waves.
Vertical profiles of horizontal velocity obtained during May and June
with a free-falling instrument using a electric field sensing
technique \cite{Sanford75} provided, for the first time, direct
estimates of the high vertical wavenumber structure of the ocean
internal wavefield.  These data are dominated by near-inertial
frequencies, and a rotary decomposition in the vertical wavenumber
domain \cite{LS75} reveals a large excess of clockwise ($cw$) phase
rotation with depth.  Clockwise phase rotation with depth is a
signature of downward energy propagation for near-inertial waves.

Despite the evidence of excess downward energy propagation that was
interpreted in terms of atmospheric generation \citep{L76}, these data
were assumed to be representative of the background internal wavefield.  
A subset obtained as a time series of $4 \frac{1}{2}$ days provided the 
basis for a revision to the isotropic Garrett and Munk spectral model, GM75, 
with high wavenumber asymptote of $m^{-5/2}$.  

The MODE profile data were included in our original study \citep{LPT}
reporting the covariability of frequency-wavenumber power laws.
Further investigation suggests the characterization of these data as
representing the background wavefield is problematic.  Several
coherent wavepackets dominate the high wavenumber energy content and
these packets can be interpreted as being strained by the mesoscale
eddy field and propagating into a critical layer \citep{P08a}.  Thus
wave-mean interactions acting in conjunction with the downward
propagating waves may be responsible for creating structure in the
vertical wavenumber domain uncharacteristic of the background
spectrum.\newline

\noindent{{\bf IWEX} - $k^{-2.4 \pm 0.4}$ {\rm and} $\sigma^{-1.75}$}

 
The Internal Wave Experiment (IWEX) represents an early attempt to
estimate a vector wavenumber-frequency spectrum with a minimum of
assumptions.  Current and temperature data were obtained with 17
Vector Averaging Current Meters (VACMs) and 3 Geodyne 850 current
meters from a taut, three-dimensional, trimoored array.

\begin{figure}[t]\vspace{-0.75cm}
\noindent\includegraphics[width=20pc]{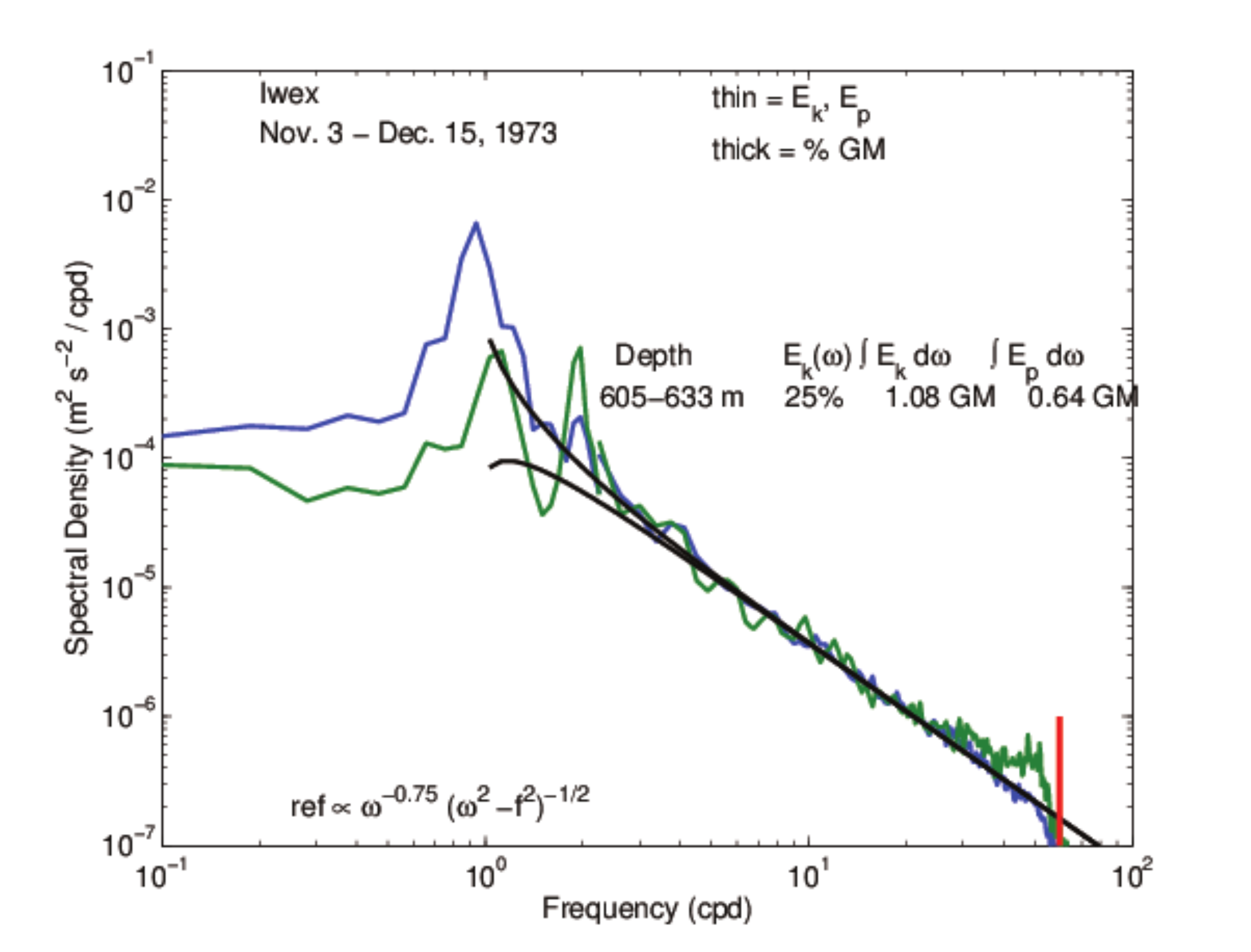}\vspace{-0.50cm}
\caption{ Iwex frequency spectra of horizontal kinetic energy and potential energy (blue and green lines).  
Black curves represent fits of (\ref{shape}) with $r=1.75$.  The thick 
vertical line represents the buoyancy frequency cut-off.  \label{Iwex-freq} }
\end{figure}

\cite{M78} assumed the spectrum could be factorized as:
$$
E^{\sigma}(k, \varphi, \sigma) = E^{\sigma}(\sigma)A(k;\sigma)S(\varphi; \sigma)
$$ with $\sigma=\pm$ denoting the sign of vertical energy propagation,
$A$ the normalized horizontal wavenumber ($k$) distribution and $S$ a
normalized azimuthal ($\varphi$) distribution.  Their horizontal wavenumber
distribution was assumed to have a parametric representation of:
$$ A(k ; \sigma) \propto [1 + (\frac{k - k_p}{k_*})^s]^{-t/s}$$ 
for $k > k_p$ and $A=0$ otherwise.  The parameters are horizontal
wavenumber scale $k_*$, low wavenumber cut-off $k_p$, shape factor
$s$, and high wavenumber slope $t$.  An inverse analysis of the
spatial lag cross spectra (the lag-coherence is proportional to the
Fourier transform of the energy spectrum) was performed to estimate
the various parameters in their proposed spectrum.  In particular,
\cite{M78} find that the horizontal wavenumber energy spectrum depends
upon wavenumber as $k^{-2.4 \pm 0.4}$, independent of frequency.  For
continuum ($f^2 \ll \sigma^2 \ll N^2$) frequencies the horizontal
and vertical wavenumber power laws are identical.  Thus $E(m) \propto
m^{-2.4 \pm 0.4}$.  The observed frequency spectra of both velocity
and temperature are characterized by the power law $\sigma^{-1.75}$,
\cite{Briscoe75b} (Fig. \ref{Iwex-freq}).

The \cite{M78} analysis includes an estimate of finestructure
contamination.  That is, their horizontal wavenumber spectrum includes
only contributions which are consistent with linear internal waves.
An increasing contamination with increasing frequency (decreasing
horizontal scale) is apparent.  As they make no distinction between
permanent finestructure contamination and Doppler shifted internal
waves, much of their finestructure contamination could be internal
waves [e.g., \cite{SandP91, P03}].  Thus the horizontal wavenumber domain power law estimate may be biased too steep.

Several features of the IWEX data set reoccur in many of the other
data sets analyzed here.  The first is that estimates of amplitude
based upon fits to the high frequency part of the spectrum tend to be
significantly lower than the GM model whereas estimates of total
energy tend to be more nearly equal to the total energy in the GM model.  The
inference is that the inertial cusp of the GM model is not
sufficiently strong to describe the oceanic peak.  While this has been
noted before (e.g., \cite{Fu81}), the IWEX analysis indicates the
excess near-inertial energy has smaller characteristic vertical
scales: At super-tidal frequencies the peak wavenumber $k_p$
corresponds to the first vertical mode.  At near-inertial frequencies
$k_p$ corresponds to the third vertical mode, indicating significantly
reduced near-inertial energy in the first two modes.  This pattern is
consistent with the vertical profile data presented here which depict
ratios of $E_k$ to $E_p$ at 100 (1000) m. vertical wavelengths that
are significantly larger (smaller) than in the GM model, for which
$E_k / E_p = 3$.

\cite{M78} avoid interpreting the IWEX spectrum in the context of how
it might relate to the background oceanic spectrum.  This is not
surprising given the short duration of the data set and the relative
lack of quality data sets at that time.  Here, however, we promote the
IWEX spectrum as a more realistic representation of the Sargasso Sea
background spectrum than the GM model.\newline

\noindent{{\bf LOTUS} - $\sigma^{-1.75}$--$\sigma^{-1.85}$}

One goal of the Long-Term Upper-Ocean Study (LOTUS) was to document
variability in the internal wavefield over several seasons in order to
investigate its association with a variety of forcing mechanisms and
environmental conditions.  Meteorological data were obtained with a
surface mooring and buoy.  Only data from subsurface moorings are
described here.  The region is notable for extreme air-sea buoyancy
exchange resulting in the production of Eighteen Degree Water,
e.g., \cite{Kwon04}

\begin{figure}[h]\vspace{-0.50cm}
\noindent\includegraphics[width=20pc]{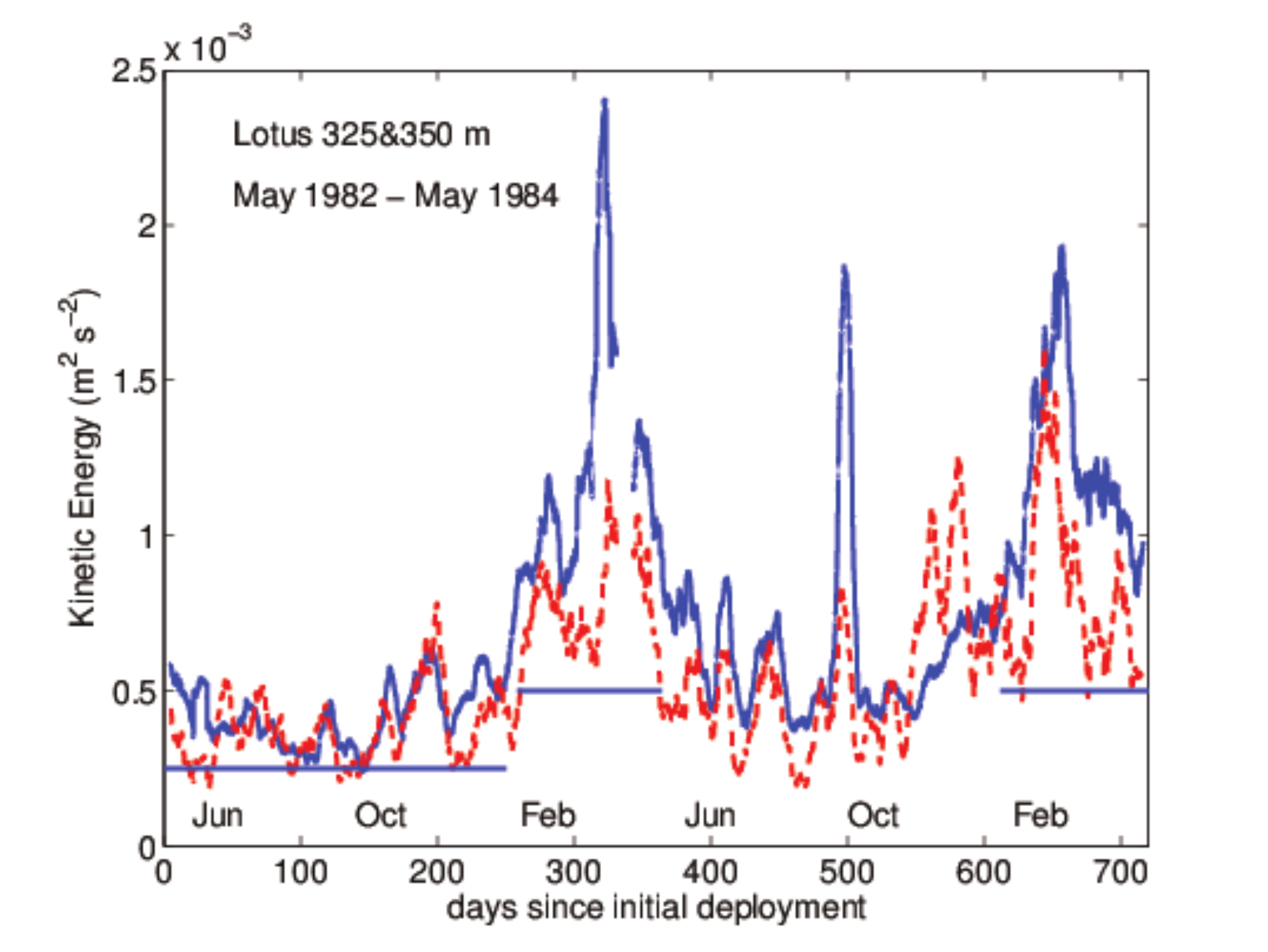}\vspace{-0.50cm}
\caption{Lotus time series of
high-frequency energy from 325 (year-1) and 350 (year-2) m water
depth.  This depth range is occupied by 18 Degree Water and represents
a local minimum in the stratification rate.  Mixed layer depths were
observed to be smaller than 300 m over the duration of the data set.
The solid line represents 10 times the high frequency energy.  The dashed line represents 
the entire internal wave band energy estimate.  The horizontal lines indicate the time periods over which spectra were averaged into bins of high and low energy states of Fig. \ref{Lotus-freq}. \label{Lotus-time} }
\end{figure}

\cite{BW84} document an annual cycle of high frequency internal wave
energy at 200-500 m depths from the first year's data.  The energy was
noted to vary from a half or a third to 2 or three times the mean at
each depth.  That signal is repeated in the second year's data (Fig
\ref{Lotus-time}).  Apart from the seasonal cycle, internal wave energy
varies on time scales of several weeks.  The relation between this
short time scale variability and low frequency (sub-inertial) shear
has been noted since MODE (see \cite{RJ79} and references therein).

In this study of how nonlinearity may shape and form the internal wave
spectrum, there is a potential link between spectral amplitude and
spectral shape.  The Lotus spectra in the 325-350 m depth range were
averaged over the time periods indicated in Fig. \ref{Lotus-time}.
Spectra from the more energetic time periods are slightly less steep
($\sigma^{-1.75}$ vs. $\sigma^{-1.85}$) for a factor of 2.5 difference
in spectral level.  If there is a dynamical link between spectral
level and spectral shape, that link is subtle relative to the gamut of
variability discussed here.

\begin{figure}[h]\vspace{-0.75cm}
\noindent\includegraphics[width=20pc]{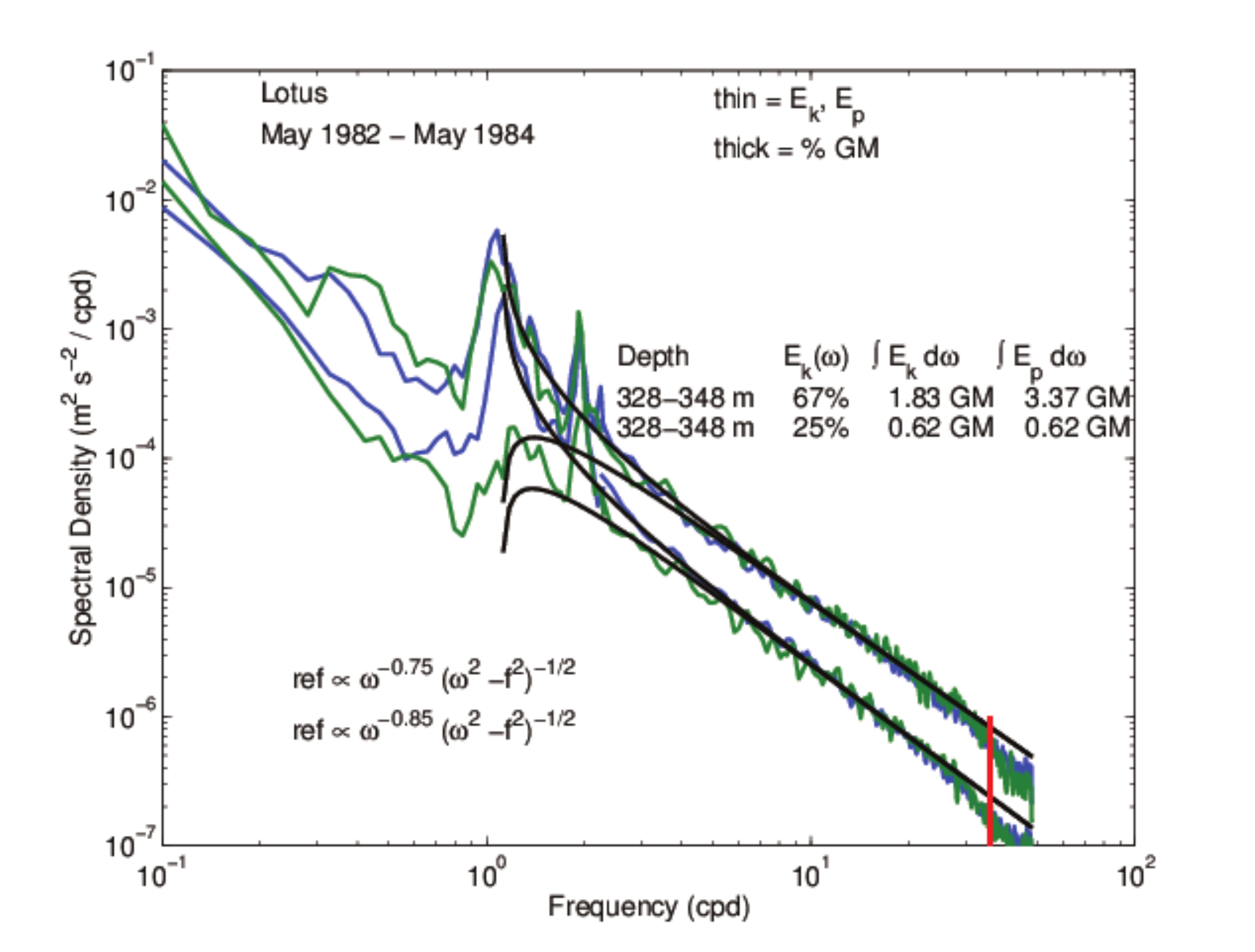}\vspace{-0.75cm}
\caption{Lotus frequency spectra of
horizontal kinetic energy and potential energy (blue and green lines) for the
high and low energy states depicted in Fig. \ref{Lotus-time}.  Black
curves represent fits of (\ref{shape}) with $r=1.75$ and $r=1.85$.
The thick vertical line represents the buoyancy frequency cut-off.
Temporal variability is dominated by variability in the overall
amplitude of the spectra rather than the shape (power law). \label{Lotus-freq} }
\end{figure}

\noindent{{\bf FASINEX} -  $m^{-2.3}$ {\rm and} $\sigma^{-1.85}$ }

\begin{figure}\vspace{-0.75cm}
\noindent\includegraphics[width=20pc]{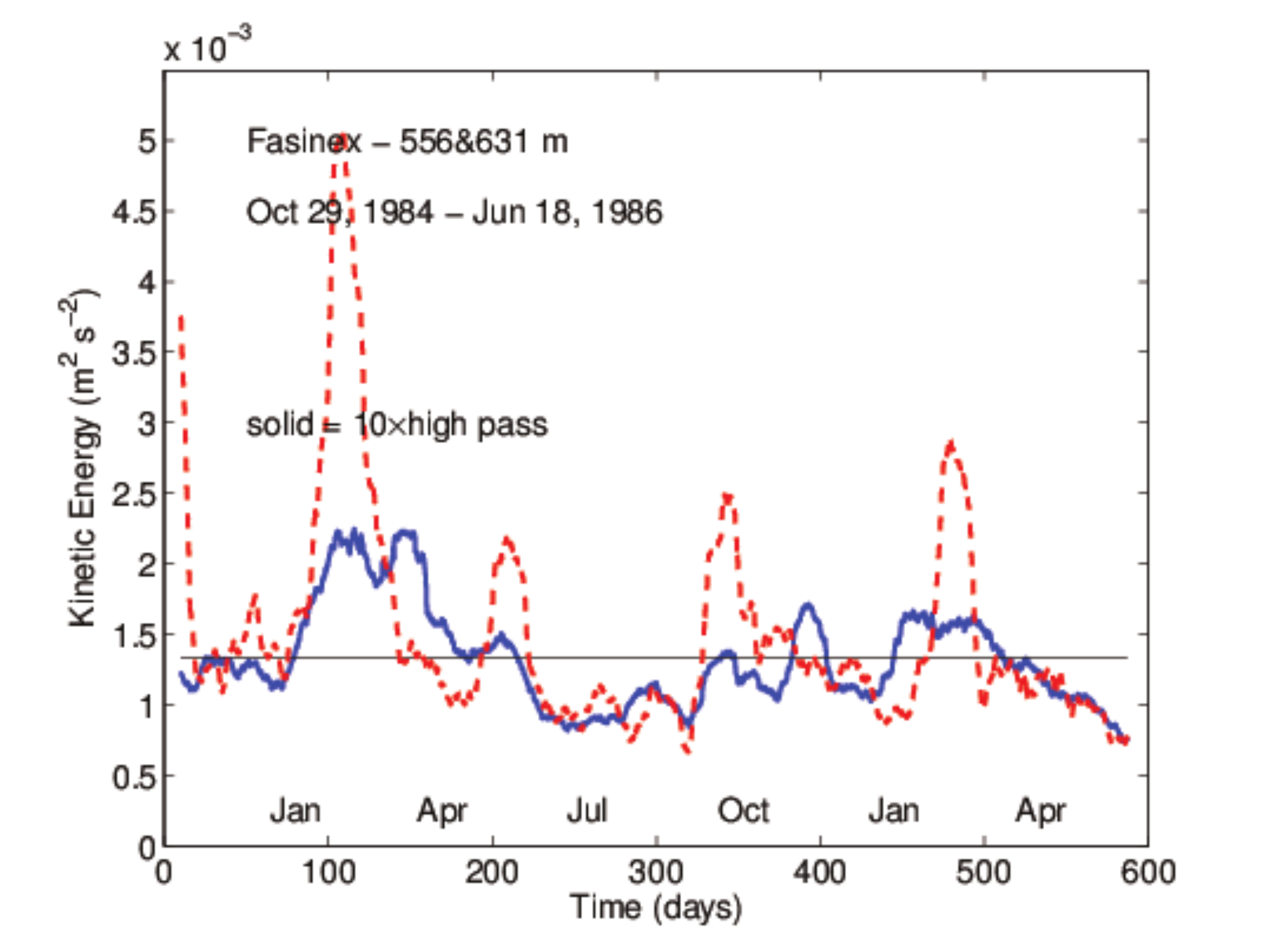}\vspace{-0.75cm}
\caption{FASINEX time series of
high-frequency energy from 556 and 631 m water
depth.   The solid line represents 10 times the high frequency energy.  The dashed line represents 
the entire internal wave band energy estimate.  \label{Fasinex-time} }
\end{figure}

\begin{figure}\vspace{-0.25cm}
\noindent\includegraphics[width=20pc]{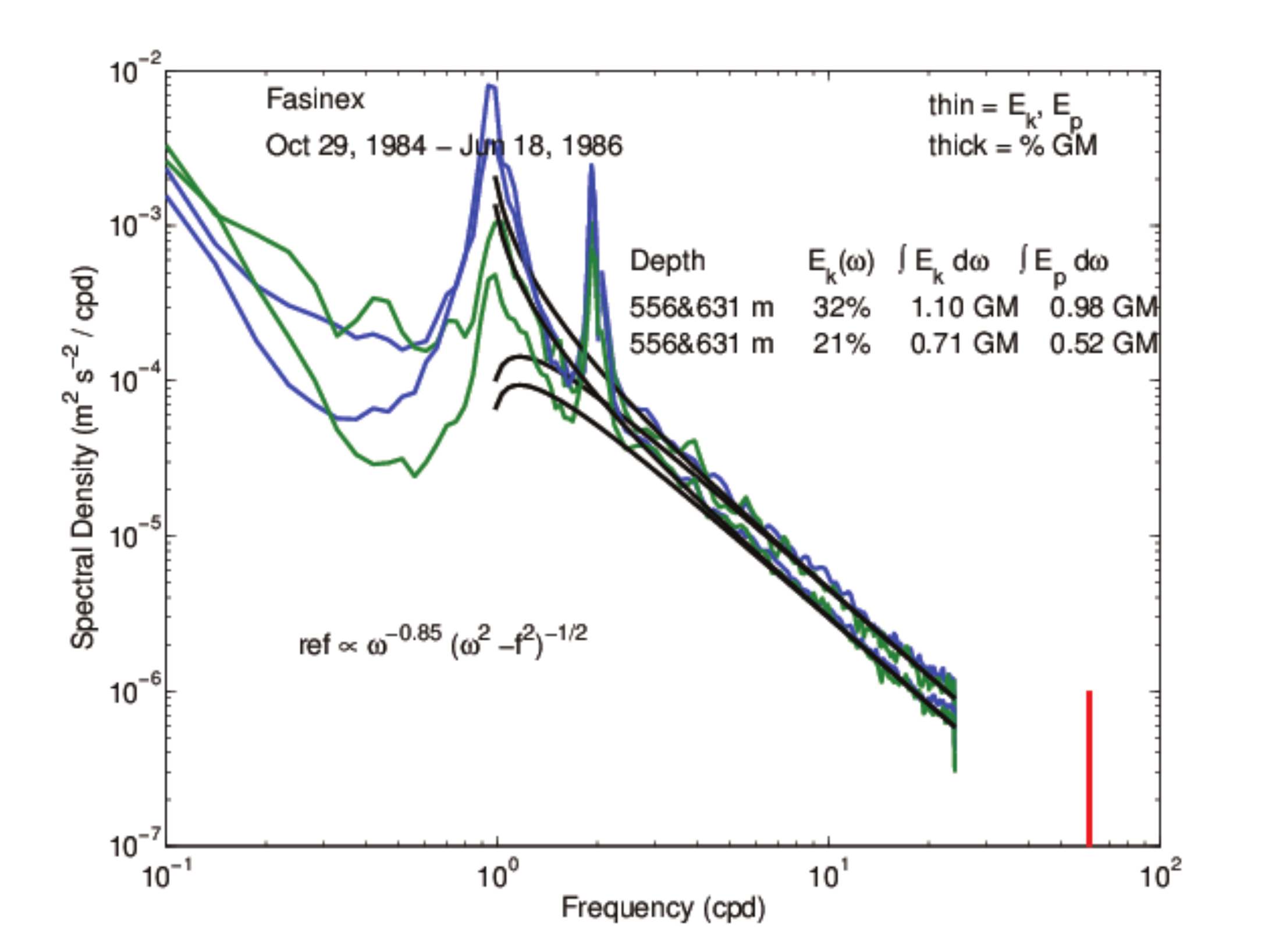}\vspace{-0.75cm}
\caption{FASINEX frequency spectra
(blue and green lines) of horizontal kinetic energy and potential energy from
the main thermocline (600 m).  Black curves represent fits of
(\ref{shape}) with $r=1.75$.  The thick vertical line represents the
buoyancy frequency cut-off.  \label{Fasinex-freq}  }
\end{figure}

The Frontal Air-Sea Interaction Experiment (FASINEX) was designed to
investigate the response of the upper ocean to atmospheric forcing in
the presence of oceanic fronts.  An array of surface and subsurface
moorings with VMCMs, VACMs and Profiling Current Meters (PCMs) was
deployed in the subtropical convergence zone of the Northwest Atlantic
( approximately $27^{\circ}$ N, $70^{\circ}$ W) from January to June
of 1986, \cite{W91} and \cite{E91}.  The moored data in this study are
taken from a long-term, subsurface mooring at $28^{\circ}$ N.
These data document an annual cycle in internal wave energy that is substantially 
reduced from the Lotus time series 500 km north (Figure \ref{Fasinex-time}).  
Frequency spectra (Fig. \ref{Fasinex-freq}) at depths of 556 and 631 m
are defined by a power law $\sigma^{-r}$ of $r=1.85$ and amplitudes
significantly smaller than the GM model.

\begin{figure}\vspace{-0.75cm}
\noindent\includegraphics[width=20pc]{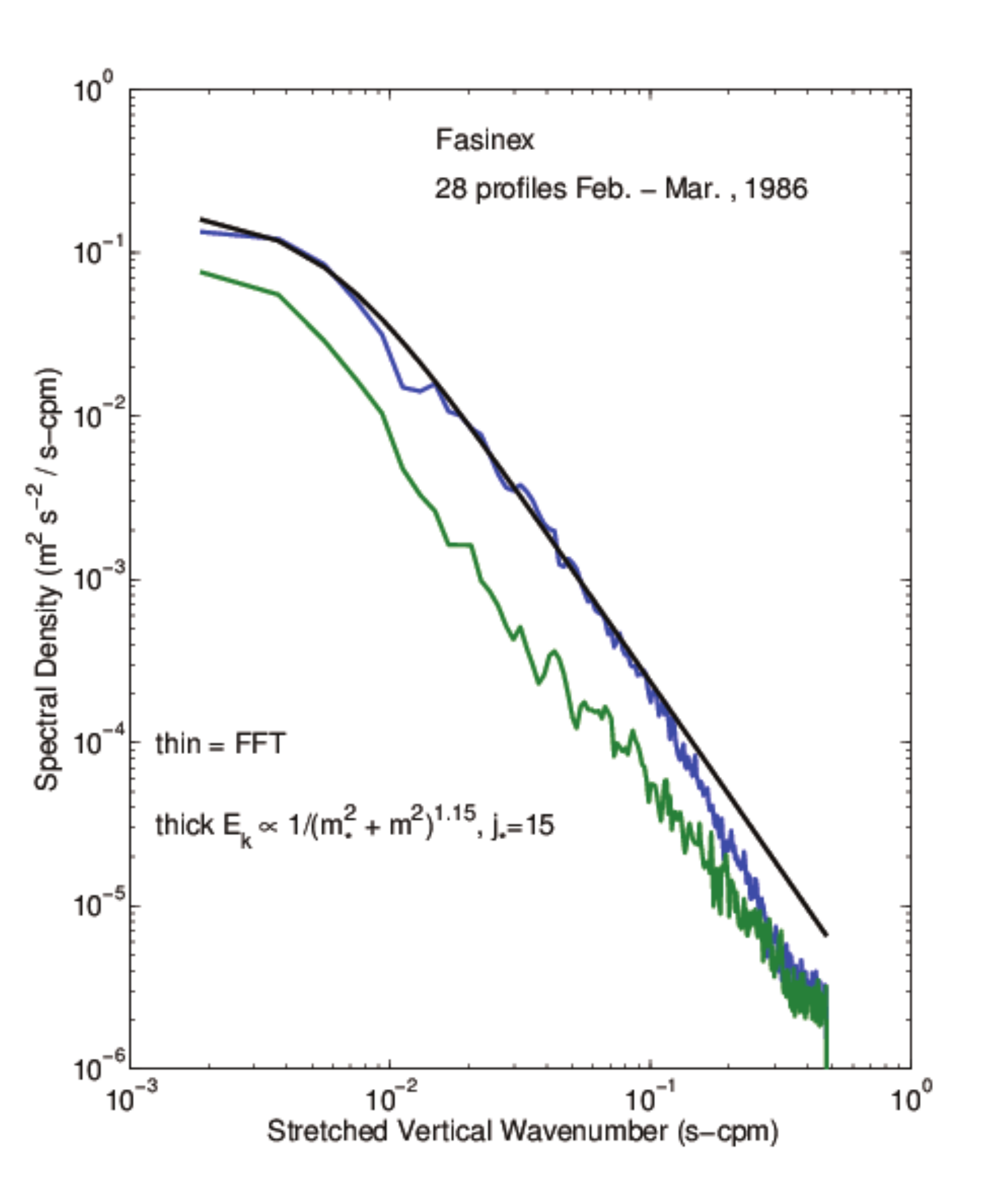}\vspace{-0.75cm}
\caption{FASINEX-vertical wavenumber
spectra of horizontal kinetic and potential energy, $N$-scaled and
stretched under the WKB approximation to $N_o = 3$ cph.  These data
were obtained 100-200 km north of the FASINEX moored array.  Sampling
was intermittent in both space and time.  Note that the low-wavenumber
spectral estimates have typically smaller ratios of $E_k$ and
$E_p$. \label{Fasinex-vert} }
\end{figure}

Vertical profiles of horizontal velocity and density were obtained
during February-March using the High Resolution Profiler (HRP),
\cite{POTS}.  The vertical profiles, obtained primarily $1^{\circ}$
north of the moored array as part of a spatial survey, revealed a
complex pattern of variability associated with the frontal velocity
structure in the upper 250 m.  Here we report results concerning data
from depths of 250-1000 m.  Those data are fit with a spectrum having
an asymptotic roll-off of $m^{-2.3}$ (Fig. \ref{Fasinex-vert}).

\cite{WandME91} document a slight excess of clockwise phase rotation
with depth over counter-clockwise shear variance, a signature of
excess downward propagating near-inertial energy.  The quoted power
law in the vertical wavenumber domain is at a depth for which $N^2$
varies by less than a factor of 2, so that, biases associated with
linear wave propagation in nonuniform $N$ are likely small.  The instrument does not, 
however, return robust velocity estimates at large vertical scales and hence diagnosing interactions with the thermocline scale geostrophic shear is problematic.  

\subsubsection{The western Tropical Atlantic \label{Tropics} } 

\noindent{{\bf SFTRE and PolyMode IIIc} -  $m^{-2.4}$ {\rm and} $\sigma^{-1.9}$ }

Extant frequency domain data from thermocline regions in the western Tropical Atlantic
are limited to those obtained as part of the PolyMode program.  The
motivation for PolyMode Array III was exploration of low frequency
variability in what were perceived as dynamically distinct regions of
the North Atlantic.  Cluster IIIc was placed in the eastward flowing 
North Equatorial Current with the intent of examining the low
frequency variability for characteristics of eddy generation by
baroclinic instability, \cite{Keffer83}.  The cluster is situated over
the northwestern extension of the Demerara Abyssal Plain.
Locally the bottom is well sedimented and relatively flat, though rough topography lies immediately to the east.  Data
presented here are an average of the three southeastern moorings
(80-81-82).

\begin{figure}\vspace{-0.75cm}
\noindent\includegraphics[width=20pc]{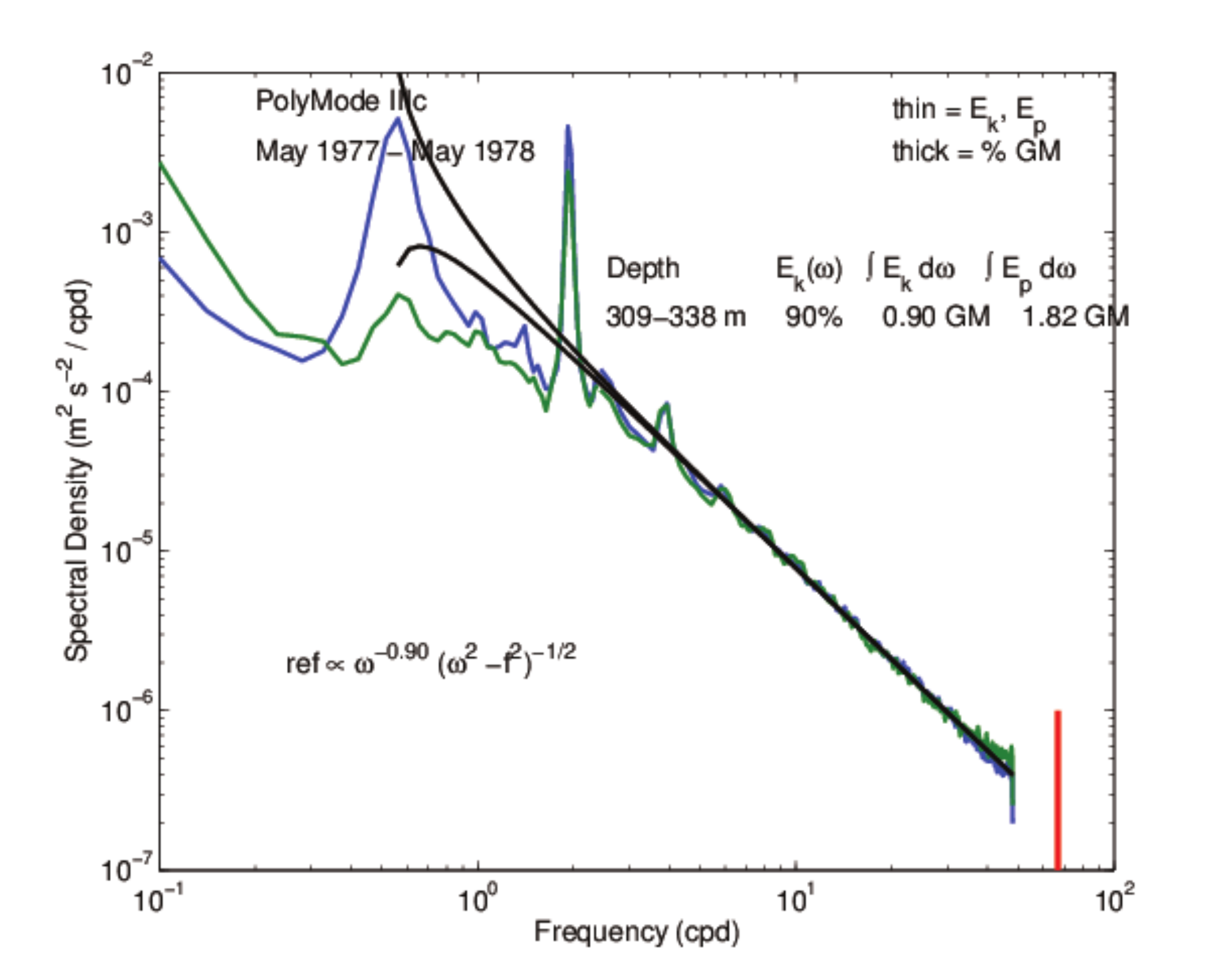}\vspace{-0.75cm}
\caption{PolyMode IIIc frequency spectra
(blue and green lines) of horizontal kinetic energy and potential energy from
the main thermocline (300 m).  Black curves represent fits of
(\ref{shape}) with $r=1.90$.  The thick vertical line represents the
buoyancy frequency cut-off.  \label{Tropics-freq} }
\end{figure}

\begin{figure}\vspace{-0.75cm}
\noindent\includegraphics[width=20pc]{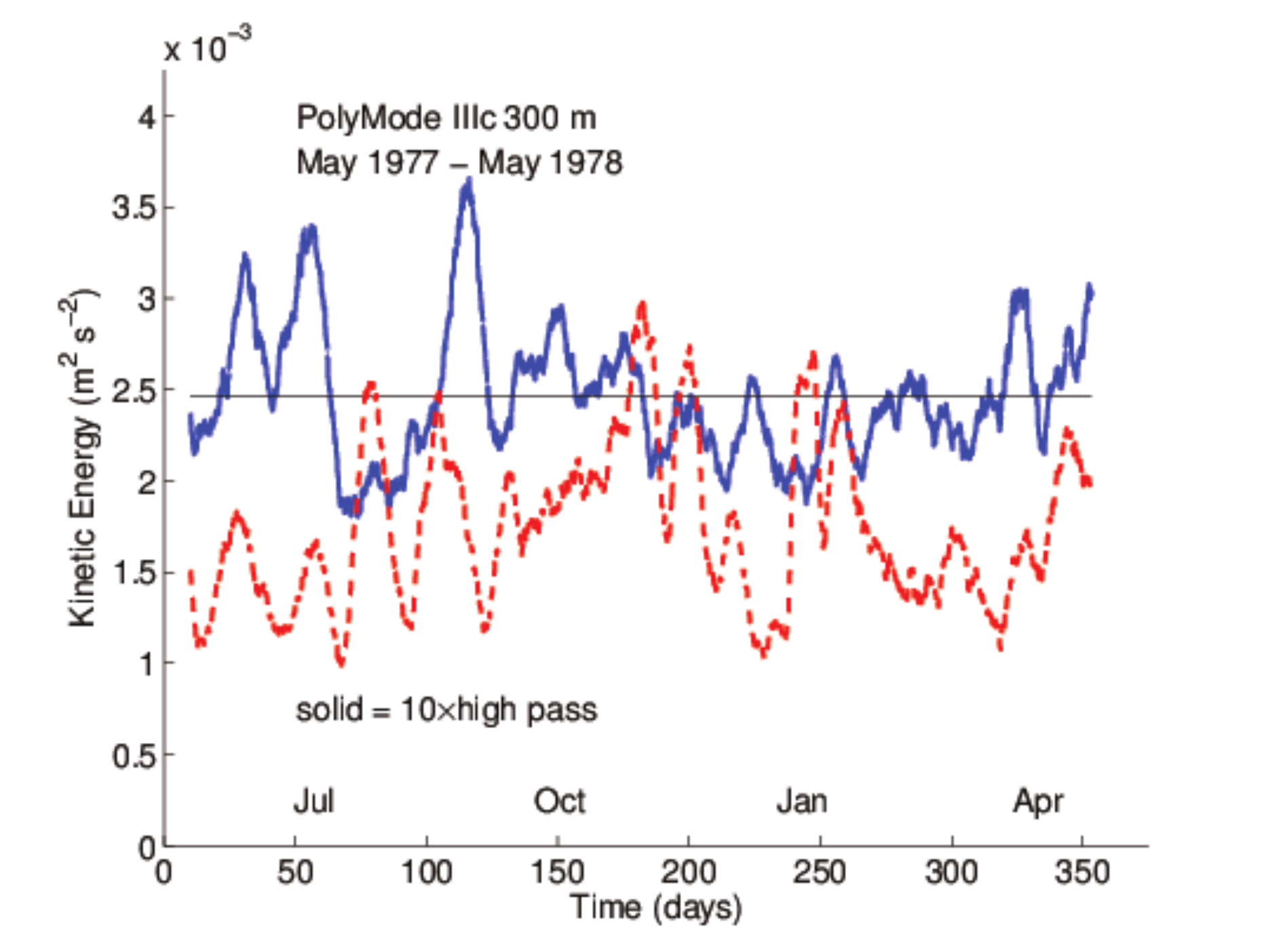}\vspace{-0.75cm}
\caption{PolyMode IIIc time series of
high-frequency and internal wave-band energy from nominal 300 m water
depth.  The solid line represents 10 times the high frequency energy.  The dashed line represents 
the entire internal wave band energy estimate.  \label{Tropics-time} }
\end{figure}

Frequency spectra (Fig. \ref{Tropics-freq}) roll-off less steeply than
$\sigma^{-2}$.  The simple power law characterization
(\ref{frequency_spectrum}) overestimates the observed spectral density
at frequencies smaller than semi-diurnal.  An annual cycle in either
high frequency or internal wave band energy is not apparent in the
time series (Fig. \ref{Tropics-time}).


Vertical profile (HRP) data used here were obtained during Nov. 2001
as part of a sampling survey for the Salt Finger Tracer Release
Experiment (SFTRE).  Warm, high-salinity Subtropical Underwater
overlies cooler, fresher Antarctic Intermediate Water
\citep{Schmitt05}.  The situation is unstable to the salt-fingering
form of double diffusive instability and a staircase layering of the
temperature, salinity and density profiles is present over much of the
region.  The profiles examined here were collected between
($14-16^{\circ}$ N, $50-57^{\circ}$ W) on the northern edge of the
survey region.

\begin{figure}[h]\vspace{-0.75cm}
\noindent\includegraphics[width=20pc]{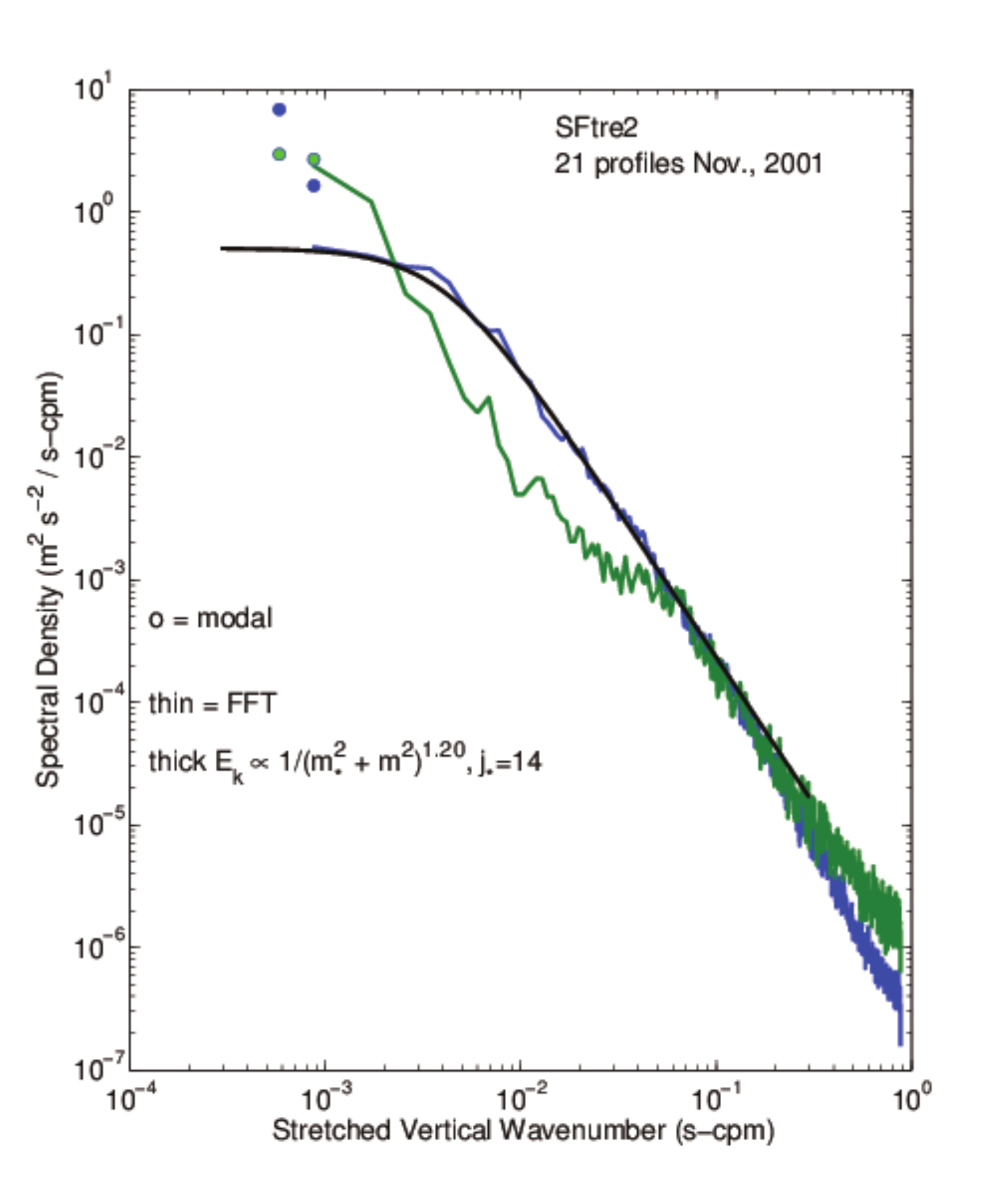}\vspace{-0.75cm}
\caption{SFtre2-vertical wavenumber
spectra of horizontal kinetic and potential energy, $N$-scaled and
stretched under the WKB approximation to $N_o = 3$ cph.  These data
were obtained on the northern part of the year-1 sampling survey grid.
Note that the low-wavenumber spectral estimates have typically smaller
ratios of $E_k$ and $E_p$. \label{Tropics-vert} }
\end{figure}

Vertical wavenumber $E_k$ spectra (Fig. \ref{Tropics-vert}) roll-off
more steeply than $m^{-2}$ at high wavenumber.  The spectra are
anomalous in that potential energy estimates exceed kinetic energy at
high wavenumber, a feature that we attribute to the staircase
features.  The parametric fit (\ref{vertical_spectrum}) produces a
low-wavenumber roll-off equivalent to mode-9.  Estimates of $E_p$
dominate $E_k$ at low wavenumber.

\subsubsection{The southeast Subtropical North Atlantic}\label{SE_ST_NA}

\noindent{{\bf NATRE and SUBDUCTION} - $m^{-2.55}$ or $m^{-2.75}$ and $\sigma^{-1.35}$}

Vertical profile (HRP) data were collected during April 1992 southwest
of the Canary Islands as part of an initial site survey for the North
Atlantic Tracer Release Experiment (NATRE).  The bulk of the data were
obtained as part of a $400\times400$ km grid centered about
(26$^\circ$N, 29$^\circ$W).  Apart from a minor seamount with a summit
at 3000 m water depth, the bottom is relatively featureless near the
survey domain.  The region is further notable for the production of a
relatively minor water mass, Madeira Mode Water \citep{Weller04}.

A rotary decomposition of the vertical profile data does not return a
consistent pattern of phase rotation with depth as a function of
wavenumber. Thus, despite the fact that the buoyancy profile in the
southeast part of the subtropical gyre decreases monotonically from
the mixed layer base through the main thermocline, a possible bias of
the spectral slope associated with a purely linear response to wave
propagation in variable $N(z)$ is unlikely.

\begin{figure*}\vspace{-0.5cm}
\noindent\includegraphics[width=39pc]{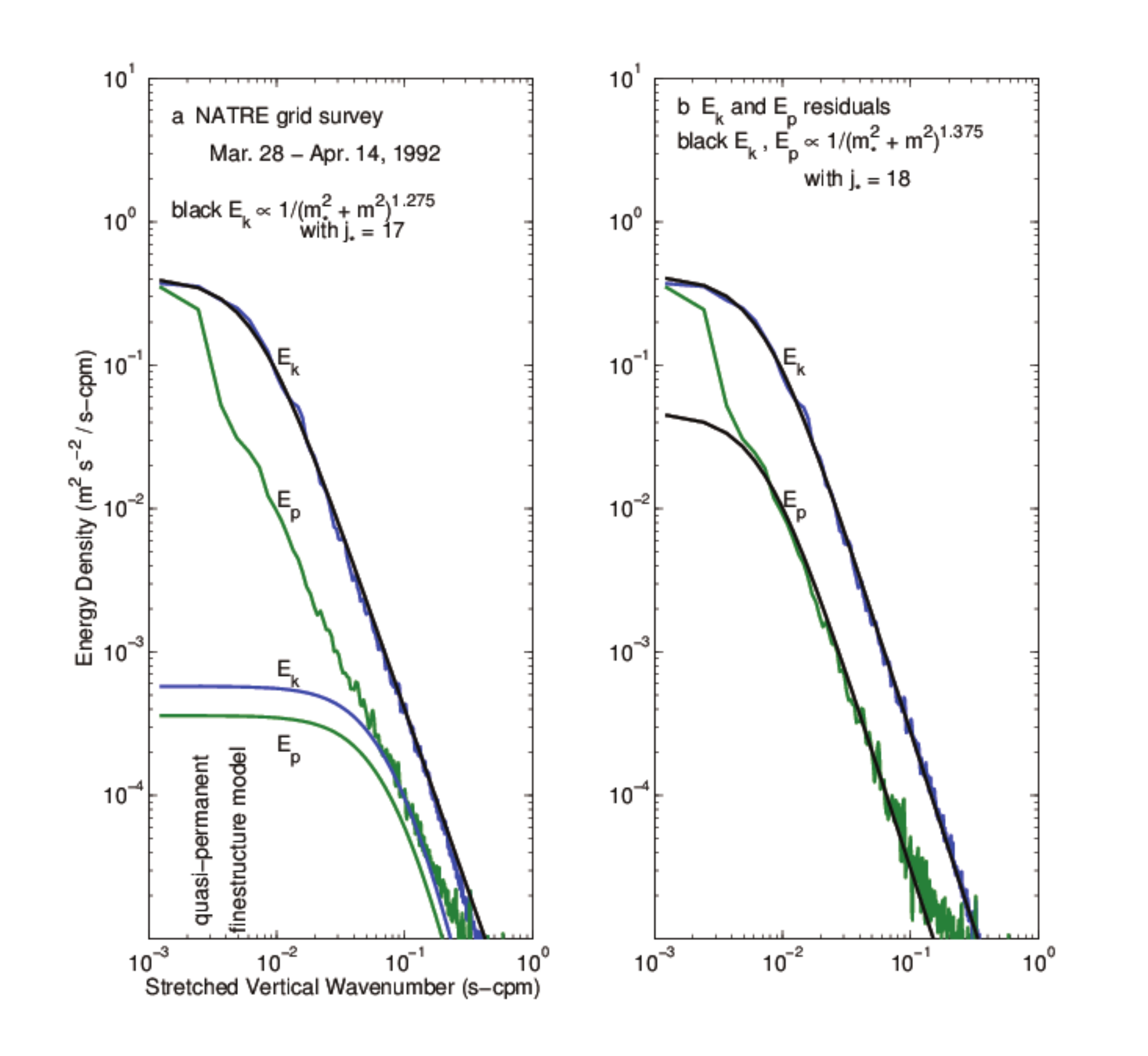}\vspace{-0.75cm}
\caption{Natre vertical wavenumber
kinetic $E_k$ and potential $E_p$ energy spectra.  These 100 profiles
were obtained as part of a 400$\times$400 km grid survey.  a) Observed
vertical spectra, $N$-scaled and stretched under the WKB approximation
to $N_o = 3$ cph, and the quasi-permanent finestructure spectrum from
\cite{Polzin03}.  b) The internal wave spectra, observed minus
quasi-permanent contributions.  The thin lines represent fits of
$1/(m_*^2 + m^2)^{11/8}$ to the spectra, with $m_*=0.0070$ cpm.  The
fit to the velocity data is obscured as it overlies the data.  Note
that the low-wavenumber spectral estimates have typically smaller
ratios of $E_k$ and $E_p$. \label{Natre_vert}  } 
\end{figure*}

The vertical wavenumber kinetic energy spectra (Fig.
\ref{Natre_vert}) are white at low wavenumber and roll-off more
steeply than $m^{-2}$ for vertical wavelengths smaller than 200 m ($E_{k} \propto m^{-2.55}$).
The observed potential energy spectra exhibit three salient features
in contrast with the kinetic energy spectrum.  First, kinetic and
potential energy are nearly equal at the largest resolved vertical
wavelengths, about 1000 meters.  The spatial survey is of sufficient
lateral extent that, even though low frequency variability is
dominated by barotropic and low mode (mode-1 and mode-2) contributions
(\cite{Wunsch97}), low frequency variability may contribute to the
observed spectra at the largest resolved scales (equivalent to about
mode-3 in Fig. \ref{Natre_vert}).  Analysis of a subset of these data
obtained on a grid with 1 nautical mile spacing returns a consistent signature:
$E_k/E_p$ ratios of 2-3 are found at the lowest two resolved vertical
wavelengths.  Second, the spectra diverge so that $E_{k}$ is about an
order of magnitude larger than $E_{p}$ at 100 m vertical wavelengths.
Third, $E_{p}$ does not roll-off as quickly at high wavenumber, as can
be inferred from increasing ratios between $E_{p}$ and $E_{k}$ at
vertical wavelengths smaller than 10 m.  \cite{P03} argue that these
increasing ratios at small scales are largely associated with an
increasing contribution of quasi-permanent finestructure (Section \ref{SlowOsc}).  
When the quasi-permanent contribution is subtracted from the
observations, both $E_{k}$ and $E_{p}$ roll-off at about the same
rate, $E_{k} \propto E_{p} \propto m^{-2.75}$.  The low wavenumber
excess of $E_{p}$ remains.  The vertical wavenumber spectrum indicates
a non-separable spectrum, with excess near-inertial content at high
wavenumber.

\begin{figure}\vspace{-0.50cm}
\noindent\includegraphics[width=20pc]{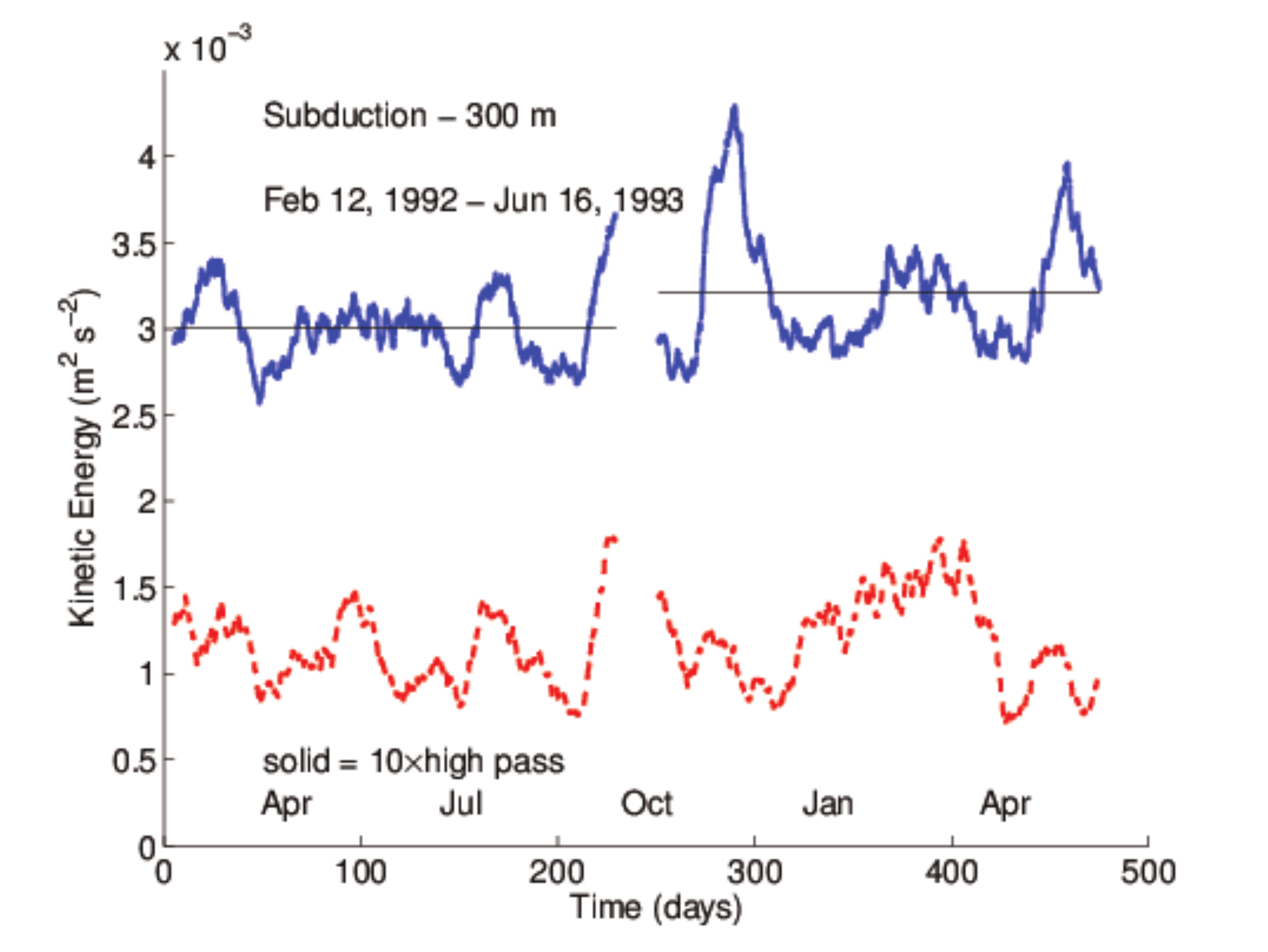}\vspace{-0.75cm}
\caption{Natre time series of
high-frequency and internal wave-band energy from nominal 300 m water
depth.  The solid line represents 10 times the high frequency energy.  The dashed line represents 
the entire internal wave band energy estimate.  \label{Natre_time} }
\end{figure}

\begin{figure}\vspace{-0.50cm}
\noindent\includegraphics[width=20pc]{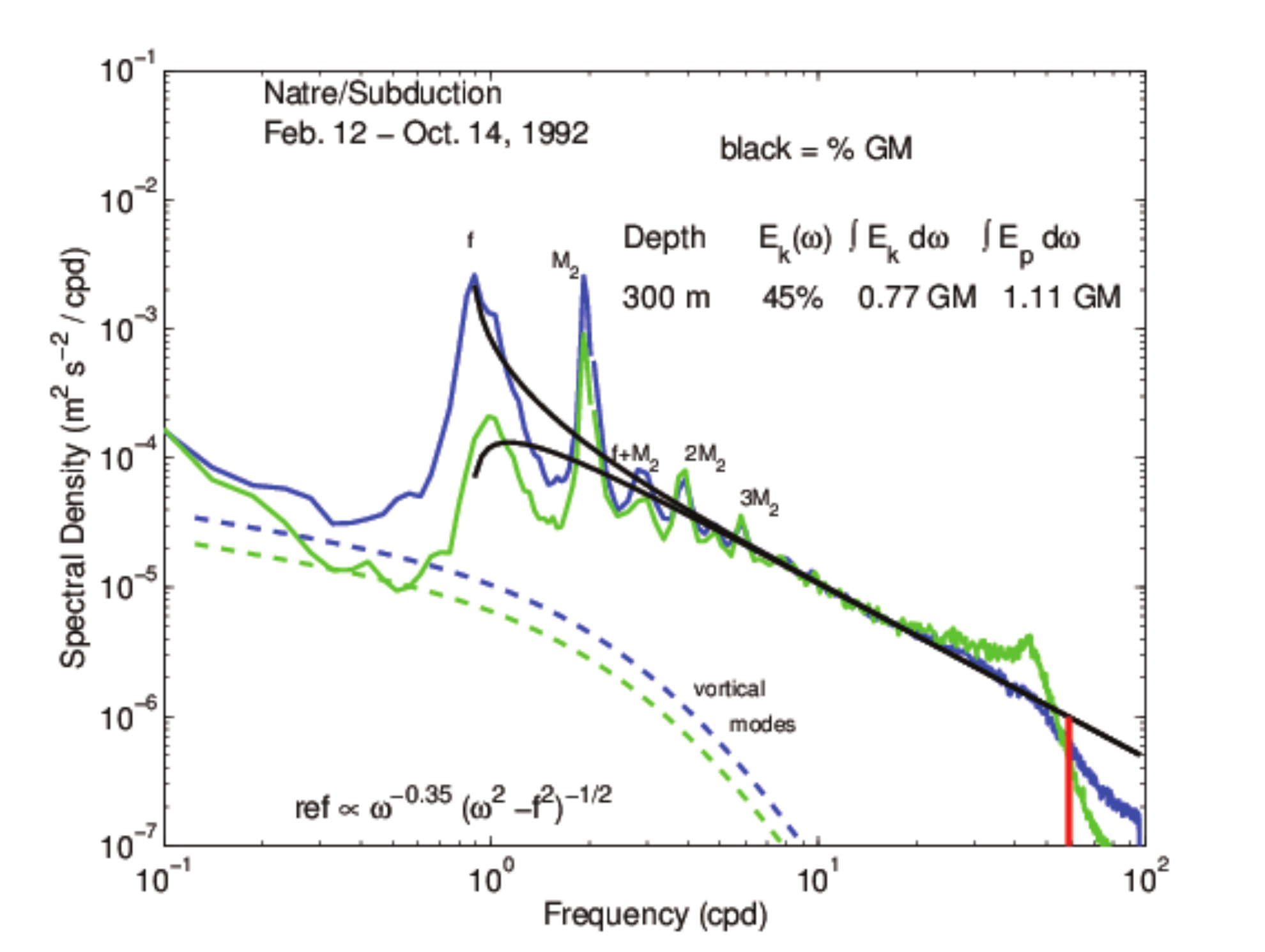}\vspace{-0.75cm}
\caption{Natre frequency spectra (thin
lines) of horizontal kinetic energy and potential energy from the main
thermocline (300 m).  Observed spectra (solid lines) with vortical
mode spectra (dashed) superimposed.  The non-propagating vortical
fluctuations are assumed to be passively advected by the mesoscale
field in this representation.  See \cite{P03}.  Inertial, semi-diurnal
and harmonic peaks are noted.  Black curves represent fits of
(\ref{shape}) with $r=1.50$.  The thick vertical line represents the
buoyancy frequency cut-off.  \label{Natre_freq} }
\end{figure}

The vertical profile data were obtained from the vicinity of the
center mooring of the Subduction array, \cite{Weller04}.  VMCMs
deployed on surface moorings document obvious inertial and tidal
peaks, and peaks at several harmonics (Fig. \ref{Natre_freq}).  The
frequency spectrum is not succinctly characterized in terms of a
single power law.  At low frequencies ($1 < \sigma < 6 $ cpd) the
spectrum exhibits a plateau (ignoring the tidal peak and
harmonics) and at higher frequencies ($6 < \sigma < 48$ cpd) rolls off as 
$\sigma^{-1.35}$.  Energy ratios [$E_k(\sigma)/E_p(\sigma)$] agree with
linear kinematics (\ref{Ek_on_Ep}), Internal wave band ratios [$\int
E_k(\sigma) d\sigma / \int E_p(\sigma)d\sigma$] are between 2-3,
consistent with a red vertical wavenumber spectrum having little
near-inertial content at low wavenumber.  Unlike the Sargasso Sea, no
seasonal cycle is apparent in either internal wave energy time series
(Fig. \ref{Natre_time}).

NATRE is special among these data sets in that it exhibits a
strikingly large $M_2$ tide and the region is a relative minimum in
eddy energy, \cite{Wunsch97}.  An obvious hypothesis is that structure
in the spectral domains reflects the detailed pathway energy takes in
draining out of the tidal peak.  A second hypothesis is suggested by
noting that the transition between the two regimes (roughly 6 cpd)
corresponds to the stratification rate near the bottom boundary ($N \cong 
3$ cpd).  Waves of higher frequency never reach the bottom and will
not be attenuated by viscous processes in the bottom boundary layer.


\subsubsection{The southeast Subtropical North Pacific}\label{PinkelsAcres}

A significant number of data sets have been obtained
offshore southern California.  The
major geographic feature here is the California Current System
consisting of a deep poleward flow along the continental slope,
equatorward flow at the shelf break and a 600-km wide
band of mesoscale variability extending from the shelf and fading into
the northeast Pacific (e.g., \cite{Ch2000}).  There are, as well,
gradients in mesoscale variability along the coast, with the
most northern of the data sets discussed here being closest to a
regional maximum in surface kinetic energy estimated from surface
drifter data \cite{Ch2000}.  The data discussed here are more than 400
km distance from the shelf break.  The buoyancy frequency [$N(z)$]
typically attains a maximum at 100 m water depth and decreases by
about a factor of three over 300 m.  This represents a large gradient.

Data from this region include the oscillating float estimates of
isopycnal displacement providing frequency spectra and vertical lag
coherence estimates reported in \cite{Cairns75} and \cite{CW76}.
These data prompted the revision of GM75 model to the GM76 form.  See
Section \ref{GM} for further discussion.

Here we utilize moored current meter data from the Eastern Boundary 
Currents program and vertical profile data from the Patches Experiment (PATCHEX; \cite{SandP91}) 
and the Surface Wave Process Program (SWAPP; \cite{Wetal91}).  PATCHEX 
was a multi-investigator study of the space and time structure of mixing events.   
SWAPP was nominally focussed upon upper ocean processes, but data 
was obtained from thermocline regions as well.  These latter two 
field programs featured yo-yo CTD and Doppler sonar data sets
obtained from the research platform FLIP.  These simultaneous vertical profiles of
velocity and density provide a unique representation of the upper
ocean wavefield and permit 2-D vertical wavenumber-frequency spectra
to be evaluated.  This has been done in both depth and density
coordinates, with the latter being consistent with and directly
comparable to the theoretical analysis presented in the companion manuscript \citep{theory}.  The use of
this coordinate system will also likely limit contamination by
quasi-permanent density finestructure.  Quantifying this is difficult,
as is quantifying the effects of Doppler shifting by sub-inertial
currents.  Results from the density data are used exclusively here as
the velocity data are subject to contamination from instrumental noise
and beam separation effects at high wavenumber and frequency, Appendix \ref{DopplerSonars}.   
A further cautionary note is that sampling during PATCHEX 
may not have been of sufficient temporal or spatial extent for the observations  
to reflect the background wavefield.  \newline

\noindent{\bf{ Eastern Boundary Currents} - $\sigma^{-2}$}

\begin{figure}\vspace{-0.50cm}
\noindent\includegraphics[width=20pc]{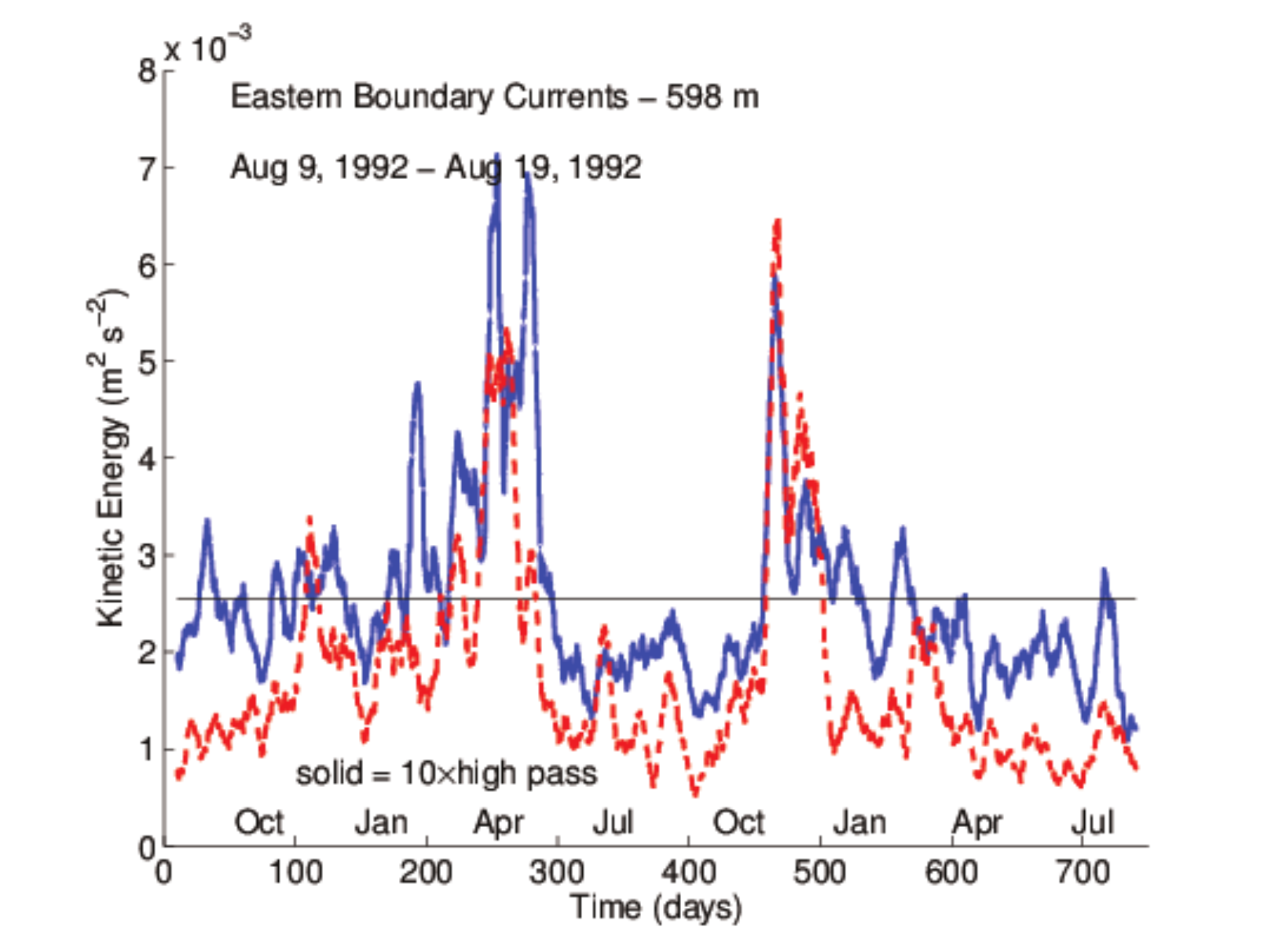}\vspace{-0.75cm}
\caption{Eastern Boundary Currents
time series of high-frequency and internal wave-band energy from 
nominal 600 m water depth.  The solid line represents 10 times the high frequency energy.  The dashed line represents 
the entire internal wave band energy estimate.  Peaks in high frequency and wave-band 
energy during April and November of 1993 coincide with the presence 
of anti-cyclonic eddies, \cite{Ch2000}.  \label{EBC-time} } 
\end{figure}
\begin{figure}\vspace{-0.50cm}
\noindent\includegraphics[width=20pc]{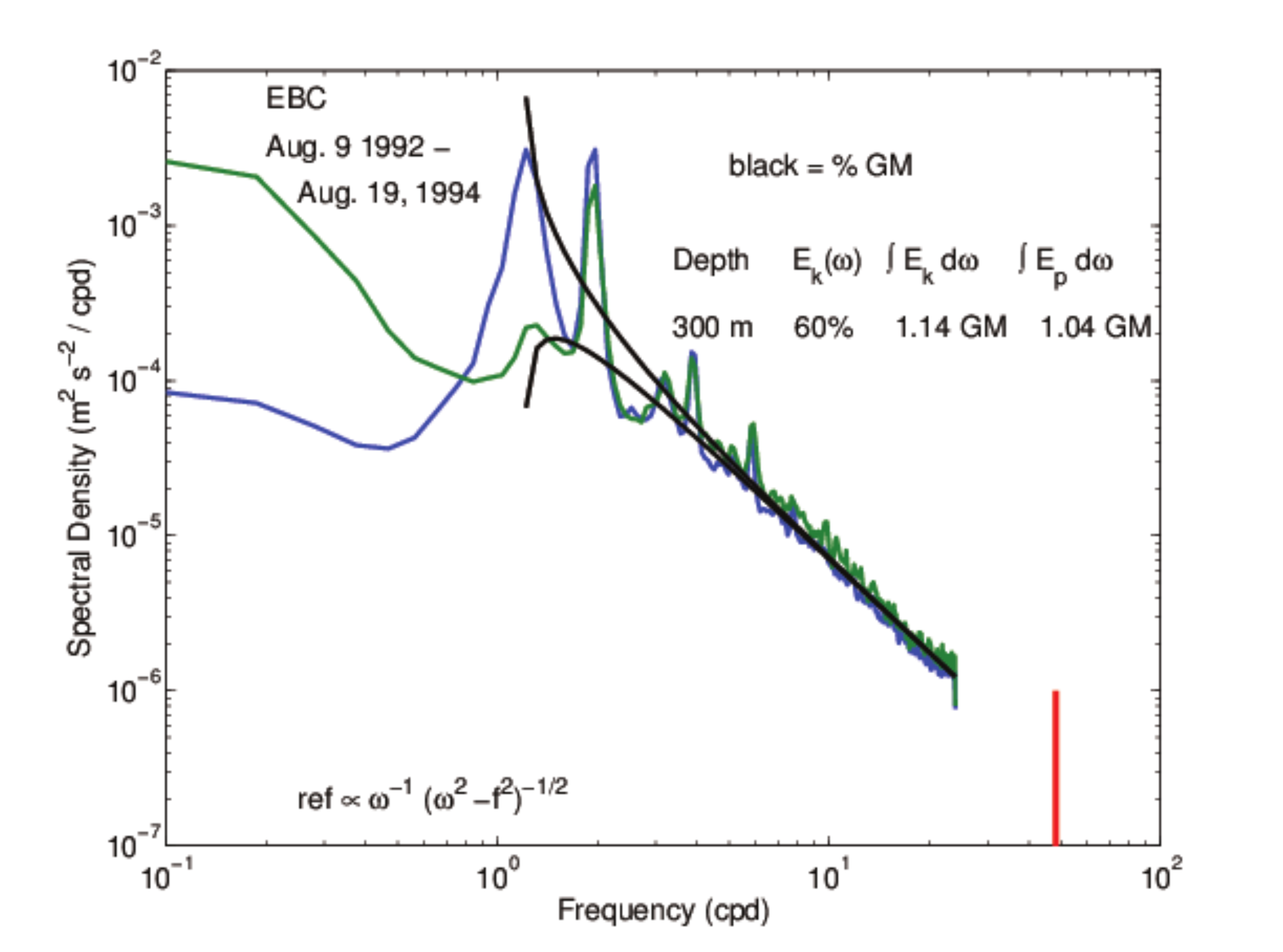}\vspace{-0.75cm}
\caption{Eastern Boundary Currents
frequency spectra (blue and green lines) of horizontal kinetic energy and
potential energy from the main thermocline (600 m).  Black curves
represent fits of (\ref{shape}) with $r=2.00$.  The thick vertical
line represents the buoyancy frequency cut-off.  \label{EBC-freq} } 
\end{figure}\noindent
A coherent moored current meter array was deployed for a two year
period (1992-1994) at the offshore edge of the California Current as
part of the Eastern Boundary Currents field program \citep{Ch2000}.
These data represent the only available long-term current meter
records in the vicinity of the other data from the northeast Pacific
discussed herein.  As those data are from short term (several week)
field programs, it is a natural question to ask whether those data are
representative of the long-term average (Fig. \ref{EBC-freq}).  The
data (both short and long term) share a common frequency dependence of
$\sigma^{-2}$ at high frequency.  Variability of both high-passed and
wave-band energy levels is associated with eddy variability rather
than a seasonal cycle (Fig. \ref{EBC-time}).\newline

\noindent{{\bf PATCHEX$^1$} - $m^{-1.9}$ }

\begin{figure}[h]\vspace{-0.75cm}
\noindent\includegraphics[width=20pc]{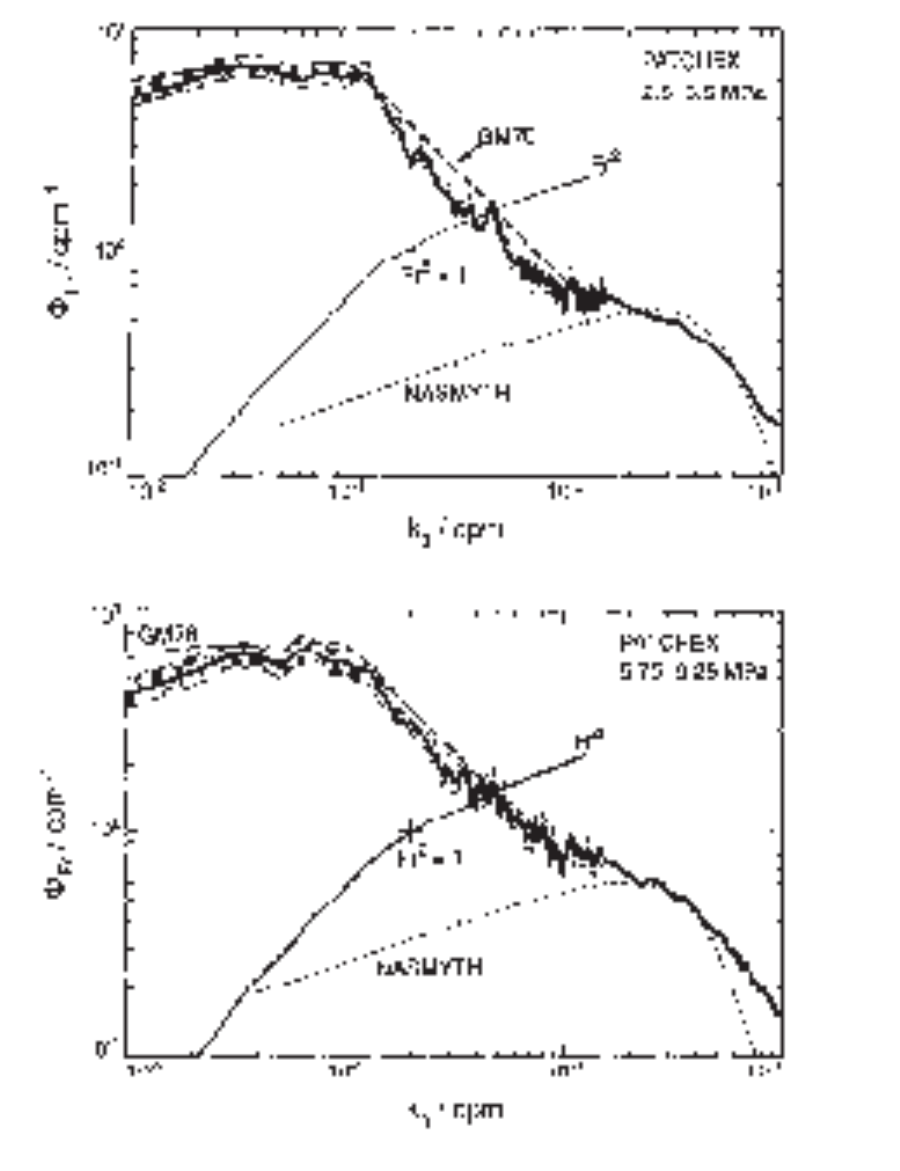}\vspace{-0.75cm}
\caption{Patchex vertical wavenumber
shear spectra using the free-fall profiler MSP.  The dashed line
represents $m^2/(m_*^2 + m^2)$, with $j_*=3$ and an $m^{-1}$
dependence at vertical wavelengths smaller than 0.1 cpm.  The domain
of interest here is $0.01 < m < 0.1$ cpm.  (Figure extracted from
\cite{G93}, their Fig. 6, permission from {\em J. Phys. Oceanogr.,}
Copyright 1993 American Meteorlogical Society). \label{Patchex_gregg}  } 
\end{figure}

Data from a freely-falling vertical profiler MSP were obtained during PATCHEX.  
\cite{GS88} note that the velocity profile data are remarkable only for 
the absence of coherent near-inertial wave packets.  \cite{G93} document 
a shear spectrum that varies from white at pressures of 250-550 db to 
slightly blue (575-925 db) over vertical wavelengths of 10-100 m (Fig. \ref{Patchex_gregg}).\newline

\noindent{{\bf PATCHEX$^2$} - $m^{-1.75}$ {\rm and} $\sigma^{-1.65} ~-~\sigma^{-2.0}$.}
During PATCHEX two CTDs were deployed from the R/P FLIP 
to obtain density profiles over 0-560
m water depth, with a cycle time of 180 s.  The density data
considered here come from the depth interval 150-406 m.  No
information regarding vertical symmetry or mean currents is available.
 
\begin{figure}\vspace{-0.50cm}
\noindent\includegraphics[width=20pc]{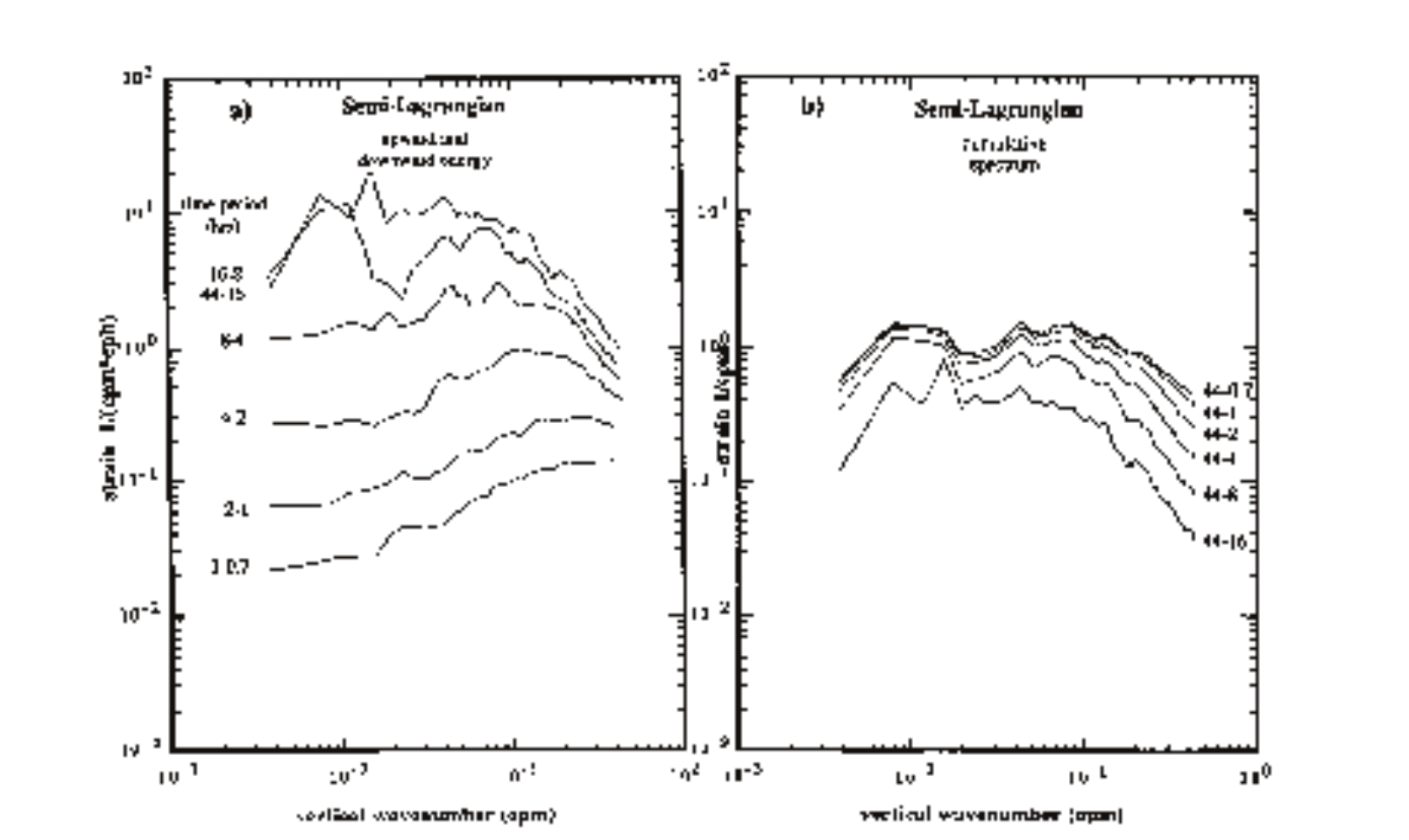}\vspace{-0.50cm}
\caption{Patchex 2-D vertical
wavenumber - frequency spectra of isopycnal displacement gradient.
The density data from which displacement is estimated were obtained
with a rapid-profiling CTD from the research platform {\em FLIP}.  The
quoted power law dependencies have been determined graphically.
(Figure extracted from \cite{SandP91}, their Fig. 6, permission from
{\em J. Phys. Oceanogr.,} Copyright 1991 American Meteorological
Society.) \label{Patchex_pinkel} } 
\end{figure}

For vertical wavenumber $1/256 < m < 1/10$ cpm, the density gradient
spectrum rolls off as $m^{-1.75}$ over $0.125 < \sigma < 1.4$ cph 
(Fig. \ref{Patchex_pinkel}) This range of wavenumbers and frequencies is
characterized by a frequency dependence of
$\sigma^{-1.65}-\sigma^{-2}$.  The cumulative spectrum (integrated
over all frequencies $f \leq \sigma \leq N$) is slightly blue for
$0.02 < m < 0.1$ cpm.  These power law relations were determined graphically with use of a straight edge.\newline

\noindent{{\bf SWAPP} - $m^{-1.9}$ {\rm and} $\sigma^{-2.0}$}

During SWAPP a pair of CTDs deployed from FLIP 
returned density profiles from ~5 to 420 m with a vertical resolution
of about 1.5 m and cycle time of 130 s for a duration of 12 days
during March, 1990.  The data cited here come from depths of 50-306 m.

\begin{figure}\vspace{-0.50cm}
\noindent\includegraphics[width=20pc]{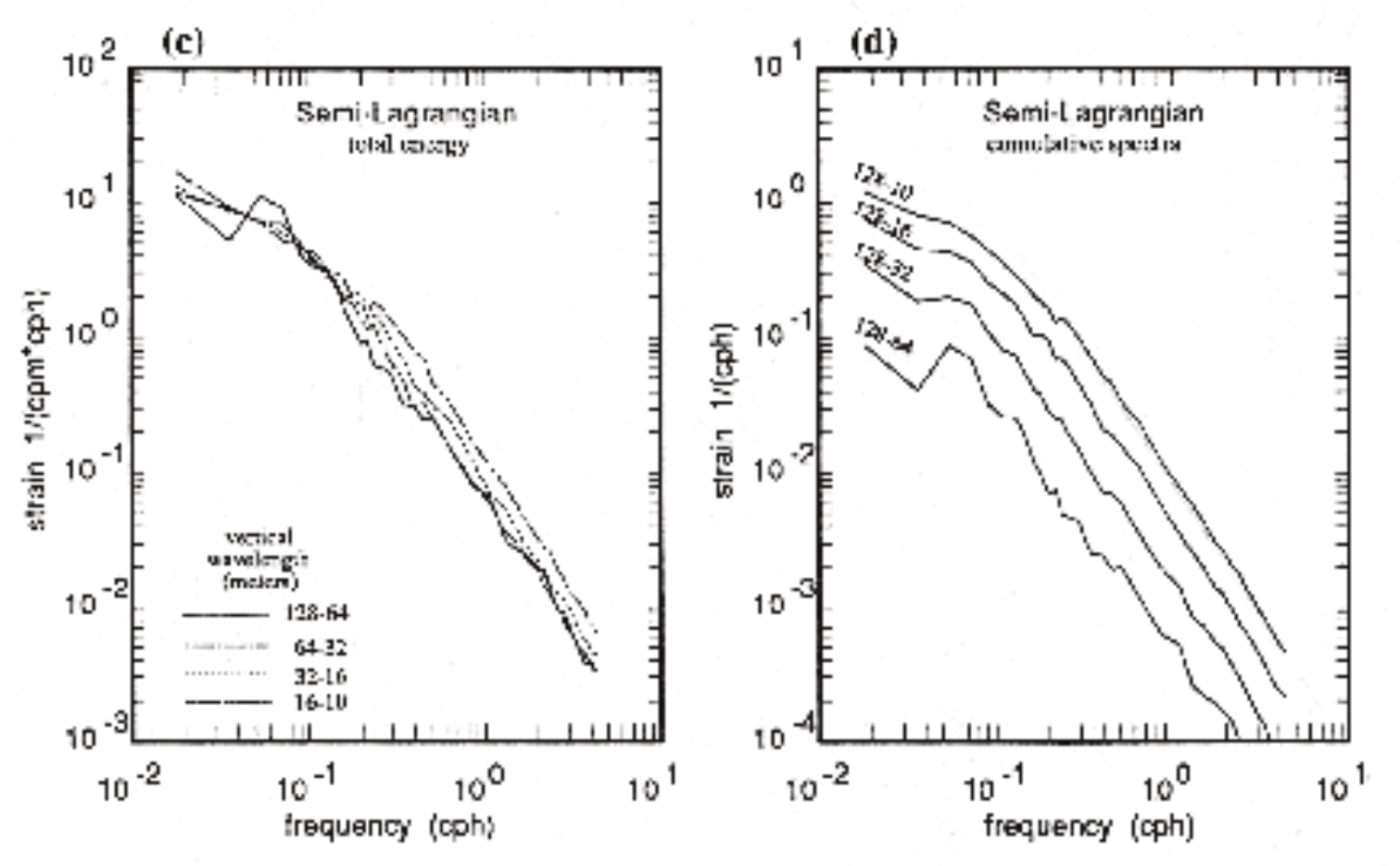}\vspace{-0.75cm}
\caption{Swapp frequency spectra of
isopycnal displacement gradients.  The density data from which
displacements are estimated were obtained with a rapid-profiling CTD
from the research platform {\em FLIP}.  The quoted power law
dependencies have been determined graphically.  (Figure extracted from
\cite{A92}, their Fig. 2.9, permission from S. Anderson.  Copyright
1992.) \label{Swapp-freq} } 
\end{figure}

\begin{figure}\vspace{-0.50cm}
\noindent\includegraphics[width=20pc]{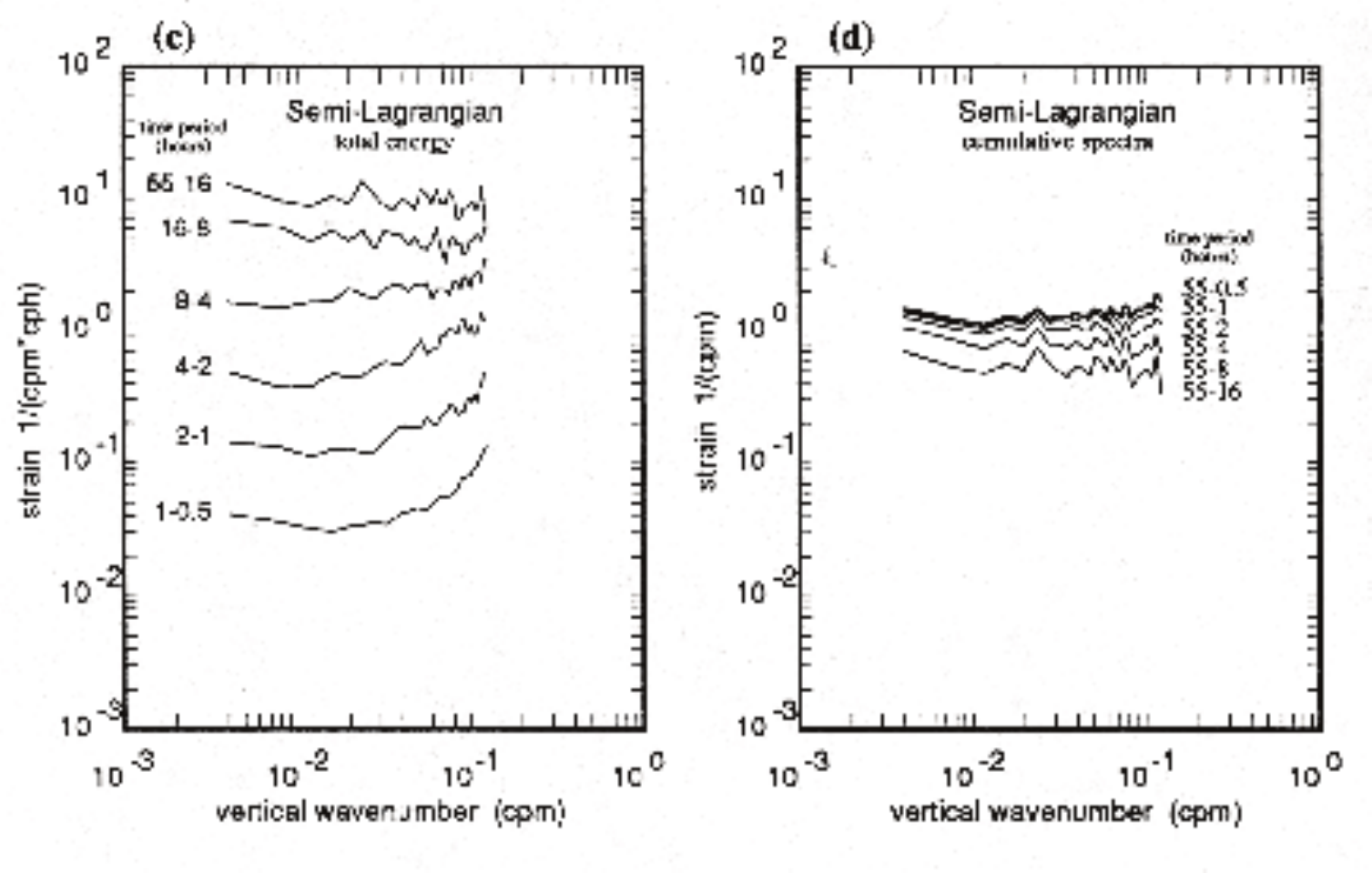}\vspace{-0.75cm}
\caption{Swapp vertical wavenumber
spectra of isopycnal displacement gradients.  The density data from
which displacement is estimated were obtained with a rapid-profiling
CTD from the research platform {\em FLIP}.  The quoted power law
dependencies have been determined graphically.  (Figure extracted from
\cite{A92}, their Fig. 2.7, permission from S. Anderson.  Copyright 1992.) \label{Swapp-vert}  } 
\end{figure}

For vertical wavenumbers $1/128 < m < 1/10$ cpm, the density gradient
spectrum rolls off as $\sigma^{-2}$ over $0.2 < \sigma < 4$ cph 
(Fig. \ref{Swapp-freq}).  For frequencies of $0.02 < \sigma < 2$ cph,
the vertical wavenumber spectrum is nearly white, $m^{0.1}$, so that
the energy spectrum rolls off as $m^{-1.9}$ (Fig. \ref{Swapp-vert}).
As with the Patchex data from FLIP, these power law relations have
simply been determined using a $\chi$-by-eye procedure.

The buoyancy frequencv [$N(z)$] attains a maximum at 100 m water depth
and decreases by about a factor of three over 300 m.  At periods
greater than 8 hours, the shear spectra indicate a dominance of
downward energy propagation, \cite{A92}.  Thus the vertical spectrum
may be influenced by purely linear kinematics, rather than
nonlinearity.  The shear spectrum is not easily characterized.  It
exhibits a peak at about 0.03 cpm associated with downward propagating
near-inertial waves, \cite{A92}.


\subsubsection{The northeast Subpolar North Pacific}

\noindent{{\bf MATE} - $m^{-2.1}$ {\rm and} $\sigma^{-1.7}$}

An extensive set of moored and profiling (both vertical and horizontal) observations was taken during June--July of 1977 as part of the Midocean Acoustic Transmission Experiment (MATE), \cite{L86}.  The experiment was located at $46^{\circ} 46^{\prime}$N, $130^{\circ} 47^{\prime}$W, midway between Cobb seamount, which rises to within 30 m of the surface, and the smaller Corn seamount which rises to depths of 1000 m.  The seamounts are separated by approximately 18 km and the local water depth is 2200 m.  The seamounts may serve as significant generators of an internal tide.  Cobb seamount has been noted to support a trapped diurnal tide \citep{CE97, LM97}.  The MATE observations focused upon subthermocline depths of 900-1300 m away from the seamounts.  

\begin{figure}\vspace{-0.50cm}
\noindent\includegraphics[width=20pc]{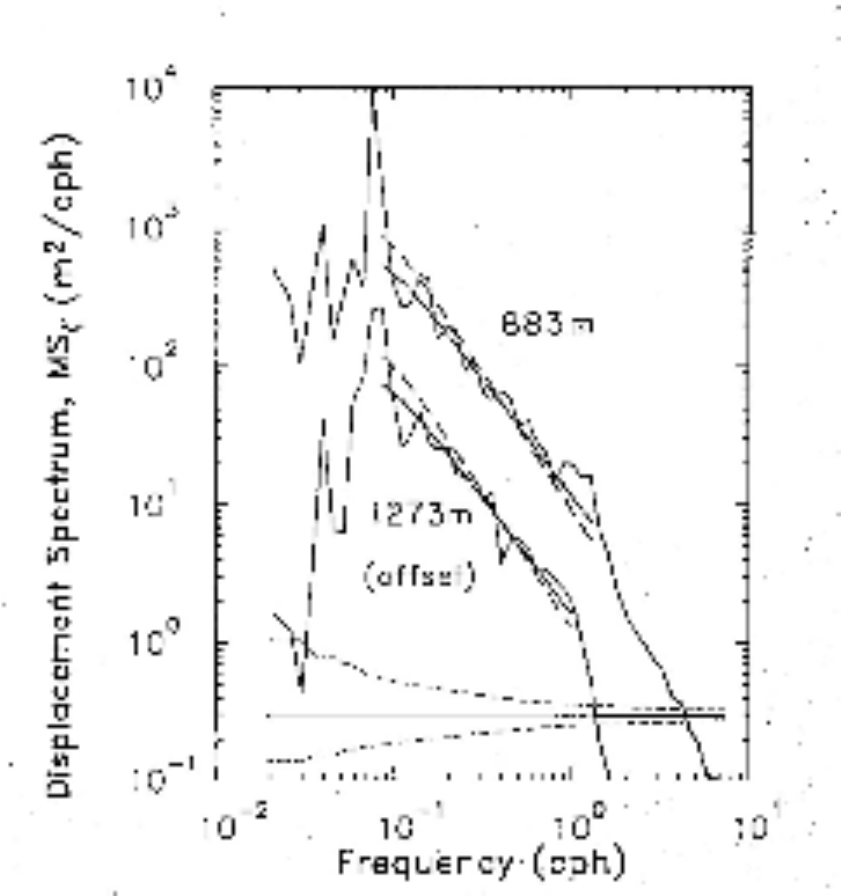}\vspace{-0.75cm}
\caption{MATE frequency spectra of vertical displacement.  Solid and dashed curves are the MATE and GM models, respectively.  (Figure extracted from \cite{L86}, their Fig. 5.  Permission from \jgr, copyright 1986 American Geophysical Union. )
\label{Mate_freq_h} }
\end{figure}

\begin{figure}\vspace{-0.50cm}
\noindent\includegraphics[width=20pc]{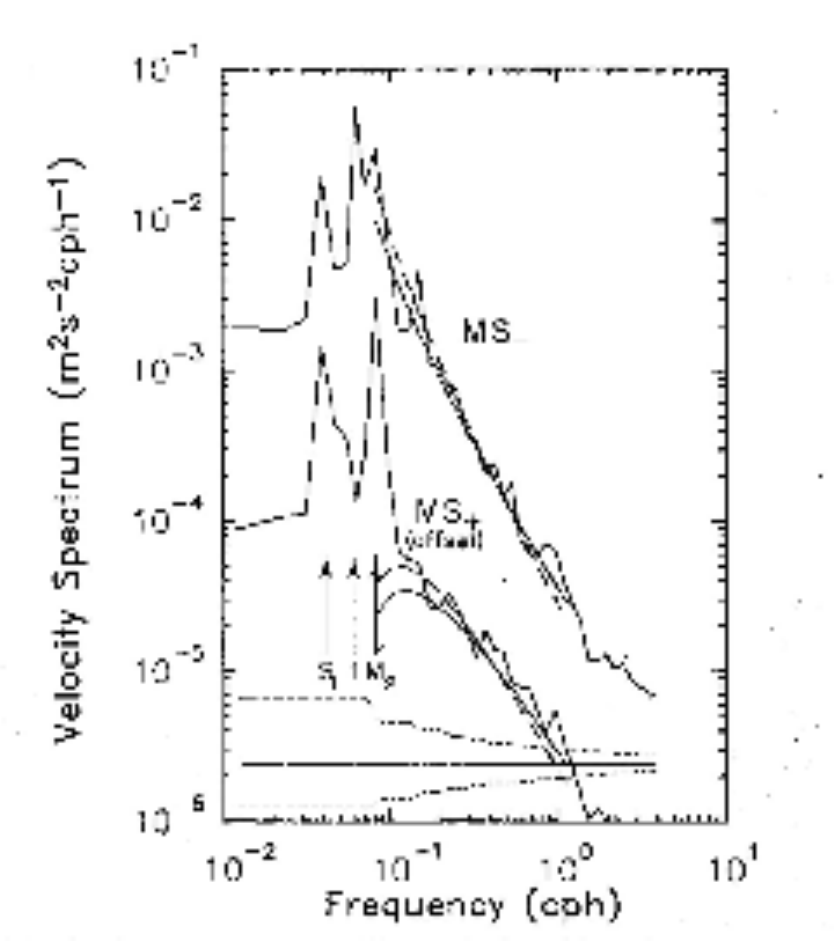}\vspace{-0.75cm}
\caption{MATE frequency spectra of horizontal kinetic energy.   Solid and dashed curves are the MATE and GM models, respectively.  (Figure extracted from \cite{L86}, their Fig. 8.  Permission from \jgr, copyright 1986 American Geophysical Union. )
\label{Mate_freq_uv} }
\end{figure}

\begin{figure}\vspace{-0.50cm}
\noindent\includegraphics[width=20pc]{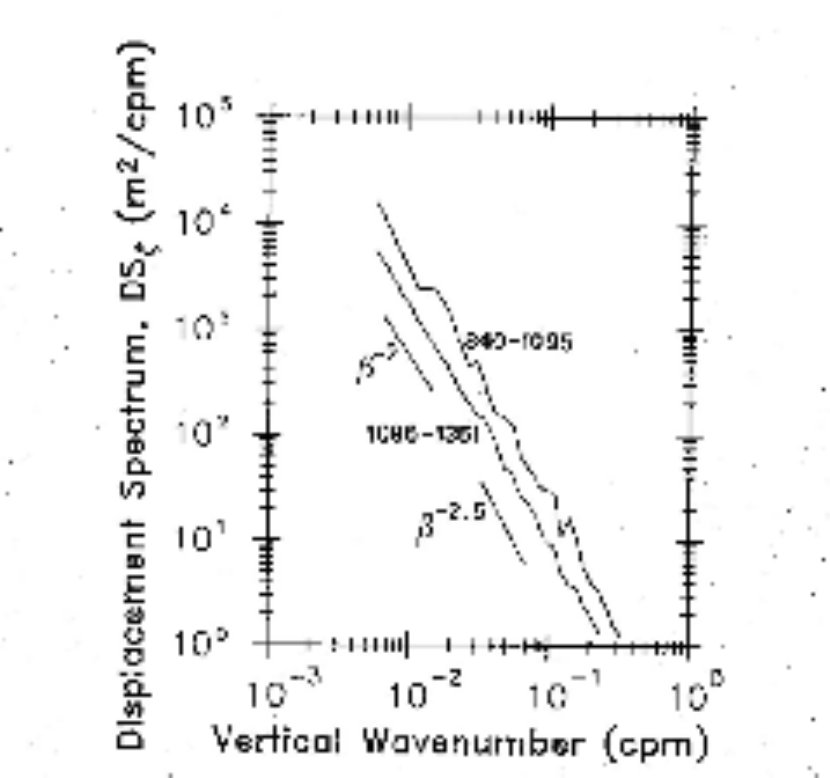}\vspace{-0.75cm}
\caption{MATE vertical wavenumber spectra of 
vertical displacement.   (Figure extracted from \cite{L86}, their Fig. 4.  Permission from \jgr, copyright 1986 American Geophysical Union. )  \label{Mate_vert}  } 
\end{figure}

\begin{figure}\vspace{-0.50cm}
\noindent\includegraphics[width=20pc]{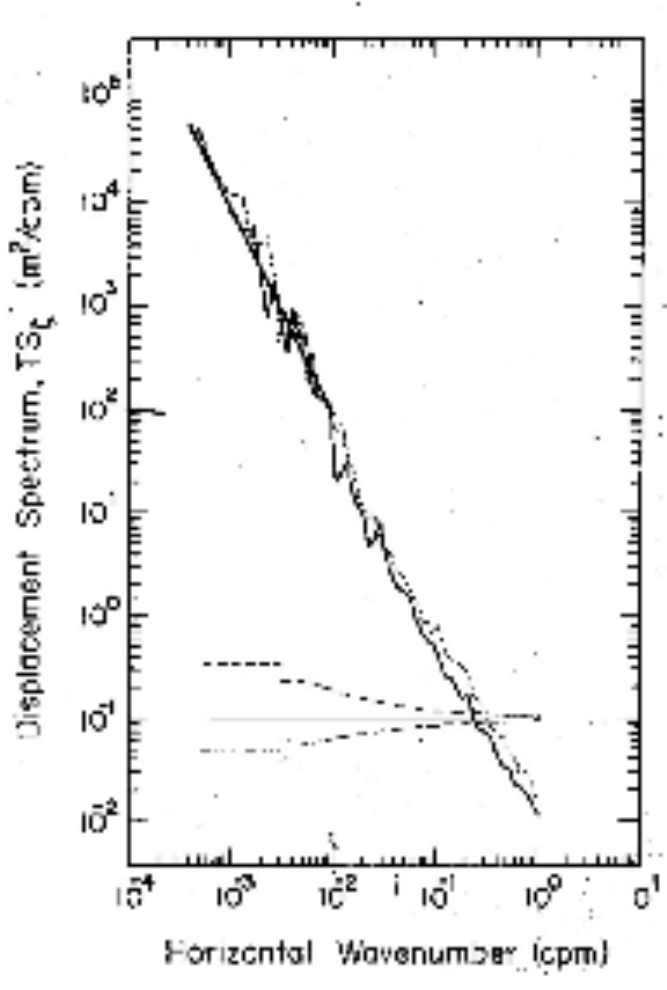}\vspace{-0.75cm}
\caption{MATE horizontal wavenumber spectra of vertical 
displacement estimated from SPRUV.   Solid and dashed curves represent are north-south and west-east runs, respectively.  (Figure extracted from \cite{L86}, their Fig. 9.  Permission from \jgr, copyright 1986 American Geophysical Union. )  \label{Mate_horz}  } 
\end{figure}

Current meter data document frequency domain power laws of $\sigma^{-1.7}$ (Fig.s \ref{Mate_freq_h} and \ref{Mate_freq_uv}).  Temperature profiles from a Bissett-Berman 9040-5 CTDSV were used along with a locally tight $\theta-S$ relation to estimate vertical displacement.  Vertical wavenumber domain power laws of $m^{-2.1}$ are quoted for $6\times10^{-3} < m < 6\times10^{-2}$ cpm (Fig. \ref{Mate_vert}).  
Horizontal wavenumber spectra of vertical displacement (again estimated from temperature) obtained from the Self-Propelled Underwater Research Vehicle (SPURV) roll off as $\mid k \mid^{-2}$ for horizontal wavenumbers $4.7\times10^{-4} < \mid k \mid < 9.0\times10^{-3}$ cpm (Fig. \ref{Mate_horz}).  A combined analysis of spatial lag coherences returns a roll off of $j_{\ast} = 6$.  


\noindent{{\bf Strex and Ocean Storms} - $m^{-2.3}$ {\rm and} $\sigma^{-2.2}$}

The northeast Subpolar North Pacific has been the subject of a number
of experiments focussing on near-inertial internal wave generation and
mixed layer processes.  The experiments [vertical profile data from the Storm Transfer
and Response Experiment (STREX, \cite{DA84}) taken during November of 1980 at 
141$^{\circ}$W, 50$^{\circ}$ N and current meter data from the Ocean Storms
Experiment \cite{DA95a} at 139.25$^{\circ}$W, 47.5$^{\circ}$ N 
are examined here] are located just south of
the climatological storm track in order to emphasize the coupling
between inertial motions in the mixed layer and an inertial time scale
anticyclonic rotation of the windstress vector associated with small
scale atmospheric disturbances.  The region is also one of relatively
low eddy energy and so minimizes the confounding effect of
mesoscale eddy -- internal wave interactions.

\begin{figure}\vspace{-0.50cm}
\noindent\includegraphics[width=20pc]{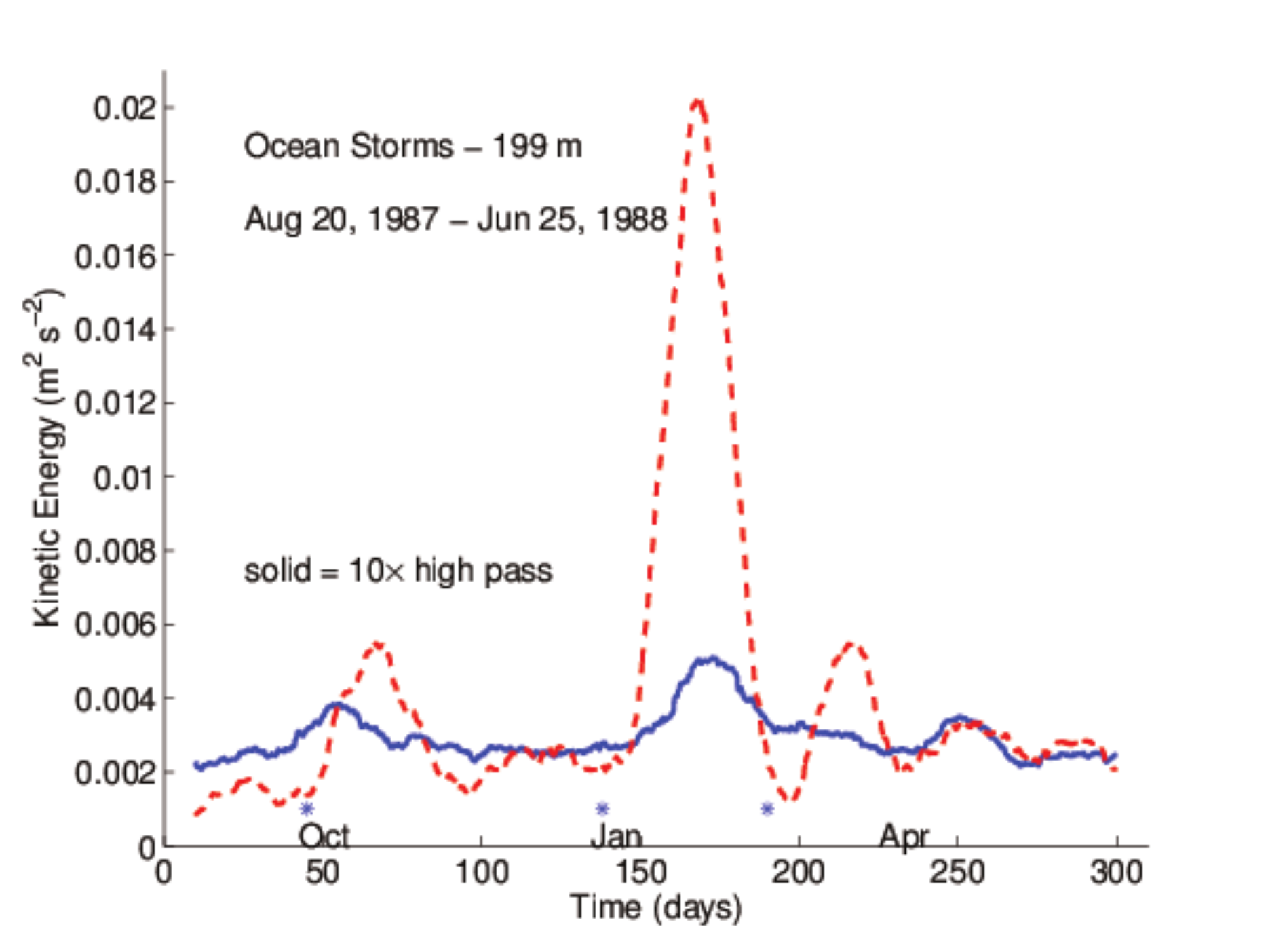}
\noindent\includegraphics[width=20pc]{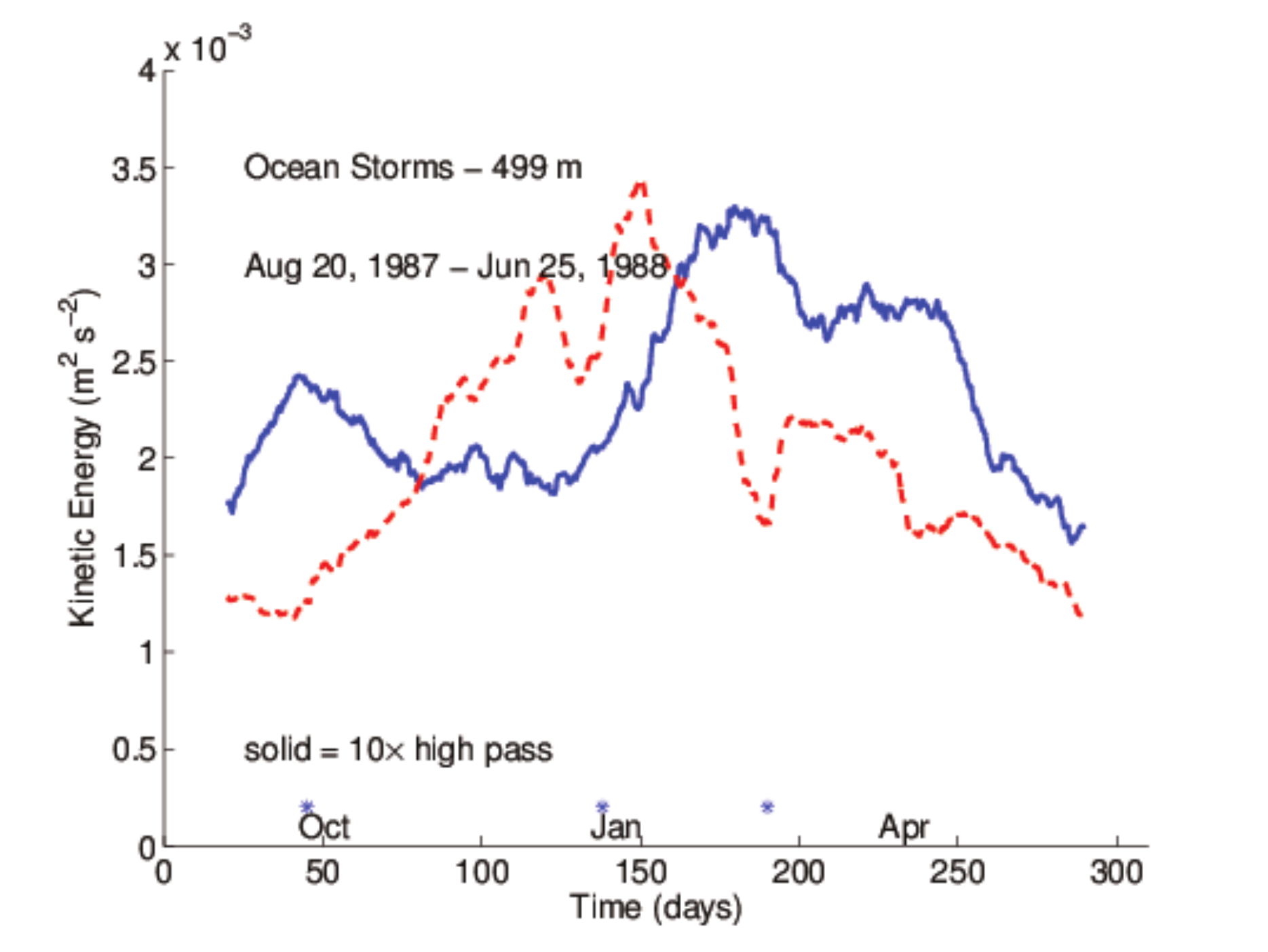}\vspace{-0.75cm}
\caption{Ocean Storms time series of
high-frequency and internal wave-band energy from nominal depths of 200 (upper panel) 
and 500 m (lower panel) water
depth.  The solid line represents 10 times the high frequency energy.  The dashed line represents 
the entire internal wave band energy estimate.  The passage of storms resulting in strong forcing of mixed layer inertial 
currents is denoted by asterisks.  \label{Storms_time} }
\end{figure}

Current meter data from 200 m water depth from the Ocean Storms Experiment clearly documents
the input of near-inertial internal wave energy associated with the
passage of individual storms (Fig. \ref{Storms_time}).  The presence of a seasonal cycle in inertial energy, though, 
is obvious only at greater depths (500 m).  In a broad survey of historical current meters, \cite{AW07} find that the seasonal cycle in 
near-inertial energy is most evident at latitudes of $25-45^{\circ}$, with decreasing seasonal modulations north of $45^{\circ}$.  
A possible interpretation is a regional emphasis on propagation.  

High frequency energy at 500 m also exhibits a seasonal cycle.  Unlike seasonal cycles in the western North Atlantic, here the high frequency energy appears to lag near-inertial energy by several 10's of days.  A possible interpretation is that the presence of such a phase lag is related to the absence of an energetic eddy field.  

The frequency spectrum fit indicates a spectrum redder than
GM.  This is paired with a vertical wavenumber spectrum of horizontal
velocity also steeper than GM (Fig. \ref{Strex_vert}).  This represents
the sole example of power laws that do not covary.

\begin{figure}\vspace{-0.50cm}
\noindent\includegraphics[width=20pc]{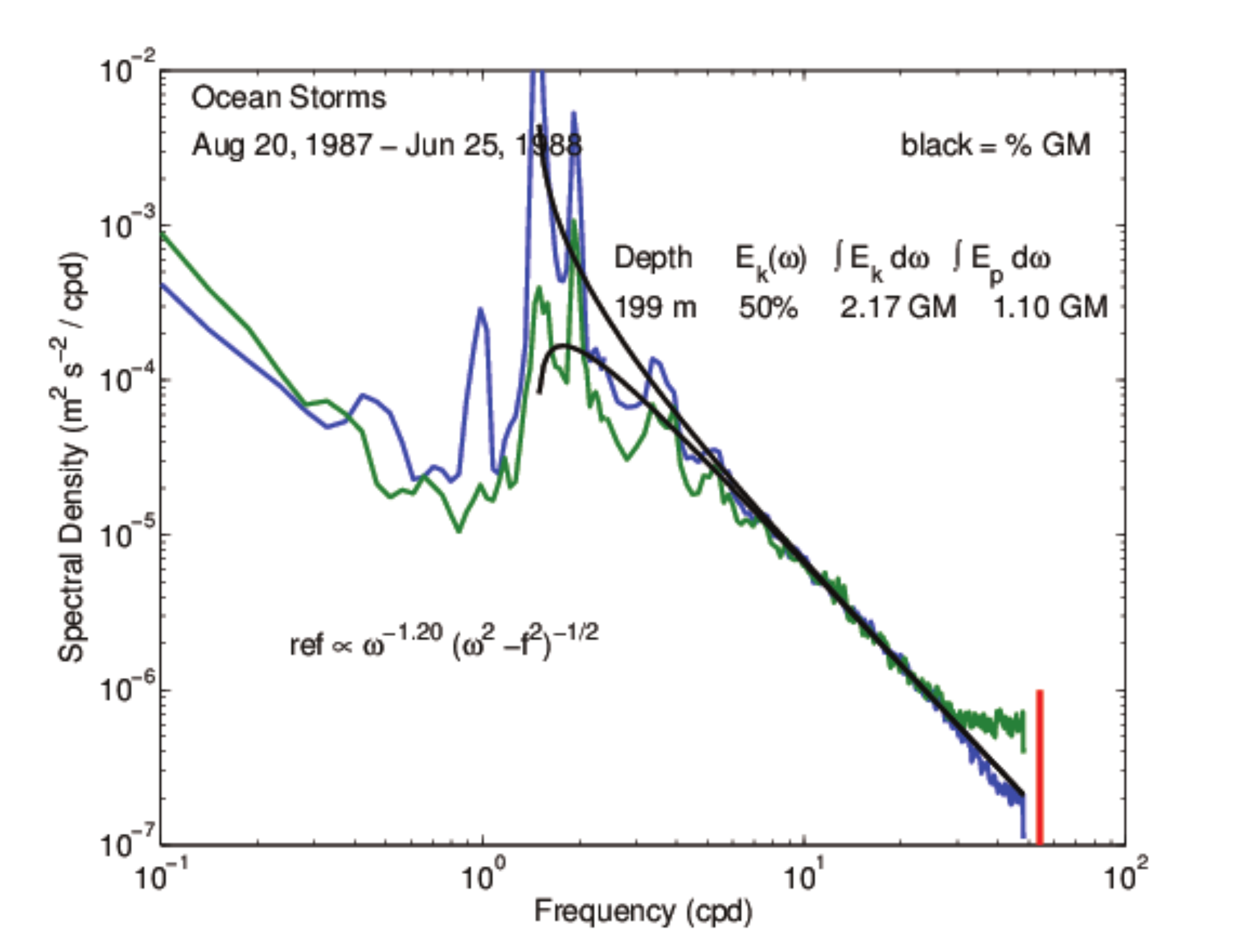}\vspace{-0.75cm}
\caption{Ocean Storms frequency spectra (blue and green lines) of horizontal kinetic energy and potential energy from the main thermocline (200 m).  Black curves represent fits of
(\ref{shape}) with $r=2.20$.  The thick vertical line represents the
buoyancy frequency cut-off.  \label{Storms_freq} }
\end{figure}
\begin{figure}\vspace{-0.50cm}
\noindent\includegraphics[width=20pc]{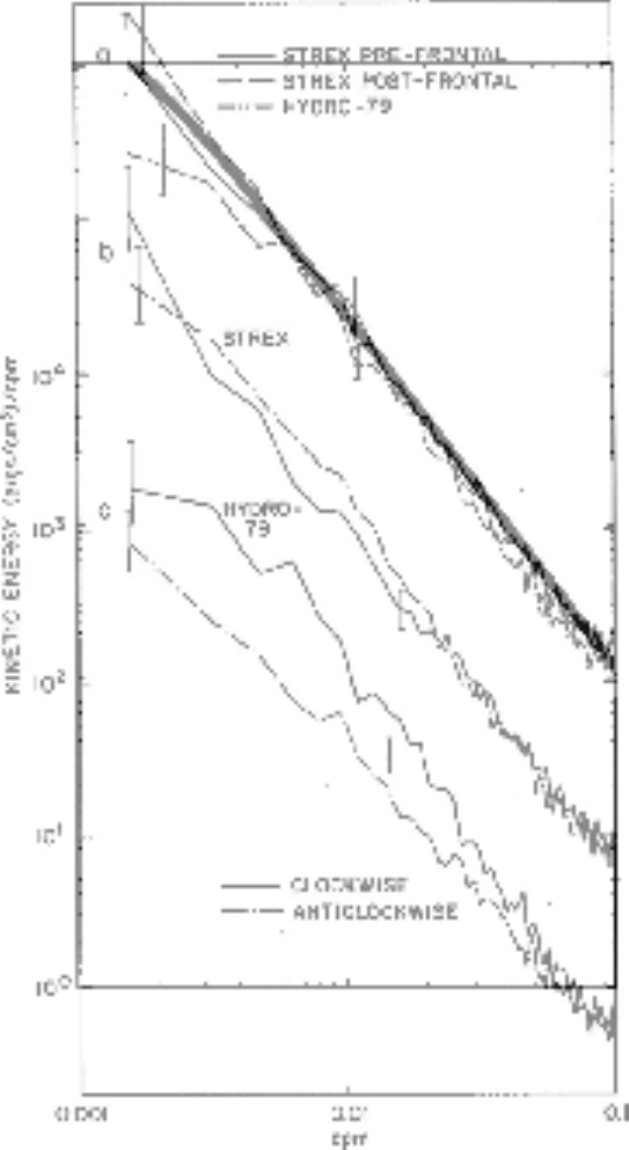}\vspace{-0.75cm}
\caption{Strex vertical wavenumber
kinetic $E_k$.  The overlay represents a fit of
$1/(m_*^2 + m^2)^{1.15}$ to the pre-frontal spectra, with $m_*=0.0070$ cpm 
(equivalent to $j_{\ast}=3$). Figure from \cite{DA84}, their Fig. 4.  Permission from \jpo, copyright American Meteorological Society 1986.   \label{Strex_vert}.  
} 
\end{figure}

\subsubsection{The Arctic}

\noindent{{\bf AIWEX} - $m^{-2.25}$ {\rm and} $\sigma^{-1.2}$ }

The Arctic Internal Wave Experiment (AIWEX) was conceived as an
attempt to study an anomalous internal wavefield.  The experiment took
place about 350 km north of Prudhoe Bay, Alaska, with sampling
extending from March to May of 1985.  Total energy levels were a
factor of 0.02 to 0.07 times smaller than in the GM model spectrum.
Possible reasons for the low energy levels are weaker winds and wind
stress, the presence of ice cover, a small barotropic tide, and a weak
circulation on basin scales, \cite{L87}.  Peaks at the $M_{2}$ tidal
frequency and its harmonics are not apparent.  There was no
discernible kinetic energy dissipation rate associated with the
wavefield, \cite{PD87}, consistent with low levels of energy and
shear.

The frequency spectra for both temperature and velocity were much
whiter than typical \cite{L87}.  For frequencies much larger than $f$,
the power law $\sigma^{-r}$ tends to $r=1.2$, \cite{L90}.

Velocity profiles were obtained with Expendable Current Profilers
(XCPs).  These data suggest a vertical wavenumber bandwidth about ten
times larger than in the GM model, \cite{DAM91}.

Several notes of caution are in order.  First, the data were not taken
as a random sample relative to the eddy field.  The Canadian Basin is
populated by relatively intense submesoscale vortices, and the AIWEX
XCP survey was intentionally taken between such features.  It is not
clear whether this leaves the data set prone to straining effects as
are evident in the MODE data set.  Secondly, \cite{DAM91} argue for
the presence of quasi-permanent velocity finestructure at 40 m
vertical wavelength with relatively small levels of quasi-permanent
density finestructure.  Their diagnostics, though, depend upon being
able to account for all noise in the XCP measurements.  Their noise
budget does not include noise induced by fall rate variations of the
XCP \citep{Sanford93}, which shows up as noise in the north-south
velocity component at 15-30 m vertical wavelength.

Finally, the spectral levels are so small that one might call into
question whether nonlinearity is important in shaping the spectrum.
Boundary dissipation has been cited as a dominant decay process rather
than interior wave breaking, \cite{Pinkel05}.
\begin{figure}\vspace{-0.50cm}
\noindent\includegraphics[width=20pc]{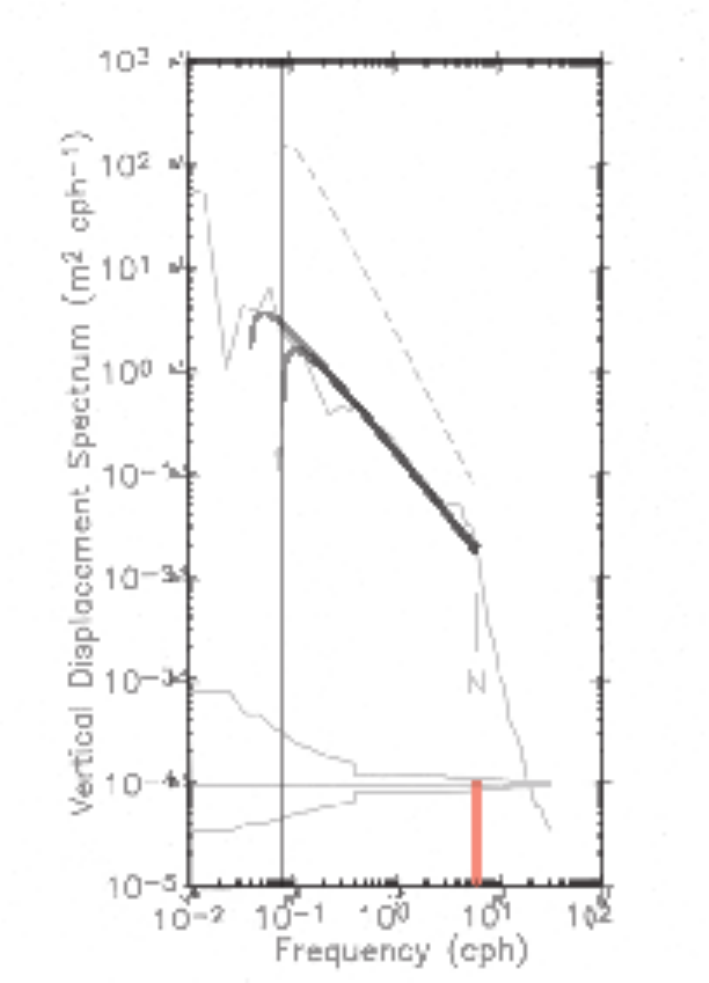}\vspace{-0.75cm}
\caption{AIwex frequency spectra of
isopycnal displacement gradients.  The dashed line represents the GM
frequency spectrum, $E(\sigma) \propto \sigma^{-1}(\sigma^2
-f^2)^{-1/2}$.  The thick line represents $E(\sigma) \propto \sigma^{-0.2}(\sigma^2
-f^2)^{-1/2}$ with with two values for $f$.  
(Figure extracted from \cite{L87}, their Fig. 2,
permission from {\em J. Geophys. Res.,} Copyright 1987 American
Geophysical Union).}
\label{AIwex-freq}  
\end{figure}
\begin{figure}\vspace{-0.50cm}
\noindent\includegraphics[width=20pc]{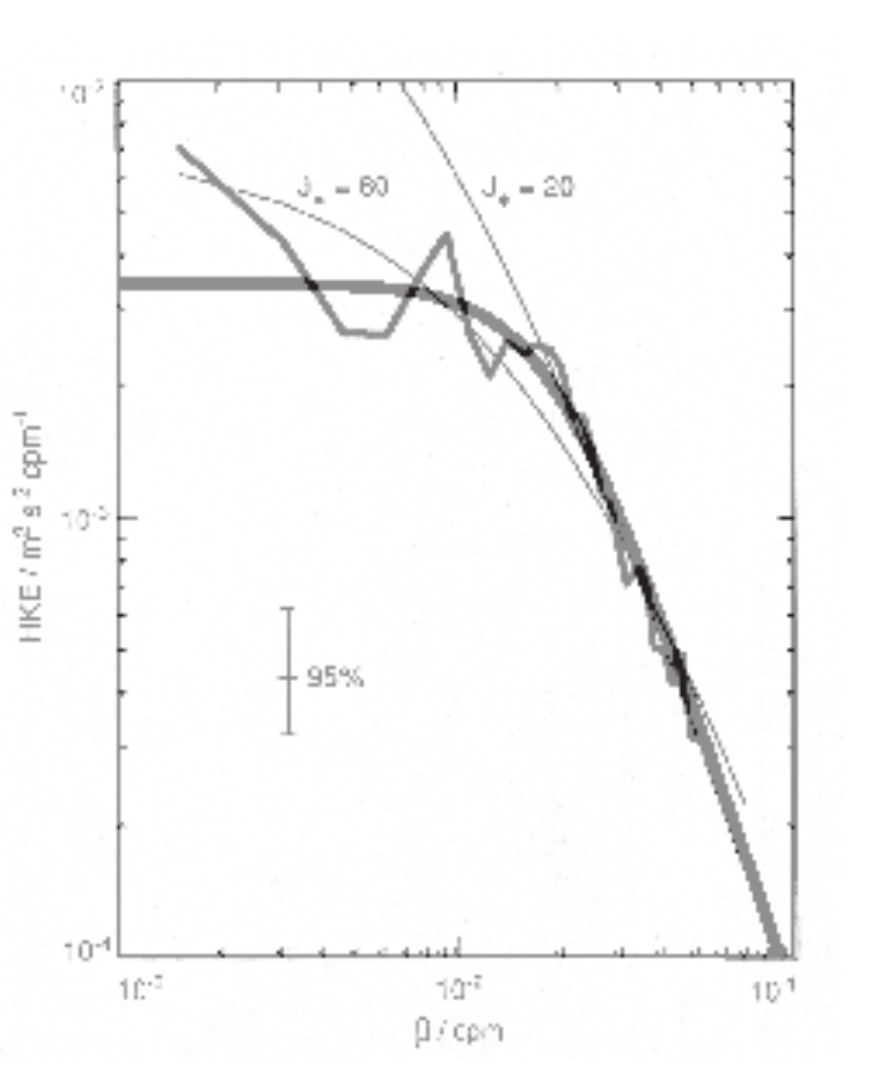}\vspace{-0.75cm}
\caption{ AIwex vertical wavenumber
spectra of horizontal velocity.  Thin lines represent fits of the GM75
spectrum, $E_k \propto (1/m_* + m)^{5/2}$, with different values of
$j_*$.  This spectrum does not approach its asymptotic roll-off
quickly.  The thick overlay is $E_k \propto (1/(m_{\ast}^3+m^3))^{0.75}$ 
with $j_{\ast}=50$ and total energy 
$0.95\times10^{-4}$ m$^2$ s$^{-2}$.  (Figure
extracted from \cite{DAM91}, their Fig. 7, permission from {\em
J. Geophys. Res.,} Copyright 1991 America Geophysical Union). }
\label{AIwex-vert}  
\end{figure}

\section{Summary}\label{Summary}

Despite the fact that major deviations in the model parameters have
been noted near boundaries \citep{WW79} and near the equator
\citep{E85}, it is astonishing that the characteristic shape, and after
buoyancy scaling, level, are noted to be nearly universal in the
literature.  Variability in the spectral
characteristics of the deep ocean internal wavefield {\bf is} apparent.  That
variability, though, tends to be subtle.

A number of patterns emerge from our analysis:  
\begin{itemize}
\item Geography: The variability of spectral shape is geographically
related, rather than a function of spectral level.  Measurements from
Site-D exhibit characteristics similar to those from the eastern Subtropical  North
Pacific and have power laws in good agreement with the GM76 model.
Both are eddy-rich regions.  Observations from the eastern North Atlantic
and the Arctic differ most significantly from the GM model.  These
regions have anomalously weak eddy fields and the internal wave
spectral levels are lower.  Spectral shapes from the Sargasso Sea 
and the eastern boundary of the Subpolar North Pacific are
intermediate in character.

\item Seasonality:  A seasonal cycle, most prominent in high frequency 
energy, is apparent in the western Subtropical North Atlantic.  A seasonal cycle is also apparent in the eastern Subpolar North Pacific.  Their character, though, differs.  The high frequency signal in the North Pacific lags the near-inertial cycle, while near-inertial and high frequency seasonal variability are in phase in the North Atlantic.  The eddy fields in the two regions are distinctly different.  

\item Amplitude: The GM spectral model has an energy of $E_o =$ 30
cm$^2$ s$^{-2}$.  With the stipulated frequency distribution
(\ref{shape}), this tends to overestimate observed frequency spectra
from the deep ocean for $\sigma \gg f$.  This pattern is evident in
tables presented by \cite{Wunsch76, WW79, Fu81, NBP86}.  Despite this
overestimate, the total energy in the internal wave band agrees 
much better with $E_o$.  The difference is additional energy in the
tides and at near-inertial frequencies.

\item Separability: The observed deep ocean internal wavefield is not,
in general, separable.  When available, vertical wavenumber spectra of
potential energy tend to have larger characteristic vertical scales
than vertical wavenumber spectra of kinetic energy.  Fits to the
kinetic energy spectra produce roll-offs ranging from $3 < j_{\ast} <
15-20$, with {\bf lower amplitude frequency spectra} being
characterized by {\bf larger values of} $j_{\ast}$.  Fits to the potential
energy spectra produce $j_*= 1-2$.  Non-separability is in the sense
of a high-wavenumber near-inertial addition to a separable spectrum.
Indications of this are also apparent in the IWEX spectrum.  The issue
of separability appears to have been considered only in the continuum
limit ($f \ll \sigma \ll N$) by \cite{GM72}.

\item Covariability: Frequency spectra with power laws whiter than the
GM model ($\sigma^{-2}$) tend to have vertical wavenumber power laws
that are redder than the GM model ($m^{-2}$).

\item The mid-frequency dip.  Many of the frequency spectra show a
departure from the smooth spectral models at frequencies between 1.5
$f$ and 4-5 cph (e.g., Fig. \ref{Tropics-freq}).  Apart from tides and tidal/inertial peak harmonics,
the observed spectra tend to be at somewhat lower levels than the
model fits in this frequency band.  \cite{L02} notes a similar pattern and suggests it is
linked to internal wave dynamics at frequencies less than
semi-diurnal.

\item Bandwidth:  A recurring pattern is one of the near-inertial field having a larger vertical wavenumber bandwidth than the rest of the internal wavefield.  Data sets were selected on the basis that they represent open ocean conditions and are not in the near-field of sites associated with topographic generation processes.  We have consequently sought to relate the near-inertial bandwidth to interior processes.  An unsettling possibility is that topographic influences may not be appreciated.  

\end{itemize}

The GM76 model is a reasonable description of the winter time internal
wave spectrum at Site-D.  The representativeness of the GM model for 
the background state here is not surprising as the GM frequency
spectrum is no more and no less than a curve fit to data from Site-D.  But what
may not be appreciated is that those original data were obtained
during the dead of winter.  At other times and in other places,
differences between this simplistic model and observations are quite apparent.  

Departures from the GM model represent clues about the relative importance of the
various possible sources and how those source inputs are modified by
dissipation, wave/wave and wave-mean interactions.  How those clues can be 
brought together is discussed in the following section.  

\section{Spectral Balances and Radiation Balance Equations}\label{Theory}

In the previous section we documented variability in the background internal wave spectrum.  
Our hypothesis is that such background states are in near equilibrium between forcing, dissipation and energy transfers.  In this section we review what we perceive to be the key elements underlying such a hypothesis.  We do not attempt an exhaustive review nor do we intend to elevate any result beyond the status of plausibility.  As a prelude to that discussion, we present a 35 year old paradigm for the spectral balances as a strawman.  What we will find in the ensuing discussion is that key components of the strawman require reconsideration. 

\begin{figure}\vspace{-0.50cm}
\noindent\includegraphics[width=20pc]{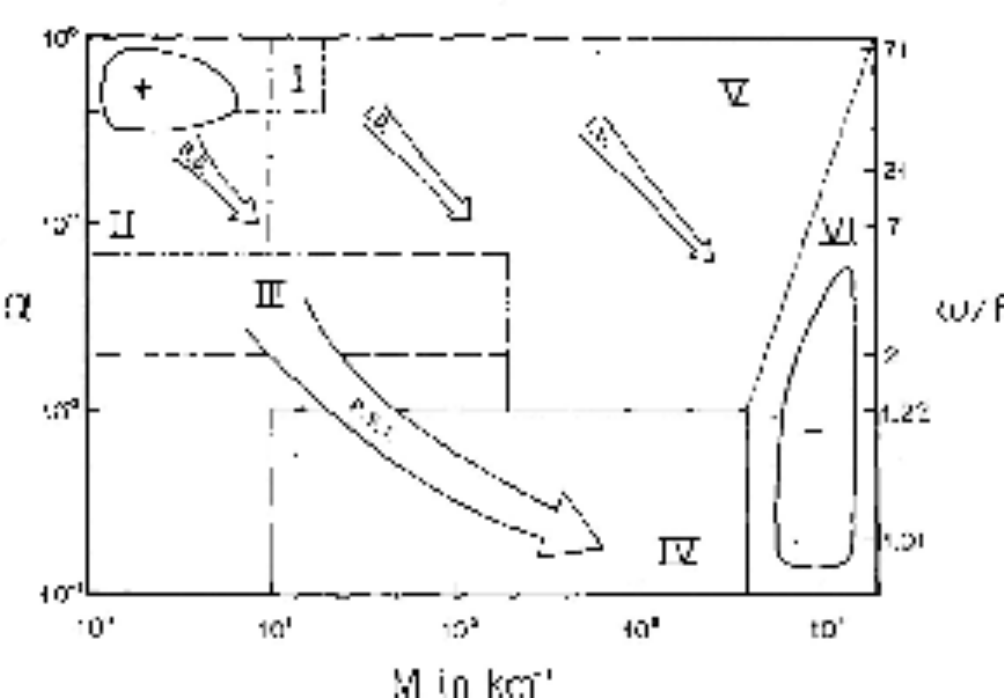}\vspace{-0.75cm}
\caption{  The \cite{M77} model for energy flow in the deep sea internal wavefield.  
\label{McComas} }
\end{figure}

Figure \ref{McComas} from \cite{M77} presents one version of the internal wave energy balance in the spectral domain.  A similar diagram can be inferred from \cite{MO75}.  Energy is gained in region I from surface waves at low vertical wavenumber and high frequency.  In the overlapping Region II (denoted S.D.), energy is input through interaction with the mesoscale eddy field and moved to region III.  Nonlinear interactions transfer energy from Region III to Region IV via the Parametric Subharmonic Instability.  Region V is in equilibrium with respect to nonlinear transfers associated with the Induced Diffusion mechanism and transports energy to Region VI, where dissipation occurs.  

We use this strawman to draw focus onto the following questions:
\begin{itemize}
\item do the observed spectra represent approximate steady states?  
\item are the sources appropriately aligned along or within the spectral domain and of sufficient strength as to balance nonlinear transfers?
\item does variability of the spectral parameters relate to variability in the forcing? 
\end{itemize}  

Intrinsic difficulties with the \cite{M77} diagram are that the source distributions are likely to be significantly different than stated and there are significant questions about the characterization of nonlinear transports associated with the weakly nonlinear theory that provides the named mechanisms above.   

We proceed by noting that a quantitative explanation of the observed spectral variability requires a corresponding analysis in the spectral domain.  In the case of media with weak inhomogeneities the spectral evolution is described by a kinetic equation that has both transfer terms associated with nonlinearity and a transport term 
that describes the motion of wave packets through the momentum-coordinate space. The
transport term is a wave analog of Liouville's theorem or the continuity equations for a distribution function in statistical mechanics \citep{LL80, PL81}.  The kinetic equation with the transport term is
called a radiative balance equation, and is given by \cite{LNW_03}:
\begin{eqnarray}
&& \frac{\partial n(\bf{p})}{\partial t} + {\bf \nabla}_{\bf r} \cdot [(\overline{{\bf u}}+{\bf C_g}) n({\bf p})] + 
{\bf \nabla}_{\bf p}  {\bf \cdot} [{\bf {\cal R}} n(\bf{p})]  = \nonumber\\
&& S_o({\bf p}) - S_i({\bf p}) + T_r({\bf p}). 
\label{rbe}
\end{eqnarray}
The factor ${\bf C_g=\nabla_{\bf p} \omega}$ represents the group velocity, $\overline{{\bf u}}$ subinertial currents, ${\bf {\cal R}} = -\nabla_{{\bf r}} (\omega + {\bf p} \cdot \overline{{\bf u}})$ refractive effects associated with spatially inhomogeneous stratification and subinertial currents, $T_r$ transfers of action, $S_o$ interior sources and $S_i$ sinks (dissipation).  Gradient operators in the spatial and spectral domains are ${\bf \nabla}_{\bf r}$ and ${\bf \nabla}_{\bf p}$, respectively. 

\begin{figure*}
\noindent\includegraphics[width=10cm]{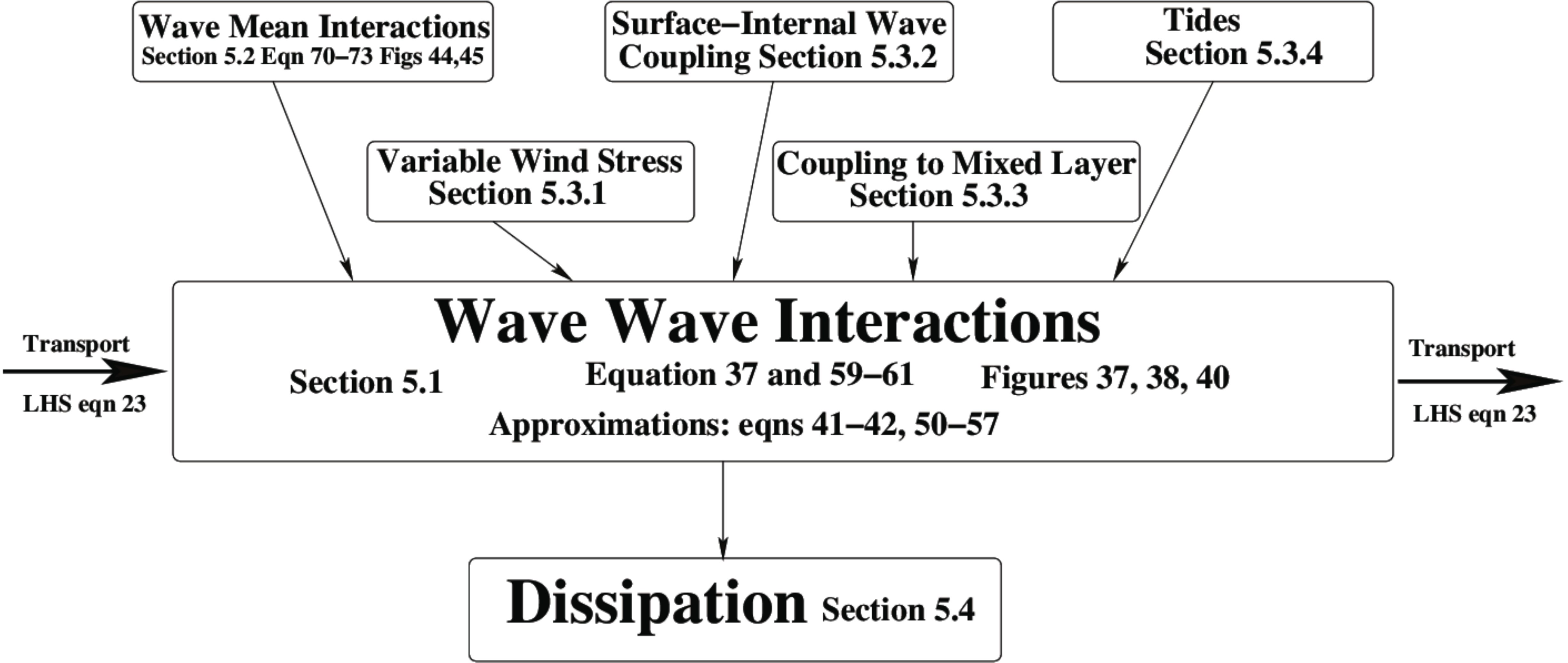}
\caption{ A schematic Table of Contents for Section \ref{Theory}.  
The radiative balance equation describes the transfer energy in the spectral and spatial domain.  
Major processes are identified and classified with section, equations and figure numbers.  }
\label{YLschematic} 
\end{figure*}

Our underlying hypothesis is that the observed variability can be understood as spatially local stationary states of the radiation balance equation (\ref{rbe}).  Thus we consider representations of nonlinear transfers $T_r({\bf p})$ in Section \ref{NonlinearitySection}, a source function $S_o({\bf p})$ characterization of wave-mean interactions \{corresponding to terms ${\bf \nabla}_{\bf r} \cdot [(\overline{{\bf u}}+{\bf C_g}) n({\bf p})] + {\bf \nabla}_{\bf p}  {\bf \cdot} [{\bf {\cal R}} n(\bf{p})] \}$ in Section \ref{WaveMean}, external forcing $S_o({\bf p})$ in Section \ref{Forcing} and dissipation $S_i({\bf p})$ in Section \ref{Dissipation}.  Possible issues related to propagation are addressed in Section \ref{Propagation}.  See Fig. \ref{YLschematic} for a schematic table of contents.  

%

\subsection{Nonlinearity $T_r({\bf p})$ in (\ref{rbe}) \label{NonlinearitySection} }

We began this investigation with the hypothesis that the observations, particularly 
the signature of covariable power laws, represent the end product of a nonlinear 
equilibration process.  Nonlinear interactions between internal waves is a fascinating subject
with a significant literature spanning 50 years.  An inclusive and insightful review is given
in \cite{M86}.  Our discussion below draws from this and includes more recent work.  

Our starting place is weak nonlinearity as described by a kinetic equation.  A kinetic equation is a closed equation for the time evolution of the wave action spectrum in a system of weakly interacting waves.  It is usually derived as a central result of wave turbulence theory.  The concepts of wave turbulence theory provide a fairly general framework for studying the statistical steady states in a large class of weakly interacting and weakly nonlinear many-body or many-wave systems.  In its essence, classical wave turbulence theory~\citep{ZLF} is a perturbation expansion in the amplitude of the nonlinearity, supposing linear  plane wave solutions of $a {\rm e}^{i({\bf r} \cdot {\bf p} -\sigma t)}$ at leading order and a slow amplitude modulation at the next order of the expansion, $a(\tau) {\rm e}^{i({\bf r} \cdot {\bf p} -\sigma t)}$, by resonant interactions.  This modulation leads to a redistribution of the spectral energy density among space- and time-scales.  Below we sketch the derivation of such an equation following \cite{ZLF}.  

Nonlinearities in the equations of motion are quadratic, so assuming the nonlinearity to be weak, first order expressions in those equations contain product terms of two waves having wavenumber-frequency $({\bf p}_1, \sigma_1)$ and $({\bf p}_2, \sigma_2)$.  For example, 
\begin{eqnarray}
{\bf u} \cdot \nabla u & \propto & e^{i({\bf p}_1 \cdot {\bf r} - \sigma_1 t)}~e^{i({\bf p}_2 \cdot {\bf r} - \sigma_2 t)} \nonumber \\ 
&& ~~~~~~~~~ = e^{i[({\bf p}_1+{\bf p}_2)\cdot {\bf r} - (\sigma_1 + \sigma_2) t]}~.  
\end{eqnarray}
If the combined wavenumber ${\bf p}_1+{\bf p}_2$ and frequency $\sigma_1 + \sigma_2$ match those associated with a free wave, resonance occurs and the amplitude of the third wave grows linearly in time as long as its amplitude remains small.  In general, the three wave resonance conditions can be written:    
\begin{equation}
{\bf p} = {\bf p_1} \pm {\bf p_2} ~{\rm  \; and \;}~ \sigma = \sigma_1 \pm \sigma_2 
\label{resonant_manifold}
\end{equation}
Solutions to (\ref{resonant_manifold}) are referred to as the resonant manifold and the three waves form a resonant triad.  The possibility of solutions to (\ref{resonant_manifold}) depends simply upon the geometry of the dispersion surfaces \citep{P60}.  Such three wave solutions are possible for internal waves but not, for example, surface gravity waves in deep water.

The evolution of wave amplitude $a_{\bf p}$ follows most directly from Hamilton's equation: 
\begin{equation}
i\frac{\partial}{\partial t} a_{{\bf p}} = \frac{\delta {\cal H}}{\delta a_{{\bf p}}^{\ast}} 
\label{fieldequation}
\end{equation}
with Hamiltonian ${\cal H}$ that is nominally the sum of kinetic and potential energies, or Lagrange's equation: 
\begin{equation}
\frac{d}{dt} \frac{\partial {\cal L}}{\partial \dot{a}_{{\bf p}}} - \frac{\partial {\cal L}}{\partial a_{{\bf p}}} = 0
\label{Lfieldequation}
\end{equation}
with Lagrangian ${\cal L}$ that is the difference of kinetic and potential energies.  

Usually, but not always, one needs to adopt a linearization to obtain the Hamiltonian ${\cal H}$ in (\ref{fieldequation}) or the Lagrangian ${\cal L}$ in (\ref{Lfieldequation}).  The difficulty is that, in order to utilize (\ref{fieldequation}) or (\ref{Lfieldequation}), the Hamiltonian or Lagrangian must {\bf first} be
constructed as a function of the generalized coordinates and momenta ($a_{\bf p}$).  
It is not always possible to to do so {\bf explicitly}, in which case one must set up a perturbation expansion which imposes a small amplitude limitation upon the result.  See Section \ref{RepDep} for further detail.    

The Hamiltonian of a system with quadratic nonlinearity can be expressed as \citep{ZLF}:  
\begin{eqnarray}
{\cal H} &=& \int d{\bf p} \, \sigma_{{\bf p}} |a_{{\bf p}}|^2 \nonumber \\
&+& \int d{\bf p}_{012}
\LSBA \delta_{{\bf p}+{\bf p}_1+{\bf p}_2} (U_{{\bf p},{\bf p}_1,{\bf p}_2} 
a_{{\bf p}}^{\ast} a_{{\bf p}_1}^{\ast} a_{{\bf p}_2}^{\ast} + \mathrm{c.c.}) \nonumber \\
&+& \delta_{-{\bf p}+{\bf p}_1+{\bf p}_2} (V_{{\bf p}_1,{\bf p}_2}^{{\bf p}}
a_{{\bf p}}^{\ast} a_{{\bf p}_1} a_{{\bf p}_2} + \mathrm{c.c.}) \RSBA 
\label{HAM}
\end{eqnarray} 
after a Fourier decomposition.  The interaction coefficients $U$ and $V$ are extended algebraic expressions involving wave amplitude $a_{\bf p}$, frequency $\sigma_{\bf p}$ and momentum ${\bf p}$ of the three waves.   

Having obtained the Hamiltonian (\ref{HAM}) one introduces wave action as
\begin{equation}
n_{\bf p} = \langle a_{\bf p}^* a_{\bf p} \rangle, 
\label{WaveAction}
\end{equation} 
where $\langle\dots\rangle$ means the averaging over statistical ensemble of many realizations of the internal waves.
To derive the time evolution of $n_{\bf p}$ the amplitude equation
(\ref{fieldequation}) with Hamiltonian (\ref{HAM}) is multiplied by $a_{\bf p}^*$, the
amplitude evolution equation for $a_{\bf p}^{\ast}$ is multiplied by $a$, the two equations are differenced and 
the result averaged  $\langle\dots\rangle$ to obtain:  
\begin{eqnarray}
\frac{\partial n_{\bf p} }{\partial t} &=& \Im
\int  \LSBA V_{{\bf p}_1{\bf p}_2}^{{\bf p}} J^{\bf p}_{{\bf p}_1 {\bf p}_2} \delta({\bf p}-{\bf p}_1-{\bf p}_2) 
\nonumber\\
   &&
    -V_{{\bf p} {\bf p}_1}^{{\bf p}_2} J^{{\bf p}_2}_{{\bf p} {\bf p}_1} \delta({\bf p}_2-{\bf p}-{\bf p}_1)  \nonumber\\
   &&
    -V_{{\bf p} {\bf p}_2}^{{\bf p}_1} J^{{\bf p}_1}_{{\bf p} {\bf p}_2} \delta({\bf p}_1-{\bf p}_2-{\bf p}) 
\RSBA d{\bf p}_1 d{\bf p}_2,
\label{TwoPoints}
\end{eqnarray}
where we introduced a triple correlation function
\begin{eqnarray} J^{{\bf p}}_{{\bf p}_1 {\bf p}_2}\delta({\bf p}_1-{\bf p}-{\bf p}_2)\equiv
\langle a_{\bf p}^* a_{{\bf p}_1} a_{{\bf p}_2} \rangle.\label{TrippleCorrelator}
\end{eqnarray} 
If we were to have non-interacting fields, i.e., fields with
$V^{{\bf p}}_{{\bf p}_1 {\bf p}_2}$ being zero, this triple correlation function would be
zero.  We therefore invoke a perturbation expansion in smallness of interactions to
calculate the triple correlation at first order.
The first order expression for
$\partial n_{\bf p}/\partial t$ thus requires computing $\partial J^{{\bf p}}_{{\bf p}_1
{\bf p}_2}/\partial t$ to
first order. To do so we take definition (\ref{TrippleCorrelator}) and
use (\ref{fieldequation}) with Hamiltonian (\ref{HAM}) and apply
 $\langle\dots\rangle$ averaging. We get
\begin{eqnarray}
\LSBA i\frac{\partial}{\partial t} &+&
 (\sigma_{{\bf p}_1}-\sigma_{{\bf p}_2}-\sigma_{{\bf p}_3})
\RSBA J^{{\bf p}_1}_{{\bf p}_2 {\bf p}_3} \nonumber\\
&=& \int \left[ -\frac{1}{2} (V^{{\bf p}_1}_{{\bf p}_4 {\bf p}_5})^* J^{{\bf p}_4 {\bf p}_5}_{{\bf p}_2
{\bf p}_3}\delta({\bf p}_1-{\bf p}_4-{\bf p}_5)
\right. \nonumber\\ &&\left.
              +(V^{{\bf p}_4}_{{\bf p}_2 {\bf p}_5})^* J^{{\bf p}_1 {\bf p}_5}_{{\bf p}_3
{\bf p}_4}\delta({\bf p}_4-{\bf p}_2-{\bf p}_5)
\right. \nonumber\\ &&\left.
+V^{{\bf p}_4}_{{\bf p}_3 {\bf p}_5} J^{{\bf p}_1 {\bf p}_5}_{{\bf p}_2 {\bf p}_4}\delta({\bf p}_4-{\bf p}_3-{\bf p}_5) \right]
d{\bf p}_4 d{\bf p}_5.
\label{TrippleCorrelatorddt}
\end{eqnarray}
Here we introduced the quadruple correlation function
\begin{eqnarray} J^{{\bf p}_1 {\bf p}_2}_{{\bf p}_3 {\bf p}_4}\delta({\bf p}_1+{\bf p}_2-{\bf p}_3-{\bf p}_4)\equiv
\langle a_{{\bf p}_1}^*  a_{{\bf p}_2}^*   a_{{\bf p}_3} a_{{\bf p}_4} \rangle.
\label{FourCorrelator} \end{eqnarray}
The next step is to assume Gaussian statistics, and to express $J^{{\bf p}_1
{\bf p}_2}_{{\bf p}_3 {\bf p}_4}$
as a product of two two-point correlators as
$$J^{{\bf p}_1 {\bf p}_2}_{{\bf p}_3 {\bf p}_4} = n_{{\bf p}_1} n_{{\bf p}_2} \Big[
\delta({\bf p}_1-{\bf p}_3)\delta({\bf p}_2-{\bf p}_4) +
\delta({\bf p}_1-{\bf p}_4)\delta({\bf p}_2-{\bf p}_3)\Big].$$
Then
\begin{eqnarray}\label{TripplePointGauss}
\LSBA
i\frac{\partial }{\partial t} &+&  (\sigma_{{\bf p}_1}-\sigma_{{\bf p}_2}-\sigma_{{\bf p}_3})
\RSBA J^{{\bf p}_1}_{{\bf p}_2{\bf p}_3} \nonumber \\
&=& (V^{{\bf p}_1}_{{\bf p}_2 {\bf p}_3})^* \left(n_1 n_3 + n_1 n_2 - n_2 n_3 \right).
\end{eqnarray}
Time integration of the equation for $J^{{\bf p}_1}_{{\bf p}_2{\bf p}_3}$ will contain fast oscillations due to the initial value of $J^{{\bf p}_1}_{{\bf p}_2{\bf p}_3}$ and slow evolution due to the nonlinear wave interactions. Contributions from the first term will rapidly decrease with time, so neglecting these terms we get
\begin{eqnarray}\label{TheMeat}
J^{{\bf p}_1}_{{\bf p}_2{\bf p}_3}
= \frac{
(V^{{\bf p}_1}_{{\bf p}_2 {\bf p}_3})^* \left(n_1 n_3 + n_1 n_2 - n_2 n_3 \right)}
{\sigma_{{\bf p}_1}-\sigma_{{\bf p}_2}-\sigma_{{\bf p}_3} + i \Gamma_{{\bf p}_1{\bf p}_2{\bf p}_3}}.
\end{eqnarray}
Here we introduced the nonlinear damping of the waves $\Gamma_{{\bf p}_1{\bf p}_2{\bf p}_3}$ 
that relates to the breadth of the resonant manifold.
We will elaborate on $\Gamma_{{\bf p}_1{\bf p}_2{\bf p}_3}$ in Section
\ref{ResonanceBroadening}.
We now  substitute (\ref{TheMeat}) into (\ref{TwoPoints}),
assume for now that the damping of the wave is small, and use
\begin{equation}
\lim_{\Gamma\to 0} \Im \left[\frac{1}{\Delta+i\Gamma} \right]= - \pi
\delta(\Delta).
\label{DeltaFunction}
\end{equation}
We then obtain the three-wave kinetic equation~\citep{ZLF,NoisyNazarenko,LLNZ}:
\begin{eqnarray}
&&T_r({\bf p}) \equiv \frac{d n_{{\bf p}}}{dt} = \nonumber \\
&&4\pi \!\! \int \!\! |V_{{\bf p}_1,{\bf p}_2}^{{\bf p}}|^2 f_{p12} \,
\delta_{{{\bf p} - {\bf p}_1-{\bf p}_2}} \delta({\sigma_{{\bf p}}
-\sigma_{{{\bf p}_1}}-\sigma_{{{\bf p}_2}}})
d {\bf p}_{12}
\nonumber \\
&&-4\pi\!\! \int \!\! 
|V_{{\bf p}_2,{\bf p}}^{{\bf p}_1}|^2 f_{12p}\, \delta_{{{\bf p}_1 - {\bf p}_2-{\bf p}}}  
\delta({{\sigma_{{\bf p}_1} -\sigma_{{\bf p}_2}-\sigma_{{\bf p}}}})
d {\bf p}_{12}
\nonumber \\
&&-4\pi\!\! \int \!\! 
|V_{{\bf p},{\bf p}_1}^{{\bf p}_2}|^2 f_{2p1}\, \delta_{{{\bf p}_2 - {\bf p}-{\bf p}_1}} 
\delta({{\sigma_{{\bf p}_2} -\sigma_{{\bf p}}-\sigma_{{\bf p}_1}}})
d {\bf p}_{12}
 \, ,\nonumber\\
&& {\rm with } \; f_{p12} = n_{{\bf p}_1}n_{{\bf p}_2} -
n_{{\bf p}}(n_{{\bf p}_1}+n_{{\bf p}_2}) \, .
\label{KineticEquation}
\end{eqnarray}   
The $\delta$ functions require the wavenumbers and frequencies sum to zero and 
ensure that pseudomomentum [${\bf p} n({\bf p})$] and energy [$\sigma_{\bf p} n({\bf p})$] spectral densities are conserved.  The Resonant Interaction Approximation (RIA) is represented by the reduction of (\ref{TheMeat}) to the resonant manifold (\ref{resonant_manifold}) and $\delta$ function representation in (\ref{DeltaFunction}).  

The typical assumptions needed for the derivation of
kinetic equations are:
\begin{itemize}
\item Weak nonlinearity,
\item Gaussian statistics of the interacting wave field in wavenumber space and
\item Resonant wave-wave interactions
\end{itemize}
\noindent In a more systematic derivation, such as \cite{LN}, the kinetic equation can be obtained using only an assumption of weak nonlinearity.  A theory of weak interaction has its limitations and this motivates discussions of renormalization attempts to account for increasing nonlinearity (\ref{Renormalization}) and ray-tracing descriptions of wave-wave interactions (Section \ref{Eikonal}).  

\subsubsection{Resonant Interactions: Scale Invariant Solutions}\label{ScaleInvariance}
Following Kolmogorov's viewpoint on energy cascades in isotropic Navier--Stokes turbulence, one may look for statistically stationary states using scale-invariant solutions\footnote{The scale invariance assumption is, formally, a generic assumption about dependent and independent variables in the kinetic equation.  See, for example, \cite{BO}, their Section 1.7} to the kinetic equation~(\ref{KineticEquation}). The solution may occur in an inertial subrange of wavenumbers and frequencies that are far from those where forcing and dissipation act, and also far from characteristic scales of the system, including the Coriolis frequency resulting from the rotation of the Earth, the buoyancy frequency due to stratification and the ocean depth.  Under these assumptions, the dispersion relation and the interaction matrix elements are locally scale-invariant and solutions of (\ref{KineticEquation}) take the form
\begin{eqnarray}
   n({\bf k}, m)=  |{\bf k}|^{-a}  |m|^{-b}.
\label{PowerLawSpectrum}
\end{eqnarray}
Values of $a$ and $b$ such that the right-hand side of (\ref{KineticEquation}) vanishes identically
correspond to steady solutions of the kinetic equation, and hopefully also to statistically steady states of the ocean's wave field.  

To connect with the observations, the corresponding vertical wavenumber - frequency spectrum of energy is obtained by transforming $ n_{{\bf k},m}$ from the three dimensional wavenumber space
$({\bf k},m)$ to the two dimensional vertical wave-number-frequency space $(\sigma,m)$ and
multiplying by frequency.  In the high frequency-high wavenumber limit,
$$ E(m,\sigma) \propto m^{2-a-b} \sigma^{2-a} ~~ .  $$ 
The frequency-wavenumber power laws inferred from the observations
($m^{-q}\sigma^{-r}$) are thus related as:
$$r=a-2, \; q=a+b-2.$$

\begin{figure}
 \begin{center}
\includegraphics[width=20pc]{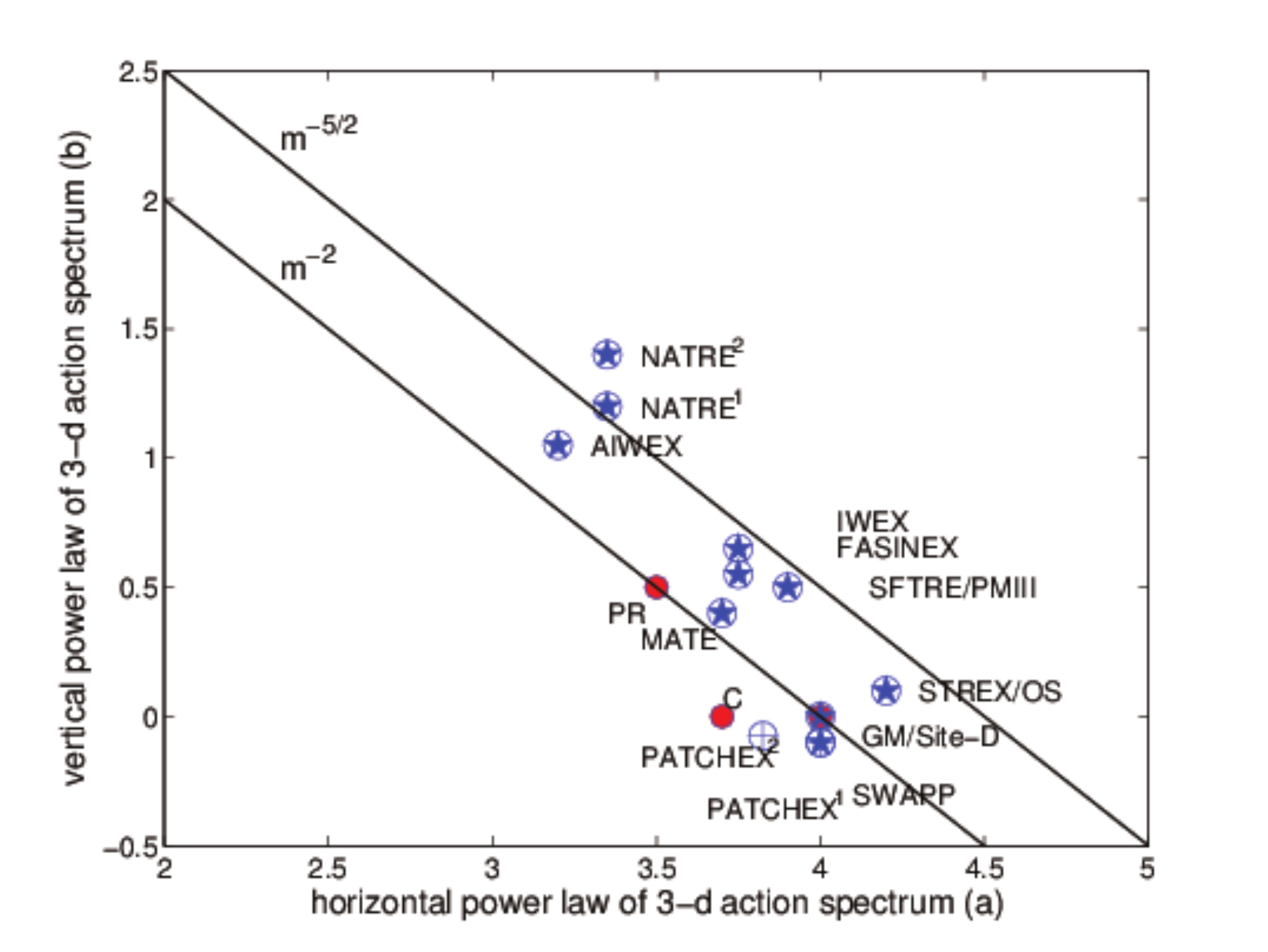}
 \end{center}
\caption{The observational points.  The filled circles represent the Pelinovsky--Raevsky (PR) spectrum [$(a,b) = (3.5,0.5)$], the convergent numerical solution [$(a,b) = (3.7,0.0)$] determined in \citet{theory} and the GM spectrum [$(a,b) = (4.0,0.0)$].  Circles with stars represent estimates based upon one-dimensional spectra from the western North Atlantic south of the Gulf Stream (IWEX, FASINEX and SFTRE/PMIII), the eastern North Pacific (STREX/OS and PATCHEX$^1$), the western North Atlantic north of the Gulf Stream (Site-D), the Arctic (AIWEX) and the eastern North Atlantic (NATRE$^1$ and NATRE$^2$).  There are two estimates obtained from two-dimensional data sets from the eastern North Pacific (SWAPP and PATCHEX$^2$) represented as circles with cross hairs.  NATRE$^1$ and NATRE$^{2}$ represent fits to the observed spectra and observed minus vortical mode spectra, respectively.  Lines of constant vertical wavenumber power laws for the observed
energy spectrum run diagonally downward.  Lines of constant frequency power laws are vertical.  }
\label{fig:onlydata}
\end{figure}

Parametric fits to the observations (Fig. \ref{fig:onlydata}) produce a clustering about two regions of parameter space with several notable outliers.  Observations from Site-D and from the subtropical North Pacific tend to cluster about the GM76 power laws ($m^{-2}\sigma^{-2}$).  These data were obtained in regions poleward of the critical latitude (28.9) at which the $M_2$ semidiurnal tide can decay into near-inertial frequencies through a parametric subharmonic instability.  The MATE data from the subpolar North Pacific and observations from south of the Gulf Stream are grouped together with power laws describing spectra distinctly redder (whiter) in the vertical wavenumber (frequency) domain.  This grouping includes data both poleward and equatorward of the semidiurnal critical latitude.  Observations from the eastern North Atlantic (NATRE), the Arctic (AIWEX) and, tentatively, the subpolar North Pacific (Ocean Storms) appear as outliers.   

\subsubsubsection{Nonrotating ($f=0$) limit}
Unlike Kolmogorov turbulence, the exponents which give steady solutions to (\ref{KineticEquation}) can not be determined by dimensional analysis alone. This is the case owing to multiple characteristic length scales in anisotropic systems, e.g., \citep{P04a}.  The anisotropic analysis is much more complicated and as a prelude to that discussion we enquire whether the improper integrals in (\ref{KineticEquation}) converge.  This is related to the question of locality of the interactions: a convergent integral characterizes the physical scenario where interactions of neighboring wavenumbers dominate the evolution of the wave spectrum, while a divergent one implies that distant, nonlocal interactions in the wavenumber space dominate. 

It turns out that the internal-wave collision integral diverges\footnote{A contrary assessment is given in \cite{LPT}.  That result is associated with a sign error in the numerical evaluation of (\ref{KineticEquation}).} for {\bf almost} all values of $a$ and $b$. In particular, the collision integral has an infra-red (IR) divergence at zero, i.e., $|\bf{k}_1|$ or $|\bf{k}_2|$ $\to 0$ and an ultra-violet (UV) divergence at infinity, i.e., $|\bf{k}_1|$ and $|\bf{k}_2|$ $\to \infty$.
A detailed analyses performed in \citep{theory} returns the convergence conditions in the IR limit, ${\bf k}_1 \;\mathrm{or}\; {\bf k}_2 \to 0$ of: 

 \begin{eqnarray}
 &
a+b/2-7/2<0 \;\; \mathrm{and} \;\; -3 < b < 3, 
\nonumber
\\
&
a-4<0 \;\; \mathrm{and} \;\; b=0,              
\nonumber
\\
&
a-7/2<0 \;\; \mathrm{and} \;\; b=1 ,           
\nonumber
\\
&
a+b-5<0 \;\; \mathrm{and} \;\; b>3 ,            
\nonumber
\\
\mathrm{or} \nonumber\\
&
a-5<0 \;\; \mathrm{and} \;\; b<-3 .              
\end{eqnarray}

\begin{figure}\vspace{-0.75cm}
\begin{center}
\includegraphics[width=20pc]{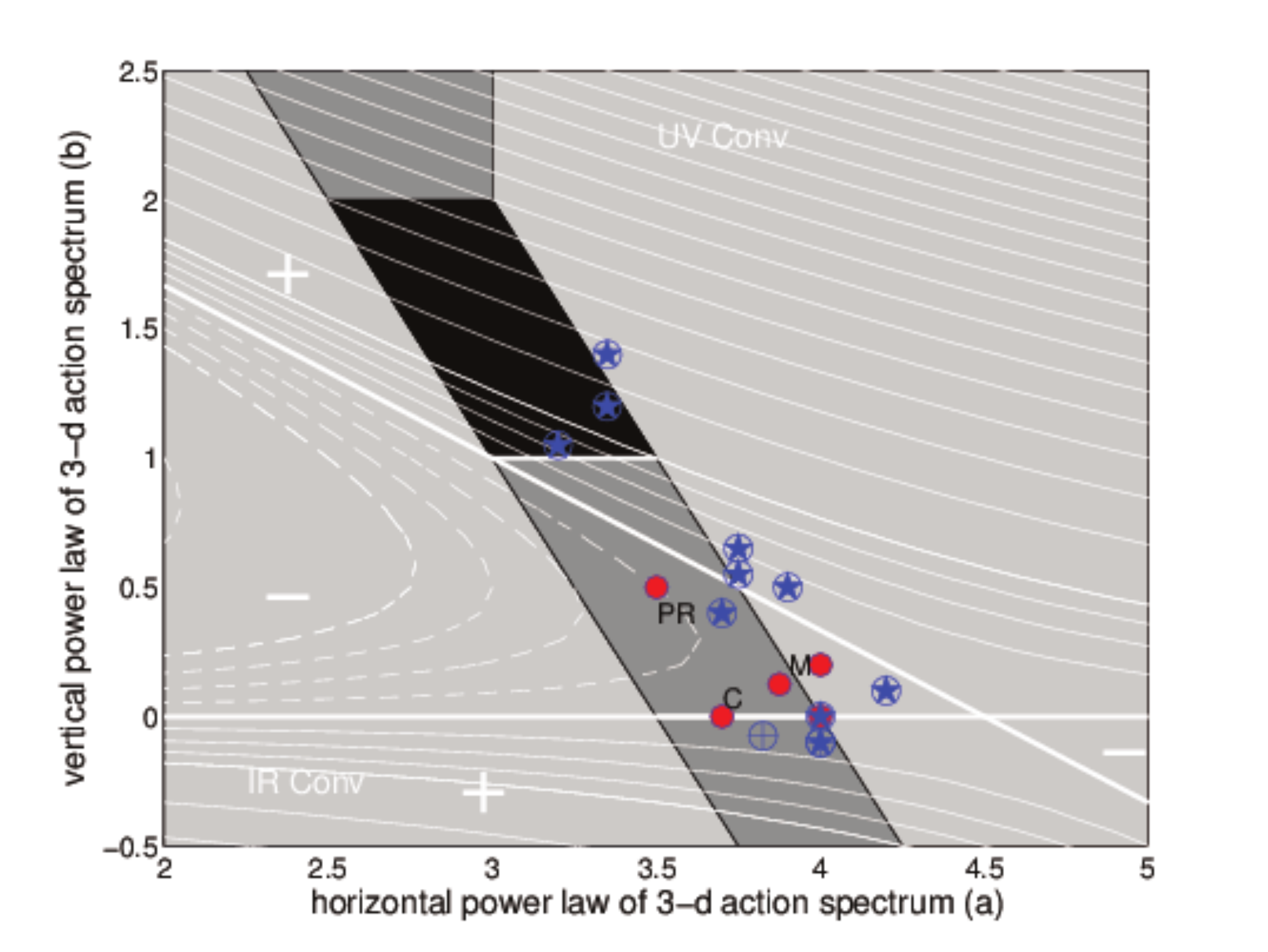}
 \end{center}
\caption{The observational points and the theories \citep{theory}.  The filled circles represent scale-invariant stationary states identified by \cite{PelinovskyRaevsky}, a convergent numerical solution determined in \cite{theory}, the GM spectrum and two possible dynamic balances identified in \cite{MMb}.  Circles with stars represent power law estimates based upon one-dimensional spectra.  Circles with cross hairs represent estimates based upon two-dimensional data sets.  See Fig. \ref{fig:onlydata} for the identification of the field programs.  Light grey shading represents regions of the power-law domain for which the collision integral converges in {\bf either} the infrared {\bf or} ultraviolet limit.  The dark grey shading represents the region of the power law domain for which {\bf neither} the infrared {\bf nor} ultraviolet limits converge.  The region of black shading represents the subdomain for which the infrared and ultraviolet divergences have the same sign.  Additionally, a line segment between $3 < a < 7/2$ along $b=1$ is IR convergent and a line segment between $7/2 < a < 4$ along $b=0$ is IR {\bf and} UV convergent.  These line segments arise as higher order cancellations in the integrand.  Overlain as solid white lines are the induced diffusion stationary states.  Thin white contours are proportional to the factor $b(9-2a-3b)$ in (\ref{InducedDiffusion}) with contour intervals of $\pm(0.25, 0.50, 0.75, 1.0, 2.0, 3.0 , ...)$.  The $+-$ symbols denote the sign of the action tendency (\ref{KineticEquation}) in the $a-b$ domain.  }
\label{fig:everything}
\end{figure}

Similarly, convergence in the UV limit $\mid {\bf k}_1 \mid \;\mathrm{and}\; \mid {\bf k}_2 \mid \to \infty$ implies that
 \begin{eqnarray}
 &
a+b/2-4>0 \;\; \mathrm{and} \;\; -2<b<2, 
\nonumber\\
&
a-7/2>0 \;\; \mathrm{and} \;\; b=0 ,      
\nonumber\\
&
a-3>0 \;\; \mathrm{and} \;\; b>2 ,         
\nonumber\\
\mathrm{or} \nonumber\\
&
a+b-3>0 \;\; \mathrm{and} \;\; b<-2 .      
\nonumber\\
\end{eqnarray}
The domains of divergence and convergence are shown in Fig.~\ref{fig:everything}.

There is only one exception where the integral converges in {\bf both} limits:  the line segment along $b=0$ with $7/2 < a < 4$.  Moreover, there is one special value, $(a,b)=(3.7,0)$, in which IR and UV contributions are both convergent {\bf and} cancel.  That is, $(a,b)=(3.7,0)$ represents a stationary state state of the kinetic equation in the absence of rotation.  Note that the line $b=0$ is 'special' in that it corresponds to wave action independent of vertical wavenumbers, $\partial n({\bf p}) / \partial m = 0$.  This convergent solution lies in the general proximity of GM76 $(a,b)=(4,0)$ and observations from Site-D, SWAPP and PATCHEX.  

The possibility also exists that one can construct a {\bf divergent} solution in the following sense.  If the UV and IR nonintegrable singularities have opposite signs for some $(a,b)$, it may be possible to devise a conformal mapping that matches the oppositely signed singularities.  In this case one would have, at least on the level of a technical mathematical exercise, an approximate stationary state.  \cite{PelinovskyRaevsky, Zeitlin1992, LT} find this to be possible for $(a,b)= (3.5,0.5)$.  \cite{theory} determine that such solutions are more general, and possible approximate stationary states occupy the dark grey shaded regions of Figure \ref{fig:everything}.  The black regions are not permitted due to UV and IR divergences having similar signs.  

The parametric spectral estimates tend to have larger $a$ values than either the convergent or the \cite{PelinovskyRaevsky} solution and also tend to lie on the border of UV convergence.  One instance (NATRE) is seen to occupy the black region, but this point migrates to a state of UV convergence if an attempt is made to subtract non-wave finestructure (See Section \ref{SE_ST_NA} for details).  

Although it represents a possible stationary state, a balance between divergent and oppositely signed integrals is not a satisfactory physical situation.  What it represents is a statement that integration endpoints other than $\pm \infty$ need to be inserted in (\ref{KineticEquation}).  Obvious candidates are $f$ in the IR limit and $N$ in the UV limit.  

\subsubsubsection{$f \ne 0$:  Extreme Scale Separated Interactions}\label{ScaleSeparation}
When the kinetic equation is evaluated on the resonance surface, extreme scale separated interactions are believed to dominate the transfers \citep{theory, M86}.  Thus approaching the issue of scale invariant solutions to the kinetic equation is not as simple as a spectrally local energy cascade.  A detailed analysis of the extreme scale separated interactions is required.  Three simple interaction mechanisms were identified by \citet{MB77} in the limit of an extreme scale separation.  The limiting cases are:

\begin{itemize}
\item
the vertical backscattering of a high-frequency wave by a low frequency wave of twice the vertical wavenumber into a second high-frequency wave of oppositely signed vertical wavenumber. This type of interaction is called elastic scattering (ES). 
\item
The scattering of a high-frequency wave by a low-frequency, small-wavenumber wave into a second, nearly identical, high-frequency large-wavenumber wave. This type of interaction is called  induced diffusion (ID). 
\item
The decay of a low wavenumber wave into two high vertical wavenumber waves of approximately one-half the frequency. This is referred to as the parametric subharmonic instability (PSI). This mechanism was identified by \cite{MB77} as causing the transfer of energy from frequencies of $2f$ and low vertical wavenumber to high vertical wavenumber near inertial oscillations in the GM spectra.   
\end{itemize}

Below we review how these extreme scale separated interactions can address the observed covariability in power law fits.  

\subsubsubsection{Induced Diffusion and covariable power laws}
One can further reduce (\ref{KineticEquation}) in the ID limit to a Fokker-Plank equation \citep{Gardiner, MB77}
\begin{equation}
\frac{\partial n({\bf p)}}{\partial t} = \frac{\partial}{\partial {\bf p}_i } D_{ij} \frac{\partial}{\partial {\bf p}_j } n({\bf p}),
\label{InducedDiffusion}
\end{equation}
such that
\begin{eqnarray}
D_{ij} = 2 \displaystyle{\int \int} d^{3}{\bf p}^{\prime} d^{3} {\bf p}^{\prime\prime}\nonumber \\ 
n( {\bf p}^{\prime\prime}) {\bf p}_i^{\prime\prime} {\bf p}_j^{\prime\prime} |V_{{\bf p}^{\prime},{\bf p}^{\prime \prime}}^{{\bf p}}|^2 \delta({\bf p}-{\bf p}^{\prime}-{\bf p}^{\prime\prime}) \delta(\sigma - \sigma^{\prime} -\sigma^{\prime\prime}),
\label{Diffusivity}
\end{eqnarray}
in which ${\bf p}$ and ${\bf p}^{\prime}$ are the wavevectors of high frequency waves, ${\bf p}^{\prime \prime}$ is the wavevector of the low frequency member of the triad and $|V_{{\bf p}^{\prime},{\bf p}^{\prime \prime}}^{{\bf p}}|^2 \cong \frac{\pi}{2} \sigma^{\prime\prime} k^2$.  To obtain the Fokker-Plank equation one:  (i) neglects terms quadratic in high frequency amplitude relative to linear terms, (ii) expands $[n({\bf p} - {\bf p}^{\prime \prime}) - n({\bf p})]$ in a Taylor series, (iii) sums the formulae for the two cases $\mid {\bf p}^{\prime} \mid > \mid {\bf p} \mid$ and $\mid {\bf p}^{\prime} \mid < \mid {\bf p} \mid$, and, noting the subtractive cancelation between the two, performs a second Taylor series expansion.  \cite{MB77} present scale estimates of (\ref{InducedDiffusion}) and argue that the vertical coordinate dominates the transfers, thereby reducing the integral expression to being proportional to the vertical gradient variance (i.e., shear variance) along the resonance curve.  We reconsider these arguments in Section \ref{ScaleInvariance}.e.  

Substituting the scale invariant solution (\ref{PowerLawSpectrum}) into (\ref{InducedDiffusion}), one obtains, for the GM class of spectra in which near-inertial frequencies dominate the shear variance,  stationary states if either 
\begin{eqnarray}
9-2a-3b=0
\quad \mathrm{or} \quad
b=0 ,
\label{NewIDcurve}
\end{eqnarray}
\cite{theory} obtain this result via rigorous asymptotic arguments, independent of the reduction of the kinetic equation (\ref{KineticEquation}) to the Fokker-Plank equation (\ref{InducedDiffusion}).  \cite{MMb} interpret $b=0$ as a no action flux in vertical wavenumber domain, while $9-2a-3b=0$ is a constant action
flux solution.  Power laws $b~^{>}_{<}~ 0$ imply a down-gradient action flux to higher/lower vertical wavenumber along lines of constant horizontal wavenumber.  Associated with this is the transport of action to lower/higher frequency and the gain/loss of energy by the near-inertial field.  A positive residual to (\ref{KineticEquation}) implies an increase in high vertical wavenumber, high frequency spectral density and is consonant with a balance between resonant transfers and a high wavenumber sink.  

The situation changes in the absence of rotation ($f=0$).  In this case, similar manipulations suggest that stationary states can be obtained if either 
\begin{eqnarray}
b-1=0
\quad \mathrm{or} \quad
b=0.
\label{NewerIDcurve}
\end{eqnarray}
in which $b=1$ is the constant action flux solution of the Fokker-Plank equation.  

The majority of the observational data points cluster in two groups that are not far from the ID stationary states for scale invariant solutions (Fig. \ref{fig:everything}).  The observations exhibit a tendency to lie in regions having either small transport ($b=0$) or transport to higher vertical wavenumber and lower frequency ($b > 0$).  The observations also tend to lie in regions having a positive residual, for which $\frac{\partial n}{\partial t} > 0$, and is consistent with a balance between the ID flux convergence and an unspecified sink.  The most extreme departure from the ID stationary states is NATRE, which is south of the critical latitude for the semi-diurnal lunar tide and has an anomalously large vertical wavenumber bandwidth ($j_{\ast}=18$).  

However appealing the apparent pattern match between the observations and stationary states under the ID mechanism may be, we believe there is more to the story.  That story includes numerical evaluations of the kinetic equation which contain significant residuals inconsistent with GM76 being a stationary state.  These residuals lead us to consider how increasing levels of nonlinearity at high wavenumber could change our interpretations in Section \ref{ResonanceBroadening}.  Before doing so, though, we fill in a bit of the background by discussing the dynamic balance of \cite{MMb}.  


\subsubsubsection{The dynamic balance} 
A quantitative formulation of the internal wave energy budget is given structure by the dynamic balance put forward in \cite{MMb}.  This key study ascribes an inertial subrange character to the GM76 spectrum after integrating over the frequency domain.  The dynamic balance is constructed by dividing the frequency domain into distinct regions and characterizing each region with transfers associated with the dominant triad class.  Thus frequencies of $f < \sigma < 2f$ and vertical wavenumber $m$ gain energy via PSI transfers from frequencies of $2f < \sigma < 4f$ and lower wavenumber.  Frequencies of $2 < \sigma < 4f$ and wavenumber $m$ loose energy via PSI transfers to frequencies of $f < \sigma < 2f$ and higher wavenumber and also gain energy via ID energy spectral fluxes across $\sigma = 4f$.  Frequencies $\sigma > 4f$ and wavenumber $m$ gain or loose energy as action is transfered to higher or lower vertical wavenumber in association with ID coupling to near-inertial waves of smaller vertical wavenumber.  Finally, frequencies $\sigma = f$ and wavenumber $m$ gain or loose energy in association with ID coupling to high frequency waves $\sigma > 4f$.  These domains are apparent in the schematic from \cite{M77} (Figure \ref{McComas}) and in numerical evaluations of the kinetic equation, e.g., Figure \ref{NonlinearityParameter}.  Matching expressions for the energy spectrum and its time rate of change across the boundary $\sigma = 4f$ provides the high frequency solution:  
\begin{equation}
E_{ID}(\sigma,\beta) = E \frac{\beta_{\ast}}{\beta^2} \frac{f}{\sigma^2}[1 + \frac{27}{64x}ln\frac{\sigma}{4f}]
\label{DB1}
\end{equation}
which is nicely fit by a power law $ n({\bf p}) \propto k^{-3.875} m^{-0.125}$.  Matching along $\beta = \beta_{\ast}$ provides
\begin{equation}
E_{ID}(\sigma,\beta) = E \frac{\beta_{\ast}}{\beta^2} \frac{f}{\sigma^2}[1 - \frac{27}{64x}ln\frac{\beta}{\beta_{\ast}}]
\label{DB2}
\end{equation}
which is fit by a power law $ n({\bf p}) \propto k^{-4.0} m^{-0.20}$.  Both solutions are plotted in Fig. \ref{fig:everything}.
  
One of the attractive properties of the first solution [$(a,b)\cong(3.875,0.125)$] is that the expressions for energy transport to small vertical scale lead to estimates that are difficult to distinguish from turbulent dissipation estimates obtained as part of the PATCHEX field program \citep{GS88}.  These estimates, along with those obtained in NATRE \citep{P95}, pin down the GM76 dissipation rate at approximately $ 8\times10^{-10}$ W/kg.  See \cite{P04a} for discussion.  

There are, however, several disquieting issues.  The first is that the downscale energy transport occurs only in vertical wavenumber and thus requires an energy source at high frequency.  The second is the requirement of integrating over the frequency domain and invocation of dissipation to obtain a stationary state.  The RIA dissipation mechanism (Section \ref{Dissipation_model}) acts primarily on near-inertial shear whereas numerical evaluations of the kinetic equation (Section \ref{ScaleInvariance}.e) indicate significant non-stationary tendencies at high frequencies.  

\subsubsubsection{Numerical Evaluations of (\ref{KineticEquation})}

The analysis presented above was based upon analytic work.  Here we report numerical evaluations of the kinetic equation expressed via the Boltzman rate:
\begin{equation}
\nu_{B} =  \frac{\dot n({{\bf p}})}{2 n{({\bf p})}}~.
\label{NonlinearTime}
\end{equation}
The Boltzman rate represents the net rate of transfer for wavenumber ${\bf p}$ and is a low order measure of nonlinearity for smooth, isotropic and homogeneous spectra.  The individual rates of transfer into and out of ${\bf p}$ maybe significantly larger for spectral spikes \citep{PMW80, M86} and potentially for smooth, homogeneous but anisotropic spectra.  

A characterization of the level of nonlinearity is made by normalizing the Boltzman rate by the characteristic linear time scale $\tau^{\mathrm{L}}_{{\bf p}}=2\pi/\sigma_{{\bf p}}$:  
\begin{equation}
{\cal  \epsilon}_{{\bf p}} = \frac{2\pi \dot n_{{\bf p}}}{n_{{\bf p}} \sigma_{{\bf p}}}.
\label{NonlinearRatio}
\end{equation}
A stationary state is defined by $\epsilon_{{\bf p}}=0$.  The normalized Boltzman rate (\ref{NonlinearRatio}) serves as a low order consistency check for the various kinetic equation derivations. An $O(1)$ value of ${\cal  \epsilon}_{{\bf p}}$ implies that the derivation of the kinetic equation is internally inconsistent.  

We present evaluations of the kinetic equation as discussed in \cite{element}:  we use a numerical scheme designed for off resonant calculations using the \cite{LT2} Hamiltonian and then we take the resonant limit.  See also Section \ref{RepDep}.  We use two spectra that mimic those at Site-D [GM76 with $(a,b)=(4.0,0.0)$] and the Sargasso Sea [$(a,b)=(3.75,0.50)$, Appendix \ref{SargassoSeaSpectrum}].  

The characteristic pattern for GM76 is two positive and one negative lobes in the vertical wavenumber - frequency domain with boundaries at approximately $2f$ and $5f$ (Fig. \ref{NonlinearityParameter}).  Action loss at frequencies $2f< \sigma<5f$ is qualitatively consistent with the PSI mechanism transferring energy into near-inertial frequencies, \citep{MMa}.  The residual at $\sigma=3f$ is, however, nearly an order of magnitude larger than what would be predicted from the PSI time scale \citep{MMa}, which leads to:
\begin{equation} 
{\rm PSI:\;\;\;\;\;\;\;} \epsilon_{\bf p} = \frac{8 \pi^3}{10x} \frac{27}{32} \frac{f}{\sigma}  \frac{m}{m_c} 
\end{equation}
with $x=\sqrt{10}$ and $m_c$ a high wavenumber cutoff defined in (\ref{MC}).  Comparative figures for GM76 can be found in \cite{MB77} (Fig. 10), \cite{M77} (Fig. 3), \cite{MMb} (Figure 11) and \cite{PMW80} (Fig. 6).  

Normalized Boltzman rates within the ID regime $\sigma > 5f$ are both $O(1)$ and positive.  The fact that they are $O(1)$ is indicative that GM76 is {\bf not} a stationary state, and the fact that they are positive is troubling in that the nearby state of (\ref{DB1}) was specifically selected by \cite{MMb} as having ID transfers with $\dot{n}({\bf p}) < 0$ to match a similar signed tendency along the ID/PSI boundary $\sigma = 4f$.  To resolve this we interpret the results in terms of the convergence conditions of the collision integral.

\begin{figure*}
 \noindent\includegraphics[width=20pc]{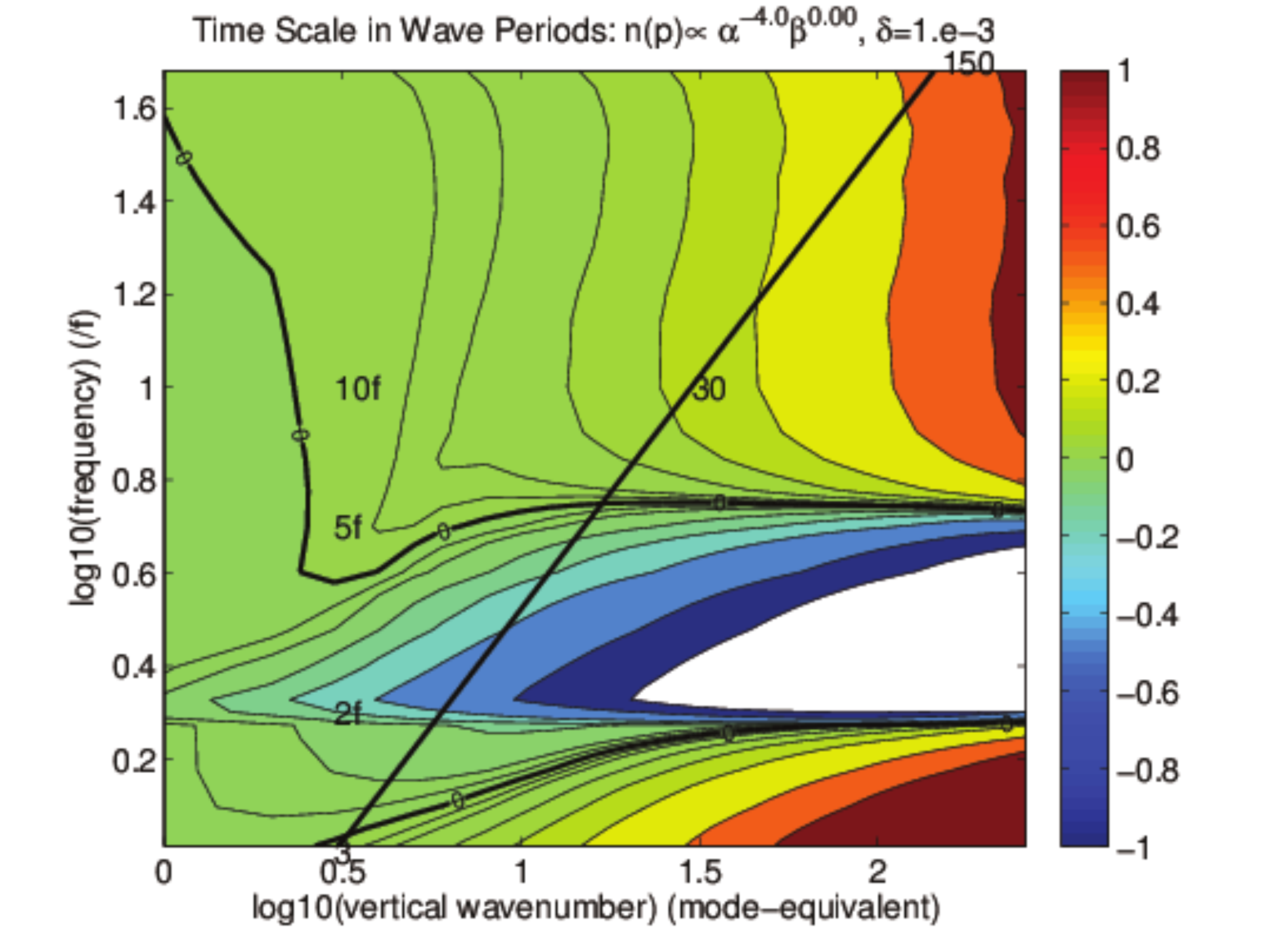}
 \noindent\includegraphics[width=20pc]{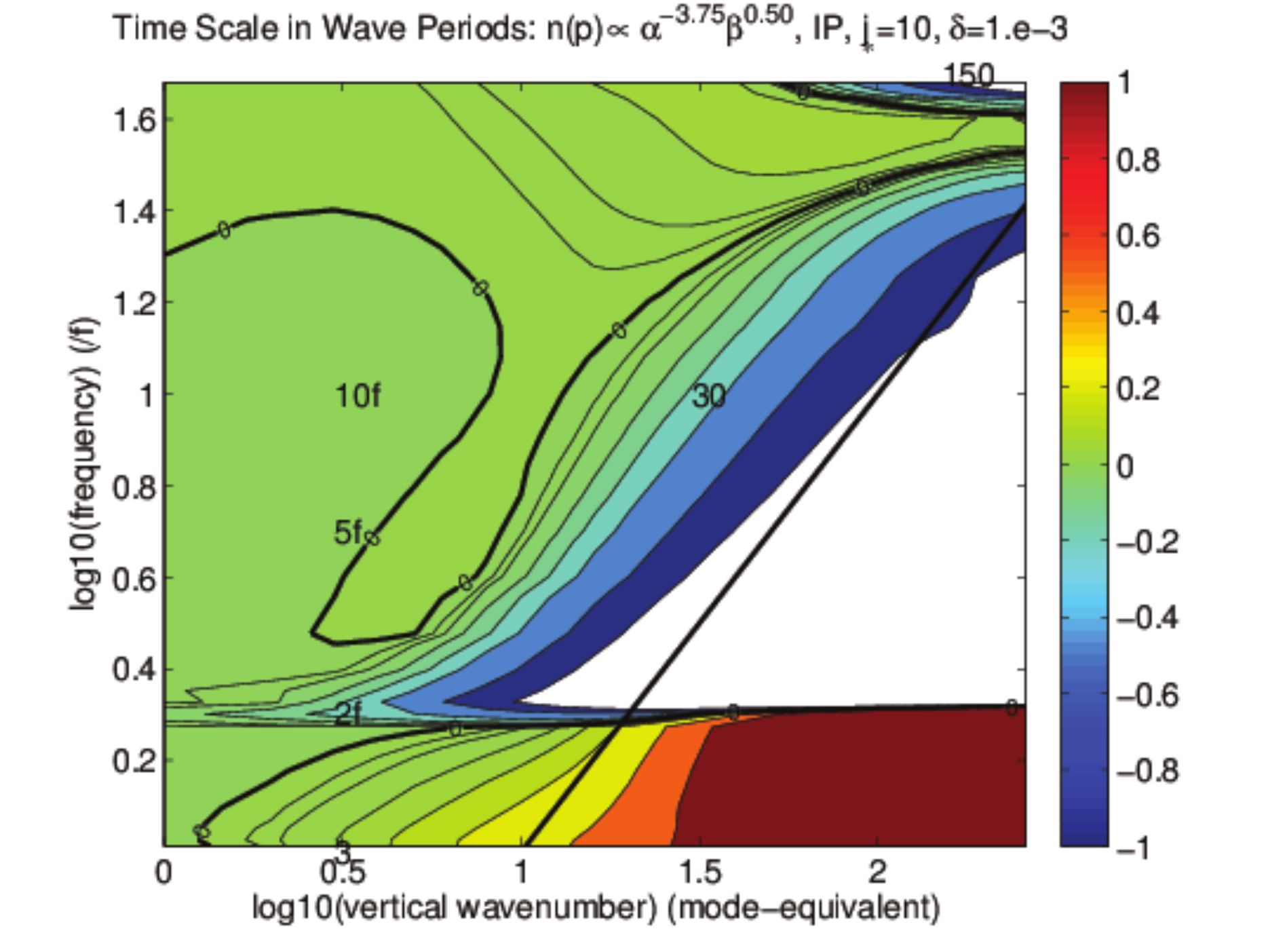}
 \noindent\includegraphics[width=20pc]{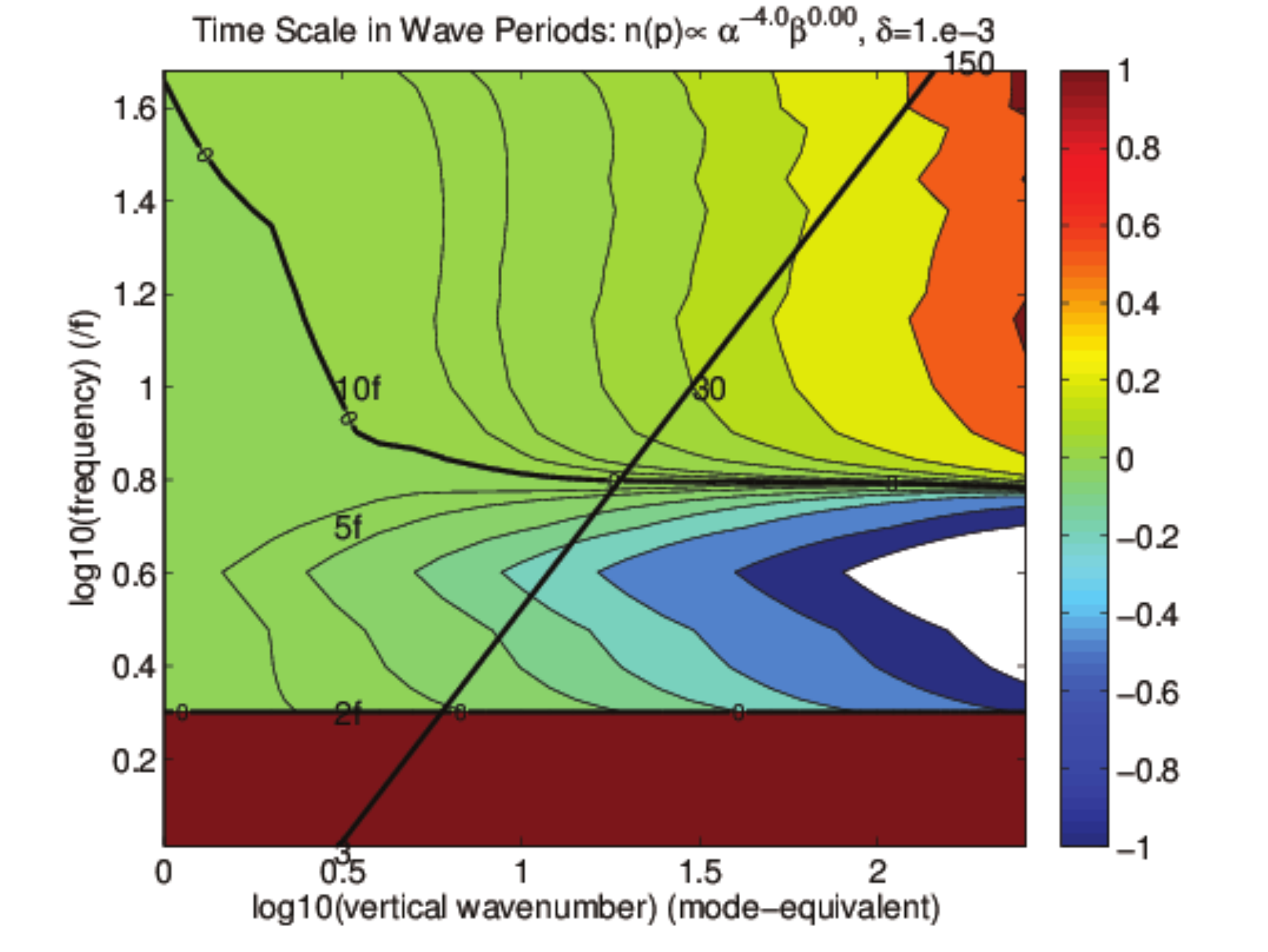}
 \noindent\includegraphics[width=20pc]{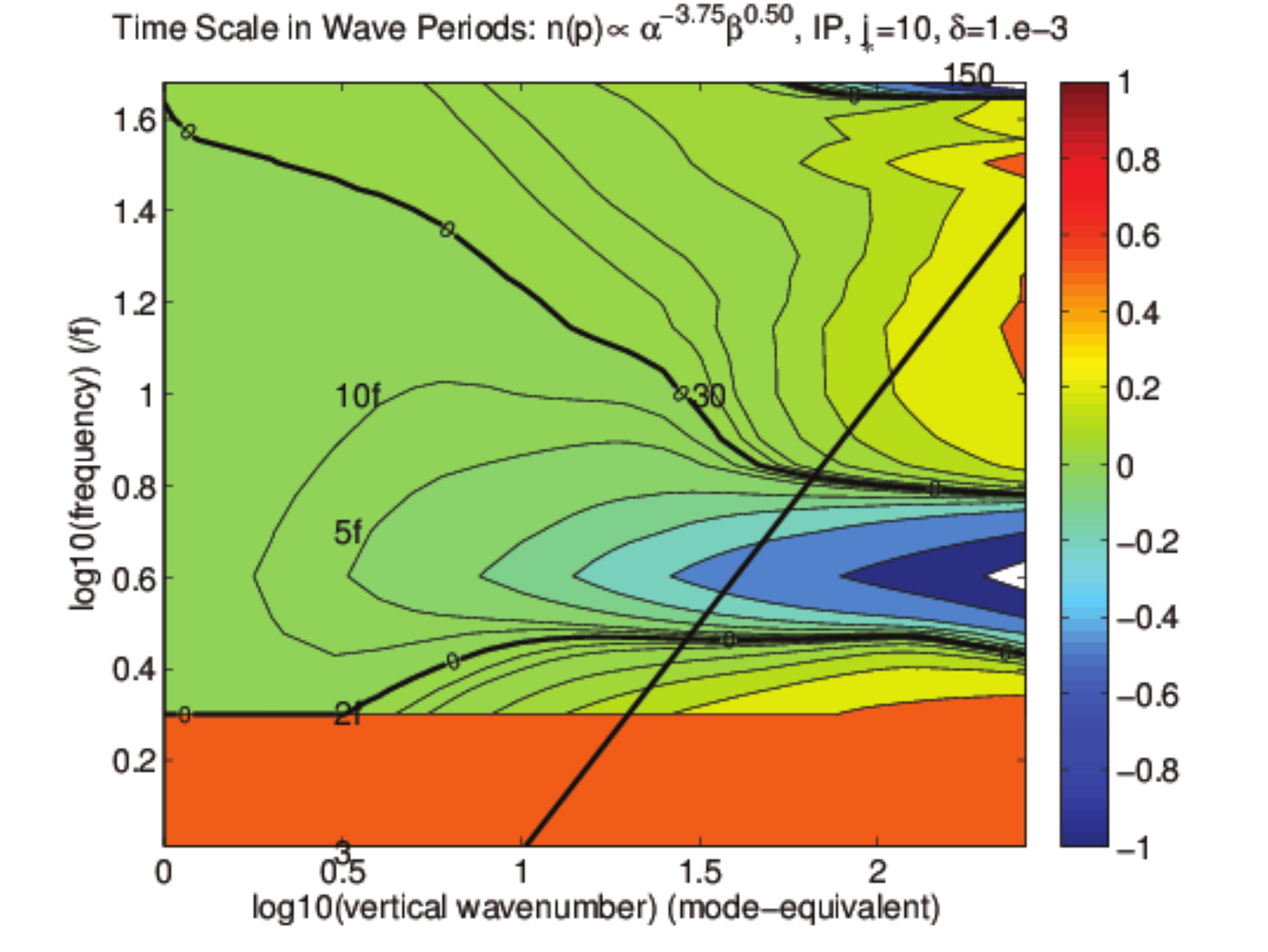}
 \noindent\includegraphics[width=20pc]{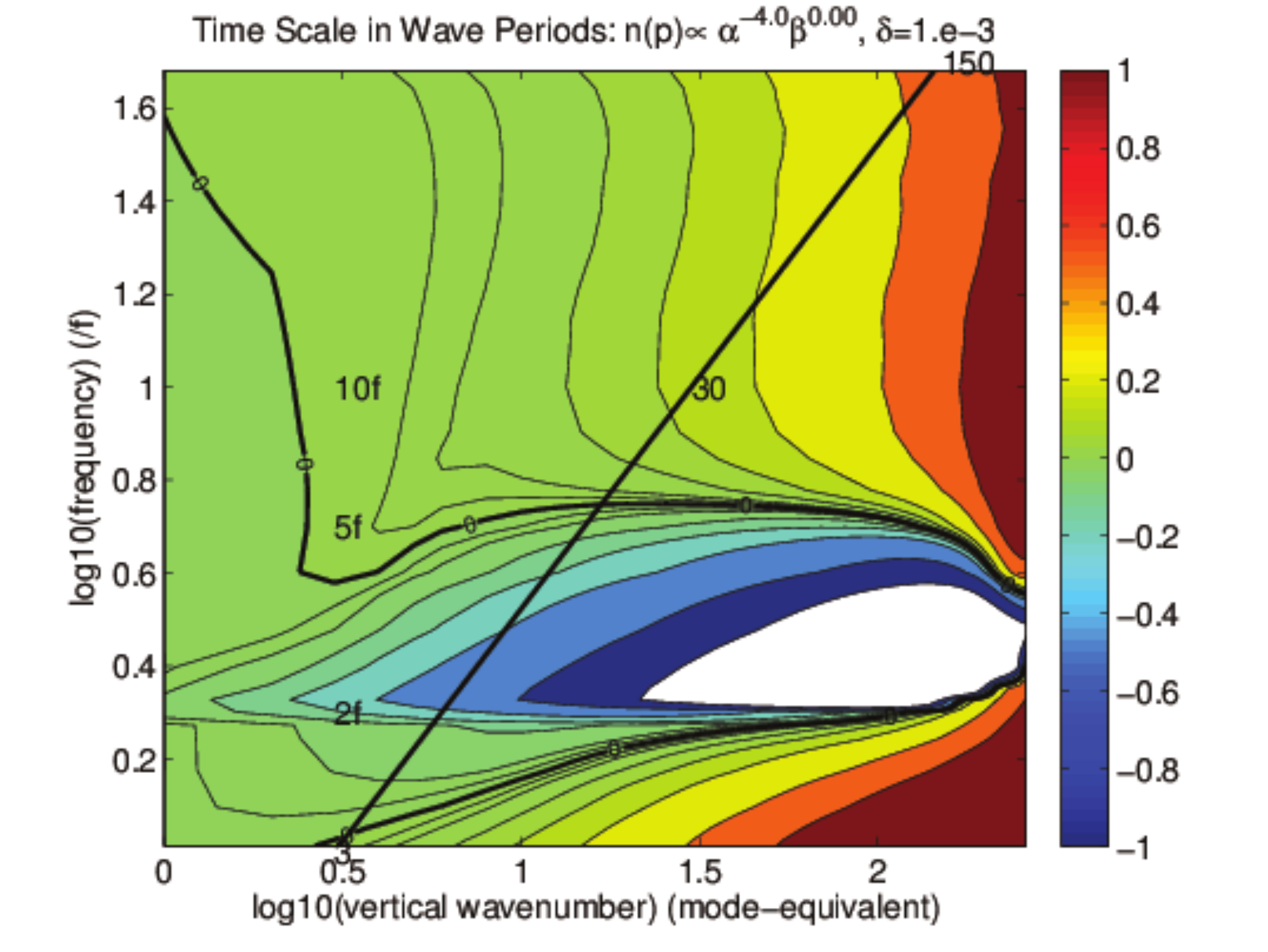}
 \noindent\includegraphics[width=20pc]{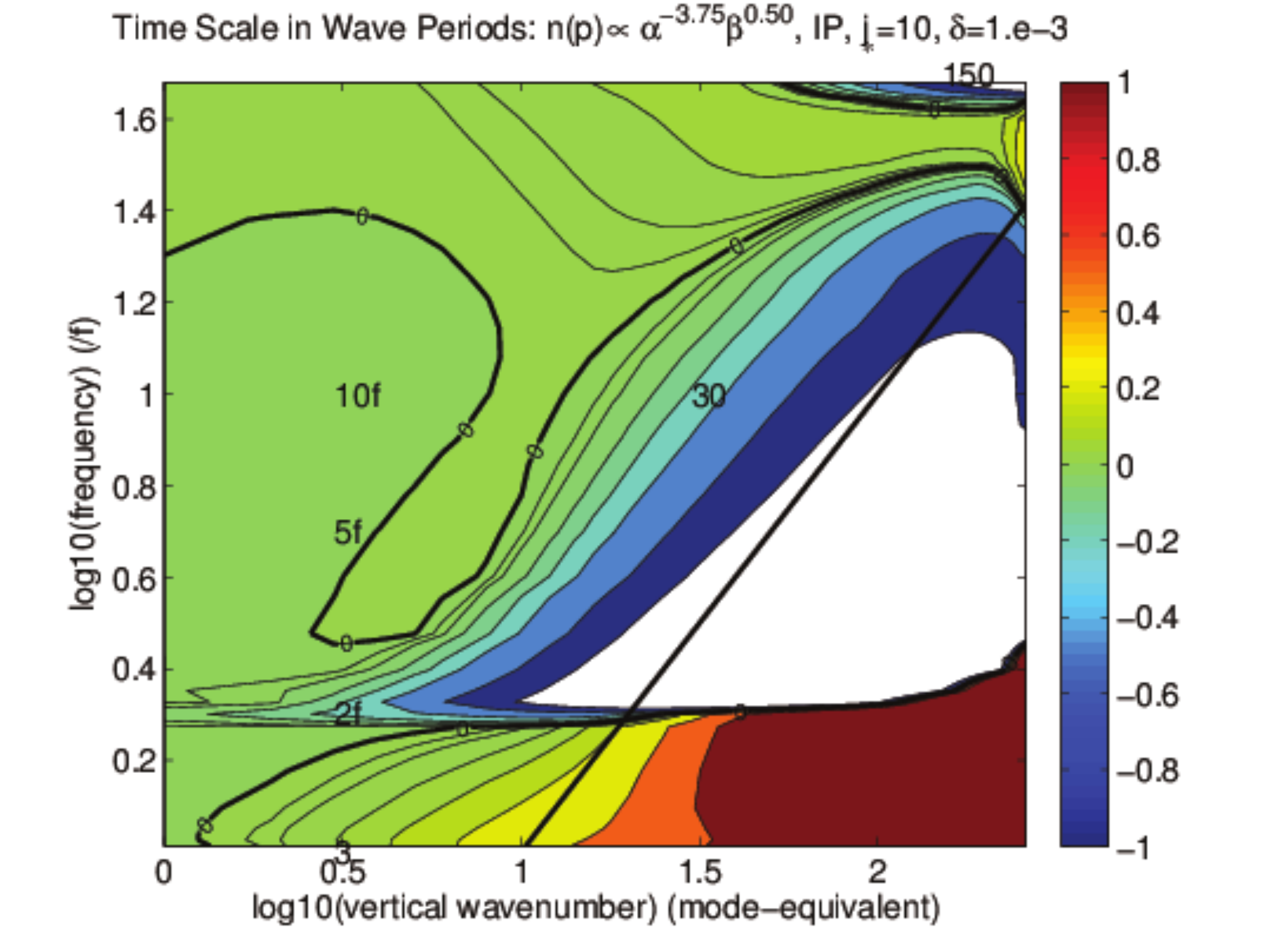}
\caption{Normalized Boltzman rates (\ref{NonlinearRatio}) calculated via (\ref{KineticEquationBroadened}) using the \cite{LT2} matrix elements with $\delta=10^{-3}$.  Left panels use the GM76 (\ref{vertical_spectrum}) spectrum, right panels use Sargasso Sea power laws (Appendix \ref{SargassoSeaSpectrum}).  Upper panels utilize a spectrum that extends to $m=8m_c$, $k=8m_c$, with $m_c$ defined in (\ref{MC}).  Middle panels:  as above, but with $E(\sigma<2f,m)=0.0$.  Lower panels:  as in the upper panels, but with restricted vertical $m < 2 m_c$ and horizontal $k < 2m_c$ wavenumbers.  Results are only plotted for $m < m_c$.  A value of $\delta=10^{-3}$ represents the resonant interaction limit.  The black diagonal lines represent the ID resonance curve ($f/m{\ast}=\sigma/m$) extending from $m_{\ast}$.  Contour values are $(0 , \pm0.01, \pm0.02, \pm0.05, \pm0.1, \pm0.2, \pm0.5, \pm1.0)$. }
\label{BoltzmanRates} 
\end{figure*}

The rigorous asymptotic analysis presented in \cite{theory} finds that both ultraviolet and infrared limits to the collision integral converge along $3.5 < a < 4$ and $b=0$.  Thus interactions along this line segment are local in the nonrotating limit.  With the addition of rotation the mathematical properties of the collision integral could be altered, but note that the line $b=0$ corresponds to an action spectrum with vanishingly small amplitude gradients in vertical wavenumber, with correspondingly small transfers that depend upon those gradients, such as the ID mechanism.  We therefore infer that the positive contours tending vertical at the highest frequencies are unrelated to extreme scale separated ID transfers.  

We double check this inference by noting that locality implies frequencies of $f < \sigma < 2f$ make little contribution to the collision integral.  If such low frequencies are excluded from the numerical evaluations, we find $\dot{n}({\bf p})$ to be virtually unchanged for GM76 at frequencies $\sigma > 4f$ (Fig. \ref{BoltzmanRates}).  Similarly, we find little change if the wavenumber domain is limited to $m < 2 \; m_c$ and $k < 2 \; m_c$.  The influence of extreme scale separated ID transfers in Fig. \ref{BoltzmanRates} would be apparent as contours of $\epsilon_{{\bf p}}$ tending to be parallel to $\sigma/m = $ constant in the vicinity of $m = m_{\ast}\sigma/f$, i.e., at small vertical wavenumbers prior to the asymptotic roll-off.  Such an influence is not apparent in any of these GM76 evaluations.  

Finally, we note that, in the asymptotic limit of high vertical wavenumber, the gradients of the 3-d action spectrum in vertical wavenumber are vanishingly small for $b=0$.  We therefore hypothesize that local interactions giving rise to these residuals are related to the gradients of the 3-d action spectrum in horizontal wavenumber.  A consequence of transfers in horizontal wavenumber is that it potentially changes the energy source at high frequency required by the vertical wavenumber Fokker-Plank equation (\ref{InducedDiffusion})!

The Sargasso Sea power law combination represents a constant action flux solution $9-2a-3b=0$ to the Fokker-Plank equation (\ref{InducedDiffusion}).  Comparative figures for the nearby state of GM75 [$(a,b)=(4.0,0.5)$] can be found in \cite{O76} (Fig. 4), \cite{MB77} (Fig. 12) and \cite{PMW80} (Fig. 6).  A bandwidth of $j_{\ast}=10$ has been chosen as being representative of the region and the amplitude has been specified to return a gradient variance at $\lambda_v = 10 $ m similar to the GM76 spectrum.  The pattern for the Sargasso Sea is, excepting $\sigma \cong N$, a consistent negative tendency for $\sigma > 2f$ and for contours to be parallel to the ID resonance condition of $m_{\ast} = f m / \sigma$.   Tendencies lead to $\epsilon_{{\bf p}} \cong O(1)$ at $m=m_c$.  Such $O(1)$ tendencies are inconsistent with the power law combination being a stationary state, but we note that this is likely an artifact of the high wavenumber, high frequency domain $(m,\sigma)=(m_c,N)$ not being within the asymptotic regime of $\sigma/m \gg f/m_{\ast}$.   

Truncating the spectral domain to exclude frequencies $\sigma < 2f$ changes the tendencies considerably, so that the sign and magnitude are similar to those of GM76.  Truncating the spectral domain to exclude vertical wavenumbers $m > 2 m_c$ changes only the results at highest vertical wavenumber.  These diagnostics suggest the Sargasso Sea results fit much better into the extreme scale separated scenario, but local interactions cannot be neglected.  

\underline{DELETE THE PARAGRAPH BELOW:  AIWEX}\newline
Note the proximity to the line segment $3 < a < 3.5$ along $b=1$ .  This line segment is an IR convergent state in the nonrotating analysis.  We have not investigated interactions in detail for this spectrum.  

\subsubsection{Resonant Interactions:  Departures from Scale Invariance}\label{The_dip}

\subsubsubsection{NATRE:  PSI decay of a tidal peak}\label{PSIdecay}

The parametric subharmonic instability described in Section \ref{ScaleInvariance}.b has also been cited as providing a pathway for the transfer of energy from low-mode internal tides \citep{HNN02, MW05}.  The question of whether such PSI transfers can explain the unique character of the NATRE spectrum is examined in \cite{Natre_lat}.  On the assumptions that $m^{\prime} \ll m,~m^{\prime \prime}$ and $\sigma^{\prime} \sim 2 \sigma \sim 2 \sigma^{\prime \prime} \sim 2 f$, the transfer integral reduces to
\begin{eqnarray}
&& T_r({\bf p}) \cong 2 \int d{\bf p}^{\prime} ~ |V_{{\bf p}^{\prime},({\bf p}^{\prime}-{\bf p})}^{{\bf p}}|^2 ~  \delta[\sigma - \sigma^{\prime} + \sigma({\bf p}^{\prime} - {\bf p})] \nonumber\\
&& \{n({\bf p}^{\prime})[n({\bf p})+n({\bf p}^{\prime} - {\bf p})] - n({\bf p})n({\bf p}^{\prime} - {\bf p})\} 
\end{eqnarray}
for the gain at high wavenumber.  The transfer function $|V_{{\bf p}^{\prime},({\bf p}^{\prime}-{\bf p})}^{{\bf p}}|^2$ in this limit is given by \citep{MMa}:
\begin{eqnarray}
|V_{{\bf p}^{\prime},({\bf p}^{\prime}-{\bf p})}^{{\bf p}}|^2 = \frac{-\pi}{8} \frac{\sigma^{\prime 3}}{\sigma (\sigma^{\prime} -\sigma)} \Big[\frac{\sigma^{\prime 2} -f^2}{\sigma^{\prime 2}}\Big]^2 \frac{\sigma^{\prime 2}-f^2}{N^2} m^{\prime 2}. \nonumber \\
\end{eqnarray}

\cite{MMa} note that with $n({\bf p}) \sim n({\bf p}^{\prime \prime}) \ll n({\bf p}^{\prime})$ the transfer integral further reduces:  
\begin{eqnarray}
&& T_r({\bf p}) \cong -\frac{\pi}{4} [n({\bf p}) + n({\bf -p})] 
\nonumber\\
&& \frac{\sigma_{T}^3}{\sigma (\sigma^{\prime} -\sigma ) } \Big[\frac{\sigma_{T}^2-f^2}{\sigma_{T}^2}\Big]^2 \frac{\sigma_{T}^2-f^2}{N^2} 
\int m^{\prime 2} n(\sigma_{T},m^{\prime}) ~ dm^{\prime} 
\nonumber\\
\label{MM}
\end{eqnarray}
with $n(\sigma,m)$ a 2-D action spectrum and semidiurnal tidal frequency $\sigma_{T}$.  
The difference between \cite{MMa} and (\ref{MM}) is that \cite{MMa} assume $n({\bf p}^{\prime})$ to be a continuous spectrum and here $n({\bf p}^{\prime})$ represents an internal tide, bandwidth limited to a narrow range of frequencies.  Assuming further that the internal tide is concentrated in mode-1 ($m_1$), the characteristic growth rate at high wavenumbers is explicitly given by 
\begin{eqnarray}
T_r({\bf p})/n({\bf p}) = \tau^{-1}({\bf p}) \cong 2\pi  \Big[\frac{\sigma_{T}^2-f^2}{\sigma_{T}^2}\Big]^2 \frac{\sigma_{T}^2-f^2}{N^2} m_1^2 E(\sigma_{T}),
\label{time_scale}
\end{eqnarray}
with $E(\sigma)$ the 1-d energy spectrum.  

The total rate of energy transfer from the semi-diurnal tide to the near-inertial field is
\begin{eqnarray}
\int \sigma T_r d^3k =  \int d^3k ~ E({\bf p}) ~ \tau^{-1} ({\bf p}) = E_{ni}/\tau,
\label{production_1}
\end{eqnarray}
with near-inertial energy $E_{ni}$ and time scale $\tau$ from (\ref{time_scale}).  This assumes that the near-inertial energy $E_{ni}$ is resonant with the tide.  One of the restrictions of the kinetic equation is apparent in (\ref{time_scale}).  The tide can not be regarded as a single plane wave.  Doing so implies $E(\sigma_{T})=E_o\delta(\sigma-\sigma_{T})$, so that the transfer rate is infinite.  Numerical simulations of PSI transfers \citep{MW05} indicate large and perhaps unrealistic transfer rates if forced with a single plane wave.  


When used as a diagnostic, \cite{Natre_lat} report that (\ref{production_1}) tends to overestimate observed dissipation rates.  An additional concern is that the Natre spectrum is unique in both its vertical wavenumber domain power law and extreme bandwidth. Implied is a scale selection process that is {\bf not} represented in (\ref{production_1}), which states that production is proportional to the shear in the tide times the near-inertial energy.  

This can be rectified by noting that the limit $n({\bf p}) ~,~ n({\bf p}^{\prime \prime}) \ll n({\bf p}^{\prime})$ does not need to be taken.  Since $n({\bf p}^{\prime \prime})=n(-{\bf p}-{\bf p}^{\prime})$, if one simply assumes that $n({\bf p}^{\prime \prime}) \cong n(-{\bf p})$, then 

\begin{eqnarray}
&& T_r({\bf p}) \cong  
\frac{\pi}{4} \frac{\sigma_{T}^3}{\sigma (\sigma_{T} -\sigma ) } \Big[\frac{\sigma_{T}^2-f^2}{\sigma_{T}^2}\Big]^2 \frac{\sigma_{T}^2-f^2}{N^2} 
\nonumber \\
&& \{-[n({\bf p}) + n({\bf -p})] \int m^{\prime 2} n(\sigma_{T},m^{\prime}) ~ dm^{\prime} 
\nonumber \\
&& + [n({\bf p}) n({\bf -p})] \frac{\sigma_{T} }{N^2}\int m^{\prime 4} ~ dm^{\prime} \}.
\label{Me}
\end{eqnarray}
Resonance constraints imply 
\begin{equation}
\int m^{\prime 4} ~ dm^{\prime} = \frac{1}{5} m^{5} \mid_{\gamma m_1}^{m},
\end{equation}
in which $\gamma$ is the minimum ratio of $m^{\prime}/m$ permitted by resonance constraints.  The factor $\gamma$ is approximately 1.5-2.0 for Natre.  The term quadratic in near-inertial amplitude represents a scale selection principle.  

A diagnostic for application to observed fields is obtained by summing the wave-action tendency equations for $n({\bf p})$ and $n({\bf p}^{\prime \prime})$, converting from a wavevector representation of the 3-D wave-action spectrum to a 2-D energy spectrum in vertical wavenumber-frequency coordinates and remembering that the observations, in distinction to the theory, reference a one-sided, rather than two-sided, spectrum.  The conversion from 3-D to 2-D assumes that the wavefield is horizontally isotropic.  The end result is that:

\begin{eqnarray}
&& T_r({\bf p}) \cong -\frac{4\gamma n({\bf p}) }{\sigma_{T}} \nonumber\\
&&\displaystyle[\displaystyle\int  m^{\prime 2} E(\sigma_{T},m^{\prime}) ~ dm^{\prime} - \frac{1}{80 \pi} 
\frac{ \sigma_{T}^2}{\sigma^2} m^3 E(\sigma, m) \displaystyle] \nonumber \\
\label{PSI_extended}
\end{eqnarray}
with 
$$\gamma = \frac{\pi}{8} \frac{\sigma_{T}^3}{\sigma (\sigma^{\prime} -\sigma ) } \Big[\frac{\sigma_{T}^2-f^2}{\sigma_{T}^2}\Big]^2 \frac{\sigma_{T}^2-f^2}{N^2}.$$ 

The total transfer from the tide to the near-inertial field is \citep{Natre_lat}:
\begin{eqnarray}
P = \int P({\bf p}) d^3 k = \int  2 \pi [\frac{\sigma_{T}^2-f^2}{\sigma_{T}^2}]^2 \frac{\sigma_{T}^2-f^2}{N^2} E({\bf p}) &&\nonumber\\
\displaystyle\times[ m^{\prime 2} E(\sigma_{T})  - \frac{m^3}{20 \pi}E(\sigma, m) \displaystyle] &&\nonumber\\ \label{production_2}
\end{eqnarray}

\cite{Natre_lat} report a close correspondence between production (\ref{production_2}) and the observed dissipation rate, and the roll-off of the vertical wavenumber spectrum as a balance between 
$$\displaystyle m^{\prime 2} E(\sigma_{T})\displaystyle$$  and 
$$\displaystyle \frac{m^3}{20 \pi}E(\sigma, m) \displaystyle.$$

\subsubsubsection{The Dip:  ID decay of a tidal peak}\label{HF_scale}

\begin{figure*}
\noindent\includegraphics[width=39pc]{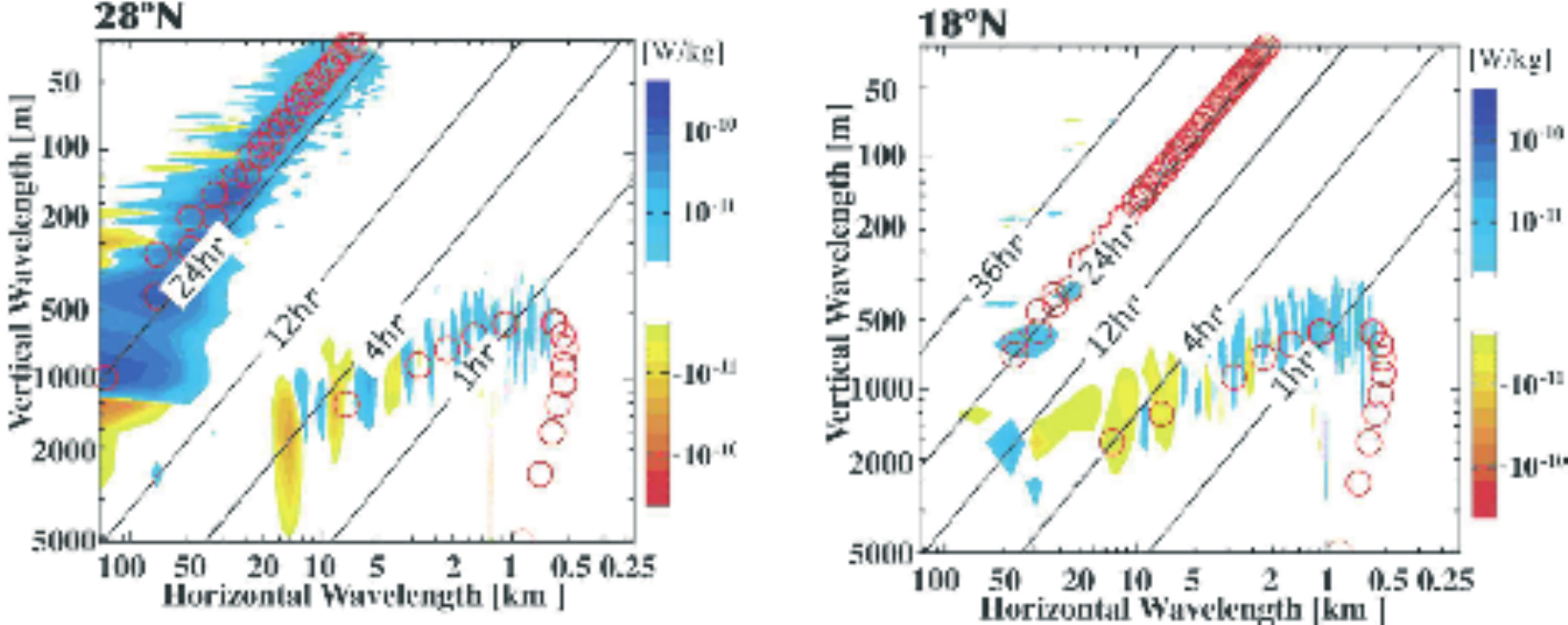} 
\caption{Nonlinear energy transfer rates from the lowest vertical wavenumber $M_2$ internal tide, with blue shades representing energy transfer from the tide to the background wavefield \citep{FHN05} .  The red circles depict the PSI (upper left) and ID (lower right) resonant curves.  Diagonal lines are curves of constant wave period.}  
\label{Toshi_Model} 
\end{figure*}

The same argument used to provide  a scale selection principle in PSI can be invoked for ID.  The term quadratic in high frequency amplitude is dropped in the Induced Diffusion approximation (\ref{InducedDiffusion}).  This term represents a consistent energy loss from the high frequency field to the low frequency tide.  In this instance the quadratic term is largest when the scale separation is smallest and the arguments used to produce the ID approximation break down.  \cite{FHN05} report an analysis of energy transfers between a tide and a background sea of internal waves initialized as the GM spectrum.  They find an energy loss from the tide to the high frequency field along the ID resonance curve at relatively high frequencies and an energy gain by the tide from the high frequency field along the ID resonance curve at relatively low frequencies (small scale separations) (Fig. \ref{Toshi_Model}).  

Such a pattern of gain and loss is consistent with the mid-frequency dip being a product of nonlinear interactions with the tide, as conjectured by \cite{L02}.  The essential question is whether this energy loss along the ID resonance curve is sufficiently large in comparison to other effects.  

\subsubsubsection{The Dip:  Departure from scale invariance}

If the pattern of loss and gain in Fig. \ref{BoltzmanRates} is time stepped in the radiation balance equation, it will, at least initially, tend to result in a pattern resembling the mid-frequency dip, e.g., Fig. \ref{Tropics-freq}.  Much depends upon the magnitude and direction of nonlinear transports and distribution of sources within the spectral domain / along the spectral boundaries.  It would help to understand how the mid-frequency dip is geographically distributed and related to power law combinations in order to understand whether the mid-frequency dip results from nonlinearity.  


\subsubsection{Resonance Broadening and Coordinate Representations}\label{ResonanceBroadening}

Numerical evaluations of (\ref{KineticEquation}), Fig. \ref{BoltzmanRates}, suggest that neither GM76 nor the Sargasso Sea spectrum is a stationary state, and this poses real problems for our hypothesis that the observed spectra are in approximate equilibrium.  A distinct possibility is that, since normalized Boltzman rates  for that calculation are $O(1)$, the character of the interaction is altered by non-resonant interactions.  We work through this issue in the following four subsections.  

\subsubsubsection{Representation Dependences}\label{RepDep}

The issue of representation (coordinate) dependent differences arises naturally in nonlinear problems.  Here it implies a question of whether a statistically stationary state in one coordinate system similarly represents a statistically stationary state in another.  

Kinetic equations have been derived in Eulerian, isopycnal and Lagrangian coordinate systems.  \cite{element} find that  Eulerian, isopycnal and Lagrangian coordinate matrix elements in the hydrostatic, non-rotating $f^2 \ll \sigma^2 \ll N^2$ limit are equivalent {\bf on the resonant manifold}.  With rotation, comparisons are restricted to isopycnal coordinate formulation of \cite{LT2} and the Lagrangian coordinate formulation of \cite{O73, M75}.  Strikingly, the $f \ne 0$, hydrostatic resonant expressions are also equivalent.  Thus representation dependent differences are not apparent at the level of the RIA.  

Some physical sense can be made by noting that the resonant interaction approximation assumes, perforce, an expansion in terms of a non-advected wavefield, with linear dispersion relation (\ref{DR}).   It is a description of the wavefield as a system of coupled oscillators with distinct resonance conditions.  

Non-resonant kinetic equations are limited to use of isopycnal coordinates.  The isopycnal Hamiltonians in \cite{LT} and \cite{LT2} are {\bf explicitly} expressed in terms of the the generalized coordinates and momenta.  The generalized coordinates and momenta are {\bf implicit} functions in the Lagrangian coordinate system formulations of \cite{O76}, \cite{MB77} and \cite{PMW80} and the Eulerian coordinate system formulation of \citep{Voronovich}.  An implicit representation requires an expansion in powers of small fluid parcel displacements in addition to an assumption of weak nonlinearity.  This small amplitude assumption represents an unconstrained approximation whose domain of validity {\em vis-a-vis} the assumption of weak nonlinearity is not well defined, \citep{M86}.  This matters when the resonances are broadened, Section \ref{Renormalization}.  In contrast, the explicit isopycnal coordinate representation of \cite{LT2} requires only an assumption of weak nonlinearity.  

The next level of approximation to the kinetic equation assumes wave amplitudes vary in time, $a=a(\tau)$, and the energy conserving delta-functions in (\ref{KineticEquation}), $\delta({\sigma_{{\bf p}} -\sigma_{{{\bf p}_1}} - \sigma_{{{\bf p}_2}} })$, consequently need to be ``broadened'' to take near resonant interactions into account:  When the resonant kinetic equation is derived, it is assumed that the amplitude of each plane wave is
constant in time, or, in other words, that the lifetime of single
plane wave is infinite. The resulting kinetic equation, nevertheless,
predicts that the wave amplitude changes. For small levels of nonlinearity this
distinction is not significant, and resonant kinetic equation
constitutes a self-consistent description. For larger values of
nonlinearity this is no longer the case, and the wave lifetime is
finite and amplitude changes need to be taken into account.
A mathematical interpretation is that resonance broadening represents a bandwidth $\Delta \sigma$ connected to the temporal localization through a Fourier uncertainty principle , i.e., the product of wave lifetime $\tau$ and bandwidth $\Delta \sigma$ is given by $\tau \Delta \sigma = 2 \pi$. This effect is larger for stronger levels of the normalized Boltzman rate (\ref{NonlinearRatio}).  The next section describes attempts to incorporate resonance broadening effects into the kinetic equation.   

\subsubsubsection{Renormalization of the Kinetic Equation}\label{Renormalization}

Inclusion of resonance broadening requires a renormalization of the
kinetic equation (\ref{KineticEquation}).  
\cite{DWW82, DWW84} tackle this starting from the Lagrangian of \cite{O76} and carrying forward their calculations in the context of the Direct Interaction Approximation (DIA).  Their nominally finite amplitude analysis still retains the small amplitude approximation inherent in the Lagrangian formulation.  
\cite{CF83} begin from the Eulerian equations of motion to derive the DIA closure equations for internal waves in two dimensions and without rotation through the application of renormalization theories originally formulated for quantum and classical statistical field theory.  A similar tack is pursued by \cite{LLNZ} and a general representation for a 3 wave system is given in terms of a Hamiltonian representation.  There the Wyld diagrammatic technique adapted
for the statistical description of a wave interaction system was used.  This
approach allows the Dyson-Wyld re-summation of the reducible infinite
class of diagrams, which presents certain parts of the nonlinear
interactions as effectively being linear on average. As a result, 
off-resonant interactions can effectively contribute to long-time
statistical averages. This result is obtained by
analytical resummation of the infinite diagrammatic series for the
Green's function and double correlator. This result
is given by
\begin{eqnarray}
\frac{d n_{{\bf p}}}{dt} = 4 \int
 |V_{{\bf p}_1,{\bf p}_2}^{{\bf p}}|^2  f_{p12} {\cal F}({\bf p} - {\bf p}_1-{\bf p}_2)
d^3 {\bf p}_{1} d^3 {\bf p}_{2} & 
\nonumber \\
-4\int
  |V_{{\bf p}_2,{\bf p}}^{{\bf p}_1}|^2 f_{12p} {\cal F}({\bf p}_1 - {\bf p}_2-{\bf p})
d^3 {\bf p}_{1} d^3 {\bf p}_{2} & 
\nonumber \\
-4\int
  |V_{{\bf p},{\bf p}_1}^{{\bf p}_2}|^2 f_{2p1}  {\cal F}({\bf p}_2 - {\bf p}-{\bf p}_1)
 d^3 {\bf p}_{1}  d^3 {\bf p}_{2}&
  ,\nonumber\\
\label{KineticEquationBroadened}
\end{eqnarray}
with ${\cal F}({\bf q})=\delta_{\bf q} {\cal L}(\sigma_{\bf q})$.  
Here ${\cal L}$ is defined as
\begin{equation}
{\cal{L}}(\Delta\sigma)  = 
\frac{\Gamma_{k12}}{(\Delta\sigma)^2 + \Gamma_{k12}^2},
\label{scriptyL}
\end{equation}
where $\Gamma_{k12}$ is the total broadening of each particular resonance and $\Delta \sigma$ 
represents the residual $\Delta \sigma = {\sigma_{{\bf p}} -\sigma_{{{\bf p}_1}} - \sigma_{{{\bf p}_2}} }$, etc..


%
The width of the resonance  $\Gamma_{k12}$ in (\ref{scriptyL}) is  given by 
$$\Gamma_{k12}=\gamma_{{\bf p}}+\gamma_{{\bf p}_1}+\gamma_{{\bf p}_2}.$$ 
It means that the total resonance broadening is the sum of individual
frequency broadening, and can be thus seen as the ``triad
interaction'' frequency.  The direct interaction approximation customarily sets $\gamma_{{\bf p}}=0$.  

The single frequency renormalization is 
calculated {\bf self-consistently} from
\begin{eqnarray}
 \gamma_{{\bf p}} & = &\nonumber \\
& \displaystyle 4 \int
 |V_{{\bf p}_1,{\bf p}_2}^{{\bf p}}|^2  (n_{{\bf p}_1}+
n_{{\bf p}_2})  {\cal F}({\bf p} - {\bf p}_1-{\bf p}_2)
d{\bf p}_{1} d{\bf p}_{2} & 
\nonumber \\
& \displaystyle -4\int
  |V_{{\bf p}_2,{\bf p}}^{{\bf p}_1}|^2 (n_{{{\bf p}}_2}  -
  n_{{{\bf p}}_1})  {\cal F}({\bf p}_1 - {\bf p}_2-{\bf p})
d{\bf p}_{1} d{\bf p}_{2} & 
\nonumber \\
& \displaystyle -4\int
  |V_{{\bf p},{\bf p}_1}^{{\bf p}_2}|^2 (n_{{{\bf p}}_1}- n_{{{\bf p}}_2})
  {\cal F}({\bf p}_2 - {\bf p}-{\bf p}_1)
d{\bf p}_{1} d{\bf p}_{2} & 
\nonumber\\
\label{Gamma}
\end{eqnarray}
Note that if the nonlinear frequency renormalization tends to zero, i.e., $\Gamma_{k12} \to 0$, ${\cal L}$ reduces to the delta function: 
$$\lim\limits_{\Gamma_{k12}\to 0} {\cal{L}}(\Delta\sigma)  =\pi 
\delta(\Delta\sigma).$$
In the limit of no broadening 
(\ref{KineticEquationBroadened}) reduces to (\ref{KineticEquation}).  

A self-consistent estimate of $\gamma_{{\bf p}} $ requires the iterative solution of (\ref{KineticEquationBroadened}) and (\ref{Gamma}) over the entire field: the width of the resonance (\ref{Gamma}) depends on the lifetime of an individual wave [from (\ref{KineticEquationBroadened})], which in turn depends on the width of the resonance (\ref{scriptyL}).  The result is a numerically intensive computation which has yet to be undertaken.  Instead, \cite{element} make the uncontrolled approximation that:
\begin{equation}
\gamma_{{\bf p}} = \varepsilon \sigma_{{\bf p}}
\label{GammaFraction}
\end{equation}
with $\varepsilon$ taken as $10^{-3}$ and larger.   

\begin{figure*}
 \noindent\includegraphics[width=20pc]{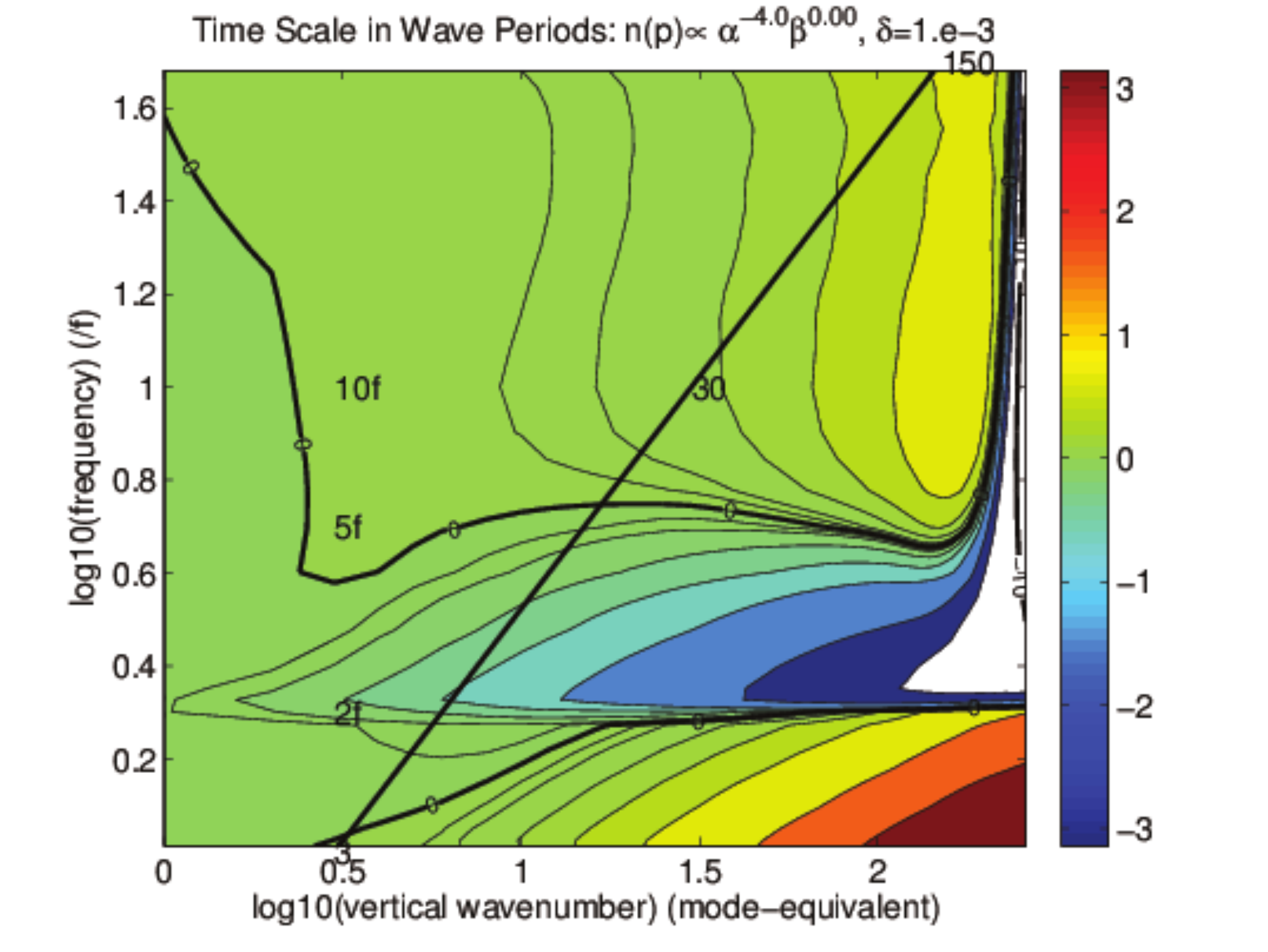}
 \noindent\includegraphics[width=20pc]{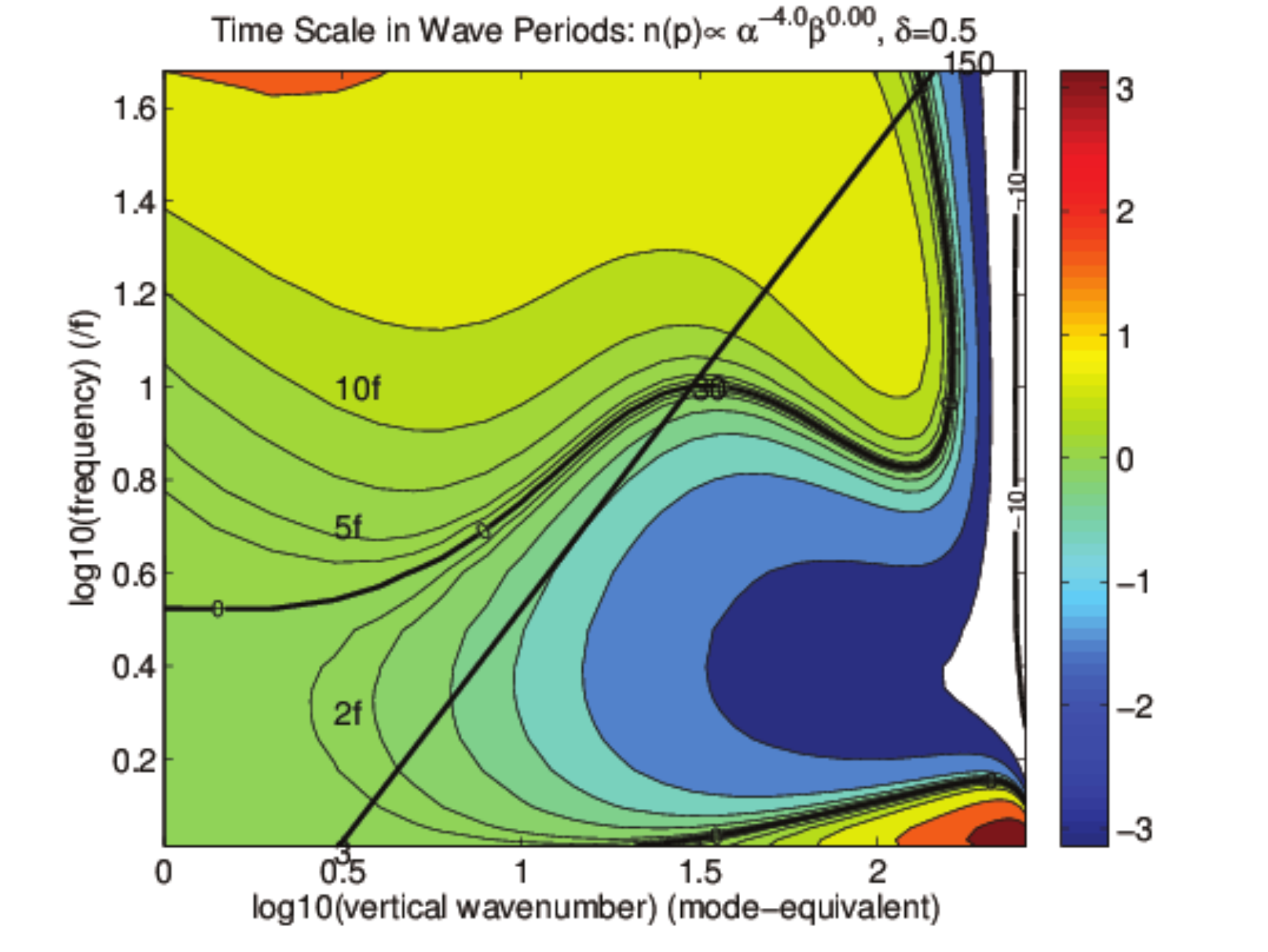}
\caption{Normalized Boltzman rates (\ref{NonlinearRatio}) for the GM76 spectrum calculated via (\ref{KineticEquationBroadened}) using the \cite{LT2} matrix elements with $\delta=10^{-3}$ (left)  and $\delta=5\times10^{-1}$ (right), \cite{element}.  A value of $\delta=10^{-3}$ represents the resonant interaction limit.  The spectra have been extended beyond $m>m_c$ by rolling off the spectrum as GM$\times m_c/m$.  }
\label{NonlinearityParameter} 
\end{figure*}

The patterns of gain and loss at $\varepsilon=1\times10^{-3}$ (Fig. \ref{NonlinearityParameter}) are quantitatively similar to previously published results including only resonant interactions, e.g., \cite{O76}.  Minimal variation is noted in evaluations using $\varepsilon=1\times10^{-3}$ and $\varepsilon=1\times10^{-2}$, implying that the $\varepsilon=1\times10^{-3}$ evaluations represent the resonant interaction limit.  Normalized Boltzman rates are $O(1)$ at high wavenumber, and in particular within the high frequency part of the spectral domain which is nominally a stationary state in the ID approximation.  \citep{element} find that normalized Boltzman rates are reduced by factors of 2-3 with $\varepsilon=0.5$ (See their Fig. 5).  Here we present results for which it is difficult to claim resonance broadening significantly decreases the normalized Boltzman rate (Fig. \ref{NonlinearityParameter}).  The difference between results presented in \cite{element} and in Fig. \ref{NonlinearityParameter} is that \cite{element} extend the GM76 spectrum to $2m_c$, zeroing the spectrum at higher vertical wavenumber.  Here we roll off the GM76 spectrum as GM76$\times m_c/m$ for $m > m_c$, zeroing the spectrum only for $m > 8m_c$.  We regard this as being a much more realistic characterization of the observations.  Clearly, the numerical evaluations are sensitive to the treatment of the spectral boundaries.  Experimentation suggests numerical evaluations of the broadened kinetic equation are also sensitive to the functional representation of $\gamma_{{\bf p}}$ other than (\ref{GammaFraction}).  

The iterative calculation needs to be carried out in order to address whether the patterns (\ref{NewIDcurve}) associated with the Induced Diffusion mechanism (Fig. \ref{fig:everything}) are more than simply fortuitous.  


\subsubsubsection{Ray-tracing models}\label{Eikonal}
The interpretation offered immediately above is one that attempts to define the breadth of the resonance associated with a forced-damped system of weakly coupled oscillators, in which 'forcing' and 'damping' can be construed as the scattering of energy into and out of $({\bf p},\sigma)$.  An alternative description is to view the wavefield as a collection of wave-packets, i.e., as particles.  In this particle perspective, the time evolution of a wave packet (a 'test' wave) is proportional to gradients of the Doppler shift as one evaluates the evolution of the wavenumber, frequency and amplitude of the wave packet along its trajectory, in contrast to the coupled oscillator construct that assumes only an amplitude modulation.  These ray methods do not make a weak interaction assumption.  They do, however, make scale separation and small amplitude assumptions.  

The ray equations \citep{Witham} are: 
\begin{equation}
{\bf {\cal R}} \equiv  \frac{d {\bf p}}{dt} = -\nabla_{\bf r} \omega -(\nabla_{\bf r} \overline{{\bf u}}) \cdot {\bf p}
\label{WavenumberTendency}
\end{equation}
for the time evolution of a wavevector following a wavepacket, and 
\begin{equation}
\frac{d {\bf r}}{dt} = \overline{{\bf u}} + \nabla_{{\bf p}} \omega 
\label{Position}
\end{equation}
for the position of the wavepacket, which varies as the sum of advection plus group velocity.  The background velocity $\overline{{\bf u}}$ is assumed to be slowly varying in comparison to the wave phase and is treated as time-dependent stochastic realizations based upon the GM model in this wave-wave interaction problem.  The intrinsic frequency 
\begin{equation}
\omega = \sigma - {\bf p} \cdot \overline{ {\bf u} }
\end{equation}
is given by a linear dispersion relation (\ref{DR}).  The amplitude of the wavepacket is assumed to be small and obeys an action conservation principle:  
\begin{equation}
\; \frac{\partial n(\bf{p})}{\partial t} + {\bf \nabla}_{\bf r} \cdot (\overline{{\bf u}}+{\bf C_g}) n({\bf p}) + 
{\bf \nabla}_{\bf p}  {\bf \cdot} {\bf {\cal R}} n(\bf{p})  = 0. 
\label{Amplitude}
\end{equation}
A rigorous derivation \citep{Witham} is based upon the concept of phase $({\bf p} \cdot {\bf r} - \sigma t)$ conservation for a single wave packet rather than energy $(\Sigma_{i=1}^{3} \sigma_i \dot{n}_i = 0)$ and momentum $(\Sigma_{i=1}^{3} {\bf p}_i \dot{n}_i = 0)$ conservation of a triad system.  The extent to which the dynamics inherent in the kinetic equation (\ref{KineticEquationBroadened}) are captured by the restricted description (\ref{WavenumberTendency})-(\ref{Position})-(\ref{Amplitude}) has not been fully elucidated.  

The ray equations do not constitute a transport representation in (\ref{rbe}), which attempts to describe the average evolution of a field, rather than the evolution of a single wavepacket.  To this end, \cite{FHW85} report that test wave spectra exhibit a tendency to relax to one-dimensional vertical wavenumber spectra proportional to $m^{-2}$ (GM76) and one-dimensional horizontal wavenumber spectra proportional to $k_h^{1/2}$ times GM76, independent of whether the background was consistent with the GM model.  This is the solution to the Kinetic Equation (\ref{KineticEquation}) defined by \cite{PelinovskyRaevsky} and that which appears in the dimensional analysis of \cite{P04a}.  The model diagnostics of \cite{FHW85} indicate action transport to higher frequency which is inconsistent with the induced diffusion model of action transport to higher vertical wavenumber without transport in horizontal wavenumber, i.e., transport to lower frequency.  These tendencies are captured by a cascade representation of $T_r$ found in \cite{P04a}:
\begin{equation}
\omega T_r({\bf p}) = \frac{\partial}{\partial m} \LSBA 0.2 m^4 N^{-1} \LSBA \frac{\omega^2-f^2}{N^2-\omega^2}\RSBA^{1/2} E(m) E({\bf p}) \RSBA ~. 
\label{Cascade} 
\end{equation}
The cascade closure has been used in combination with PSI forcing of the inertial field  \citep{Natre_lat} to obtain a link between power law dependencies in the vertical wavenumber and frequency domains.  This closure is also used to discuss the near-boundary decay problem in \cite{P04b}.  It is equivalent to the ray-tracing based formulas employed in diagnostic studies of dissipation parameterizations \citep{G89, P95, G03}.    

The most important point is the frequency domain weighting of energy transports to small vertical scales in (\ref{Cascade}):
\begin{equation}
\int_f^N \sigma T_r(m,\sigma) d\sigma \propto \int_f^N \LSBA \frac{\omega^2-f^2}{N^2-\omega^2} \RSBA^{1/2} E(\omega) d\omega 
\end{equation}
Despite the small spectral density at high frequency, high frequencies make a significant contribution to the total spectral transports.  

\subsubsubsection{Numerical Simulations}

Direct numerical simulations of the equations of motion are not limited by the dynamical assumptions inherent in either the weakly nonlinear or eikonal representations.  They are, however, subject to computational restrictions.  

\cite{WD97} present spin-down simulations based upon the GM76 spectrum with varying amplitude.  They regard their results as being consistent with the magnitude ($\epsilon = 7\times10^{-10}$ W/kg) and scaling of \cite{G89}, $\epsilon \sim E^2N^2$.  \cite{P95} and \cite{P04a} argue that theoretical estimates based upon the eikonal model of \citep{HWF} and the dynamic balance of \cite{MMb} are indistinguishable from the observational results.  Thus, both observations, \cite{HWF}, \cite{MMb} and \cite{WD97} are consistent.  The domain considered by \cite{WD97} consists of a rectangular box $80\times10\times1$ km on a side with resolved wavelengths of 1 km in the horizontal and 50 m in the vertical.  Note that this domain does not include regions in Fig. \ref{BoltzmanRates} exhibiting large values of the normalized Boltzman rate.  Interactions in the resolved domain may be dominated by PSI transfers as discussed in \cite{MMb}.  

Forced non-rotating simulations are presented in \cite{F03}.  The computational domain is a box of horizontal size $100\times100\times128$ m height.  The forcing is isotropic in wavenumber and peaks at a horizontal wavelength of 25 m.  The forcing is controlled so that amplitudes are consistent with GM76 and the resulting dissipation is a significant fraction of that associated with GM76.  While we are uncertain about how to relate these forced simulations to the wave-wave interaction problem, we believe finding significant dissipation at high frequencies is credible [see (\ref{Cascade}) and Section \ref{Dissipation_model}].  Also notable are findings that wavenumber-local 
interactions dominate the transfers and that energy transport to higher horizontal wavenumbers (Section \ref{ScaleInvariance}.e) is a robust feature of the  simulations whereas energy transport in vertical wavenumber is strongly dependent upon spectral shape.  

Quantitative estimates of spectral transports in high resolution and realistically forced direct numerical simulations is a high priority, especially in conjunction with efforts to understand the limitations of a broadened kinetic equation (\ref{KineticEquationBroadened}).  


\subsubsection{Does the Continuum Exist?}\label{PinkelFantasy}

Eulerian frequency spectra at high vertical wavenumber are contaminated by vertical Doppler shifting:  near-inertial frequency energy is Doppler shifted to higher frequency at approximately the same vertical wavelength.  High frequency, small horizontal scale internal waves are also easily advected past a similar Eulerian observer by the horizontal velocity of near-inertial waves.  Such differences motivate our concern with different coordinate systems in Section \ref{RepDep}.  

\begin{figure}\vspace{-0.75cm}
\noindent\includegraphics[width=20pc]{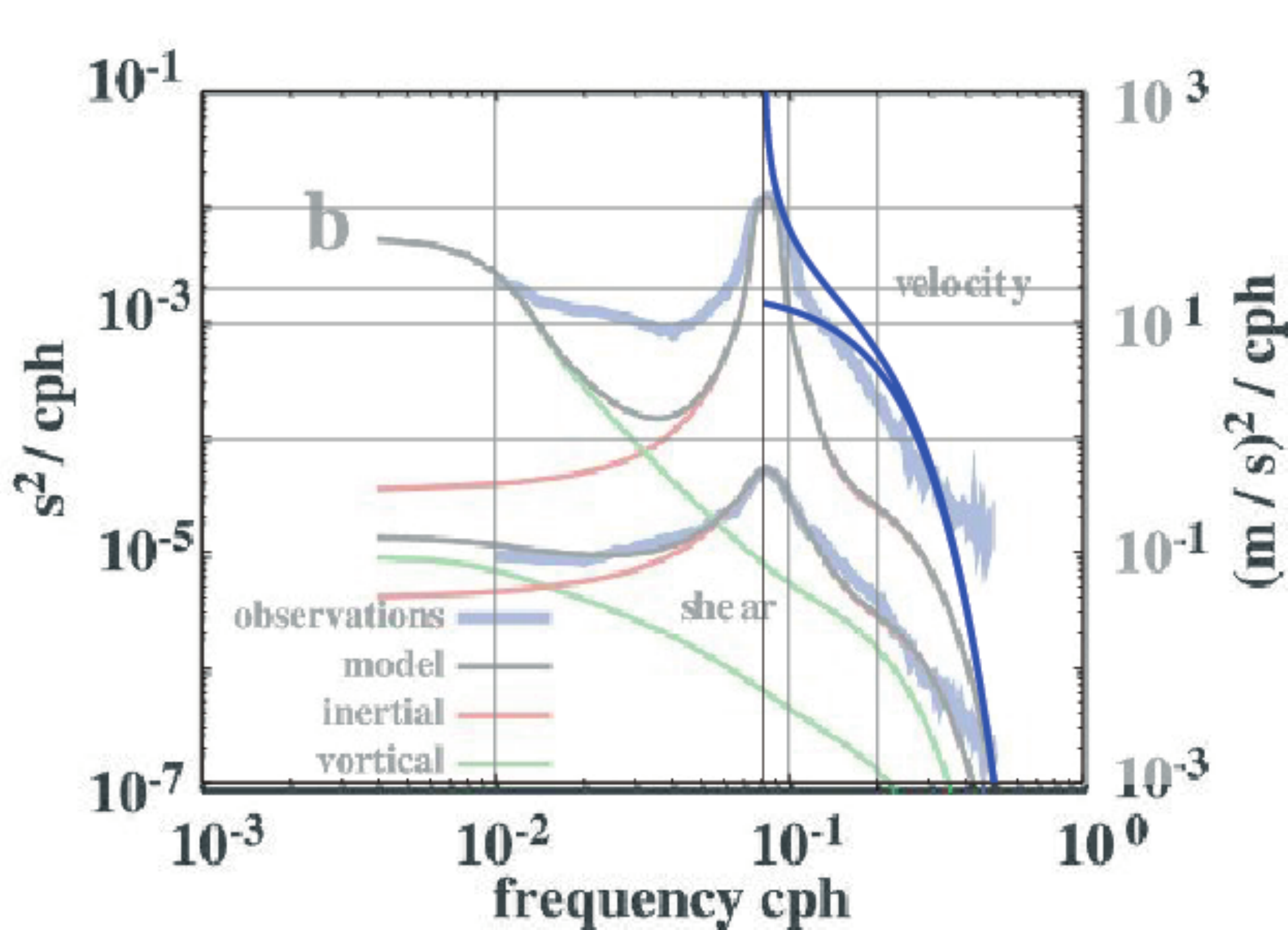}\vspace{-0.75cm}
\caption{Fig. 4b from \cite{Pinkel08} with a displacement variance overlay and associated velocity spectrum, assuming the Gaussian distribution given by 
\cite{Pinkel08} with 2 hour lag correlation and 1.25 m$^2$ displacement variance.  
The associated velocity spectrum (upper curve within the overlay) lies above the observed velocity 
spectrum.  A buoyancy frequency of $N= 6$ cph has been assumed.  
\label{Pinkel_Sheba} }
\end{figure}

\cite{Pinkel08} posits that the apparently continuous wavenumber-frequency spectrum at high vertical wavenumber results from the advective smearing of variance contained in a few discrete spectral lines.  A ``simple'' model is introduced to quantify the effects of lateral advection, and both random and deterministic vertical advection on these spectral lines.  Two data sets, including one from the Arctic (the Surface Heat Budget of the Arctic [SHEBA], Fig. \ref{Pinkel_Sheba}) were used to demonstrate the model.  

Model implementation includes arbitrary choices of functional forms for lag coherence estimates and choosing model parameters to maximize the high frequency shear variance which would be interpreted as resulting from Doppler smearing.  Further observational constraints, such as requiring the vertical displacement spectrum to be consistent with the observations, could have been implemented.  As presented, the model's vertical displacement spectrum appears about a factor of two larger than expected (Fig. \ref{Pinkel_Sheba}).  

The SHEBA spectrum is quite different from the AIWEX spectrum.  The SHEBA frequency domain power laws are much steeper.  This is possibly related to time dependence  in forcing or nonlinearity (the SHEBA spectrum was obtained in August-September 1998, AIWEX was conducted in March-May 1985 and \cite{Pinkel05} notes significant temporal variability in the stratification with the recent buildup of a seasonal thermocline prior to the SHEBA observations) and/or to spatial variability (the reported SHEBA observations were obtained atop the continental shelf whereas the AIWEX data represent deep ocean conditions).  The AIWEX data \cite{L90} appear to argue strongly for a high frequency, high wavenumber continuum.  

While there assuredly is some contamination of the high vertical wavenumber continuum by Doppler shifting, the goodness of fit in \cite{Pinkel08} can not be interpreted as evidence for the nonexistence of the continuum.  If robust, the goodness of fit simply implies that the velocity observations are consistent with the proposed model.  Further consistency tests based upon both density and velocity fields could demonstrate otherwise.  Note that the 2-D PATCHEX and SWAPP spectral parameters (Section \ref{PinkelsAcres}) were extracted from $E_p$ rather than $E_k$ spectra in an attempt to limit the effects of Doppler shifting and instrument response functions.  

\subsubsection{A Summary and Future Challenges}


The characterization of nonlinear interactions is a complicated process and therefore reductions to simplified forms for various limiting cases are highly desirable.  The reduction of the kinetic equation in the ID limit to a Fokker-Plank equation in vertical wavenumber is fundamental in this regard:  the stationary states of this equation collapse much of the observed variability in the spectral parameters $(a,b)$ (Fig. \ref{fig:everything}).  There are, however, significant questions regarding the validity of the underlying assumption of resonant interactions, especially in the ID limit at small vertical scales, that prompt one to wonder if the pattern match is merely fortuitous.  At these small vertical scales, diagnostic studies that compare simplified representations of the downscale transport with estimates of turbulent dissipation often settle upon the ray-tracing model (\ref{Cascade}) as providing the theoretical framework for the finescale internal wavefield.  However, the simplified representations derived within the RIA provide predictions that are similar both in magnitude and in scaling \citep{P04b}.  Thus one {\bf might} be prompted to suggest that such misgivings are simply due to offenses against one's theoretical sensibilities.  

There is, however, a very substantive issue that cries out for explanation.  Associated with the ID mechanism and $b>0$ are energy transports to higher vertical wavenumber and lower frequency.  A stationary state under the ID mechanism and $b>0$ thus requires either:
\begin{itemize}
\item An external energy source at high frequencies
External forcing is discussed in Section \ref{WaveMean} and \ref{Forcing}
\item An alteration of spectral transports associated with finite amplitude effects that acts to reduce transports through the frequency domain.  Section \ref{ResonanceBroadening} was dedicated to the issue of resonance broadening.  
\item A reassessment of resonant transfers.  We find that residuals in the resonant calculation for GM76 give rise to normalized Boltzman rates $\epsilon_{\bf p}$ that are $O(1)$.  We argue that these residuals arise from inherently local interactions and hypothesize that they are associated with spectral transports in the horizontal wavenumber domain.  It is possible that such residual interactions determine the transport of GM76.  
\item An effective source associated with nonhydrostatic effects.  
Only the numerical evaluations of \cite{O76} have included nonhydrostatic effects.  Those results are silent on this issue.  
\end{itemize} 

\subsection{Wave-Mean Interactions\label{WaveMean}}
\subsubsection{Spectral Patterns}

Much of the analysis in Section \ref{NonlinearitySection} was predicated upon the assumption of an inertial subrange; that there exists a range of wavenumbers and frequencies removed from the influence of forcing and dissipation, and far from the characteristic spatial and temporal scales of the system, that the spectrum can be characterized as being scale invariant.  We have noted the difficulties for this scenario with respect to the importance of extreme scale separated interactions in Section \ref{ScaleInvariance}.  Here we note that wave-mean interactions pose a special problem for the scale invariant solution in that wave-mean interactions can be cast as a source representation $S_o({\bf p})$ acting throughout the spectral and spatial domains rather than being located at the boundaries.  

A large body of work focusses upon the simplified cases of waves interacting with parallel shear flows \citep{MMY05, R80} or vortices \citep{K85}, i.e., symmetric background flows.  This conceptual paradigm may be made much more general and useful by allowing the background flow to have spatial gradients in both horizontal coordinates.  While the assumption of a spatially symmetric background flow makes for a more tractable theoretical problem, it also introduces a hidden constraint in which the vertical flux of streamwise pseudo-momentum is nondivergent except at a critical layer.  In the general problem of internal waves interacting with a mesoscale eddy field having 3-dimensional structure, a wave packet can freely exchange wave psuedo-momentum for eddy potential vorticity, \cite{BM05}.  This is restricted from happening in a symmetric background flow.  

The essence of the interaction in three dimensions is the filamentation of an internal wave by the horizontal rate of strain and vertical propagation into a critical layer.  This can be directly seen from a linearized internal wave energy equation for wave propagation in a quasigeostrophic background \citep{P08b}: 
\begin{eqnarray}
(\frac{\partial}{\partial t} + \overline{u} \cdot \nabla_h ) (E_k + E_p) + 
\nabla \cdot \overline{ \pi^{\prime\prime} \bf{ u}^{\prime\prime} } = \nonumber && \\
- [ \overline{u^{\prime\prime} u^{\prime\prime}} -\overline{v^{\prime\prime} v^{\prime\prime}} ] S_n/2
-\overline{ u^{\prime\prime} v^{\prime\prime} } S_s && \nonumber \\
-[ \overline{u^{\prime\prime}w^{\prime\prime}} - \frac{f}{N^2}\overline{b^{\prime\prime}v^{\prime\prime}} ] \overline{u}_z
- [ \overline{v^{\prime\prime}w^{\prime\prime}} + \frac{f}{N^2}\overline{b^{\prime\prime}u^{\prime\prime}} ] \overline{v}_z && 
\label{Energy}
\end{eqnarray}
with kinetic [$E_k = (u^{\prime\prime 2} + v^{\prime\prime 2} + w^{\prime\prime 2})/2$] and 
potential [$E_p = (N^{-2} b^{\prime\prime 2})/2$] energies.  Temporal variability
and advection of internal wave energy by the geostrophic velocity field are balanced by
wave propagation and energy exchanges between the quasigeostrophic and 
internal wave fields.  Nonlinearity and dissipation are assumed to be higher order effects.  
Spatial gradients of the vertical velocity $\overline{w}$ do not appear as $\overline{w}_z$ is small [order Rossby number squared] in the quasigeostrophic approximation.  The thermal wind relation has been invoked to cast the vertical Reynolds stress and horizontal buoyancy flux as the rate of work by an effective vertical stress acting on the vertical gradient of horizontal momentum.  These two terms in the effective stress will cancel each other in the limit that $\omega \rightarrow f$, \cite{RJ79}.  Finally, $S_s \equiv {\overline v}_x + {\overline u}_y$ represents the shear component and $S_n \equiv {\overline u}_x - {\overline v}_y$ the normal component of the rate of strain tensor.  In the limit of an extreme scale separation between wave and background flow, horizontal group velocities are small and an internal wave haves much like a passive tracer, \citep{BM05, P08a} as the rate of strain cascades the wave to higher horizontal wavenumber.  As it does so, the horizontal components of the internal wave stress result in work {\bf against} the mean flow gradients and {\bf gain} of wave energy.  In the situation in which a wave is propagating vertically into a critical layer, the vertical stresses are such that the wave {\bf looses} energy to the eddy field.  

The coupling of mesoscale eddies and internal waves and construction of a $S_o({\bf p})$ representation was explicitly addressed in \cite{M76}.  \cite{M76}'s analysis was based upon a perturbation expansion of the internal wave radiation balance equation (\ref{rbe}).  Internal waves were assumed to be of small amplitude and have small spatial scales relative to a geostrophically balanced background that evolves over a much longer time scale.  Spatial gradients in the background were assumed to be sufficiently weak that wave-mean interactions affect wave propagation only through an advective Doppler shift.  In this limit of small amplitude waves interacting with a quasi-qeostrophic eddy field, exchanges of energy, momentum and vorticity are reversible unless an external force is invoked which acts in phase on the wave-mean induced perturbations.  In \cite{M76}, the external force is identified as the tendency for nonlinear interactions to relax the wave-mean induced perturbations to an isotropic state.  It is this relaxation that creates a permanent exchange of pseudo-momentum for potential vorticity.  In order to close the problem, \cite{M76} invoked a relaxation time approximation and further assumed the relaxation time scale to be constant.  Estimates of the eddy-internal wave coupling strength were not supported by observations \citep{RJ79, Bow} and this line of investigation was dropped.  With regards to the strawman in Fig. \ref{McComas}, the consequence is that wave-mean interactions are not available as a source to supply energy depleted by PSI transfers.  

\cite{M76} was written from the perspective that the thermocline is characterized by a diapycnal
diffusivity of $K_{\rho}=1\times 10^{-4}$ m$^{2}$ s$^{-1}$, which
implies a time scale of 5-10 days for nonlinear interactions to drain
energy out of the background internal wavefield.  Several decades of
research has since demonstrated that the background diffusivity is
$K_{\rho}= 5\times 10^{-6}$ m$^{2}$ s$^{-1}$, with corresponding time
scale of 50-100 days.  The difference is crucial, as M\"uller assumed
that the eddy-internal wave interaction was local in the spatial
domain.  With O(50-100) day relaxation time scales, wave propagation
effects become important: larger scale internal waves can propagate
through an eddy-wave interaction event with minimal {\bf permanent} exchange
of pseudo-momentum and vorticity.  Fortunately, \cite{M76} tells us how to account for such propagation effects.  
Accounting for propagation effects and use of a more realistic prescription of the nonlinear time scale returns estimates of the coupling that are in much better agreement with the observations, \citep{P08c}.  

The radiation balance equation source function:
\begin{equation}
\sigma S_o({\bf p}) = S_{o}^{h}  + S^{v}_{o} 
\end{equation}
can be characterized with the following relaxation time scale formulas: 
\begin{eqnarray}
S_{o}^{h}(\sigma,m) &=& -\frac{1}{8} \frac{\sigma^2-f^2}{\sigma^2} \frac{m^2}{m^2+k_h^2} \frac{\sigma k_h \tau_R}{1+(\tau_{R}/\tau_{p})^2}\nonumber\\ 
&& \times \frac{\partial n_3^{(0)}}{\partial k_h}~(\overline{u}_x^2 + \overline{u}_y^2 + \overline{v}_x^2 + \overline{v}_y^2)
\label{Source-horz} \\
S^{v}_{o}(\sigma,m) &=& \frac{1}{2} \frac{\sigma^2-f^2}{\sigma^2} \frac{\sigma k_h^2}{m} \frac{m^2}{m^2+k_h^2} \frac{\tau_R}{1+(\tau_{R}/\tau_{p})^2} \nonumber\\ 
&& \times\frac{\partial n_3^{(0)}(k_h,m)}{\partial m}~(\overline{u}_z^2 + \overline{v}_z^2).
\label{Source-vert}
\end{eqnarray}
with relaxation time scale $\tau_{R}$ and propagation time scale (wave group velocity divided by eddy length scale) $\tau_{p}$.  Evaluations of the relaxation time-scale formulas are presented in Fig. (\ref{Source-Function}).  We use two spectra  intended to mimic those at Site-D (GM76) and the Sargasso Sea [(\ref{vertical_spectrum}) and (\ref{frequency_spectrum}) with ($s=2,t=2.3,j_{\ast}=15,r=0.8,q=0.5,N=2.6 {\rm cph},E=0.7E^{GM}$)].  Mean flow gradient statistics are taken from the LDE data \citep{P08b}:  ($ H$, $L$, $\overline{u}_z^2 + \overline{v}_z^2$, $\overline{u}_x^2 + \overline{u}_y^2 + \overline{v}_x^2 + \overline{v}_y^2$)=(700 m, 100 km, $4.1\times10^{-8}$ s$^{-2}$, $5.8\times10^{-12}$ s$^{-2}$).  Preliminary estimates from Site-D array data indicate that this is an appropriate characterization there.  The relaxation time scale is specified using (\ref{Cascade}).  

The coupling is a complicated  function of the spectral parameters: spectral amplitude, high 
wavenumber/frequency power law specifications and low wavenumber roll-off; and external parameters, i.e., mesoscale velocity gradient variances and characteristic spatial scales (Fig.s \ref{Eddies_Atlantic} and \ref{Eddies_Pacific}).  Without having executed a complete parameter regime survey, we rationalize the results as follows.

In the continuum frequency band, differences between the vertical source functions for GM76 and the Sargasso Sea result from the GM76 3-d action spectrum having minimal gradients of action in the vertical wavenumber domain, i.e., $b=0$.  For $b \ne 0$, the zero-contour demarcating source/sink tendencies is dictated by $m_{\ast}$:  larger vertical wavelengths are sources of energy for the internal wavefield.  The magnitude of the positive region relative to the high wavenumber negative region is dictated by the ratio of relaxation to propagation time scales, $\tau_r \colon \tau_p$.  The possibility of a positive feedback mechanism dictating $m_{\ast}$ is obvious.  

In the near-inertial band, vertical transfers are sinks and horizontal transfers are sources, consistent with a vertical critical layer and horizontal filamentation scenario.  The {\bf net} transfer obviously is a function of the horizontal vs vertical gradient 
\samepage{
\begin{figure*}[H]
\begin{tabular}[t]{lr}
\begin{minipage}[t]{20pc}
variances of the mesoscale eddy field, i.e., it's aspect ratio.  The Sargasso Sea estimates imply an eddy aspect ratio slightly less than $f/N$ and this may be typical \citep{AandF}.  We have also included non-hydrostatic conditions in these calculations.  There is a signature of a significant
\end{minipage} &
\begin{minipage}[t]{20pc}\noindent
 source at the buoyancy frequency in the vertical source function that is largest for $b \ne 0$.  This could potentially resolve issues related to ID transfers requiring a high frequency source in order to maintain a transport of action to high vertical wavenumber, Section \ref{ScaleInvariance}.d.  Some skepticism 
\end{minipage}
\end{tabular} 
\noindent\includegraphics[width=20pc]{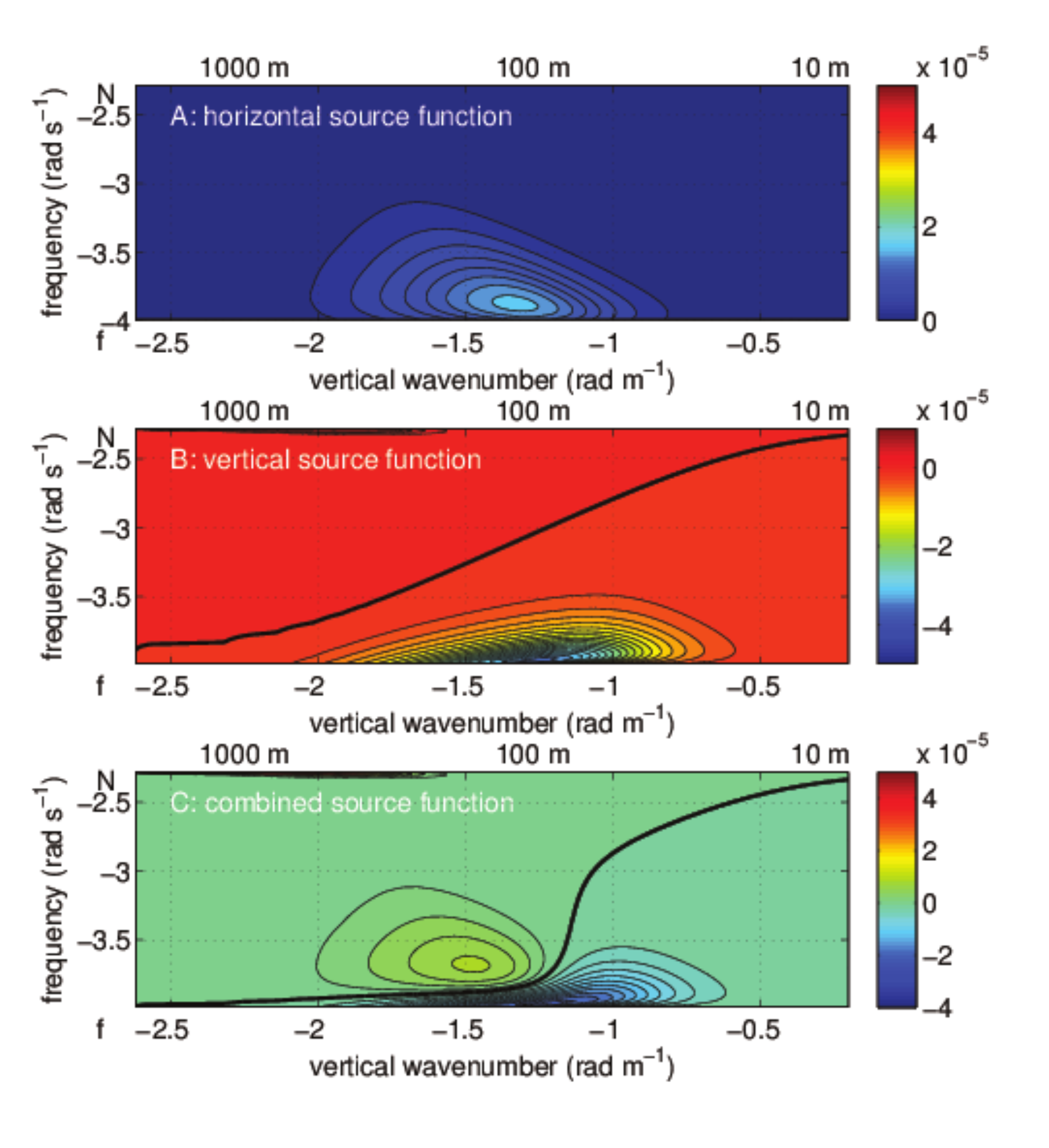}
\noindent\includegraphics[width=20pc]{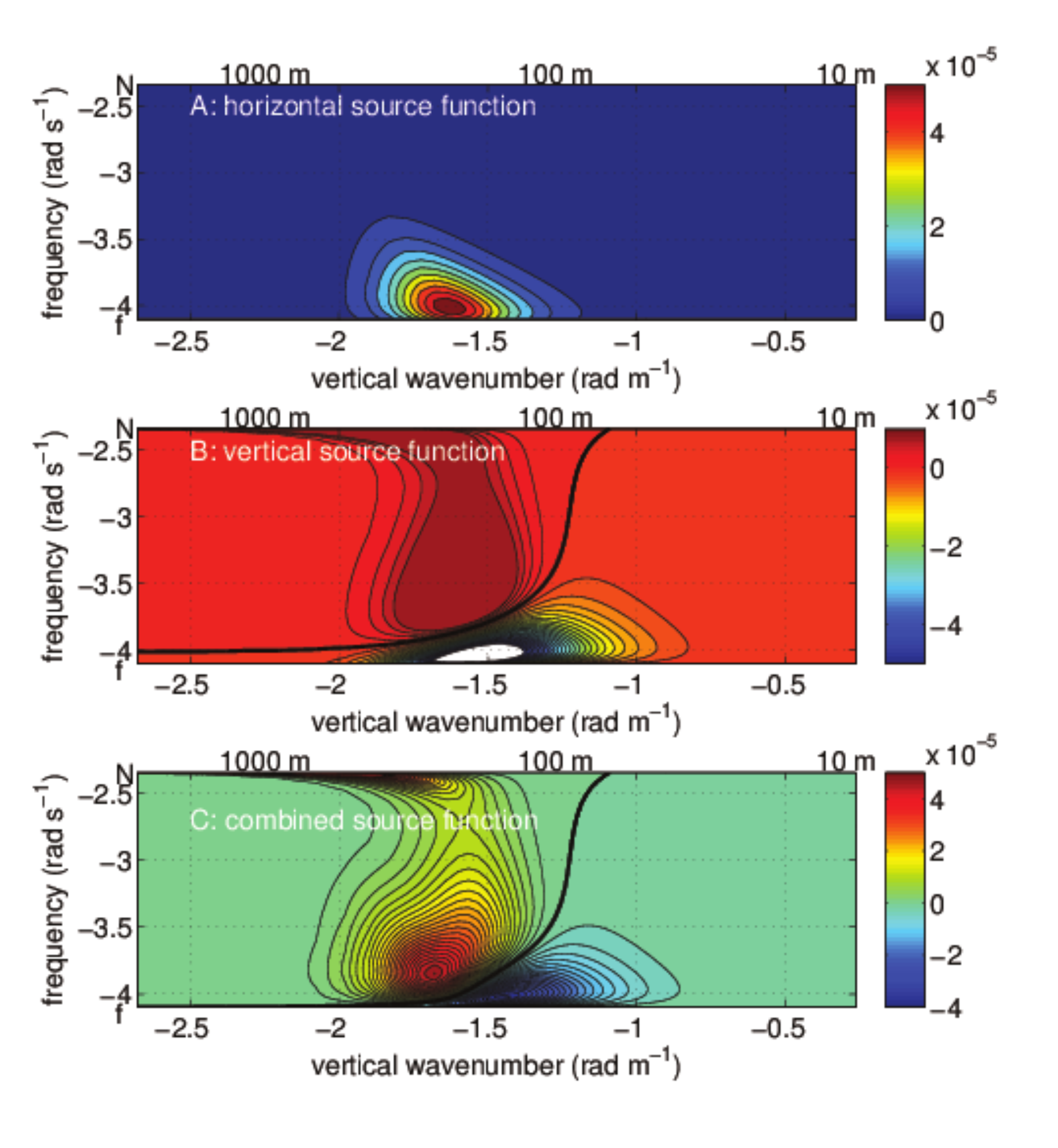}\vspace{-0.50cm}
\caption{The spectral source functions defined in (\ref{Source-horz}) and (\ref{Source-vert}).  Left-hand panels are GM76 evaluations.  Source functions in the right-hand panels use a 'Sargasso Sea' spectrum.  Panels A, B, and C represent the horizontal, vertical and combined source functions.  The thick line separates positive source values from negative.  Negative values of the source function are, effectively, energy sinks for the internal wavefield.  
\label{Source-Function} }
\end{figure*}
\begin{figure*}[H]\vspace{-0.50cm}
\noindent\includegraphics[width=20pc]{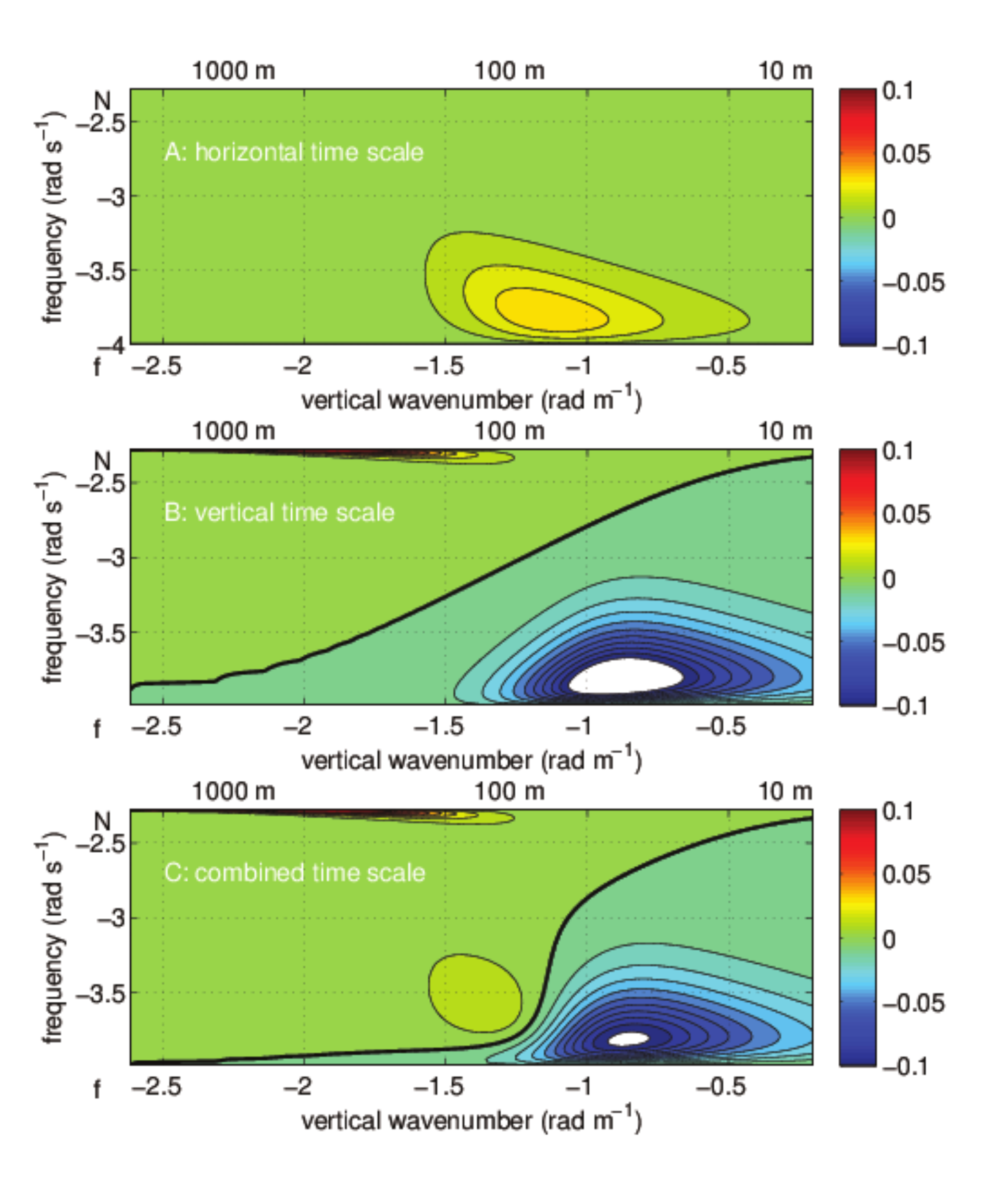}
\noindent\includegraphics[width=20pc]{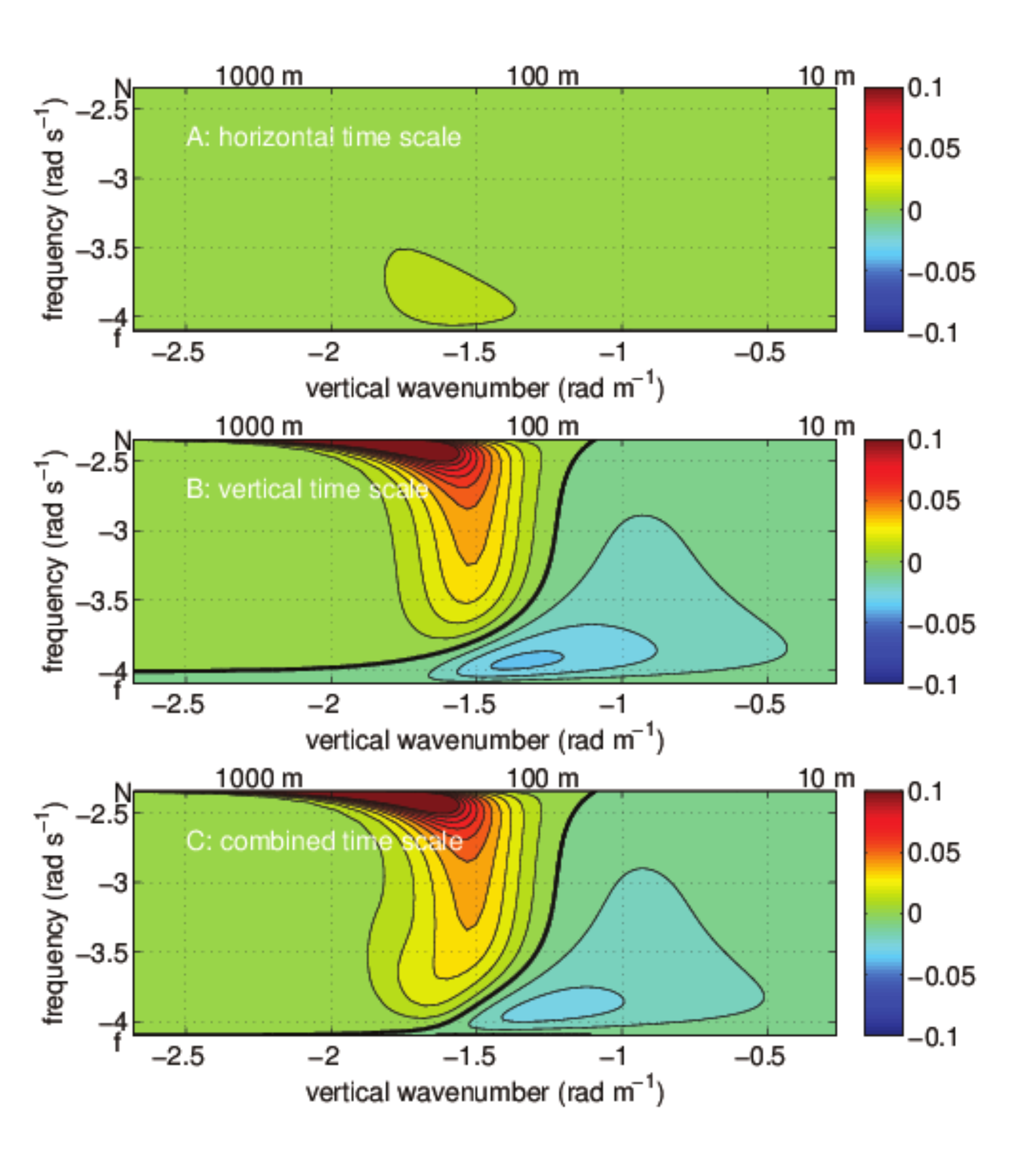}\vspace{-0.50cm}
\caption{The spectral source functions appearing in Fig. \ref{Source-Function} but expressed as a normalized transfer rate $2 \pi S_o(\sigma,m) / \sigma E(\sigma,m)$.   This normalization permits a direct comparison with Fig.s \ref{BoltzmanRates} and \ref{NonlinearityParameter}.  Contour intervals here are 0.01.  
\label{wave-mean_time-scales} }
\end{figure*} }
\clearpage
\begin{tabular*}{39pc}[t]{lr}
\begin{minipage}[b]{20pc}
is required:  such dynamics will compete with turning point effects and questions of how to specify $H$ in $\tau_p$ arise.  Apart from the buoyancy frequency, note that these wave-mean transfers are far too small to offset {\bf resonant} wave-wave interactions:  Compare Fig. \ref{BoltzmanRates} with \ref{wave-mean_time-scales}.  

\subsubsection{Budgets}

Energy exchange between the mean and wave fields results from a stress-strain relation in the horizontal.  Direct evaluation of the energy exchange using the PolyMode LDE current meter array data \citep{P08b} returns a transfer rate somewhat smaller than previously obtained \citep{Bow},  $4 \times 10^{-10}$ W/kg at 825 m depth.  If this value characterizes the thermocline region
having a characteristic depth $H=1000$ m, a net source strength of 0.4
mW m$^{-2}$ can be anticipated, Table \ref{Sources}.  
This estimate may be 
biased low, however, as both vertical and horizontal velocity
gradients increase towards the surface and the rate at which energy is
transfered is proportional to the velocity gradient variances.
The LDE was located at ($31^{\circ}$N, $70^{\circ}$W), midway between
Lotus and Fasinex.  The estimates of energy transfer through
eddy-internal wave coupling at 825 m depth do not 
\end{minipage} 
\begin{minipage}[b]{20pc}
\begin{figure}[h]
\noindent\includegraphics[width=20pc]{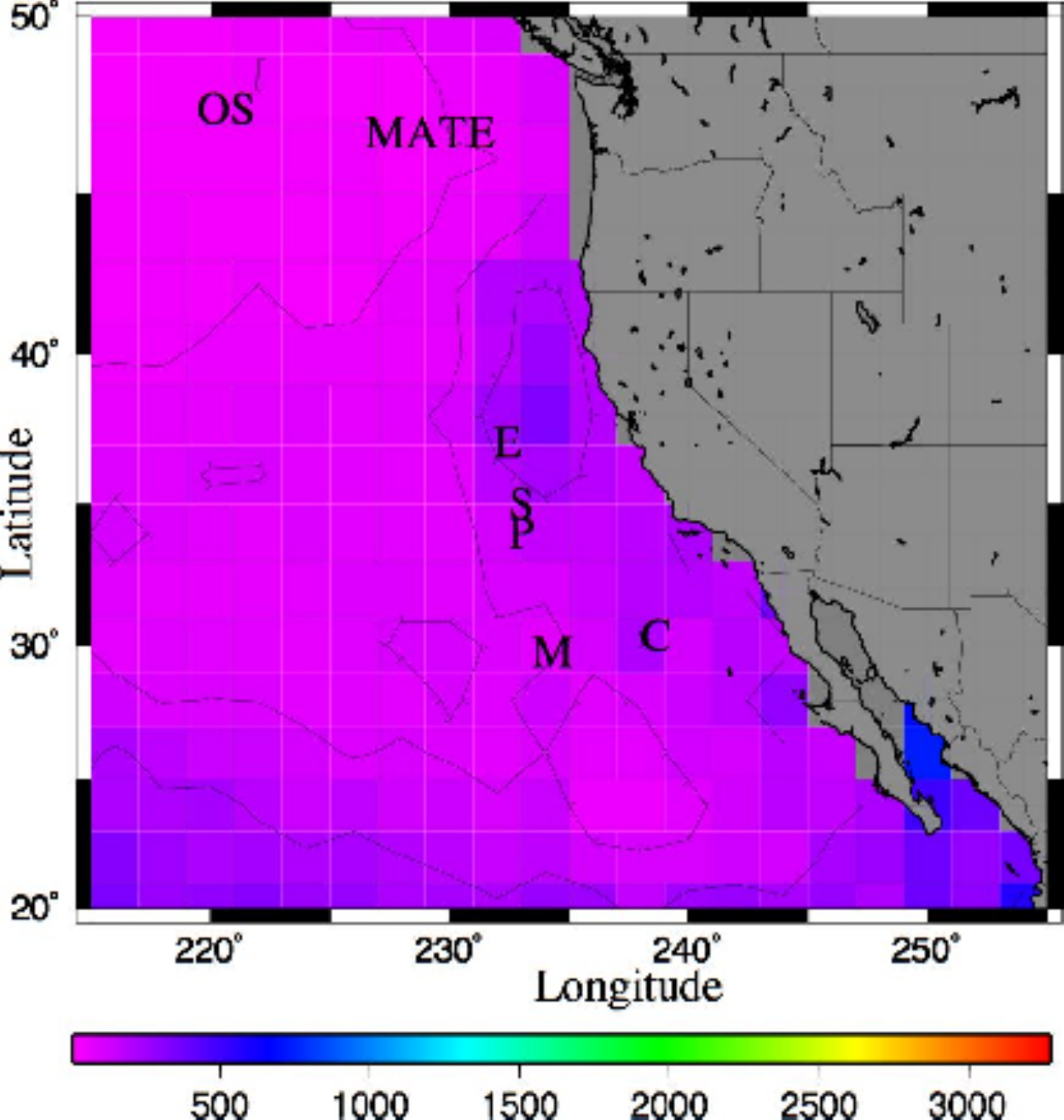}
\caption{Map of surface eddy kinetic energy in units of $10^{-4}$ m$^2$ s$^{-2}$.  Contour intervals are $[50, 100, 200, 500, 1000, 2000]\times10^{-4}$ m$^2$ s$^{-2}$.   \label{Eddies_Pacific}  }
%
\end{figure}
\end{minipage} 
\end{tabular*}  
\begin{figure*}[b]
\begin{minipage}[b]{40pc}
\noindent\includegraphics[width=39pc]{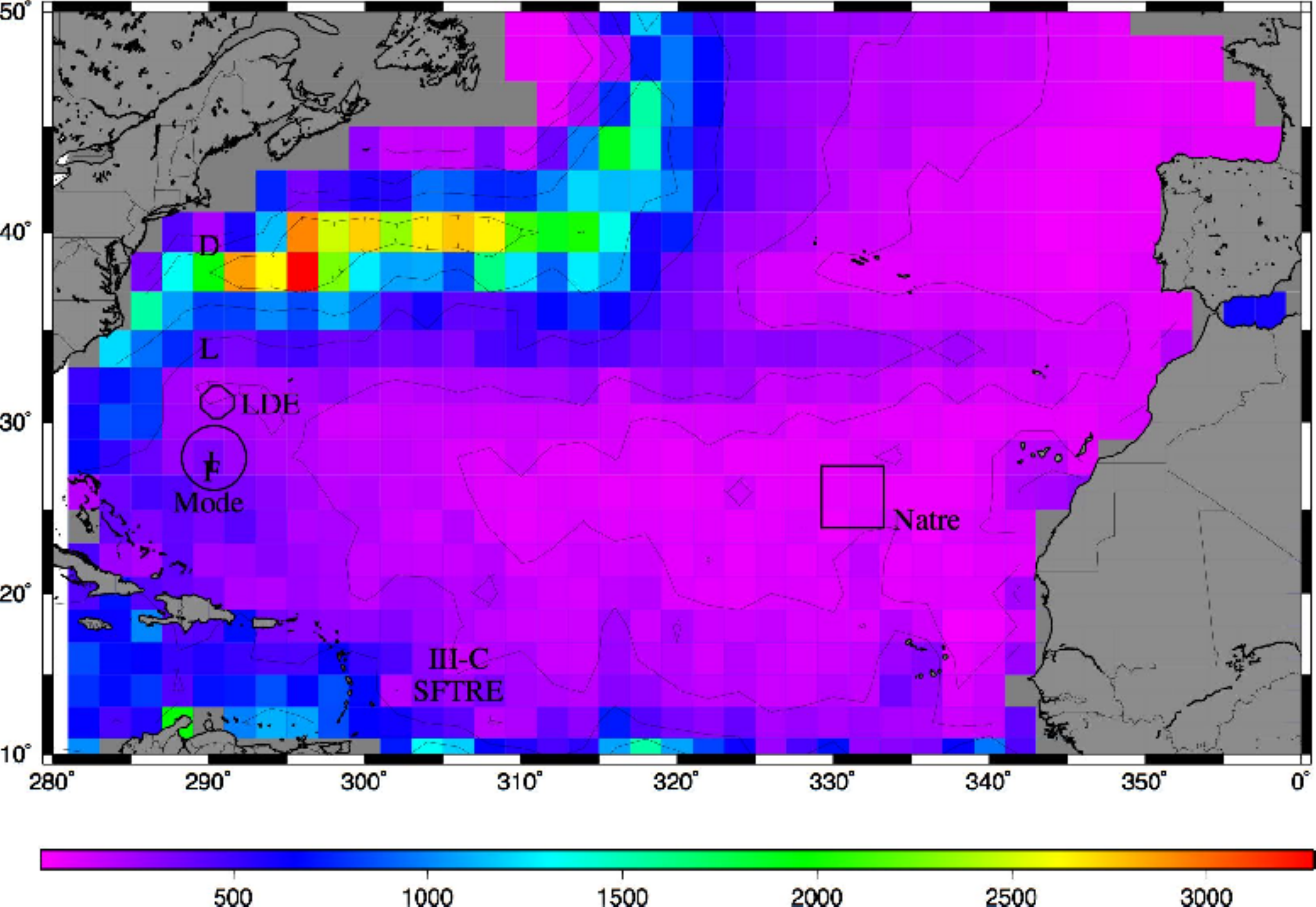}
\caption{Map of surface eddy kinetic energy in units of $10^{-4}$ m$^2$ s$^{-2}$.  Contour intervals are $[50, 100, 200, 500, 1000, 2000]\times10^{-4}$ m$^2$ s$^{-2}$  \label{Eddies_Atlantic} }
\end{minipage}
\end{figure*} 
\newpage
\begin{tabular*}{39pc}[t]{lr}
\begin{minipage}[b]{20pc}

\noindent appear to explain
the regional seasonal cycle of internal wave energy.  There is,
however, a seasonal cycle in upper ocean fronts associated with
buoyancy forcing, e.g., \cite{W91}.  This upper ocean frontal regime is
characterized by an $O(1)$ Froude (vertical gradient of horizontal
velocity divided by buoyancy frequency) and Rossby (relative vorticity divided by $f$) number
parameter regime \citep{PR92} which may support more
efficient energy transfers than the quasi-geostrophic regime at the
thermocline base.




\subsection{Forcing\label{Forcing}}

The focus of the previous subsections was the genesis of spectral structure.  In Sections \ref{Forcing}, \ref{Dissipation} and \ref{Propagation} we focus more upon (i) whether the variability in spectral parameters represents the character of the forcing at the endpoints of the cascade / spectral boundaries, and (ii) whether the forcing is consistent with the direction and magnitude of energy transfers. 

\subsubsection{Near-inertial wave generation by a variable wind stress}

Work done by the wind in generating inertial frequency mixed-layer motions has been the 
subject of several recent publications.  Nearly-global maps 
\end{minipage} 
\begin{minipage}[b]{20pc}
\begin{figure}[h]
\noindent\includegraphics[width=20pc]{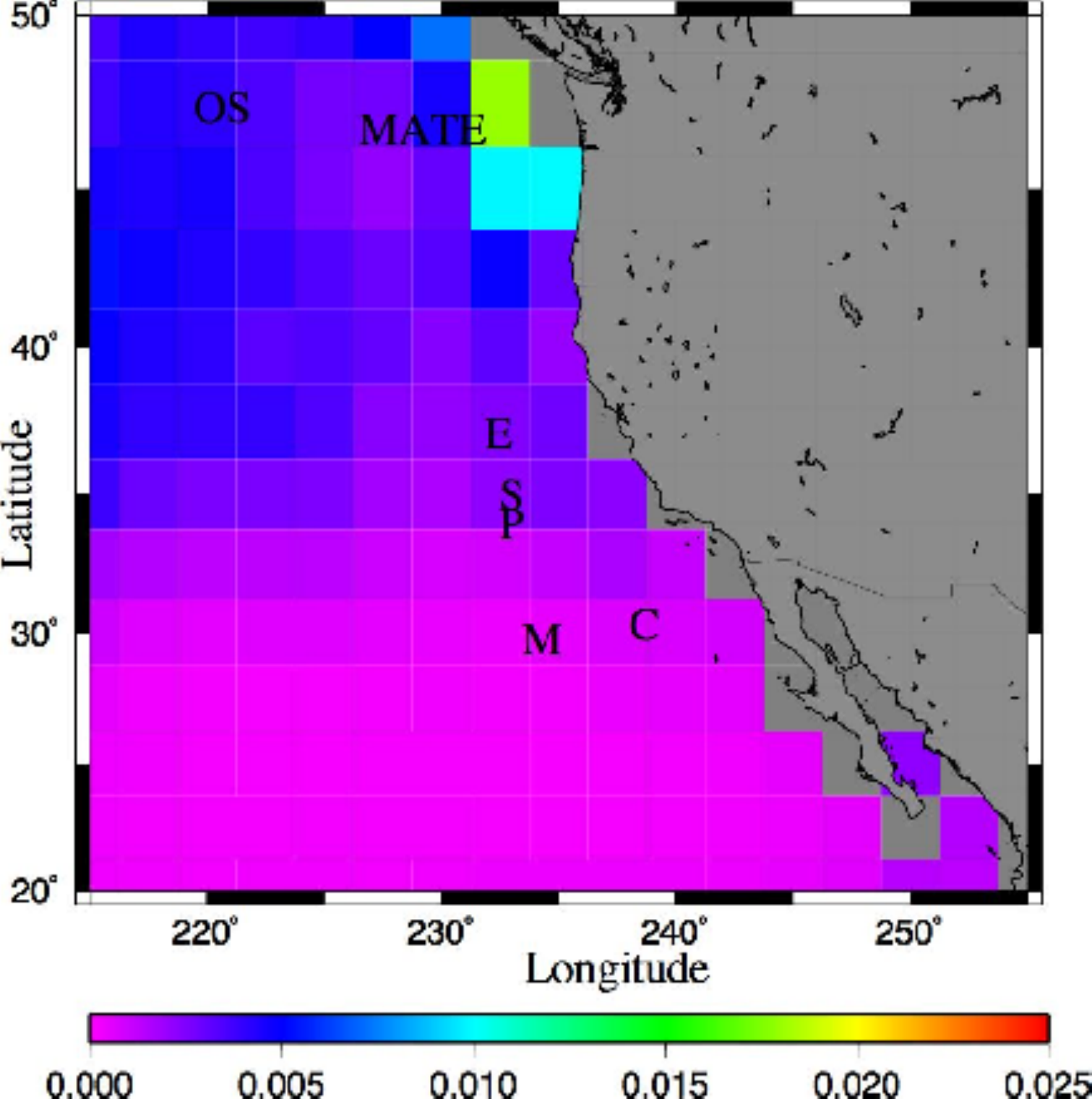}
\caption{Map of near-inertial wind input into the mixed layer in units of W/m$^2$, \citep{A03}.  }
\label{near-inertial-input-P} 
\end{figure}
\end{minipage} 
\end{tabular*}  

\begin{figure*}[b]
\begin{minipage}[b]{40pc}
\noindent\includegraphics[width=39pc]{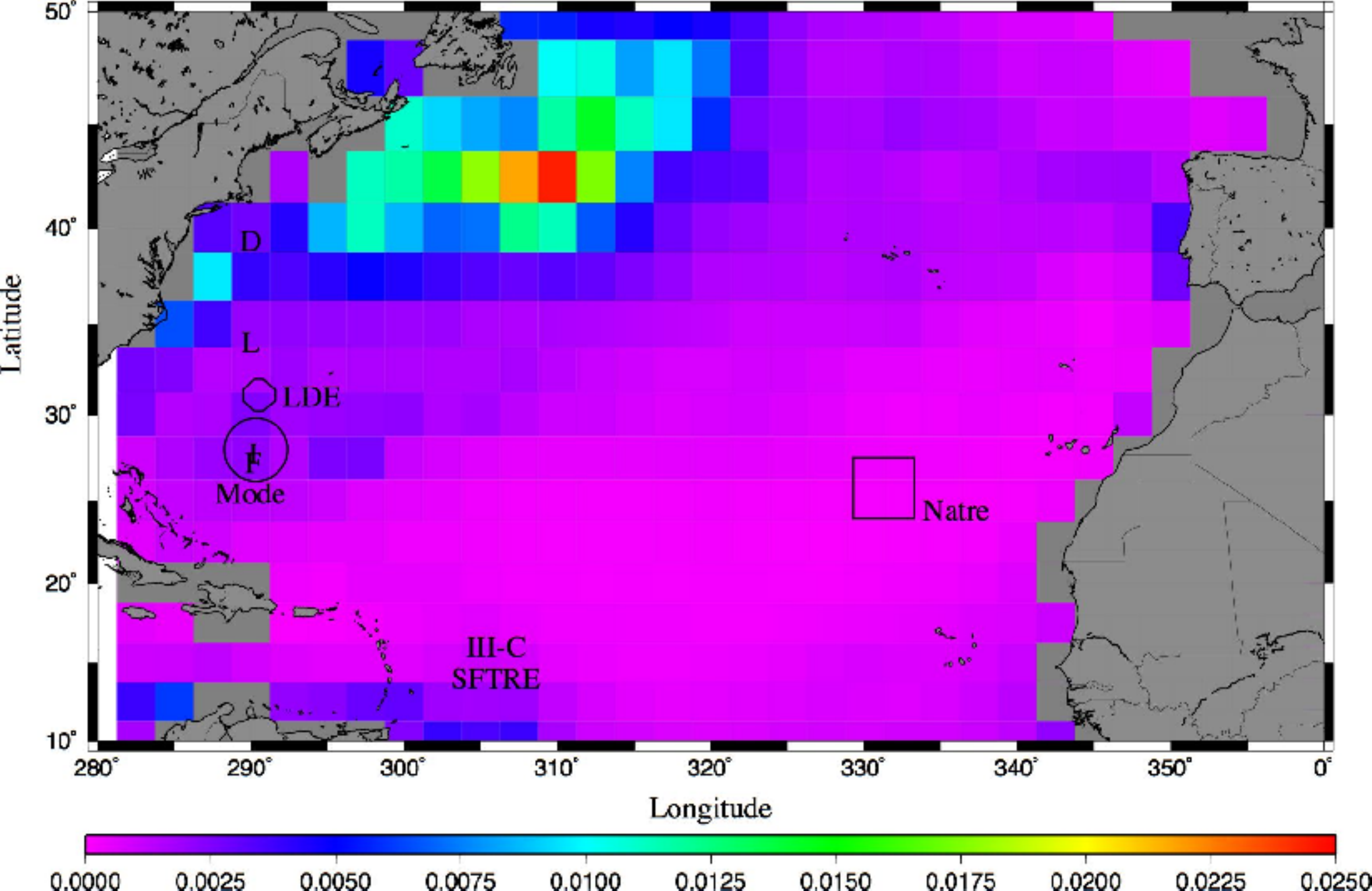}
\caption{Map of near-inertial wind input into the mixed layer in units of W/m$^2$, \citep{A03}.  
\label{near-inertial-input-A} }
\end{minipage} 
\end{figure*}
\newpage

\begin{figure*}[t]
\noindent\includegraphics[width=39pc]{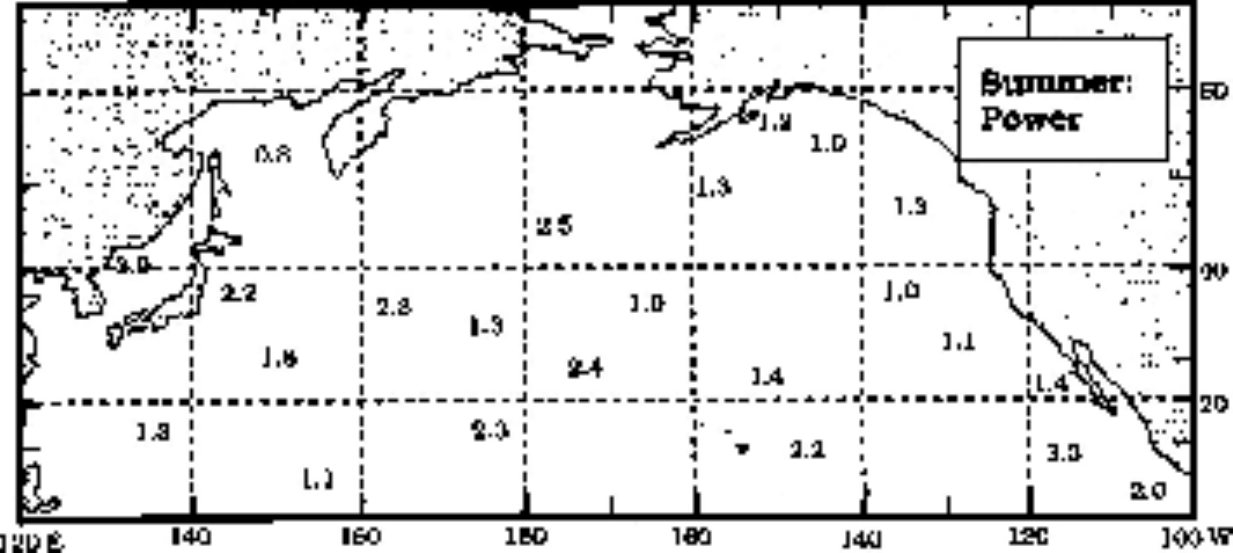}
\caption{Power {\bf lost} from the internal wavefield to the surface wavefield in units of $10^{-4}$ W/m$^2$ for summer months \citep{W94}.  
\label{windwave_Summer} }
\end{figure*}

\begin{figure*}[h]\vspace*{-1.5cm}
\noindent\includegraphics[width=39pc]{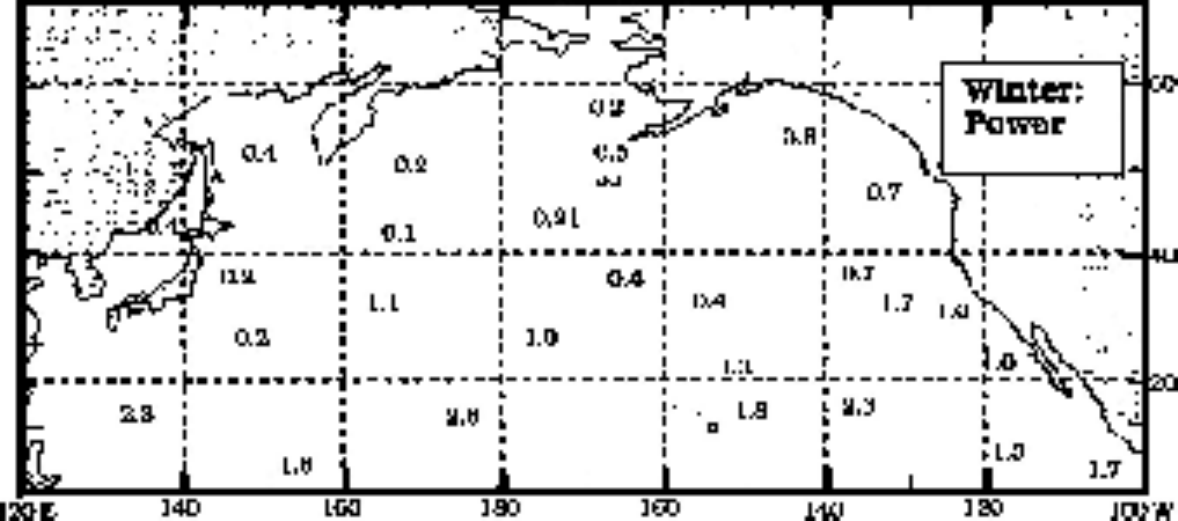}\vspace*{-0.5cm}
\caption{Power {\bf lost} from the internal wavefield to the surface wavefield in units of $10^{-4}$ W/m$^2$ for winter months \citep{W94}.  
\label{windwave_Winter} }
\end{figure*}

\begin{tabular*}{39pc}[t]{lr}
\begin{minipage}[t]{20pc}\vspace*{3.0cm}
of this wind work using the \cite{PM70}
damped-slab model are presented in \cite{A03, WH02}, Fig.s \ref{near-inertial-input-P} and \ref{near-inertial-input-A}.  The slab model is highly idealized.  Using observations of the upper ocean response to atmospheric forcing and a more sophisticated mixed layer model, \cite{PF06} describe how the slab model tends to overestimate inertial frequency wind work by a factor of two.  Their estimates for the Lotus, Fasinex and Subduction field programs are given in Table 
\ref{source_strength}.  Similar estimates for Site-D are presented in \cite{ST09}.  

\cite{PF06} also note that other modeling investigations (\cite{CL96, SSC00}) imply that only a fraction of this work ($50 \%$) may be available for near-inertial 
\end{minipage} &
\begin{minipage}[t]{20pc}\vspace*{3.0cm}
internal wave generation as the other part is lost to viscous dissipation in the mixed layer for the resonant forcing conditions that dominate the revised transfer estimates.  'Best' estimates of source strength in Table 
\ref{source_strength} incorporates this dissipation.  

The aforementioned work concerns only getting inertial energy into the mixed layer.  Radiation of this energy into the stratified interior can be represented by invoking a Fourier decomposition of the slab model and mapping onto the internal wavefield using a linear model that includes the variation of the Coriolis parameter with latitude \citep{DA95a}.  
There are, however, unresolved issues:  the linear model under estimates the initial rate at which energy leaves the mixed layer by $20-50\%$ and can not explain the evolution of the small vertical wavelength 
\end{minipage}
\end{tabular*}

\newpage

\noindent near-inertial field.  Nonlinear interactions are held to be a primary candidate to explain the later \citep{DA95b}.  For further consideration of near-inertial motions see Section \ref{Propagation}.    

\subsubsection{Surface-internal wave coupling}

Internal waves can also exchange energy with the surface wavefield, 
thereby implying a parametric dependence of source strength upon wind stress and 
possible seasonality.  In the resonant interaction approximation, a pair of surface 
waves with nearly equal wavenumber and frequency can interact with an internal wave mode 
through a difference interaction, \cite{WWC, OH78}.  Estimates of energy transfer are sensitive to the structure of the upper ocean buoyancy profile.  Estimates \citep{W94} using climatological buoyancy profiles and seasonally averaged wind estimates return transfer rates of several tenths of a mW m$^{-2}$ {\bf from} the internal wavefield {\bf to} the surface wavefield during summer months (Fig. \ref{windwave_Summer}) and at low latitudes during winter months (Fig \ref{windwave_Winter}).  The high latitude transfer rates during winter are smaller.  In contrast to these wind wave results, rapid transfers of energy {\bf from} a {\bf narrow band ocean swell} {\bf to} the internal wavefield are possible \citep{W90} and remain to be quantified.  

The coupling of surface and internal gravity waves thus does not appear to be the effective source at high frequencies and low vertical wavenumber as depicted in the strawman Fig. \ref{McComas} and required as a boundary condition on the energy transports in the dynamic balance, Section \ref{ScaleInvariance}.d.  

\subsubsection{Internal wave coupling to mixed-layer turbulence}

The coupling of surface and internal gravity waves is not the only possible source of high frequency internal wave energy.  It has long been conjectured that mixed-layer turbulence can be coupled to the internal wavefield, \cite{Bell78}.  Observations presented in \cite{WD91} reveal a narrow band internal wave response in the upper equatorial ocean at frequencies near the upper ocean stratification rate.  \cite{SD94} point towards shear instability as a generation mechanism.  It is unclear how this mechanism translates to a rotating system.  

\cite{Polton08} use large-eddy simulations of mixed-layer turbulence with a parameterization of Stokes drift forcing by a surface gravity wavefield.  This forcing sets up Langmuir circulations and a inertial current oscillation is superimposed that advects the Langmuir cells relative to the stratified interior.  Vertical circulations of the Langmuir cells excite high frequency internal waves having a phase velocity that matches the depth averaged mixed layer velocity.  

\cite{Polton08} estimate that energy and momentum are radiated at rates comparable to those associated with near-inertial wave radiation.  These waves, however, are likely trapped within the highly stratified transition layer with either implications for mixing or for acting as a source of high frequency energy that can be transported to lower frequency via wave-wave interactions.  Analysis of upper ocean data sets for a high frequency response on the time scale of the wind forcing could be quite enlightening.


\subsubsection{Tides}
\subsubsubsection{Spatial Patterns}

\begin{figure*}
\noindent\includegraphics[width=39pc]{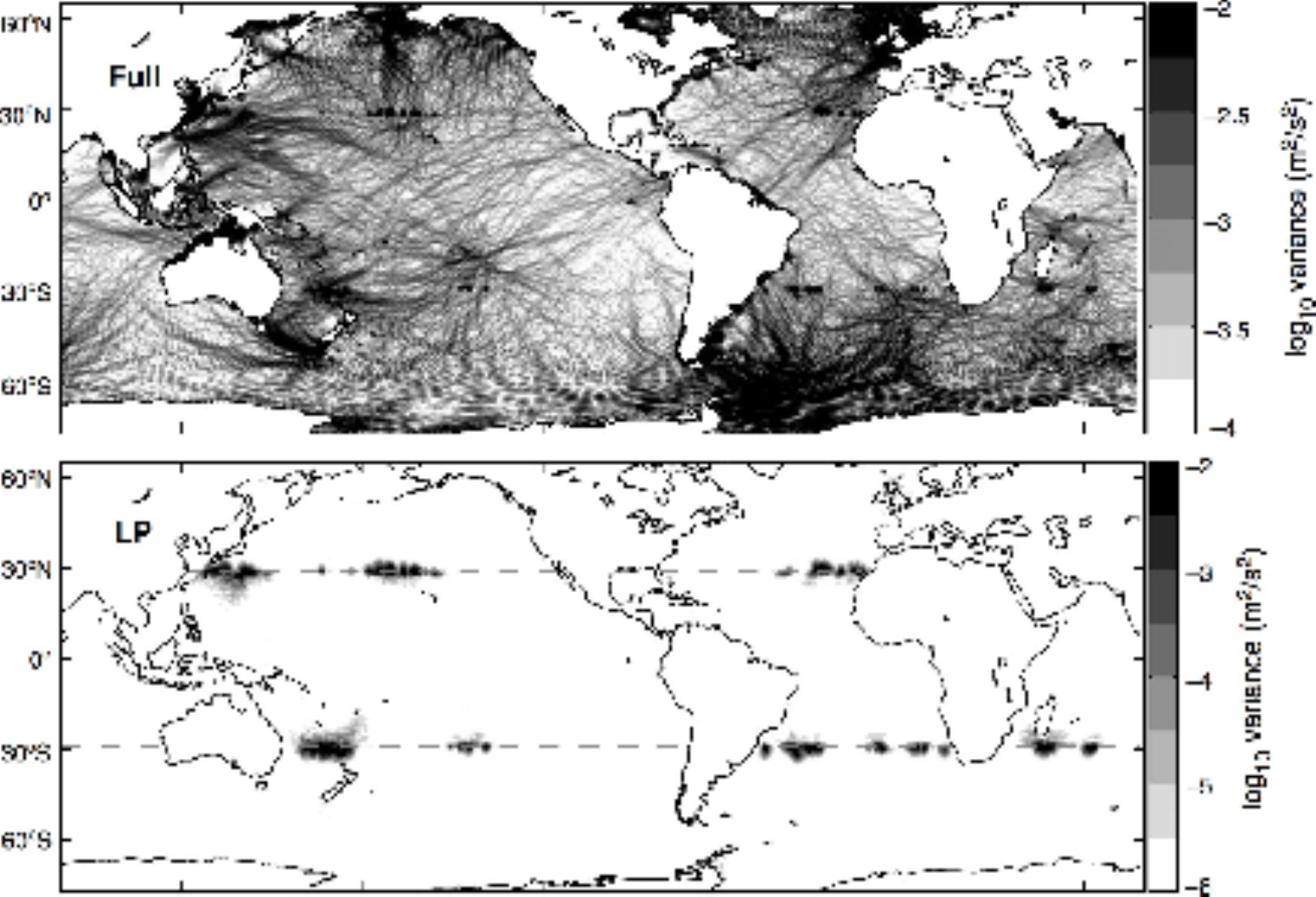}
\caption{Maps of baroclinic tidal and near-inertial energy.   From \cite{Simmons08}. }
\label{Harper} 
\end{figure*}

Numerical models have, in several instances, been used to both estimate the energy conversion from barotropic to baroclinic tides {\bf and} keep track of the nonlinear evolution of the internal tides (e.g., \cite{Simmons08}).  While such models may not have the spatial resolution required to get the details of the nonlinear transfer process right, or pick up the generation of baroclinic tides from small scale topographic roughness, 
they should give some indication of where the PSI decay can expected to be important.  The eastern Atlantic is one such region, Fig. \ref{Harper}, providing further circumstantial evidence that the Natre spectrum is forced primarily through PSI decay.  

\subsubsubsection{Budgets}
The issue of internal tide generation, propagation and decay has received a great deal of attention this 
past decade, in part due to several major field programs (the Brazil Basin Tracer Release Experiment 
\cite{L00} and the Hawaiian Ocean Mixing Experiment \cite{Rud03}).  Near global maps of the rate of conversion of barotropic tidal energy to baroclinic (internal) tidal energy have been provided by several \cite{ER03, SHA, Simmons08, NH01} groups.  As our focus is on the background internal wavefield, none of the data sets described here is from a region of significant barotropic to baroclinic conversion.  

The relevant phenomena appears to be the far-field decay of a low-mode tide via resonant interactions.  
This particular phenomena is believed to be most important when the frequency of the low-mode tide is 
slightly higher than the Coriolis frequency, e.g., \cite{FHN05}.  For $\sigma = M_2$, this corresponds to latitudes equatorward of $28.9^{\circ}$.  Many of the data sets examined here are in the vicinity of this critical latitude, particularly in the western North Atlantic.  There, however, estimates suggest a low mode internal tide with horizontal energy flux of approximately 300 W m$^{-1}$ \citep{H77}\footnote{Hendry's estimate of 300 W/m to the south east represents a {\bf short term} estimate from the MODE current meter {\bf array} centered about $28^{\circ}$N, $70^{\circ}$W.  {\bf Long term} estimates from {\bf single moorings} at $31^{\circ}$N, $70^{\circ}$W (LDE) and $34^{\circ}$N, $70^{\circ}$W (LOTUS) \citep{AZ07} are approximately 150 W/m to the north-northwest and 200 W/m to the southeast.  It is tempting to interpret energy flux estimates in Fig. 1 of \cite{Simmons08} as suggesting that the western boundary is a source of baroclinic tidal energy at these latitudes.  The lack of consistency is troubling and the issue warrants further investigation.} and analyses of relevant data from the western North Atlantic have not emphasized an interpretation of semidiurnal tidal PSI decay.  An order of magnitude estimate of source strength as the observed flux distributed over the distance from the assumed source (the Blake Escarpment some 700 km distant) is smaller ( 300 W m$^{-1}$ / 700 km) than the nominal background dissipation rate, O(1 mW m$^{-2}$), Section \ref{Dissipation}.  Moreover, the internal tide likely cannot explain the observed seasonal cycle in high frequency internal wave energy if the source has been correctly identified.  \cite{Baines82} finds that the details of the stratification profile are important in estimating the rate of conversion of barotropic tidal to baroclinic energy, but the Blake Escarpment is a relatively deep feature and extends only to 1000 m water depth, well below seasonal changes in stratification.  

Decay of a low mode-tide via resonant interactions poleward of the critical latitude would be through the induced diffusion mechanism.  \cite{OP81} argue that decay of a line spectrum with a typical energy density of 100 J m$^{-2}$ would result in loss from the tides at rates between 0.005 to 0.05 mW m$^{-2}$.  Such values will not contribute significantly to the internal wave energy balance, but note:  the calculation is based upon GM76, and with $b=0$ the estimate {\bf is} biased.  

Barotropic tidal velocities in the eastern North Atlantic are typically larger than in the west, and, consistently, available estimates of the low mode internal tide (\cite{Rossby88, SP91, AZ07}) are typically larger.  However, such estimates exhibit significant spatial and temporal variability.  Preliminary estimates of internal tide parameters from the Subduction array's center mooring assign rms buoyancy scaled velocities $u_o,v_o$ of (0.02,0.02) m s$^{-1}$ and 4 m displacement to mode-1, and implying an energy flux of approximately 1 kW m$^{-1}$.  This is a surface mooring and contamination of energy flux estimates by vertical motion in response to ambient currents is a definite concern.  \cite{AZ07} present energy flux estimates of roughly 750 W/m for this mooring.  This energy flux is directed to the northeast, seemingly at odds with both near-field and far-field sources (a minor seamount that rises to depths of 3000 m immediately to the east, the Great Meteor seamount to the north and the European continental shelf to the northeast).  Energy flux vectors in \cite{Simmons08} are to the west.  

An order of magnitude estimate of the flux divergence (Table \ref{Sources}) can be obtained by dividing 1 kW m$^{-1}$ by the length scale characterizing the spatial variability in the internal tide amplitude, roughly 1,000 km.

\begin{table*}
\caption{Estimates of source strengths.  Best estimates of the internal wave source associated with wind work are a fraction (50 \%) of the wind work.  The internal wave source associated with the decay of a low-mode tide is obtained by dividing the energy flux estimate by a length scale.  Those length scales are ill constrained.  \label{source_strength} }
\begin{center}
\begin{tabular}{|c|c|c|c|c|c|c|}
\hline
program & latitude & longitude & mechanism & start & stop & source strength \\
\hline
Lotus & 34$^{\circ}$N & 70$^{\circ}$W & wind work & May 1982 & Oct. 1982 & 1 mW m$^{-2} \times$ 50\% \\
LDE & 31$^{\circ}$N & 70$^{\circ}$W & eddy-wave coupling & May 1978 & July 1979 & $>$ 0.4 mW m$^{-2}$ \\
Mode & 28$^{\circ}$N & 70$^{\circ}$W & tide & March 1973 & July 1973 & 300 W m$^{-1}$ / 700 km \\
Fasinex & 27$^{\circ}$N & 70$^{\circ}$W & wind work & Jan. 1986 & May 1986 & 0.4 mW m$^{-2} \times$  50\% \\
Subduction & 25.5$^{\circ}$N & 29$^{\circ}$W& wind work & April 1992 & Oct. 1992 & 0.05 mW m$^{-2} \times$  50\% \\
Subduction & 25.5$^{\circ}$N & 29$^{\circ}$W& tides & April 1992 & Oct. 1992 &  1 kW m$^{-1}$ / 1000 km\\
Site-D & 39$^{\circ}$N & 69$^{\circ}$W& wind work & November 2001 & April 2006 &  1.0 mW m$^{-2} \times$ 50\% \\
\hline
\end{tabular}
\end{center}
\label{Sources}
\end{table*}

\subsection{Dissipation\label{Dissipation}}

Our underlying hypothesis is that the observed variability can be understood as spatially local stationary states of the radiation balance equation (\ref{rbe}).  In this context, the role of dissipation is possibly two fold.  First, in a local stationary balance, forcing balances dissipation.  Second, dissipation may play a role in actively shaping high wavenumber spectra.  We take up these issues in turn below.  

\subsubsection{Budgets\label{DissipationBudgets}}

Internal wave energy can be dissipated in the interior through wave breaking and at the boundaries as work done against viscous stresses.  

Depth integrated dissipation rates in the open ocean will typically exceed dissipation in the bottom boundary layer, \cite{DA82, Fu81}.  The contribution of the bottom boundary layer is mitigated by buoyancy scaling of internal wave velocity and low stratification in the abyss.  Away from regions of topographic generation, dissipation associated with interior wave breaking is dominated by thermocline and upper ocean regions having large stratification rates.  One atypical example in which boundary layer dissipation can dominate is under ice cover in the Canadian Basin of the Arctic Ocean.  There, the internal wavefield is anomalously weak and buoyancy scaling implies internal wave velocities are largest near the upper boundary.  A second atypical example is in regions of large geostrophic currents in which wave dissipation can be modeled through a linearization of the bottom stress \citep{P08b}.  

%

Interior wave breaking is usually conceived of as a high wavenumber sink associated with instability mechanisms, e.g., \cite{P96}.  That sink represents a production term ${\cal P}$ in a turbulent mechanical energy budget:  
\begin{equation} 
\sigma S_{i} \equiv {\cal P} = {\cal B} - \epsilon
\end{equation}
being balanced by a buoyancy flux ${\cal B}$ and dissipation $\epsilon$ and a stochastic production/dissipation balance.  The correlations inherent in both production and buoyancy flux terms are difficult to measure directly.  Rather, specialized instrumentation and platforms are used to estimate the rate of dissipation $\epsilon$.  Full-depth estimates of such microstructure data are limited.  In its stead, use has been made of finestructure parameterizations based upon (\ref{Cascade}).  These assign a value of  
\begin{eqnarray}
\cal{P} & = & 8\times10^{-10} \; {\rm W / kg} \nonumber \\
& {\rm and} & \nonumber \\
K_{\rho} & = & {\cal B}/N^2 = 5\times10^{-6} \; {\rm m}^2 {\rm s}^{-1} \nonumber 
\end{eqnarray}
to GM76 at 30 degrees latitude and a buoyancy frequency of 3 cph.  

With either fine- or microstructure method, a major caveat is that one needs to understand how (not just whether) the sampling of the wavefield is biased.  For example, while one might hope to make sense of observations in NATRE as being almost exclusively forced by a PSI decay \citep{Natre_lat}, analysis of the diapycnal dispersion of an anthropogenic tracer suggests a long term average of a factor of two larger.  The difference may actually be related to the spatial variability of the tide.  Experience \citep{P09} suggests that issues of biasing in regions of rough topography are much more problematic than the factor of two in NATRE.  

\subsubsection{Spectral Patterns: models\label{Dissipation_model}}

In order to reconcile the $O(1)$ high wavenumber normalized Boltzman rates with the claim that GM76 represents a stationary state, \cite{MMb} invoke a spectral representation that assumes dissipation events have space-time scales much smaller than the scales of the internal waves and that the turbulence can be parameterized by an eddy viscosity $\nu$, resulting in
\begin{equation}
\sigma S_i({\bf p}) = \frac{\partial E({\bf p})}{ \partial t} = - \nu m^2 E({\bf p}) . 
\label{DissipationFunction}
\end{equation}
Similar representations are examined in greater detail in \cite{NM05}.  

Some generalization of this model may be required if it is necessary to invoke dissipation in order to obtain a stationary state.  Dissipation events will typically have vertical scales smaller than the nominal 10 meter cut-off, gauged either by correlation length scales of turbulent dissipation \citep{Gregg93} or the exclusive contribution of sub-10 meter wavelength shear to the occurrence of supercritical Richardson number events \citep{P96} (with supercritical Richardson number being indicative of shear instability), so the {\bf spatial} scale separation between waves and turbulence could be justified.  But the application of phenomenological mixing models indicates that dissipation events have lifetimes in excess of a buoyancy period \citep{P96} so that a {\bf temporal} scale separation is problematic.  In general, \cite{P96} finds such phenomenological mixing models to be useful diagnostic tools, but difficult to develop into prognostic models because of this finite lifetime.  

A key point in a generalized model is that the finestructure parameterization (\ref{Cascade}) weights the contribution of each frequency to dissipation as a product of vertical gradient variance and aspect ratio and thus the contribution of high frequency waves.  The dissipation function (\ref{DissipationFunction}) weights each frequency only as the vertical gradient variance and so emphasizes the near-inertial that dominate the vertical shear.  Phenomenological mixing models \citep{P96} have been shown to be in agreement with (\ref{Cascade}).  Observations presented in \cite{AP00} also appear to support the role of high frequency variability.  

A qualitative distinction between (\ref{DissipationFunction}) and (\ref{Cascade}) may be drawn by noting that (\ref{DissipationFunction}) is likely to be appropriate for a system of coupled oscillators.  It may not make sense for a description of the internal wavefield as a system of wavepackets in which dissipation events are associated with the transport of individual wave packets to breaking scales, as invoked in the ray tracing models, e.g., \cite{HWF}.  The ray-tracing model does not need a dissipation function because the wave packets are followed as particles to their annihilation.  

We regard the impact of dissipation on the internal wave spectral domain as an open question awaiting rigorous analysis.  


\subsubsection{Spectral Patterns:  The mid-frequency dip and bottom boundary conditions}

Many of the frequency spectra are poorly characterized by a simple power law fit near 
semidiurnal frequencies (e.g., Fig. \ref{Tropics-freq}).  Apart from the tidal peak, observed spectral levels tend to be 
smaller than the power law characterization.  \cite{L02} forwards the hypothesis that this 
mid-frequency dip is related to nonlinearity in how energy is transfered to higher and 
lower frequency out of the semidiurnal peak.  Here we forward an additional hypothesis 
regarding such departures.  With a buoyancy profile that decreases towards the bottom 
boundary, only the low frequency portion of the wavefield reaches the bottom, where 
wave energy can be dissipated as work done against viscous stresses and vertical 
scale transformations associated with reflection or scattering from a nonuniform 
bathymetry can act.  Calculations have already been made to assess the rate at which 
energy is redistributed by such scale transformations for the GM spectrum, e.g., \citep{MX92, Eriksen}. 
With an eye toward explaining spatial variability of the background spectrum, one can also ask the question of how the equilibrated spectrum could be shaped by the cumulative 
effects of scale transformations acting through the bottom boundary condition and nonlinearity acting through radiation balance equation's  source function. 
This is a topic for future research, which requires a relatively sophisticated treatment of (\ref{rbe}).  

\subsection{Propagation\label{Propagation}}

The ability of low mode internal waves to propagate long distances from their generation sites is loosely constrained in the observational record and is a serious issue for internal wave energy balance studies.  The most significant processes are tidal constituents and near-inertial waves.  These internal waves do not play a major role in this Regional study of the high wavenumber / high frequency spectral variability.  They do, however, act as sources for the internal wave continuum, whether it be through extreme scale separated interactions (Section \ref{ScaleInvariance}.c) or via a PSI decay process of tidal energy into near-inertial waves that then is linked to the continuum (Section \ref{PSIdecay}).  

We anticipate that variability in the tidal and inertial peaks is linked to variability in the continuum.  \cite{Fu81} documents variability of the inertial peak in the Western North Atlantic and obtains the following Regional classification scheme:  Class 1 spectra have prominent inertial peaks.  Such spectra occur in records obtained from the Mid-Atlantic Ridge in the vicinity of $28^{\circ}$ N and occur in conjunction with relatively white frequency spectra.  We anticipate that the inertial peaks here are related to a PSI decay process (Section \ref{The_dip}.a).  Class 2 spectra have less prominent peaks and are found (a) in the upper ocean at depths less than 2000 m, (b) in the deep ocean above rough topography off the Mid-Atlantic Ridge axis and away from $28^{\circ}$ N and (c) in the deep ocean under the Gulf Stream.  We anticipate issues of internal lee-wave generation and wave-mean interactions (Section \ref{WaveMean}) to be significant.  Finally, Class 3 spectra have the smallest inertial peaks and are found above smooth topography.  \cite{Fu81} argues that Class 1 and Class 2 inertial peaks are associated with local processes and Class 3 inertial peaks are consistent with propagation effects as waves initialized with the GM76 spectrum reflect from their turning latitudes.  Such quantitative estimates depend upon the vertical modal distribution of energy, and the regional variability of $m_{\ast}$ is not well constrained in that study.  \cite{Fu81}'s study included the MODE Central mooring instrumented at [500 (class 2a), 1500(class 2a), 4000(class 3)] m depth, whose location was the site of moorings during IWEX and FASINEX, Section \ref{SargassoSea}; and PolyMode IIIc mooring 82 [172(class 2a), 322(class 2a)  522 (class 2a), 2446 (class 2b), 3946 (class 2b)], Section \ref{Tropics}.  The vertical profile data sets at these locations in our study are focused upon depths less than 2000 m.  \cite{L76}, however, comments upon differences between vertical profiles obtained over rough and smooth topography in the vicinity of MODE Central.    

The last decade has seen the development of General Circulation Models that include both tidal forcing and wind products with sufficient temporal resolution to force inertial motions \citep{Simmons08, Arbic10}.  The models, however, obey their own dynamics:  near-inertial forcing will be influenced by the mixed layer scheme, absence of an eddy field, presence/absence of a realistic internal wavefield, topographic representation, sub-grid closures, etc.  Unraveling how such model fields relate to the ocean will be a complicated, but useful aid in interpreting the oceanic internal wavefield.   

The issue of propagation demands careful assessment.  As long as the roles of the bottom boundary condition and refraction in the mesoscale eddy field are not appreciated, the dominant balance of terms in an energy budget can seriously be misconstrued.  See, for example, the difference of opinion about the Brazil Basin data set in \cite{SG02} and \cite{P04b} that appears to be resolved in favor of a high efficiency of topographic scattering in \cite{P09}.  

Similarly, \cite{AZ07} estimate significant horizontal energy fluxes in the near-inertial band north of the Gulf Stream and regard those flux estimates as symptomatic of the ability of low-mode waves to propagate large distances.  The issue of mooring motion contaminating the horizontal energy flux estimates bears further investigation before those flux estimates can be considered robust.

\section{Conclusion}\label{Conclusions}

The ingredients that shape the deep ocean energy spectrum have been known for some time.  Fig. \ref{McComas} from \cite{M77} presents one version of how these ingredients can be combined.  In that recipe, the primary sources are at high frequency and a transfer of energy to higher vertical wavenumber is associated with the ID mechanism.  Parametric spectral fits summarized in Fig. \ref{fig:everything} are highly suggestive that the ID mechanism plays an active role in determining the observed power laws.  However, the details are not consistent with such an interpretation.  

We now believe that near-inertial and tidal sources dominate high frequency sources of internal wave energy, and this posses a real challenge to the induced diffusion mechanism:  Energy transports associated with induced diffusion are to higher vertical wavenumber and lower frequency for vertical wavenumber domain power laws $b \ge 0$.  That requires a source of high frequency energy.  Possible sources are coupling of internal waves with surface swell, internal wave coupling to mixed layer turbulence and a weaker version of internal wave coupling to the eddy field.  These possible sources are not well constrained, either observationally or theoretically.  

A second facet is that transports under the resonant interaction approximation are not completely understood.  Hitherto neglected local interactions may play a significant role in determining nonlinear transports.  

A third facet is that {\bf resonant} kinetic equation evaluations in Fig. \ref{NonlinearityParameter} indicate GM76 is far from being a stationary state, seemingly at odds with GM76 representing an ID no flux state.  
Self-consistent solutions to a {\bf broadened} kinetic equation could be much closer to being stationary and exhibit energy transfers appropriate for low frequency sources.  A closely related result is that ray-tracing diagnostics imply a transport of action to higher frequency.  This transport could provide a pathway to supply energy to the high wavenumber continuum.  A first principles derivation of the spectral transports associated with the ray-tracing models would be enlightening. 

Despite such uncertainty, nonlinearity is clearly an organizing principle.  Power laws associated with parametric spectral representations lie close to the Induced Diffusion stationary states of the resonant kinetic equation describing the lowest order nonlinear transfers.  The one exception appears to be a wavefield set up by the decay of a semidiurnal internal tide through the parametric subharmonic instability.  

While our ideas of the ingredients and their geographic distribution has evolved, our knowledge of how to create the observed spectrum (the recipe) has remained relatively static.  We are confident that such recipes are within reach, but it will take a combination of observations, analytic work in the context of a radiation balance equation and realistic direct numerical simulations.  Theoretical studies coupled to observational programs, especially those that would document 2-D spectra, transfer rates, the spectral character of the forcing functions and seek to define the role of the geostrophic flow field are a high priority.  Resonant calculations show strong non-stationary tendencies, and this issue needs to be either resolved or circumvented before addressing the contributions of other ingredients to the recipes.      

{\bf Regionality}

Variability in the spectral characteristics of the deep ocean internal wavefield documented in Section \ref{obs} and consideration of regional variability in the forcing fields in Section \ref{Theory} leads us to propose the following strawmen for further investigation:  

\begin{itemize}
\item The Natre region is a minimum for eddy energy in the North Atlantic and wind work at near-inertial frequencies is small.  Equatorward of the critical latitude for $M_2$ tides, PSI decay of the low-mode internal tide represents the major forcing of the internal wavefield.  The Natre spectrum is an outlier with extremely high bandwidth, steep vertical wavenumber and shallow frequency spectral slopes.  A preliminary recipe for the Natre region is described in Section \ref{The_dip}.a.  This recipe invokes a heuristic cascade closure for nonlinearity that has its roots in the ray-tracing diagnostics.   

\item The contrasts between the wavefields north and south of the Gulf Stream are distinct and appear to have a parallel with differences between subtropical and subpolar gyres in the North Pacific along its eastern boundary.  The later case deserves further investigation to confirm spectral parameters graphically extracted from the literature.  Our leading hypothesis is that the differences can be directly related to the relative roles of near-inertial wind forcing and interactions with the mesocale eddy field via (\ref{Source-horz}) and (\ref{Source-vert}). 
We have not yet translated this hypothesis into a recipe.   

\item  The tentative signature of a phase lag between seasonal cycles in high frequency and near-inertial frequency waves in the eddy desert of the North Pacific may be a signature of nonlinear transfers in the absence of wave-mean coupling.  

\end{itemize}

The future will certainly bring many further exciting developments,
and the synthesis of theoretical, observational and numerical results
yet to be obtained.





%
%
%
%
%
%
\appendix\label{Appendix}

\section{Meta data}

Meta data for the data sets used in this study can be found in Tables \ref{Mooring-Meta} and \ref{Profiler-Meta}.  

\begin{table*}{\tiny 
\caption{Meta data for moored current meters used in this study. \label{Meta_freq} }
\begin{center}
\begin{tabular}{|c|c|c|c|c|c|c|c|c|c|c|}
\hline
place & latitude & longitude & water depth & sensor depth & inst. type & code & start & stop  & $N$ cph & $\theta_z$ C/m\\
\hline
Site-D & 39 $17.9^{\prime}$ & 70 $05.6^{\prime}$ & 2640 m & 106 & VACM & WHOI 2203 & Feb. 26, 1967& Mar. 25, 1967 & 3.03 & \\
& & & & 511 & & WHOI 2204 & & & 1.48$^{*}$ & \\
& & & & 1013 & & WHOI 2205 & & & 0.66$^{*}$ & \\
& & & & 1950 & & WHOI 2206 & & & 0.58$^{*}$ & \\
Shelf-Slope& 39 $36.6^{\prime}$ & 70 $56.5^{\prime}$ & 2305 m & 305 & VACM & WHOI 5881 & Feb. 10, 1976 & Aug. 8, 1976 & 2.82$^{*}$ & -0.0235$^{*}$ \\
Primer & 39 $05.2^{\prime}$ & 69 21.4$^{\prime}$ & 2990 m & 391 & VACM & WHOI 9872 & Dec. 8, 1995 & Dec. 4, 1997 & 2.39$^{*}$ & -0.0170$^{*}$ \\
\hline
Iwex & 27 $43.9^{\prime}$ & 69 50.95$^{\prime}$ & 5453 m & 604 & VACM & WHOI 515A4 &  Nov. 3, 1973 & Dec. 15, 1973 & 2.55 & -0.0175 \\
 & & & & 604 & VACM & WHOI 515B4 &  Nov. 3, 1973 & Dec. 15, 1973 & 2.55 & -0.0177\\
 & & & & 633 & VACM & WHOI 515C5 &  Nov. 3, 1973 & Dec. 15, 1973 & 2.60 & -0.0157\\
\hline
Lotus & 34 1.2 & 70 1.45 & 5366 & 328 & VACM & WHOI 7666 & May 11, 1982 & April 12, 1983 & 1.49 & -0.00356 \\
 & 33 58.55 & 70 0.36 & 5366 & 348 & VACM & WHOI 7886 & April 14, 1983 & May 1, 1984 & 1.49 & -0.00356 \\
\hline
PolyMode III-c & 15 23.40 & 53 55.20 & & 319 & VACM & NOVA 80 & May 11, 1977 & May 1, 1978 & 2.78 & -0.03096 \\
& 15 11.50 & 53 12.30 & & 309 & VACM & NOVA 81 & May 12, 1977 & Dec. 20, 1977 & & \\
& 15 02.10 & 54 12.90 & & 338 & VACM & NOVA 82 & May 13, 1977 & May 1, 1978 & & \\
\hline
Fasinex & 27 58.9 & 69 58.8 & & 556 & VACM & WHOI 8293 & Oct. 29, 1984 & June 18, 1986 & 2.48 & -0.0178 \\
 & & & & 631& VACM & WHOI 8294 & & & 2.61 & -0.0202 \\
\hline
Subduction & 25 31.90 & 28 57.2 & & 300 & VMCM & WHOI & Feb. 12, 1992 & Oct. 14, 1992 & 2.44 & -0.02244 \\
Subduction & 25 31.90 & 28 57.2 & & 300 & VMCM & WHOI & Oct. 15, 1992 & June 16, 1993 & 2.44 & -0.02244 \\
\hline
EBC & 37 6.7 & 127 32.1 & 4752 & 598 & Aanderaa & rcm02674 & Aug. 9, 1992 & Aug. 19, 1994 & 2.01 & 0.00455 \\
\hline
Ocean Storms & 47 25.4 &  139 17.8 & 4224 & 199 & VMCM & rcm07420 & Aug. 20, 1987 & June 25, 1988 & 2.25 & 0.006 \\
\hline
AIWEX & 74 &  143-144 & 3700 & 250 & SBE-3 & - & Mar. 20, 1985 & Apr. 5, 1985 & - & 0.02 \\
\hline
\end{tabular}
\end{center}
}
\label{Mooring-Meta}
\end{table*}%

\begin{table*}
\caption{Meta data for the vertical profiling instrumentation used in this study \label{Meta_vert} }
\begin{center}
\begin{tabular}{|c|c|c|c|c|c|c|}
\hline
place & latitude & longitude & water depth & inst. type & start & stop \\
\hline
Station-W(3) & 39 & 70 & 3000 & MMP & Jun. 20, 2003 & Aug. 5, 2003 \\
Station-W(3) & 39 & 70 & 3000 & MMP & Jan. 17, 2004 & Mar. 2, 2004 \\
Mode & 28 & 69 40${^\prime}$ & 5440 & EMVP & Jun. 11, 1973 & Jun. 15, 1973\\
Fasinex & 27-29 & 67 $30^{\prime}$ - 70 $30^{\prime}$ & $\cong$ 5200 & HRP & Feb. 17, 1986 & Mar. 5 1986\\
Natre & 23 $56^{\prime}$ - 27 $32^{\prime}$ & 26 $45^{\prime}$ - 30 $43^{\prime}$ & $\cong$ 5500 & HRP & Mar. 28, 1992 & Apr. 14, 1992\\
SFtre2 & 14-16 & 50-57 & $\cong$ 5400 & HRP & Nov. 14, 2001 & Nov. 24, 2001\\
Patchex$^1$ & 34 & 127 & $\cong$ 4700 & MSP & Oct. 17, 1986 & Oct. 24, 1986 \\
Patchex$^2$ & 34 & 127 & $\cong$ 4700 & FLIP & Oct. 1986 & 7.5 days \\
Swapp & 35 8.2$^{\prime}$ & 126 59.0$^{\prime}$ & $\cong$ 4700 & FLIP & Feb.-Mar. 1990 & 18 days \\
STREX & 50 & 140 & $\cong$ 4000 & XCP & Nov. 8 1980 & Nov. 22, 1980 \\
AIWEX & 74 0$^{\prime}$ - 74 12$^{\prime}$  & 144 $0^{\prime}$ - 145 $0^{\prime}$& $\cong$ 3750 & XCP & April 3, 1985 & April 14, 1985 \\
\hline
\end{tabular}
\end{center}
\label{Profiler-Meta}
\end{table*}%

\section{Instrumentation}

A terse presentation of many of these data was made in \cite{LPT},
simply summarizing the available power law estimates as points in a
frequency-wavenumber domain.  That presentation was limited to spectra
appearing in published literature.  The intent of this work was to
present pertinent data in a common framework, limiting the potential
pitfalls associated with an irregular analysis.  This does not limit
uncertainties associated with the use of different instrument systems.
Details of an instrumental and technical nature that impact the
interpretation of the data are collected herein.

\subsection{Moored Current Meters}\label{CurrentMeters}
Regarding the interpretation of moored current meter data as internal
waves, there are two primary sets of issues.  The first set is
referred to as finestructure contamination.  Here the presumption is
that the data record represents signal and that departures from linear
internal wave kinematics are associated with either: (a)
quasi-permanent finestructure (either density or velocity) being
advected past the sensor (e.g., \cite{P03}, \ref{SlowOsc}), (b) self-advection within
the wavefield (e.g., \cite{SandP91}, \ref{PinkelFantasy}), or (c) attempting to estimate
density perturbations using temperature data only and invoking a
stable relation between density, temperature and salinity.  This later
assumption fails in regions that have significant large-scale
gradients of temperature and salinity on isopycnals
(e.g., \cite{FandP05}).  See the cited references for further discussion of 
these issues.  The second set of issues are instrumental in
nature: (a) drag associated with flow past a mooring will induce
movement (e.g., \cite{Foff67}), (b) typical sampling rates are not sufficient to resolve internal wave
spectrum out to the buoyancy frequency in the upper ocean, and (c)
instrument response issues.  

\begin{figure}
\noindent\includegraphics[width=20pc]{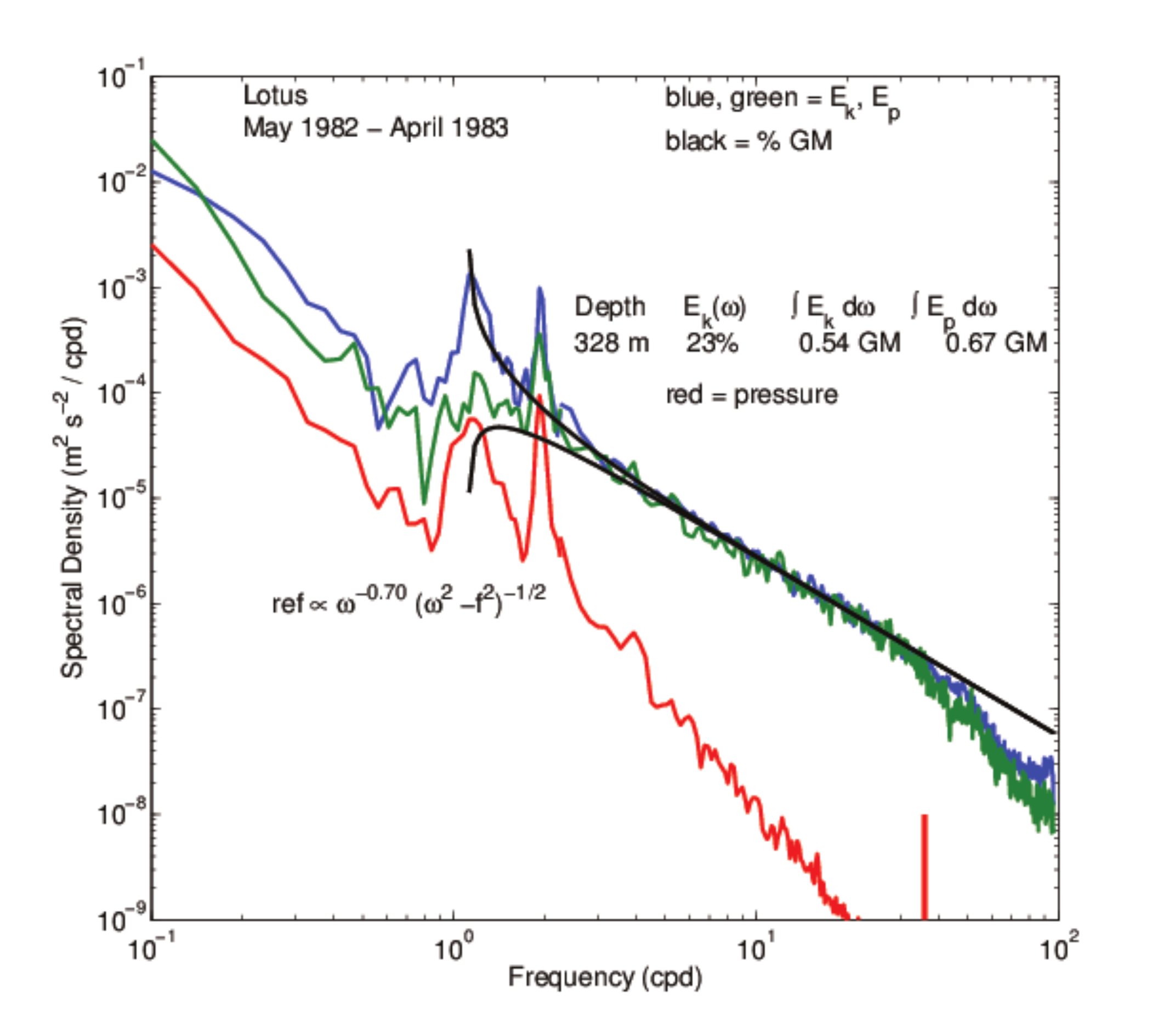}
\caption{LOTUS frequency spectra of horizontal kinetic energy (blue line), 
potential energy (green line), and pressure $p$ interpreted as a vertical displacement, $N^2 p^2 /2$.  This interpretation implies possible contamination by mooring motion is more problematic at inertial and tidal frequencies than within the internal wave continuum.  
\label{pressure_as_diagnostic} }
\end{figure}

Regarding (a), there are typically 3
classes of moorings: deep, intermediate and surface.  A deep mooring
will utilize glass balls as buoyancy elements distributed along the
mooring cable.  The primary buoyancy element for an intermediate
mooring will be a large syntactic sphere at the upper terminus.  This
mooring type utilized at WHOI starting in the early 1980's.  It has less over all drag and consequently 
provides a more stable platform.  Finally, surface moorings are
loosely tethered in order to accommodate the often sizable surface
motions associated with surface waves.  As a result, the surface buoy
inscribes a watch circle roughly equal to the water depth
\citep{Trask}.  Vertical motion of the instruments can be diagnosed if
pressure sensors are included, e.g., Fig. \ref{pressure_as_diagnostic}.  Experience 
suggests inertial and tidal frequencies suffer greater contamination by mooring motion 
than super-tidal frequencies.  Significant discussions appear in \cite{Fu81} and \cite{RJ79}.  

Regarding (b), insufficient sampling rates rates will result in the
aliasing of high frequency signals back into the resolved frequency
domain.  Ordinarily, the frequency spectra will be sufficiently red
that the aliased energy is relatively small.  However, {\bf if} there
is a substantial bump at the buoyancy frequency associated with
turning point dynamics (e.g., \cite{Desaubies75}), the aliased energy
{\bf may} be significant.  Ascertaining whether this is the case
requires information on the vertical wavenumber content of the
wavefield and details of the vertical structure of the buoyancy
profile.  Without such information and in the absence of pressure
records, it is difficult to assert that the departure of the observed
high frequency spectra in Fig. \ref{D-freq}, for example, are noise
rather than signal.

Regarding (c), the principle moored current meters used here are the
Vector Averaging Current Meter (VACM) and Vector Measuring Current
Meter (VMCM).  The VACM employs a Savonius rotor and the VMCM uses
sets of propellors.  The VMCM \citep{WellerDavis} was designed to have a cosine response
to eliminate `pumping' associated with the motion of a surface
mooring.  VACMs are standard for subsurface moorings.  VACMs have a
finite stall speed of about 2 cm s$^{-1}$ \citep{LuytenStommel}.  Good directional data is
believed to be obtained at speeds smaller than this.  The resulting
`noise' is not well defined.  VMCM data are calibrated assuming no
significant stalling.  See \cite{HoggFrye} for a more recent discussion of current meter performance.  

\subsection{Doppler Sonars}\label{DopplerSonars}

Doppler sonars do not measure horizontal velocity as a point but determine Doppler shifts as a function of range along multiple acoustic beams.  These Doppler shifts are interpreted as slant velocities of suspended acoustic backscattering targets moving with the water relative to the instrument platform.  Individual estimates of ocean velocity are highly uncertain:  hundreds of pings are typically used to average incoherent noise from the system.  Estimates of signal-to-noise ratios appear in \cite{A92} and an appendix of \cite{AP00}.  The insidious complication of package motion and beam separation effects are considered in \cite{P02}.  

\subsection{Vertical Profiling Instrumentation}\label{Pro}
Most of the vertical wavenumber domain information considered here was
obtained with vertical profiling instrumentation.  Estimates of the
vertical structure of the velocity field can be obtained by using a
number of different sensors, each of which with its own strengths and
weaknesses.

Sanford's ElectroMagnetic Velocity Profiler (EMVP, \cite{Sanford75})
senses the voltage drop associated with an electrical current
(seawater) in a magnetic field (the Earth's, in this case).  The
measurement is uncertain to within a conductivity weighted mean, so
that other sensors need to be incorporated to provide estimates of the
absolute velocity field.  Noise levels are typically around 0.5 cm
s$^{-1}$, which is sufficient to start resolving vertical wavelengths
smaller than 10m.  Expendable Current Profilers (XCPs) operate on the
same principle, but may have somewhat larger noise levels (1.0 cm
s$^{-1}$).

Data from several acoustically tracked dropsonde's have been analyzed but not presented.  With
this instrument system, the horizontal position is estimated from
range information provide by an acoustical net and depth information
from an onboard CTD (Conductivity-Temperature-Depth sensor suite).
Oceanic velocity estimates assume that the package tracks the flow
field as it descends.  This method provides estimates of the absolute
velocity, but the small scale noise is relatively high.  Acoustic
tracking provides positions to within 1 m.  A typical descent rate of 1
m s$^{-1}$ implies that the background internal wavefield having
vertical wavelengths smaller than several hundred meters will be
obscured by noise.

A third method is to utilize an acoustic travel time sensor to
estimate relative flow past a freely-falling vehicle [e.g., the High
resolution Profiler \citep{Schmitt88} and the Multi-Scale
Profiler \citep{Winkel96}].  Estimates of the oceanic velocity profile
are deduced from these relative velocity data and a model of how the
vehicle responds to the relative flow.  The method is capable of
resolving oceanic shear at 1-m scales.  The limitations are at larger
wavelengths.  The offsets (zeros) of the acoustic travel time sensor
are typically determined {\em in situ} and may even be temperature and
pressure dependent.  Thus the resulting profiles are uncertain to a
linear trend.  Other information, such as provided by an electric
field sensor or matching with a shipboard ADCP record, are required to
provide the largest scale information.

Acoustic travel time sensors are used in the McClane Moored Profiler
(MP, \cite{Doherty99}).  This instrument autonomously samples relative
velocity, temperature, conductivity and pressure while transiting a
mooring cable.  Estimates of relative velocity are thus absolute.
Noise levels in the velocity record are nominally estimated at 0.5 cm
$s^{-1}$.  Biases related to drifting zeros and mooring oscillations
are not presently well defined.

Velocity data obtained with Doppler sonars typically estimate
horizontal velocity from back-to-back acoustic beams.  In so doing,
one assumes that the velocity field is horizontally uniform.  This
assumption breaks down as the beam separation increases and will
preferentially contaminate high frequency, small vertical scale
signals.  \cite{P02}.

\section{Processing}\label{Processed}
In the absence of a mean flow, the linear internal wave equation is
(e.g., \cite{Gill}):
\begin{equation}
(\partial_t^2 + f^2)\partial_z^2 w+[N^2(z)+\partial_t^2](\partial_x^2 + \partial_y^2) w = 0
\label{repeat}
\end{equation}
for arbitrary stratification profile $N^2(z)$ and vertical velocity
$w$.  If the stratification profile varies much more slowly than the
wave phase, a WKB approximation for vertically propagating waves
provides the approximate solution:
\begin{equation}
w \propto N(z)^{-1/2} e^{i \int N(z) dz} 
\end{equation}
and so the effects of a variable buoyancy profile can be accounted
for by stretching the depth coordinate by $N$ and scaling the
horizontal velocities by $N^{1/2}/N_o^{1/2}$, in which $N_o$ is a
reference stratification.  The value $N_o=$ 3 cph is often used.  The
use of the WKB approximation requires $\omega^2 \ll N^2$.  If this
relation is not satisfied, solutions can be found by
treating (\ref{repeat}) as an eigen value problem with appropriate
boundary conditions.  If $\omega^2 \ll N^2$ but the wave-phase is not
slowly varying, the boundary conditions are that $w=0$ at the top
($z=0$) and bottom ($z=H$), which then implies the horizontal
velocities $[u(z), v(z)]$ are proportional to:
\begin{equation}
N(z)^{1/2} cos(n \pi \int_0^z N(z^{\prime}) dz^{\prime}/ \int_0^H
N(z^{\prime}) dz^{\prime}),
\end{equation}
for integer values of n.  For data sets that document the velocity
profile over the entire water column, the first three modes have been
estimated by using a linear regression.  Vertical wavenumber spectra 
are calculated by using a cubic spline to interpolate onto a uniform 
grid having approximately the same resolution as the unstretched data.  

Current meter data used here have sampling intervals of 3.75 minutes to 
1 hour.  Frequency spectra have been calculated using transform intervals 
of $10 \frac{2}{3}$ and $\frac{8}{3}$ days.  Spectra are displayed using the 
later at super-tidal frequencies ($\sigma > 2.4$ / cpd) and the former at 
lower frequencies.  The intent is to give finer resolution to inertial and tidal peaks 
and decreased variability at higher frequency.  

\section{A Sargasso Sea spectrum}\label{SargassoSeaSpectrum}

We have characterized the background internal wave spectrum in the Sargasso Sea as:
\begin{eqnarray}
&& E(m,\sigma)=\frac{2A}{\pi m_{\ast}} \LSBA 1+\frac{49f^2}{[1764(\sigma-f)^2+f^2]} \RSBA \frac{1}{\sigma^{1.75}} \nonumber \\
&& \frac{1}{[1+(m/m_{\ast})^{2}]^{1.125}}, 
\end{eqnarray}
with $m_{\ast} = 10 \pi / 1300$ m$^{-1}$ and normalization constant $A$ chosen to return $E=0.56\times30\times10^{-4}$ m$^2$ s$^{-2}$.  This places the gradient variance at approximately $\frac{3}{4} \int_0^{2\pi/10 {\rm m}} 2 m^2E(m) dm = \frac{2\pi}{10} N^2$.  Power laws are characteristic of the Sargasso Sea Data sets (Section \ref{SargassoSea}) are selected specifically to lie on the Induced Diffusion constant flux solution (Section \ref{ScaleInvariance}.c).  The inertial peak was selected with the constraint that $E_k/E_p = 4.4$ based upon moored data.  We have deferred from invoking non-separable conditions via $m_{\ast}=m_{\ast}(\sigma)$ that are a documented part of the observational record \citep{M78} and appear as $E_k(m=0.01 {\rm cpm})/E_p(m=0.01 {\rm cpm})$ in excess of this value \citep{P03}.    

%
%
%
%

\begin{acknowledgments}

The authors gratefully acknowledge the efforts of many talented
researchers who were instrumental in collecting, processing and archiving the data
presented here.  Special thanks are extended to John Toole for providing yet 
unpublished MP data from Site-D, Tom Sanford for providing access to a 
number of  his EMVP and AVP data sets, and to Carl Wunsch for the PolyMode IIIc current meter data set.  We greatfully acknowledge funding provided by a Collaborations in 
Mathematical Geosciences (CMG) grant from the National
Science Foundation.  

\end{acknowledgments}

%
%
%
%
%
%
%
%
%
%


\bibliographystyle{agufull04} 

%
%

\end{article} 




%
%
%
%
%
%


\end{document}